\documentclass[12pt,reqno,oneside]{amsbook}
\usepackage{amsmath,amsthm}
\usepackage{eucal}
\usepackage{eufrak}
\newcommand{\ar}{\renewcommand{\arraystretch}{1}} 
\DeclareMathAlphabet{\bb}{U}{msb}{m}{n}
\gdef\C{\bb C}
\gdef\dZ{\bb Z}
\gdef\A{\bb A}

\gdef\dS{\bb S}
\gdef\R{\bb R}
\gdef\K{\bb K}
\gdef\BH{\bb H}
\gdef\F{\bb F}
\gdef\dO{\bb O}

\DeclareMathOperator{\End}{End}
\DeclareMathOperator{\spin}{{\bf Spin}}
\DeclareMathOperator{\pin}{{\bf Pin}}
\DeclareMathOperator{\fD}{\mathfrak{D}}
\DeclareMathOperator{\Id}{Id}

\DeclareMathOperator{\Aut}{Aut}
\DeclareMathOperator{\Ker}{Ker}

\DeclareMathOperator{\bA}{{\bf A}}

\DeclareMathOperator{\Sym}{Sym}

\newcommand{\s}{\!}

\newcommand{\re}{\mbox{\rm Re}\,}
\newcommand{\im}{\mbox{\rm Im}\,}
\newcommand{\Om}{{\bf\Omega}} 
\newcommand{\cA}{\mathcal{A}}
\newcommand{\cB}{\mathcal{B}}

\newcommand{\cD}{\mathcal{D}}
\newcommand{\cE}{\mathcal{E}}
\newcommand{\cN}{\mathcal{N}}
\newcommand{\cL}{\mathcal{L}}

\newcommand{\M}{{\bf\sf M}}
\newcommand{\bsA}{{\bf\sf A}}
\newcommand{\sA}{{\sf A}}
\newcommand{\sB}{{\sf B}}
\newcommand{\sI}{{\sf I}}
\newcommand{\sW}{{\sf W}}
\newcommand{\sE}{{\sf E}}
\newcommand{\sC}{{\sf C}}
\newcommand{\sT}{{\sf T}}

\newcommand{\sP}{{\sf P}}
\newcommand{\sU}{{\sf U}}
\newcommand{\sQ}{{\sf Q}}
\newcommand{\sAut}{{\sf Aut}}
\newcommand{\bi}{{\bf i}}
\newcommand{\bj}{{\bf j}}
\newcommand{\bk}{{\bf k}}
\newcommand{\bx}{{\bf x}}

\newcommand{\bZ}{{\bf Z}}
\newcommand{\bF}{{\bf F}}
\newcommand{\bE}{{\bf E}}
\newcommand{\bH}{{\bf H}}
\newcommand{\fM}{\mathfrak{M}}
\newcommand{\fG}{\mathfrak{G}}
\newcommand{\fC}{\mathfrak{C}}
\newcommand{\fR}{\mathfrak{R}}
\newcommand{\fH}{\mathfrak{H}}
\newcommand{\fO}{\mathfrak{O}}
\newcommand{\hs}{\hspace{0.2cm}}
\newcommand{\Lip}{\boldsymbol{\Gamma}}

\newcommand{\cl}{C\kern -0.2em \ell}

\newcommand{\p}{\prime}
\newcommand{\e}{\mbox{\bf e}}

\newcommand{\ld}{\left[}
\newcommand{\rd}{\right]}
\newcommand{\lf}{\left\{}
\newcommand{\rf}{\right\}}
\newtheorem{theorem}{Theorem}
\newtheorem{cor}{Corollary}

\newtheorem{prop}{Proposition}

\begin{document}
\frontmatter
\thispagestyle{empty}
\setcounter{page}{1}
\phantom{CC}
\vspace{0.88truein}
\vspace{2.0cm}
\vspace{1.5cm}
\vspace{1.0cm}

\centerline{\bf GROUP\, THEORETICAL\, DESCRIPTION\, OF}\vspace{0.1cm}
\centerline{\bf SPACE\, INVERSION,\, TIME REVERSAL\, AND\, CHARGE}\vspace{0.1cm}
\centerline{\bf CONJUGATION}
\vspace*{0.035truein}
\vspace*{0.37truein}
\centerline{V.~V.~Varlamov}
\newpage
\mainmatter
\pagenumbering{roman}
\markboth{\contentsname}{\contentsname}
\tableofcontents
\markboth{\contentsname}{\contentsname}
\newpage
\pagenumbering{arabic}
\chapter*{Introduction}
Importance of discrete transformations is well--known, many textbooks
on quantum theory began with description of the discrete symmetries, and
famous L\"{u}ders--Pauli $CPT$--Theorem is a keystone of quantum field
theory. 
Besides, a fundamental notion of antimatter immediately relates with the
charge conjugation $C$. The requirement of invariance concerning the
each of the discrete transformations gives rise to certain consequences
which can be verified in experience. So, from the invariance concerning
the $CPT$-transformation it follows an identity of the masses and full life
times of particles and antiparticles, from the invariance concerning the
time reversal $T$ we have certain relations between the forward and
reverse reaction cross-sections. In turn, from the invariance concerning
the charge conjugation $C$ it follows an absence of the reactions
forbidden by the conservation law of charge parity, and from the
invariance concerning the space inversion $P$ and time reversal $T$
it follows an absence of the electric dipol moment of particles.
As follows from experience, for the processes defined by strong and 
electromagnetic interactions there exists an invariance with respect to
the all discrete transformations. In contrast to strong and electromagnetic
interactions, for the weak interaction, as shown by experiment, 
there is no invariance concerning space inversion $P$, but there is invariance
with respect to $CP$--transformation. Moreover, there are experimental
evidences confirming $CP$--violation (a decay of the neutral $K$--mesons).
It is clear that the analysis of the discrete symmetries allows
to reveal the most profound structural characteristics of the matter.

However, usual practice of definition of the discrete symmetries
from the analysis of relativistic wave equations does not give a full and
consistent theory of the discrete transformations. In the standard approach,
except a well studied case of the spin $j=1/2$ (Dirac equation), a situation
with the discrete symmetries remains vague for the fields of higher spin
$j>1/2$. It is obvious that a main reason of this is an absence of a fully
adequate formalism for description of higher--spin fields (all widely
accepted higher--spin formalisms such as Rarita--Schwinger approach \cite{RS41},
Bargmann--Wigner \cite{BW48} and Gel'fand--Yaglom \cite{GY48} multispinor
theories, and also Joos--Weinberg $2(2j+1)$--component formalism 
\cite{Joo62,Wein} have many intrinsic contradictions and difficulties).
The first attempt of going out from this situation was initiated by
Gel'fand, Minlos and Shapiro in 1958 \cite{GMS}. In the
Gel'fand--Minlos--Shapiro approach the discrete symmetries are represented
by outer involutory automorphisms of the Lorentz group (there are also other
realizations of the discrete symmetries via the outer automorphisms, see
\cite{Mic64,Kuo71,Sil92}).
At present the
Gel'fand--Minlos--Shapiro ideas have been found further development in the
works of Buchbinder, Gitman and Shelepin \cite{BGS00,GS00}, where the
discrete symmetries are represented by both outer and inner automorphisms of
the Poincar\'{e} group.

Discrete symmetries $P$ and $T$ transform (reflect) space and time
(two the most fundamental notions in physics), but in the Minkowski
4--dimensional space--time continuum \cite{Min} space and time are not
separate and independent. By this reason a transformation of one (space or
time) induces a transformation of another. Therefore, discrete symmetries
should be expressed by such trnsformations of the continuum, that transformed 
all its structure totally with a full preservation of discrete nature. The
only possible candidates on the role of such transformations are
automorphisms. In such a way the idea of representation of the discrete
symmetries via the automorphisms of the Lorentz group (`rotation' group of the
4--dimensional continuum) is appearred in the Gel'fand--Minlos--Shapiro
approach, or via the automorphisms of the Poincar\'{e} group (motion
group of the 4--dimensional continuum) in the Buchbinder--Gitman--Shelepin
approach.

In 1909, Minkowski showed \cite{Min} that a causal structure of the world
is described by a 4--dimensional pseudo--Euclidean geometry. In accordance
with \cite{Min} the quadratic form $x^2+y^2+z^2-c^2t^2$ remains invariant
under the action of linear transformations of the four variables $x,y,z$ and 
$t$,
which form a general Lorentz group $G$. As known, the general Lorentz group
$G$ consists of an own Lorentz group $G_0$ and three reflections
(discrete transformations) $P,\,T,\,PT$, where $P$ and $T$ are space and
time reversal, and $PT$ is a so--called full reflection. The discrete
transformations $P,\,T$ and $PT$ added to an identical transformation
form a finite group. Thus, the general Lorentz group may be represented by
a semidirect product $G_0\odot\{1,P,T,PT\}$. Analogously, an orthogonal
group $O(p,q)$ of the real space $\R^{p,q}$ is represented by the semidirect
product of a connected component $O_0(p,q)$ and a discrete subgroup.

In 1957, Shirokov pointed out \cite{Shi57} that an universal covering of the
inhomogeneous Lorentz group has eight inequivalent realizations. Later on,
in the eighties this idea was applied to a general orthogonal group
$O(p,q)$ by D\c{a}browski \cite{Dab88}.
As known, the orthogonal
group $O(p,q)$ of the real space $\R^{p,q}$ is represented by the semidirect
product of a connected component $O_0(p,q)$ and a discrete subgroup
$\{1,P,T,PT\}$.
Further, a double covering of the orthogonal group $O(p,q)$ is a
Clifford--Lipschitz group $\pin(p,q)$ which is completely constructed within
a Clifford algebra $\cl_{p,q}$. In accordance with squares of elements of the
discrete subgroup ($a=P^2,\,b=T^2,\,c=(PT)^2$) there exist eight double
coverings (D\c{a}browski groups \cite{Dab88}) of the orthogonal group
defining by the signatures $(a,b,c)$, where $a,b,c\in\{-,+\}$. Such in brief is
a standard description scheme of the discrete transformations.
However, in this scheme there is one essential flaw. Namely, the
Clifford--Lipschitz group is an intrinsic notion of the algebra $\cl_{p,q}$
(a set of the all invertible elements of $\cl_{p,q}$), whereas the discrete
subgroup is introduced into the standard scheme in an external way, and the
choice of the signature $(a,b,c)$ of the discrete subgroup is not
determined by the signature of the space $\R^{p,q}$. Moreover, it is suggest
by default that for any signature $(p,q)$ of the vector space there exist
the all eight kinds of the discrete subgroups. It is obvious that a
consistent description of the double coverings of $O(p,q)$ in terms of
the Clifford--Lipschitz groups $\pin(p,q)\subset\cl_{p,q}$ can be obtained
only in the case when the discrete subgroup $\{1,P,T,PT\}$ is also defined
within the algebra $\cl_{p,q}$. Such a description has been given in the
works \cite{Var99,Var00,Var03}, where the discrete symmetries are
represented by fundamental automorphisms of the Clifford algebras.
So, the space inversion $P$, time reversal $T$ and their
combination $PT$ correspond to an automorphism $\star$
(involution), an antiautomorphism $\widetilde{\phantom{cc}}$ (reversion) and
an antiautomorphism $\widetilde{\star}$ (conjugation), respectively.
Group theoretical structure of the discrete transformations is a central
point in this work. The fundamental automorphisms of the Clifford algebras
are compared to elements of the finite group formed by the discrete
transformations. In its turn, a set of the fundamental automorphisms,
added by an identical automorphism, forms a finite group 
$\Aut(\cl)$, for which in virtue of the Wedderburn--Artin Theorem
there exists a matrix representation. In such a way, an isomorphism
$\{1,P,T,PT\}\simeq\Aut(\cl)$ plays a central role.
It is shown that
the division ring structure of $\cl_{p,q}$ imposes hard restrictions on
existence and choice of the discrete subgroup, and the signature
$(a,b,c)$ depends upon the signature of the underlying space $\R^{p,q}$.
Moreover, this description allows to incorporate the
Gel'fand--Minlos--Shapiro automorphism theory into 
Shirokov--D\c{a}browski scheme and further to unite them on the basis
of the Clifford algebra theory.

In the present work such an unification is given. First of all,
Clifford algebras are understood as `algebraic coverings' of 
finite--dimensional representations of the proper Lorentz group $\fG_+$.
Clifford algebras $\C_n$ over the field $\F=\C$ are
associated with complex finite--dimensional representations $\fC$ of the
group $\fG_+$. It allows to define a new class of the finite--dimensional
representations of $\fG_+$ (quotient representations) corresponded to the
type $n\equiv 1\pmod{2}$ of the algebras $\C_n$. In its turn, representation
spaces of $\fC$ are the spinspaces $\dS_{2^{n/2}}$ or the minimal left
ideals of the algebras $\C_n$. In virtue of this the
discrete symmetries representing by spinor representations of the
fundamental automorphisms of $\C_n$ are defined for both complex and real
finite--dimensional representations of the group $\fG_+$.

\begin{sloppypar}
The aim of present research is a construction of the full and consistent
theory of the discrete transformations for both space--time and spaces of
any dimension and signature. At this point the main subject is a group
theoretical consideration of the problem. Namely, space--time is considered
as a space of the fundamental representation of the Lorentz group, and
multidimensional spaces are understood as spaces of the finite--dimensional
representations of the same group. In such a way, the main goal of the
research is the construction of a theory of the discrete symmetries on 
spaces of the finite--dimensional representations of the Lorentz group
(correspondingly, Poincar\'{e} group). According to Wigner, a quantum system,
described by an irreducible representation of the Poincar\'{e} group,
is called an elementary particle. Therefore, discrete transformations,
defined for some irreducible representation of the Lorentz
(Poincar\'{e}) group, form discrete symmetries of some elementary particle.
Usually, the spin of the elementary particle is associated with the
weight of the irreducible representation, but, as noted previously,
at present time there exists no fully adequate higher--spin formalism.
The construction of such a formalism is a main problem of quantum field
theory. It stands to reason that in the future higher--spin formalism the
discrete symmetries should be represented by the automorphisms of the
Lorentz group (Poincar\'{e} group, Clifford algebra), therefore,
the present work should be considered as a certain step in this
direction.\end{sloppypar}

Other goal, a full realization of which comes beyond the framework of the
present research, is an unification of space--time and internal symmetries
of elementary particles. As known, the standard approach in this direction
is a search of some inification group which includes the Poincar\'{e}
group and the group (or, groups) of internal symmetries
($SU(2)$, $SU(3)$ and so on) as a subgroup. However, in this approach
a physical sense of the unification group is vague. As an alternative to
the standard approach we will consider in the present work a complex
algebraic structure associated with the system of finite--dimensional
irreducible representations of the Lorentz group. A bundle
`finite--dimensional representation of the Lorentz group $+$ complex
Clifford algebra' serves as a basis for the unification of space--time and
internal symmetries. It is well known that an octonion algebra, triality
and spinor group
$\spin(8)$, defined a basis of the algebraic description of quark symmetries
(G\"{u}naydin and G\"{u}rsey works), admit a full description within
eight--dimensional Clifford algebras.
The octonion algebra $\dO$, or the algebra of Cayley numbers \cite{Cal},
is also the hypercomplex system, but in contrast to Clifford algebras
this system is non--associative. Nevertheless, there exists an octonion
representation of the Clifford algebras
\cite{Oku91a,Oku91b,Oku95a,Oku95b,MaSh93}. Further, in view of the fact that
an automorphism group of the algebra
$\dO$ coincides with an exceptional group $G_2$ containing the group
$SU(3)$ as a subgroup, G\"{u}naydin and G\"{u}rsey 
\cite{GG73,Gur75,Gur79,GT96} established a relationship between
$SU(3)$--color quark symmetry and the algebraic structure of the fields
constructed on the octonions. As known, there exist only four division
algebras (Frobenius Theorem)
$\R,\,\C,\,\BH,\,\dO$. The first three algebras ($\R$ and $\C$ are the fields
of real and complex numbers, $\BH$ is a quaternion algebra) are the simplest
Clifford algebras. As noted previously, the octonion algebra
$\dO$ admits a representation within eight--dimensional Clifford algebras.
Moreover, Dixon
\cite{Dix} introduced the algebra $\A=\R\otimes\C\otimes\BH\otimes\dO$ and
showed that the group $SU(3)\times SU(2)\times U(1)$ of
so--called Standard Model appears as a subgroup of the automorphism group
of $\A$. In such a way, we can consider the Clifford algebra theory as an
unification base, in terms of which different algebraic structures related
with the internal summetries of elementary particles are united.
\begin{sloppypar}
Clifford algebras form a theoretical basis of the research. It is well
known that Clifford algebras present one of the most basic mathematical
constructions in quantum field theory.
In connection with this it is in order to give a brief review of the
Clifford algebra theory and its relations with theoretical physics.
From historical point of view these algebras have essentially geometric
origin \cite{3}, because they are the synthesis of Hamilton quaternion
calculus \cite{Ham} and Grassmann {\it Ausdehnungslehre}
\cite{Grass}, and by this reason they called by Clifford as
{\it geometric algebras} \cite{Cliff2}. Further, Lipschitz
\cite{Lips} showed that Clifford algebras closely related with the study
of the rotation groups of multidimensional spaces. After fundamental works of
Cartan \cite{Car08}, Witt \cite{Wit37} and Chevalley \cite{Che54}
the Clifford algebra theory takes its modern form.
\cite{Cru91,Port95,Lou97}. In spite of the fact that Clifford algebras
(or systems of hypercompex numbers) are known in mathematics from the
middle of XIX century a broad application of the hypercomplex numbers
in physics began at the end of twenties of XX century after the famous
Dirac work on the electron theory
\cite{Dir28}. On the other hand, long before the appearance of quantum
mechanics Clifford algebras are presented in physics. Namely,
a mathematical apparatus of classical electrodynamics (vector analysis),
created by Gibbs and Heavyside, takes its beginning from the Hamilton
quaternion calculus, and, as noted previously, the quaternion algebra is one
of the simplest Clifford algebras. In virtue of this there exists a
quaternionic description of classical electrodynamics \cite{Mer35}).
By analogy with this there is a quaternionic description of quantum
mechanics (so--called quaternionic quantum mechanics \cite{Adl95}).
\end{sloppypar}\begin{sloppypar}
An isomorphism $\{1,P,T,PT\}\simeq\Aut(\cl)$ allows to use methods of the
Clifford algebra theory at the study of a group theoretical structure
of the discrete transformations. First of all, it allows to classify the
discrete groups into Abelian 
$\dZ_2\otimes\dZ_2$, $\dZ_4$ and non--Abelian
$D_4$, $Q_4$ finite groups, and also to establish a dependence between the
finite groups and signature of the spaces in case of real numbers.
\end{sloppypar}

A full system $\fM=\fM^+\oplus\fM^-$ of the finite--dimensional
representations of the group $\fG_+$ allows to define in the Chapter 4
the Atiyah--Bott--Shapiro periodicity \cite{AtBSh} on the Lorentz group.
In case of the field $\F=\C$ we have modulo 2 periodicity on the
representations $\fC$, $\fC\cup\fC$, that allows to take a new look at the
de Broglie--Jordan neutrino theory of light \cite{Bro32,Jor35}. In its turn,
over the field $\F=\R$ we have on the system $\fM$ the modulo 8 periodicity
which relates with octonionic representations of the Lorentz group and
the G\"{u}naydin--G\"{u}rsey construction of the quark structure in terms
of an octonion algebra $\dO$ \cite{GG73}. In essence, the modulo 8
periodicity on the system $\fM$ gives an another realization of the
well--known Gell-Mann--Ne'emann eightfold way \cite{GN64}. It should be
noted here that a first attempt in this direction was initiated by
Coquereaux in 1982 \cite{Coq82}.

Other important discrete symmetry is the charge conjugation $C$. In contrast
with the transformations $P$, $T$, $PT$ the operation $C$ is not
space--time discrete symmetry. This transformation is firstly appearred
on the representation spaces of the Lorentz group and its nature is
strongly different from other discrete symmetries. By this reason in the
Chapter 4 the charge conjugation $C$ is represented by a
pseudoautomorphism $\cA\rightarrow\overline{\cA}$ which is not fundamental
automorphism of the Clifford algebra. All spinor representations of the
pseudoautomorphism $\cA\rightarrow\overline{\cA}$ are given in
Theorem \ref{tpseudo}.

Quotient representations of the group $\fG_+$ compose the second half
$\fM^-$ of the full system $\fM$ and correspond to the types $n\equiv 1\pmod{2}$
($\F=\C$) and $p-q\equiv 1,5\pmod{8}$ ($\F=\R$). An explicit form
of the quotient representations is given in
Theorem \ref{tfactor}. In the Chapter 4 the first simplest physical
field (neutrino field), corresponded to a fundamental representation
$\fC^{1,0}$ of the group $\fG_+$, is studied within a quotient
representation ${}^\chi\fC^{1,0}_c\cup{}^\chi\fC^{0,-1}_c$. Such a
description of the neutrino was firstly given in the work \cite{Var99},
but in \cite{Var99} this description looks like an exotic case, whereas
in the present work it is a direct consequence of all mathematical
background developed in the previous chapters. It is shown also that
the neutrino field $(1/2,0)\cup(0,1/2)$ can be defined in terms of a
Dirac--Hestenes spinor field \cite{Hest66,Hest67}, and the wave function
of this field satisfies the Weyl--Hestenes equations 
(massless Dirac--Hestenes equations).
\chapter{Clifford algebras}
\section{Definition of the Clifford algebra}
In this section we will consider some basic facts about Clifford algebras
and Clifford--Lipschitz groups which we will widely use below.
Let $\F$ be a field of characteristic 0 $(\F=\R,\,\F=\C)$, where
$\R$ and $\C$ are the fields of real and complex numbers, respectively.
A Clifford algebra $\cl$ over a field $\F$ is an algebra with
$2^n$ basis elements: $\e_0$
(unit of the algebra) $\e_1,\e_2,\ldots,\e_n$ and products of the one--index
elements $\e_{i_1i_2\ldots i_k}=\e_{i_1}\e_{i_2}\ldots\e_{i_k}$.
Over the field $\F=\R$ the Clifford algebra denoted as $\cl_{p,q}$, where
the indices
$p$ and $q$ correspond to the indices of the quadratic form
\[
Q=x^2_1+\ldots+x^2_p-\ldots-x^2_{p+q}
\]
of a vector space $V$ associated with $\cl_{p,q}$. The multiplication law
of $\cl_{p,q}$ is defined by a following rule:
\begin{equation}\label{e1}
\e^2_i=\sigma(p-i)\e_0,\quad\e_i\e_j=-\e_j\e_i,
\end{equation}
where
\begin{equation}\label{e2}
\sigma(n)=\left\{\begin{array}{rl}
-1 & \mbox{if $n\leq 0$},\\
+1 & \mbox{if $n>0$}.
\end{array}\right.
\end{equation}
The square of a volume element $\omega=\e_{12\ldots n}$ ($n=p+q$) plays an
important role in the theory of Clifford algebras,
\begin{equation}\label{e3}
\omega^2=\left\{\begin{array}{rl}
-1 & \mbox{if $p-q\equiv 2,3,6,7\pmod{8}$},\\
+1 & \mbox{if $p-q\equiv 0,1,4,5\pmod{8}$}.
\end{array}\right.
\end{equation}  
A center $\bZ_{p,q}$ of the algebra $\cl_{p,q}$ consists of the unit $\e_0$ 
and the volume element $\omega$. The element $\omega=\e_{12\ldots n}$ 
belongs to a center when $n$ is odd. Indeed,
\begin{eqnarray}
\e_{12\ldots n}\e_i&=&(-1)^{n-i}\sigma(q-i)\e_{12\ldots i-1 i+1\ldots n},
\nonumber\\
\e_i\e_{12\ldots n}&=&(-1)^{i-1}\sigma(q-i)\e_{12\ldots i-1 i+1\ldots n},
\nonumber
\end{eqnarray}
therefore, $\omega\in\bZ_{p,q}$ if and only if $n-i\equiv i-1\pmod{2}$, 
that is, $n$ is odd. Further, using (\ref{e3}) we obtain
\begin{equation}\label{e4}
\bZ_{p,q}=\left\{\begin{array}{rl}
\phantom{1,}1 & \mbox{if $p-q\equiv 0,2,4,6\pmod{8}$},\\
1,\omega & \mbox{if $p-q\equiv 1,3,5,7\pmod{8}$}.
\end{array}\right.
\end{equation}

An arbitrary element $\cA$ of the algebra $\cl_{p,q}$ is represented by a
following formal polynomial
\begin{gather}
\cA=a^0\e_0+\sum^n_{i=1}a^i\e_i+\sum^n_{i=1}\sum^n_{j=1}a^{ij}\e_{ij}+
\ldots+\sum^n_{i_1=1}\cdots\sum^n_{i_k=1}a^{i_1\ldots i_k}\e_{i_1\ldots i_k}+
\nonumber\\
+\ldots+a^{12\ldots n}\e_{12\ldots n}=\sum^n_{k=0}a^{i_1i_2\ldots i_k}
\e_{i_1i_2\ldots i_k}.\nonumber
\end{gather}

In Clifford algebra $\cl$ there exist four fundamental automorphisms.\\[0.2cm]
1) {\bf Identity}: An automorphism $\cA\rightarrow\cA$ and 
$\e_{i}\rightarrow\e_{i}$.\\
This automorphism is an identical automorphism of the algebra $\cl$. 
$\cA$ is an arbitrary element of $\cl$.\\[0.2cm]
2) {\bf Involution}: An automorphism $\cA\rightarrow\cA^\star$ and 
$\e_{i}\rightarrow-\e_{i}$.\\
In more details, for an arbitrary element $\cA\in\cl$ there exists a
decomposition
$
\cA=\cA^{\p}+\cA^{\p\p},
$
where $\cA^{\p}$ is an element consisting of homogeneous odd elements, and
$\cA^{\p\p}$ is an element consisting of homogeneous even elements,
respectively. Then the automorphism
$\cA\rightarrow\cA^{\star}$ is such that the element
$\cA^{\p\p}$ is not changed, and the element $\cA^{\p}$ changes sign:
$
\cA^{\star}=-\cA^{\p}+\cA^{\p\p}.
$
If $\cA$ is a homogeneous element, then
\begin{equation}\label{auto16}
\cA^{\star}=(-1)^{k}\cA,
\end{equation}
where $k$ is a degree of the element. It is easy to see that the
automorphism $\cA\rightarrow\cA^{\star}$ may be expressed via the volume
element $\omega=\e_{12\ldots p+q}$:
\begin{equation}\label{auto17}
\cA^{\star}=\omega\cA\omega^{-1},
\end{equation}
where
$\omega^{-1}=(-1)^{\frac{(p+q)(p+q-1)}{2}}\omega$. When $k$ is odd, 
for the basis elements 
$\e_{i_{1}i_{2}\ldots i_{k}}$ the sign changes, and when $k$ is even, the sign
is not changed.\\[0.2cm]
3) {\bf Reversion}: An antiautomorphism $\cA\rightarrow\widetilde{\cA}$ and
$\e_i\rightarrow\e_i$.\\
The antiautomorphism $\cA\rightarrow\widetilde{\cA}$ is a reversion of the
element $\cA$, that is the substitution of the each basis element
$\e_{i_{1}i_{2}\ldots i_{k}}\in\cA$ by the element
$\e_{i_{k}i_{k-1}\ldots i_{1}}$:
\[
\e_{i_{k}i_{k-1}\ldots i_{1}}=(-1)^{\frac{k(k-1)}{2}}
\e_{i_{1}i_{2}\ldots i_{k}}.
\]
Therefore, for any $\cA\in\cl_{p,q}$, we have
\begin{equation}\label{auto19}
\widetilde{\cA}=(-1)^{\frac{k(k-1)}{2}}\cA.
\end{equation}
4) {\bf Conjugation}: An antiautomorpism $\cA\rightarrow\widetilde{\cA^\star}$
and $\e_i\rightarrow-\e_i$.\\
This antiautomorphism is a composition of the antiautomorphism
$\cA\rightarrow\widetilde{\cA}$ with the automorphism
$\cA\rightarrow\cA^{\star}$. In the case of a homogeneous element from
the formulae (\ref{auto16}) and (\ref{auto19}), it follows
\begin{equation}\label{20}
\widetilde{\cA^{\star}}=(-1)^{\frac{k(k+1)}{2}}\cA.
\end{equation}
\begin{theorem}\label{t1}
If $n=p+q$ is odd, then
\begin{eqnarray}
\cl_{p,q}&\simeq&\C_{p+q-1}\quad\text{if $p-q\equiv 3,7\pmod{8}$},\nonumber\\
\cl_{p,q}&\simeq&\cl_{p,q-1}\oplus\cl_{p,q-1}\nonumber\\
&\simeq&\cl_{q,p-1}\oplus\cl_{q,p-1}\quad\text{if $p-q\equiv 1,5\pmod{8}$}.
\label{e5'}
\end{eqnarray}
\end{theorem}
\begin{proof} The structure of $\cl_{p,q}$ allows to identify the Clifford
algebras over the different fields. Indeed, transitions $\cl_{p-1,q}
\rightarrow\cl_{p,q},\;\cl_{p,q-1}\rightarrow\cl_{p,q}$ may be represented
as transitions from the real coordinates in $\cl_{p-1,q},\,\cl_{p,q-1}$
to complex coordinates of the form $a+\omega b$, where $\omega$ is
an additional basis element $\e_{12\ldots n}$ (volume element). Since
$n=p+q$ is odd, then the volume element $\omega$ in accordance with
(\ref{e4}) belongs to $\bZ_{p,q}$. Therefore, we can to identify it
with imaginary unit $i$ if $p-q\equiv 3,7\pmod{8}$ and with a double unit
$e$ if $p-q\equiv 1,5\pmod{8}$. The general element of the algebra $\cl_{p,q}$
has a form $\cA=\cA^{\p}+\omega\cA^{\p}$, where $\cA^{\p}$ is a general
element of the algebras $\cl_{q,p-1},\,\cl_{p,q-1}$.
\end{proof}

Over the field $\F=\C$ there is the analogous result:
\begin{theorem}[{\rm Rashevskii \cite{Rash}}]\label{t2}
When $p+q\equiv 1,3,5,7\pmod{8}$ the Clifford algebra over the field
$\F=\C$ decomposes into a direct sum of two subalgebras:
\[
\C_{p+q}\simeq\C_{p+q-1}\oplus\C_{p+q-1}.
\]
\end{theorem}
\section{The structure of the Clifford algebras}
One of the most fundamental theorems in the theory of associative algebras
is as follows
\begin{theorem}[{\rm Wedderburn--Artin}]
Any finite--dimensional associative simple algebra $\mathfrak{A}$ over the
field $\F$ is isomorphic to a full matrix algebra $\M_n(\K)$, where $n$ is
natural number defined unambiguously, and $\K$ a division ring defined
with an accuracy of isomorphism.
\end{theorem}
In accordance with this theorem, all properties of the initial algebra
$\mathfrak{A}$ are isomorphically transferred to the matrix algebra 
$\M_n(\K)$. Later on we will widely use this theorem. In its turn, for the
Clifford algebra $\cl_{p,q}$ over the field $\F=\R$ we have an isomorphism
$\cl_{p,q}\simeq\End_{\K}(I_{p,q})\simeq\M_{2^m}(\K)$, where $m=\frac{p+q}{2}$,
$I_{p,q}=\cl_{p,q}f$ is a minimal left ideal of $\cl_{p,q}$, and
$\K=f\cl_{p,q}f$ is a division ring of $\cl_{p,q}$. The primitive idempotent
of the algebra $\cl_{p,q}$ has a form
\[
f=\frac{1}{2}(1\pm\e_{\alpha_1})\frac{1}{2}(1\pm\e_{\alpha_2})\cdots\frac{1}{2}
(1\pm\e_{\alpha_k}),
\]
where $\e_{\alpha_1},\e_{\alpha_2},\ldots,\e_{\alpha_k}$ are commuting
elements with square 1 of the canonical basis of $\cl_{p,q}$ generating
a group of order $2^k$. The values of $k$ are defined by a formula
$k=q-r_{q-p}$, where $r_i$ are the Radon--Hurwitz numbers \cite{Rad22,Hur23},
values of which form a cycle of the period 8: $r_{i+8}=r_i+4$. The values of
all $r_i$ are
\begin{center}
\begin{tabular}{lcccccccc}
$i$  & 0 & 1 & 2 & 3 & 4 & 5 & 6 & 7\\ \hline
$r_i$& 0 & 1 & 2 & 2 & 3 & 3 & 3 & 3
\end{tabular}.
\end{center}
{\bf Examples}. 1) {\it Space--time algebra}. Let us consider a minimal left
ideal of the space--time algebra
$\cl_{1,3}$. In this case the Radon--Hurwitz number is equal to
$r_{q-p}=r_2=2$ and, therefore, for a quantity of the commuting elements
of the canonical basis we have $k=1$. Thus, a primitive idempotent of the
algebra $\cl_{1,3}$ has a form
\[
e_{13}=\frac{1}{2}(1+\e_0),
\]
and the minimal left ideal is defined by an expression
\begin{equation}\label{e20'}
I_{1,3}=\cl_{1,3}\frac{1}{2}(1+\e_0).
\end{equation}
2) {\it Dirac algebra}. Using the recurrence relation $r_{i+8}=r_i+4$ we
we obtain for the Dirac algebra $\cl_{4,1}$: $k=1-r_{-3}=1-(r_5-4)=2$.
Therefore, the idempotent $e_{41}=\frac{1}{2}(1+\e_{0123})\frac{1}{2}(1+
\e_{034})\in\cl_{4,1}$ is primitive, and a minimal left ideal of the algebra
$\cl_{4,1}$ is equal to
\begin{equation}\label{e20''}
I_{4,1}=\cl_{4,1}\frac{1}{2}(1+\e_{0123})\frac{1}{2}(1+\e_{034}).
\end{equation} 
3) {\it The algebras $\cl_{0,8}$ and $\cl_{8,0}$}. For the algebra $\cl_{0,8}$ 
we have
$k=8-r_8=8-(r_0+4)=4$ and, therefore,
\begin{equation}\label{2e0}
e_{08}=\frac{1}{2}(1+\e_{1248})\frac{1}{2}(1+\e_{2358})\frac{1}{2}(1+\e_{3468})
\frac{1}{2}(1+\e_{4578}).
\end{equation}
Analogously, for the algebra $\cl_{8,0}$: $k=0-r_{-8}=0-(r_0-4)=4$ and, 
therefore, the primitive idempotents of the algebras
$\cl_{8,0}$ and $\cl_{0,8}$ are coincide, since for any element
$\e_{i_1i_2i_3i_4}\in\cl_{8,0}$ we have $(\e_{i_1i_2i_3i_4})^2=1$.
All the Clifford algebras $\cl_{p,q}$ over the field $\F=\R$ are divided
into eight different types with the following division ring structure:\\[0.3cm]
{\bf I}. Central simple algebras.
\begin{description}
\item[1] Two types $p-q\equiv 0,2\pmod{8}$ with a division ring 
$\K\simeq\R$.
\item[2] Two types $p-q\equiv 3,7\pmod{8}$ with a division ring
$\K\simeq\C$.
\item[3] Two types $p-q\equiv 4,6\pmod{8}$ with a division ring
$\K\simeq\BH$.
\end{description}
{\bf II}. Semisimple algebras.
\begin{description}
\item[4] The type $p-q\equiv 1\pmod{8}$ with a double division ring
$\K\simeq\R\oplus\R$.
\item[5] The type $p-q\equiv 5\pmod{8}$ with a double quaternionic 
division ring $\K\simeq\BH\oplus\BH$.
\end{description}
The following table (Budinich--Trautman Periodic Table \cite{BT88})
explicitly shows a distribution of the real Clifford algebras in dependence
on the division ring structure, here ${}^2\R(n)=\R(n)\oplus\R(n)$ and
${}^2\BH(n)=\BH(n)\oplus\BH(n)$.
\begin{equation}\label{Periodic}{\renewcommand{\arraystretch}{1.2}
\begin{tabular}{c|cccccccccc}
  & p & 0 & 1 & 2 & 3 & 4 & 5 & 6 & 7 & \ldots\\ \hline
q &   &   &   &   &   &   &   &   &   &\\
0 &   &$\R$&${}^2\R$&$\R(2)$&$\C(2)$&$\BH(2)$&${}^2\BH(2)$&$\BH(4)$&$\C(8)$&
$\ldots$\\
1&&$\C$&$\R(2)$&${}^2\R(2)$&$\R(4)$&$\C(4)$&$\BH(4)$&${}^2\BH(4)$&$\BH(8)$&
$\ldots$\\
2&&$\BH$&$\C(2)$&$\R(4)$&${}^2\R(4)$&$\R(8)$&$\C(8)$&$\BH(8)$&${}^2\BH(8)$&
$\ldots$\\
3&&${}^2\BH$&$\BH(2)$&$\C(4)$&$\R(8)$&${}^2\R(8)$&$\R(16)$&$\C(16)$&$\BH(16)$&
$\ldots$\\
4&&$\BH(2)$&${}^2\BH(2)$&$\BH(4)$&$\C(8)$&$\R(16)$&${}^2\R(16)$&$\R(32)$&
$\C(32)$&$\ldots$\\
5&&$\C(4)$&$\BH(4)$&${}^2\BH(4)$&$\BH(8)$&$\C(16)$&$\R(32)$&${}^2\R(32)$&
$\R(64)$&$\ldots$\\
6&&$\R(8)$&$\C(8)$&$\BH(8)$&${}^2\BH(8)$&$\BH(16)$&$\C(32)$&$\R(64)$&
${}^2\R(64)$&$\ldots$\\
7&&${}^2\R(8)$&$\R(16)$&$\C(16)$&$\BH(16)$&${}^2\BH(16)$&$\BH(32)$&$\C(64)$&
$\R(128)$&$\ldots$\\
$\vdots$&&$\vdots$&$\vdots$&$\vdots$&$\vdots$&$\vdots$&$\vdots$&$\vdots$&
$\vdots$
\end{tabular}}
\end{equation}

Over the field $\F=\C$ there is an isomorphism $\C_n\simeq\M_{2^{n/2}}(\C)$
and there are two different types of complex Clifford algebras $\C_n$:
$n\equiv 0\pmod{2}$ and $n\equiv 1\pmod{2}$.

In virtue of the Wedderburn--Artin theorem, all fundamental automorphisms
of $\cl$ are transferred to the matrix algebra. Matrix representations of the
fundamental automorphisms of $\C_n$ was first obtained by Rashevskii in 1955
\cite{Rash}: 1) Involution: $\sA^\star=\sW\sA\sW^{-1}$, where $\sW$ is a
matrix of the automorphism $\star$ (matrix representation of the volume
element $\omega$); 2) Reversion: $\widetilde{\sA}=\sE\sA^{\sT}\sE^{-1}$, where
$\sE$ is a matrix of the antiautomorphism $\widetilde{\phantom{cc}}$
satisfying the conditions $\cE_i\sE-\sE\cE^{\sT}_i=0$ and 
$\sE^{\sT}=(-1)^{\frac{m(m-1)}{2}}\sE$, here $\cE_i=\gamma(\e_i)$ are matrix
representations of the units of the algebra $\cl$; 3) Conjugation:
$\widetilde{\sA^\star}=\sC\sA^{\sT}\sC^{-1}$, where $\sC=\sE\sW^{\sT}$ 
is a matrix of
the antiautomorphism $\widetilde{\star}$ satisfying the conditions
$\sC\cE^{\sT}_i+\cE_i\sC=0$ and
$\sC^{\sT}=(-1)^{\frac{m(m+1)}{2}}\sC$.

{\bf Example}. Let consider matrix representations of the fundamental
automorphisms of the Dirac algebra $\cl_{4,1}$. In virtue of Theorem
\ref{t1} there is an isomorphism $\cl_{4,1}\simeq\C_4$, and therefore
$\cl_{4,1}\simeq\C_4\simeq\M_4(\C)$. In the capacity of the matrix
representations of the units $\e_i\in\C_4$ $(i=1,2,3,4)$ we take the
well-known Dirac $\gamma$-matrices (so-called canonical representation):
\[
\ar
\gamma_1=\begin{pmatrix}
0 & 0 & 0 & -i\\
0 & 0 & -i& 0\\
0 & i & 0 & 0\\
i & 0 & 0 & 0
\end{pmatrix},\quad\gamma_2=\begin{pmatrix}
0 & 0 & 0 & -1\\
0 & 0 & 1 & 0\\
0 & 1 & 0 & 0\\
-1& 0 & 0 & 0
\end{pmatrix},
\]
\begin{equation}\label{e20}\ar
\gamma_3=\begin{pmatrix}
0 & 0 & -i & 0\\
0 & 0 & 0 & i\\
i & 0 & 0 & 0\\
0 & -i& 0 & 0
\end{pmatrix},\quad\gamma_4=\begin{pmatrix}
1 & 0 & 0 & 0\\
0 & 1 & 0 & 0\\
0 & 0 &-1 & 0\\
0 & 0 & 0 &-1
\end{pmatrix}.
\end{equation}   
$\gamma$-matrices form the only one basis from the set of isomorphic
matrix basises of $\cl_{4,1}\simeq\C_4$. In the basis (\ref{e20}) the element
$\omega=\e_1\e_2\e_3\e_4$ is represented by a matrix $\sW=\gamma_5=\gamma_1
\gamma_2\gamma_3\gamma_4$. Since $\gamma^2_5=1$, then $\varepsilon=1\;
(\sW^{\p}=\sW)$ and the matrix
\begin{equation}\label{e21}\ar
\sW^T=\sW=\begin{pmatrix}
0 & 0 &-1 & 0\\
0 & 0 & 0 &-1\\
-1& 0 & 0 & 0\\
0 &-1 & 0 & 0
\end{pmatrix}
\end{equation}
in accordance with (\ref{auto17}) is a matrix of the automorphism $\cA
\rightarrow\cA^\star$. Further, in the matrix representation the
antiautomorphism $\cA\rightarrow\widetilde{\cA}$ is defined by the
transformation $\widetilde{\sA}=\sE\sA^T\sE^{-1}$. For the $\gamma$-matrices
 we have
$\gamma^T_1=-\gamma_1,\,\gamma^T_2=\gamma_2,\,\gamma^T_3=-\gamma_3,
\,\gamma^T_4=\gamma_4$. Further,
\begin{eqnarray}
&&\gamma_1=-\sE\gamma_1\sE^{-1},\quad\gamma_2=\sE\gamma_2\sE^{-1},\nonumber\\
&&\gamma_3=-\sE\gamma_3\sE^{-1},\quad\gamma_4=\sE\gamma_4\sE^{-1}.\nonumber
\end{eqnarray}
It is easy to verify that a matrix $\sE=\gamma_1\gamma_3$ satisfies the
latter relations and, therefore, the antiautomorphism $\cA\rightarrow
\widetilde{\cA}$ in the basis (\ref{e20}) is defined by the matrix
\begin{equation}\label{e22}\ar
\sE=\gamma_1\gamma_3=\begin{pmatrix}
0 & -1 & 0 & 0\\
1 & 0 & 0 & 0\\
0 & 0 & 0 & -1\\
0 & 0 & 1 & 0
\end{pmatrix}.
\end{equation}
Finally, for the matrix $\sC=\sE\sW^T$ of the antiautomorphism $\cA\rightarrow
\widetilde{\cA^\star}$ from (\ref{e21}) and (\ref{e22}) in accordance with
(\ref{auto19}) we obtain
\begin{equation}\label{be23}\ar
\sC=\sE\sW^T=\begin{pmatrix}
0 & 0 & 0 & 1\\
0 & 0 &-1 & 0\\
0 & 1 & 0 & 0\\
-1& 0 & 0 & 0
\end{pmatrix}.
\end{equation}

\section{Salingaros groups}
The structure of the Clifford algebras admits a very elegant description
in terms of finite groups \cite{Sal81a,Sal82,Sal84}. In accordance with
the multiplication law (\ref{e1}) the basis elements of $\cl_{p,q}$ form
a finite group of order $2^{n+1}$
\begin{equation}\label{FG}
G(p,q)=\left\{\pm 1,\,\pm\e_i,\,\pm\e_i\e_j,\,\pm\e_i\e_j\e_k,\,\ldots,\,
\pm\e_1\e_2\cdots\e_n\right\}\quad(i<j<k<\ldots).
\end{equation}

Salingaros showed \cite{Sal81a,Sal82} that there exist five distinct types
of finite groups (\ref{FG}) that arise from Clifford algebras.
In \cite{Sal81a,Sal82} they were called `vee groups' and were labelled as
\begin{equation}\label{FG2}
N_{\text{odd}},\;N_{\text{even}},\;\Omega_{\text{odd}},\;\Omega_{\text{even}},\;
S_k.
\end{equation}
The odd $N$--groups correspond to real spinors, for example, $N_1$ is related
to real 2--spinors, and $N_3$ is the group of the real Majorana matrices.
The even $N$--groups define the quaternionic groups. The $S$--groups
are the `spinor groups' ($S_k=N_{2k}\otimes\C\simeq N_{2k-1}\otimes\C$):
$S_1$ is the group of the complex Pauli matrices, and $S_2$ is the group of the
Dirac matrices. Furthermore, the $\Omega$--groups are double copies of the
$N$--groups and can be written as a direct product of the $N$--groups with
the group of two elements $\dZ_2$:
\begin{equation}\label{FG3}
\Omega_k=N_k\otimes\dZ_2.
\end{equation}

Let us consider now several simplest examples of the groups (\ref{FG2}).
First of all, a finite group corresponding to the Clifford algebra $\cl_{0,0}$
with an arbitrary element $\cA=a^0$ and division ring $\K\simeq\R$
($p-q\equiv 0\pmod{8}$) is a cyclic group $\dZ_2=\{1,-1\}$ with the
following multiplication table
\[
{\renewcommand{\arraystretch}{1.4}
\begin{tabular}{|c||c|c|}\hline
 & $1$ & $-1$\\ \hline\hline  
$1$ & $1$ & $-1$ \\ \hline
$-1$& $-1$& $1$ \\ \hline
\end{tabular}
}
\]
It is easy to see that in accordance with (\ref{FG2}) the finite group
corresponding to $\cl_{0,0}$ is $N_0$--group ($N_0=\dZ_2$).

A further Clifford algebra is $\cl_{1,0}$: $\cA=a^0+a^1\e_1,\,\e^2_1=1$,
$\K\simeq\R\oplus\R$, $p-q\equiv 1\pmod{8}$. In this case the basis elements
of $\cl_{1,0}$ form the Gauss--Klein four--group 
$\dZ_2\otimes\dZ_2=\{1,-1,\e_1,-\e_1\}$. The multiplication table of
$\dZ_2\otimes\dZ_2$ has a form
\[
{\renewcommand{\arraystretch}{1.4}
\begin{tabular}{|c||c|c|c|c|}\hline
  & $1$ & $-1$ & $\e_1$ & $-\e_1$ \\ \hline\hline
$1$ & $1$ & $-1$ & $\e_1$ & $-\e_1$ \\ \hline
$-1$ & $-1$ & $1$ & $-\e_1$ & $\e_1$ \\ \hline
$\e_1$ & $\e_1$ & $-\e_1$ & $1$ & $-1$ \\ \hline
$-\e_1$ & $-\e_1$ & $\e_1$ & $-1$ & $1$ \\ \hline
\end{tabular}
}
\]
In accordance with (\ref{FG3}) we have here a first $\Omega$--group:
$\Omega_0=N_0\otimes\dZ_2=N_0\otimes N_0=\dZ_2\otimes\dZ_2$.

The algebra $\cl_{0,1}$ with $\cA=a^0+a^1\e_1$, $\e^2_1=-1$
($\K\simeq\C,\,p-q\equiv 7\pmod{8}$) corresponds to the complex group
$\dZ_4=\{1,-1,\e_1,-\e_1\}$ with the multiplication table
\[
{\renewcommand{\arraystretch}{1.4}
\begin{tabular}{|c||c|c|c|c|}\hline
  & $1$ & $-1$ & $\e_1$ & $-\e_1$ \\ \hline\hline
$1$ & $1$ & $-1$ & $\e_1$ & $-\e_1$ \\ \hline
$-1$ & $-1$ & $1$ & $-\e_1$ & $\e_1$ \\ \hline
$\e_1$ & $\e_1$ & $-\e_1$ & $-1$ & $1$ \\ \hline
$-\e_1$ & $-\e_1$ & $\e_1$ & $1$ & $-1$ \\ \hline
\end{tabular}
}
\]
It is easy to see that in accordance with Salingaros classification this
group is a first $S$--group: $S_0=\dZ_4$.

The three finite groups considered previously are Abelian groups. All other
Salingaros groups (\ref{FG2}) are non--Abelian. The first non--Abelian
Salingaros group correspond to the algebra $\cl_{2,0}$ with an arbitrary
element $\cA=a^0+a^1\e_1+a^2\e_2+a^{12}\e_{12}$ and $\e^2_1=\e^2_2=1,\,
\e_{12}^2=-1$ ($p-q\equiv 2\pmod{8},\,\K\simeq\R$). The basis elements of
$\cl_{2,0}$ form a dihedral group $D_4=\{1,-1,\e_1,-\e_1,\e_2,-\e_2,
\e_{12},-\e_{12}\}$ with the table
\[
{\renewcommand{\arraystretch}{1.4}
\begin{tabular}{|c||c|c|c|c|c|c|c|c|}\hline
  & $1$ & $-1$ & $\e_1$ & $-\e_1$ & $\e_2$ & $-\e_2$ & $\e_{12}$ & $-\e_{12}$
\\ \hline\hline
$1$ & $1$ & $-1$ & $\e_1$ & $-\e_1$ & $\e_2$ & $-\e_2$ & $\e_{12}$ & $-\e_{12}$
\\ \hline
$-1$ & $-1$ & $1$ & $-\e_1$ & $\e_1$ & $-\e_2$ & $\e_2$ & $-\e_{12}$ & $\e_{12}$
\\ \hline
$\e_1$ & $\e_1$ & $-\e_1$ & $1$ & $-1$ & $\e_{12}$ & $-\e_{12}$ &$\e_2$&$-\e_2$
\\ \hline
$-\e_1$& $-\e_1$& $\e_1$ & $-1$ & $1$ &$-\e_{12}$ &$\e_{12}$ & $-\e_2$&$\e_2$
\\ \hline
$\e_2$ & $\e_2$ & $-\e_2$&$-\e_{12}$ &$\e_{12}$ & $1$ &$-1$& $-\e_1$ &$\e_1$
\\ \hline
$-\e_2$& $-\e_2$& $\e_2$ &$\e_{12}$ & $-\e_{12}$& $-1$ &$1$ & $\e_1$ & $-\e_1$
\\ \hline
$\e_{12}$& $\e_{12}$&$-\e_{12}$&$-\e_2$ &$\e_2$ &$\e_1$ &$-\e_1$ &$-1$ &$1$
\\ \hline
$-\e_{12}$&$-\e_{12}$&$\e_{12}$&$\e_2$ & $-\e_2$&$-\e_1$&$\e_1$ &$1$ &$-1$
\\ \hline
\end{tabular}
}
\]
It is a first $N_{\text{odd}}$--group: $N_1=D_4$. It is easy to verify
that we come to the same group $N_1=D_4$ for the algebra $\cl_{1,1}$ with the
ring $\K\simeq\R$, $p-q\equiv 0\pmod{8}$.

The following non--Abelian finite group we obtain for the algebra
$\cl_{0,2}$ with a quaternionic ring $\K\simeq\BH$, $p-q\equiv 6\pmod{8}$.
In this case the basis elements of $\cl_{0,2}$ form a quaternionic group
$Q_4=\{\pm 1,\pm\e_1,\pm\e_2,\pm\e_{12}\}$ with the multiplication table
\[
{\renewcommand{\arraystretch}{1.4}
\begin{tabular}{|c||c|c|c|c|c|c|c|c|}\hline
  & $1$ & $-1$ & $\e_1$ & $-\e_1$ & $\e_2$ & $-\e_2$ & $\e_{12}$ & $-\e_{12}$
\\ \hline\hline
$1$ & $1$ & $-1$ & $\e_1$ & $-\e_1$ & $\e_2$ & $-\e_2$ & $\e_{12}$ & $-\e_{12}$
\\ \hline
$-1$ & $-1$ & $1$ & $-\e_1$ & $\e_1$ & $-\e_2$ & $\e_2$ & $-\e_{12}$ & $\e_{12}$
\\ \hline
$\e_1$ & $\e_1$ & $-\e_1$ & $-1$ & $1$ & $\e_{12}$ & $-\e_{12}$ &$-\e_2$&$\e_2$
\\ \hline
$-\e_1$& $-\e_1$& $\e_1$ & $1$ & $-1$ &$-\e_{12}$ &$\e_{12}$ & $\e_2$&$-\e_2$
\\ \hline
$\e_2$ & $\e_2$ & $-\e_2$&$-\e_{12}$ &$\e_{12}$ & $-1$ &$1$& $\e_1$ &$-\e_1$
\\ \hline
$-\e_2$& $-\e_2$& $\e_2$ &$\e_{12}$ & $-\e_{12}$& $1$ &$-1$ & $-\e_1$ & $\e_1$
\\ \hline
$\e_{12}$& $\e_{12}$&$-\e_{12}$&$\e_2$ &$-\e_2$ &$-\e_1$ &$\e_1$ &$-1$ &$1$
\\ \hline
$-\e_{12}$&$-\e_{12}$&$\e_{12}$&$-\e_2$ & $\e_2$&$\e_1$&$-\e_1$ &$1$ &$-1$
\\ \hline
\end{tabular}
}
\]
It is a first $N_{\text{even}}$--group: $N_2=Q_4$.

Now we can to establish a relationship between the finite group and
division ring structures of $\cl_{p,q}$. It is easy to see that the five
distinct types of Salingaros groups correspond to the five division rings
of the real Clifford algebras as follows
\begin{eqnarray}
N_{\text{odd}}&\leftrightarrow&\cl_{p,q},\;p-q\equiv 0,2\pmod{8},\;\K\simeq\R;
\nonumber\\
N_{\text{even}}&\leftrightarrow&\cl_{p,q},\;p-q\equiv 4,6\pmod{8},\;
\K\simeq\BH;\nonumber\\
\Omega_{\text{odd}}&\leftrightarrow&\cl_{p,q},\;p-q\equiv 1\pmod{8},\;
\K\simeq\R\oplus\R;\nonumber\\
\Omega_{\text{even}}&\leftrightarrow&\cl_{p,q},\;p-q\equiv 5\pmod{8},\;
\K\simeq\BH\oplus\BH;\nonumber\\
S_k&\leftrightarrow&\cl_{p,q},\;p-q\equiv 3,7\pmod{8},\;\K\simeq\C.\nonumber
\end{eqnarray}
Therefore, the Periodic Table (\ref{Periodic}) can be rewritten in terms
of finite group structure
\[
{\renewcommand{\arraystretch}{1.2}
\begin{tabular}{c|cccccccccc}
  & p & 0 & 1 & 2 & 3 & 4 & 5 & 6 & 7 & \ldots\\ \hline
q &   &   &   &   &   &   &   &   &   &\\
0 &   &$N_1$&$\Omega_0$&$N_1$&$S_1$&$N_4$&$\Omega_4$&$N_6$&$S_3$&
$\ldots$\\
1&&$S_0$&$N_1$&$\Omega_1$&$N_3$&$S_2$&$N_6$&$\Omega_6$&$N_8$&
$\ldots$\\
2&&$N_2$&$S_1$&$N_3$&$\Omega_3$&$N_5$&$S_3$&$N_8$&$\Omega_8$&
$\ldots$\\
3&&$\Omega_2$&$N_4$&$S_2$&$N_5$&$\Omega_5$&$N_7$&$S_4$&$N_{10}$&
$\ldots$\\
4&&$N_4$&$\Omega_4$&$N_6$&$S_3$&$N_7$&$\Omega_7$&$N_9$&
$S_5$&$\ldots$\\
5&&$S_2$&$N_6$&$\Omega_6$&$N_8$&$S_4$&$N_9$&$\Omega_9$&
$N_{11}$&$\ldots$\\
6&&$N_5$&$S_3$&$N_8$&$\Omega_8$&$N_{10}$&$S_5$&$N_{11}$&
$\Omega_{11}$&$\ldots$\\
7&&$\Omega_5$&$N_7$&$S_4$&$N_{10}$&$\Omega_{10}$&$N_{12}$&$S_6$&
$N_{13}$&$\ldots$\\
$\vdots$&&$\vdots$&$\vdots$&$\vdots$&$\vdots$&$\vdots$&$\vdots$&$\vdots$&
$\vdots$
\end{tabular}}
\]
Further, in accordance with (\ref{e4}) a center $\bZ_{p,q}$ of the algebra
$\cl_{p,q}$ consists of the unit if $p-q\equiv 0,2,4,6\pmod{8}$ and the
elements $1,\;\omega=\e_{12\ldots n}$ if $p-q\equiv 1,3,5,7\pmod{8}$.
Let $\bZ(p,q)\subset\cl_{p,q}$ be a center of the finite group (\ref{FG}).
In such a way, we have three distinct realizations of the center
$\bZ(p,q)$:
\begin{eqnarray}
\bZ(p,q)&=&\{1,-1\}\simeq\dZ_2\;\;\text{if}\;p-q\equiv 0,2,4,6\pmod{8};
\nonumber\\
\bZ(p,q)&=&\{1,-1,\omega,-\omega\}\simeq\dZ_2\otimes\dZ_2\;\;\text{if}\;
p-q\equiv 1,5\pmod{8};\nonumber\\
\bZ(p,q)&=&\{1,-1,\omega,-\omega\}\simeq\dZ_4\;\;\text{if}\;
p-q\equiv 3,7\pmod{8}.\nonumber
\end{eqnarray}
The Abelian groups $\bZ(p,q)$ are the subgroups of the Salingaros groups
(\ref{FG2}). Namely, $N$--groups have the center $\dZ_2$,
$\Omega$--groups have the center $\dZ_2\otimes\dZ_2$, and $S$-group has
the center $\dZ_4$.

In the following sections we will consider finite groups that arise from
the fundamental automorphisms of the Clifford algebras. It will be shown
that these groups (both Abelian and non--Abelian) are the subgroups of the
Salingaros groups. Moreover, it will be shown also that the finite groups
of automorphisms, $\Aut(\cl_{p,q})$, form a natural basis for description
of the discrete symmetries in quantum field theory.
\section{Clifford--Lipschitz groups}
The Lipschitz group $\Lip_{p,q}$, also called the Clifford group, introduced
by Lipschitz in 1886 \cite{Lips}, may be defined as the subgroup of
invertible elements $s$ of the algebra $\cl_{p,q}$:
\[
\Lip_{p,q}=\left\{s\in\cl^+_{p,q}\cup\cl^-_{p,q}\;|\;\forall x\in\R^{p,q},\;
s\bx s^{-1}\in\R^{p,q}\right\}.
\]
The set $\Lip^+_{p,q}=\Lip_{p,q}\cap\cl^+_{p,q}$ is called {\it special
Lipschitz group} \cite{Che55}.

Let $N:\;\cl_{p,q}\rightarrow\cl_{p,q},\;N(\bx)=\bx\widetilde{\bx}$.
If $\bx\in\R^{p,q}$, then $N(\bx)=\bx(-\bx)=-\bx^2=-Q(\bx)$. Further, the
group $\Lip_{p,q}$ has a subgroup
\begin{equation}\label{Pin}
\pin(p,q)=\left\{s\in\Lip_{p,q}\;|\;N(s)=\pm 1\right\}.
\end{equation}
Analogously, {\it a spinor group} $\spin(p,q)$ is defined by the set
\begin{equation}\label{Spin}
\spin(p,q)=\left\{s\in\Lip^+_{p,q}\;|\;N(s)=\pm 1\right\}.
\end{equation}
It is obvious that $\spin(p,q)=\pin(p,q)\cap\cl^+_{p,q}$.
The group $\spin(p,q)$ contains a subgroup
\begin{equation}\label{Spin+}
\spin_+(p,q)=\left\{s\in\spin(p,q)\;|\;N(s)=1\right\}.
\end{equation}
It is easy to see that the groups $O(p,q),\,SO(p,q)$ and $SO_+(p,q)$ are
isomorphic, respectively, to the following quotient groups
\[
O(p,q)\simeq\pin(p,q)/\dZ_2,\quad
SO(p,q)\simeq\spin(p,q)/\dZ_2,\quad
SO_+(p,q)\simeq\spin_+(p,q)/\dZ_2,
\]
\begin{sloppypar}\noindent
where the kernel $\dZ_2=\{1,-1\}$. Thus, the groups $\pin(p,q)$, $\spin(p,q)$
and $\spin_+(p,q)$ are the double coverings of the groups $O(p,q),\,SO(p,q)$
and $SO_+(p,q)$, respectively.\end{sloppypar}

Further, since $\cl^+_{p,q}\simeq\cl^+_{q,p}$, then
\[
\spin(p,q)\simeq\spin(q,p).
\]
In contrast with this, the groups $\pin(p,q)$ and $\pin(q,p)$ are 
non--isomorphic. Denote $\spin(n)=\spin(n,0)\simeq\spin(0,n)$.
\begin{theorem}[{\rm\cite{Cor84}}]\label{t3}
The spinor groups
\[
\spin(2),\;\;\spin(3),\;\;\spin(4),\;\;\spin(5),\;\;\spin(6)
\]
are isomorphic to the unitary groups
\[
U(1),\;\;Sp(1)\sim SU(2),\;\;SU(2)\times SU(2),\;\;Sp(2),\;\;SU(4).
\]
\end{theorem} 
In accordance with Theorem \ref{t1} and decompositions (\ref{e5'})
over the field $\F=\R$ the algebra $\cl_{p,q}$ is isomorphic to a direct
sum of two mutually annihilating simple ideals $\frac{1}{2}(1\pm\omega)
\cl_{p,q}$: $\cl_{p,q}\simeq\frac{1}{2}(1+\omega)\cl_{p,q}\oplus\frac{1}{2}
(1-\omega)\cl_{p,q}$, where $\omega=\e_{12\ldots p+q},\,p-q\equiv 1,5
\pmod{8}$. At this point,the each ideal is isomorpic to $\cl_{p,q-1}$ or
$\cl_{q,p-1}$. Therefore, for the Clifford--Lipschitz groups we have the
following isomorphisms
\begin{eqnarray}
\pin(p,q)&\simeq&\pin(p,q-1)\bigcup\pin(p,q-1)\nonumber\\
&\simeq&\pin(q,p-1)\bigcup\pin(q,p-1).
\end{eqnarray}
Or, since $\cl_{p,q-1}\simeq\cl^+_{p,q}\subset\cl_{p,q}$, then 
according to (\ref{e11})
\[
\pin(p,q)\simeq\spin(p,q)\bigcup\spin(p,q)
\]
if $p-q\equiv 1,5\pmod{8}$.

Further, when $p-q\equiv 3,7\pmod{8}$ from Theorem \ref{t1} it follows
that $\cl_{p,q}$ is isomorphic to a complex algebra $\C_{p+q-1}$. Therefore,
for the $\pin$ groups we obtain
\begin{eqnarray}
\pin(p,q)&\simeq&\pin(p,q-1)\bigcup\e_{12\ldots p+q}\pin(p,q-1)\nonumber\\
&\simeq&\pin(q,p-1)\bigcup\e_{12\ldots p+q}\pin(q,p-1)\label{e13}
\end{eqnarray}
if $p-q\equiv 1,5\pmod{8}$ and correspondingly
\begin{equation}\label{e14}
\pin(p,q)\simeq\spin(p,q)\cup\e_{12\ldots p+q}\spin(p,q).
\end{equation}
In case of $p-q\equiv 3,7\pmod{8}$ we have isomorphisms which are analoguos
to (\ref{e13})-(\ref{e14}), since $\omega\cl_{p,q}\sim\cl_{p,q}$.
Generalizing we obtain the following
\begin{theorem}\label{t4}
Let $\pin(p,q)$ and $\spin(p,q)$ be the Clifford-Lipschitz groups of the
invertible elements of the algebras $\cl_{p,q}$ with odd dimensionality,
$p-q\equiv 1,3,5,7\pmod{8}$. Then
\begin{eqnarray}
\pin(p,q)&\simeq&\pin(p,q-1)\bigcup\omega\pin(p,q-1)\nonumber\\
&\simeq&\pin(q,p-1)\bigcup\omega\pin(q,p-1)\nonumber
\end{eqnarray}
and
\[
\pin(p,q)\simeq\spin(p,q)\bigcup\omega\spin(p,q),
\]
where $\omega=\e_{12\ldots p+q}$ is a volume element of $\cl_{p,q}$.
\end{theorem}
In case of low dimensionalities from Theorem \ref{t3} and Theorem
\ref{t4} it immediately follows
\begin{theorem}\label{t5}
For $p+q\leq 5$ and $p-q\equiv 3,5\pmod{8}$,
\begin{eqnarray}
\pin(3,0)&\simeq&SU(2)\cup iSU(2),\nonumber\\
\pin(0,3)&\simeq&SU(2)\cup eSU(2),\nonumber\\
\pin(5,0)&\simeq&Sp(2)\cup eSp(2),\nonumber\\
\pin(0,5)&\simeq&Sp(2)\cup iSp(2).\nonumber
\end{eqnarray}
\end{theorem}
\begin{proof}\begin{sloppypar}\noindent
Indeed, in accordance with Theorem \ref{t4} $\pin(3,0)\simeq\spin(3)
\cup\e_{123}\spin(3)$. Further, from Theorem \ref{t3} we have
$\spin(3)\simeq SU(2)$, and a square of the element $\omega=\e_{123}$ is
equal to $-1$, therefore $\omega\sim i$. Thus, $\pin(3,0)\simeq SU(2)\cup
iSU(2)$. For the group $\pin(0,3)$ a square of $\omega$ is equal to $+1$,
therefore $\pin(0,3)\simeq SU(2)\cup eSU(2)$, $e$ is a double unit.
As expected, $\pin(3,0)\not\simeq\pin(0,3)$. The isomorphisms for the
groups $\pin(5,0)$ and $\pin(0,5)$ are analogously proved.\end{sloppypar}
\end{proof}

On the other hand, there exists a more detailed version of the $\pin$--group
(\ref{Pin}) proposed by D\c{a}browski in 1988 \cite{Dab88}. In general,
there are eight double coverings of the orthogonal group 
$O(p,q)$ \cite{Dab88,BD89}:
\[
\rho^{a,b,c}:\;\;\pin^{a,b,c}(p,q)\longrightarrow O(p,q),
\]
where $a,b,c\in\{+,-\}$. As known, the group $O(p,q)$ consists of four
connected components: identity connected component $O_0(p,q)$, and three
components corresponding to parity reversal $P$, time reversal
$T$, and the combination of these two $PT$, i.e., $O(p,q)=(O_0(p,q))\cup
P(Q_0(p,q))\cup T(O_0(p,q))\cup PT(O_0(p,q))$. Further, since the
four--element group (reflection group) $\{1,\,P,\,T,\,PT\}$ is isomorphic to
the finite group $\dZ_2\otimes\dZ_2$ 
(Gauss--Klein veergruppe \cite{Sal81a,Sal84}), then
$O(p,q)$ may be represented by a semidirect product $O(p,q)\simeq O_0(p,q)
\odot(\dZ_2\otimes\dZ_2)$. The signs of $a,b,c$ correspond to the signs of the
squares of the elements in $\pin^{a,b,c}(p,q)$ which cover space reflection
$P$, time reversal $T$ and a combination of these two
$PT$ ($a=-P^2,\,b=T^2,\,c=-(PT)^2$ in D\c{a}browski's notation \cite{Dab88} and
$a=P^2,\,b=T^2,\,c=(PT)^2$ in Chamblin's notation \cite{Ch94} which we will
use below).
An explicit form of the group $\pin^{a,b,c}(p,q)$ is given by the following
semidirect product
\begin{equation}\label{Pinabc}
\pin^{a,b,c}(p,q)\simeq\frac{(\spin_0(p,q)\odot C^{a,b,c})}{\dZ_2},
\end{equation}
where $C^{a,b,c}$ are the four double coverings of
$\dZ_2\otimes\dZ_2$. 
All the eight double coverings of the orthogonal group
$O(p,q)$ are given in the following table:
\begin{center}
{\renewcommand{\arraystretch}{1.4}
\begin{tabular}{|c|l|l|}\hline
$a$ $b$ $c$ & $C^{a,b,c}$ & Remark \\ \hline
$+$ $+$ $+$ & $\dZ_2\otimes\dZ_2\otimes\dZ_2$ & $PT=TP$\\
$+$ $-$ $-$ & $\dZ_2\otimes\dZ_4$ & $PT=TP$\\
$-$ $+$ $-$ & $\dZ_2\otimes\dZ_4$ & $PT=TP$\\
$-$ $-$ $+$ & $\dZ_2\otimes\dZ_4$ & $PT=TP$\\ \hline
$-$ $-$ $-$ & $Q_4$ & $PT=-TP$\\
$-$ $+$ $+$ & $D_4$ & $PT=-TP$\\
$+$ $-$ $+$ & $D_4$ & $PT=-TP$\\
$+$ $+$ $-$ & $D_4$ & $PT=-TP$\\ \hline
\end{tabular}
}
\end{center}
Here $\dZ_4$, $Q_4$, and $D_4$ are complex, quaternion, and
dihedral groups, respectively.
According to \cite{Dab88} the group $\pin^{a,b,c}(p,q)$ satisfying the
condition
$PT=-TP$ is called {\it Cliffordian}, and respectively {\it
non--Cliffordian} when $PT=TP$. 
\section[DIRAC--HESTENES SPINOR FIELDS]
{Dirac-Hestenes spinor fields and minimal left ideals of the algebras
$\cl_{1,3}$ and $\cl_{4,1}$}
An introduction of Dirac--Hestenes fields \cite{Hest66,Hest67,Hest76} 
were caused by the following problem. As known, a 5--dimensional real space
$\R^{4,1}$, associated with the Dirac algebra
$\cl_{4,1}$, is a non--physical space, whereas in virtue of an isomorphism
$\cl_{4,1}\simeq\C_4$ the Dirac spinor $\Phi\in\End_{\C}(I_{4,1})$ describes
the wave function of the electron \cite{Dir}. Thus, with a view to remove
this disparity it needs to define the Dirac field in a physical space
$\R^{1,3}$ (Minkowski space--time) associated with the algebra $\cl_{1,3}$.
The transition from the complex space
$\C_4$ to space--time $\R^{1,3}$ was given by Hestenes and his collaborators
\cite{Hest66,Hest76,Hest90,RVR93,VR93,DHSV93,DLG93,DR}).

So, the minimal left ideals of the algebras
$\cl_{1,3}$ and $\cl_{4,1}$ have correspondingly the following form
\begin{eqnarray}
I_{1,3}&=&\cl_{1,3}e_{13}=\cl_{1,3}\frac{1}{2}(1+\gamma_0),\nonumber\\
I_{4,1}&=&\cl_{4,1}e_{41}=\cl_{4,1}\frac{1}{2}(1+\gamma_0)\frac{1}{2}
(1+i\gamma_{12}),\nonumber
\end{eqnarray}
where $\gamma_i\;(i=0,1,2,3)$ are matrix representations of the units of
$\cl_{1,3}$:
\begin{equation}\label{Gamma}
\ar\gamma_0=\begin{pmatrix}
E & 0\\
0 & -E
\end{pmatrix},\quad\gamma_1=\begin{pmatrix}
0 & \sigma_1\\
-\sigma_1 & 0
\end{pmatrix},\quad\gamma_2=\begin{pmatrix}
0 & \sigma_2\\
-\sigma_2 & 0
\end{pmatrix},\quad\gamma_3=\begin{pmatrix}
0 & \sigma_3\\
-\sigma_3 & 0
\end{pmatrix},
\end{equation}
whre $\sigma_i$ are Pauli matrices
\[
\ar\sigma_1=\begin{pmatrix}
0 & 1\\
1 & 0
\end{pmatrix},\quad\sigma_2=\begin{pmatrix}
0 & -i\\
i & 0
\end{pmatrix},\quad\sigma_3=\begin{pmatrix}
1 & 0\\
0 & -1
\end{pmatrix},
\]
$E$ is an unit matrix. Further, for the Dirac algebra take place the following
isomorphisms $\cl_{4,1}\simeq\C_4=\C\otimes\cl_{1,3}\simeq\M_2(\C_2)$ 
, $\cl^+_{4,1}\simeq\cl_{1,3}\simeq\M^{\BH}_2(\cl_{1,1})$. In virtue of an
identity $\cl_{1,3}e_{13}=\cl^+_{1,3}e_{13}$
\cite{FRO90a,FRO90b} for the minimal left ideal of the algebra
$\cl_{4,1}$ we obtain 
\begin{multline}\label{2e1}
I_{4,1}=\cl_{4,1}e_{41}=(\C\otimes\cl_{1,3})e_{41}\simeq\cl^+_{4,1}e_{41}
\simeq\cl_{1,3}e_{41}=\\
\cl_{1,3}e_{13}\frac{1}{2}(1+i\gamma_{12})=\cl^+_{1,3}e_{13}\frac{1}{2}
(1+i\gamma_{12}).
\end{multline}
Let $\Phi\in\cl_{4,1}\simeq\M_4(\C)$ be a Dirac spinor and let $\phi\in
\cl^+_{1,3}\simeq\cl_{3,0}$ be a Dirac--Hestenes spinor. Then from (\ref{2e1})
it immediately follows a relation between the spinors
$\Phi$ and $\phi$:
\begin{equation}\label{2e2}
\Phi=\phi\frac{1}{2}(1+\gamma_0)\frac{1}{2}(1+i\gamma_{12}).
\end{equation}
Since $\phi\in\cl^+_{1,3}\simeq\cl_{3,0}$, then the Dirac--Hestenes spinor
can be represented by a biquaternion number
\begin{equation}\label{2e3}
\phi=a^0+a^{01}\gamma_{01}+a^{02}\gamma_{02}+a^{03}\gamma_{03}+
a^{12}\gamma_{12}+a^{13}\gamma_{13}+a^{23}\gamma_{23}+a^{0123}\gamma_{0123}.
\end{equation}
Or in the matrix form
\begin{equation}\label{e23}\ar
\phi=\begin{pmatrix}
\phi_1 & -\phi^\ast_2 & \phi_3 & \phi^\ast_4\\
\phi_2 & \phi^\ast_1  & \phi_4 & -\phi^\ast_3\\
\phi_3 & \phi^\ast_4  & \phi_1 & -\phi^\ast_2\\
\phi_4 & -\phi^\ast_3 & \phi_2 & \phi^\ast_1
\end{pmatrix},
\end{equation}
where
\begin{eqnarray}
\phi_1&=&a^0-ia^{12},\nonumber\\
\phi_2&=&a^{13}-ia^{23},\nonumber\\
\phi_3&=&a^{03}-ia^{0123},\nonumber\\
\phi_4&=&a^{01}+ia^{02}.\nonumber
\end{eqnarray}
According to (\ref{2e1}) and (\ref{2e2}), (\ref{e23}) the minimal left ideal of
the algebra $\cl_{4,1}$ in the matrix representation has a form
\[\ar
\Phi=\begin{pmatrix}
\phi_1 & 0 & 0 & 0\\
\phi_2 & 0 & 0 & 0\\
\phi_3 & 0 & 0 & 0\\
\phi_4 & 0 & 0 & 0
\end{pmatrix}
\]
In such a way, the elements of this ideal contain four complex, or eight real,
parameters, which are just sufficient to define a Dirac spinor.

In conclusion of this section it should be noted that the Hestenes program
of reinterpretation of quantum mechanics within the field of real numbers
involves a wide variety of interesting applications of the Dirac--Hestenes
fields in geometry and general theory of relativity
\cite{HS84,RF90,RO90,RS93,RSVL,Keller,Var99a,Var99b,Var01,Var02}.
\chapter{D\c{a}browski groups}
\setcounter{theorem}{0}
\setcounter{cor}{0}
\setcounter{opr}{0}
\setcounter{lem}{0}
\setcounter{axiom}{0}
\setcounter{equation}{0}
\section{Generalization of the D\c{a}browski groups and fundamental
automorphisms of the Clifford algebras}
As noted above, there exists a close relationship between D\c{a}browski groups
$\pin^{a,b,c}(p,q)$ and discrete tansformations of the orthogonal group
$O(p,q)$ (in particular, Lorentz group $O(1,3)$) \cite{DWGK,Ch97,Ch94,AlCh94,
AlCh96}. On the other hand, discrete transformations of the group $O(p,q)$
which acting in the space $\R^{p,q}$ associated with the algebra $\cl_{p,q}$,
may be realized via the fundamental automorphisms of $\cl_{p,q}$. In essence,
the group $\pin(p,q)$ is an intrinsic notion of $\cl_{p,q}$, since
in accordance with (\ref{e10}) $\pin(p,q)\subset\cl_{p,q}$. Let us show
that the D\c{a}browski group $\pin^{a,b,c}(p,q)$ is also completely defined
in the framework of the algebra $\cl_{p,q}$, that is, there is an equivalence
between $\pin^{a,b,c}(p,q)$ and the group $\pin(p,q)\subset\cl_{p,q}$
complemented by the transformations $\cA\rightarrow\cA^\star,\,\cA\rightarrow
\widetilde{\cA},\,\cA\rightarrow\widetilde{\cA^\star}$ (in connection with
this it should be noted that the Gauss--Klein group $\dZ_2\otimes\dZ_2$
is a finite group
corresponded to the algebra $\cl_{1,0}=\Om$ \cite{Sal81a,Sal84}).
\begin{prop}\label{prop1}
Let $\cl_{p,q}$ ($p+q=2m$) be a Clifford algebra over the field $\F=\R$ and
let $\pin(p,q)$ be a double covering of the orthogonal group $O(p,q)=O_0(p,q)
\odot\{1,P,T,PT\}\simeq O_0(p,q)\odot(\dZ_2\otimes\dZ_2)$ of transformations
of the space $\R^{p,q}$, where $\{1,P,T,PT\}\simeq\dZ_2\otimes\dZ_2$ is a
group of discrete transformations of $\R^{p,q}$, $\dZ_2\otimes\dZ_2$ is the
Gauss--Klein group. Then there is an isomorphism between the group
$\{1,P,T,PT\}$ and an automorphism group $\{\Id,\star,\widetilde{\phantom{cc}},
\widetilde{\star}\}$ of the algebra $\cl_{p,q}$. In this case, space
inversion $P$, time reversal $T$ and combination $PT$ are correspond 
respectively to the fundamental automorphisms $\cA\rightarrow\cA^\star,\,
\cA\rightarrow\widetilde{\cA}$ and $\cA\rightarrow\widetilde{\cA^\star}$.
\end{prop} 
\begin{proof} As known, the transformations $1,P,T,PT$ at the conditions
$P^2=T^2=(PT)^2=1,\;PT=TP$ form an Abelian group with the following
multiplication table
\begin{center}{\renewcommand{\arraystretch}{1.4}
\begin{tabular}{|c||c|c|c|c|}\hline
    & $1$ & $P$ & $T$ & $PT$\\ \hline\hline
$1$ & $1$ & $P$ & $T$ & $PT$\\ \hline
$P$ & $P$ & $1$ & $PT$& $T$\\ \hline
$T$ & $T$ & $PT$& $1$ & $P$\\ \hline
$PT$& $PT$& $T$ & $P$ & $1$\\ \hline
\end{tabular}
}
\end{center}
Analogously, for the automorphism group $\{\Id,\star,\widetilde{\phantom{cc}},
\widetilde{\star}\}$ in virtue of the commutativity $\widetilde{(\cA^\star)}=
(\widetilde{\cA})^\star$ and the 
involution conditions $(\star)^2=(\widetilde{\phantom{cc}
})^2=\Id$ a following multiplication table takes place
\begin{center}{\renewcommand{\arraystretch}{1.4}
\begin{tabular}{|c||c|c|c|c|}\hline
        & $\Id$ & $\star$ & $\widetilde{\phantom{cc}}$ & $\widetilde{\star}$\\ \hline\hline
$\Id$   & $\Id$ & $\star$ & $\widetilde{\phantom{cc}}$ & $\widetilde{\star}$\\ \hline
$\star$ & $\star$ & $\Id$ & $\widetilde{\star}$ & $\widetilde{\phantom{cc}}$\\ \hline
$\widetilde{\phantom{cc}}$ & $\widetilde{\phantom{cc}}$ &$\widetilde{\star}$
& $\Id$ & $\star$ \\ \hline
$\widetilde{\star}$ & $\widetilde{\star}$ & $\widetilde{\phantom{cc}}$ &
$\star$ & $\Id$\\ \hline
\end{tabular}
}
\end{center}
The identity of the multiplication tables proves the isomorphism of the
groups $\{1,P,T,PT\}$ and $\{\Id,\star,\widetilde{\phantom{cc}},\widetilde{
\star}\}$.
\end{proof}
Further, in the case of anticommutativity $PT=-TP$ and $P^2=T^2=(PT)^2=\pm 1$
an isomorphism between the group $\{1,P,T,PT\}$ and an automorphism group
$\{\sI,\sW,\sE,\sC\}$, where $\sW,\sE$ and $\sC$ in accordance with 
(\ref{auto17}),
(\ref{20}) and (\ref{auto19}) are the matrix representations of the
automorphisms $\cA\rightarrow\cA^\star,\,\cA\rightarrow\widetilde{\cA}$
and $\cA\rightarrow\widetilde{\cA^\star}$, is analogously proved.\\[0.4cm]
{\bf Example}. According to (\ref{e20}), (\ref{e21}), (\ref{e22}) and
(\ref{be23}) the matrix representation of the fundamental automorphisms
of the Dirac algebra $\C_4$ is defined by the following expressions:
$\sW=\gamma_1\gamma_2\gamma_3\gamma_4,\,\sE=\gamma_1\gamma_3,\,
\sC=\gamma_2\gamma_4$.
The multiplication table of the group $\{\sI,\sW,\sE,\sC\}\sim
\{\sI,\gamma_1\gamma_2
\gamma_3\gamma_4,\gamma_1\gamma_3,\gamma_2\gamma_4\}$ has a form 
\begin{equation}\label{e30}{\renewcommand{\arraystretch}{1.4}
\begin{tabular}{|c||c|c|c|c|}\hline
    & $\sI$ & $\gamma_{1234}$ & $\gamma_{13}$ &
$\gamma_{24}$ \\ \hline\hline
$\sI$ & $I$ & $\gamma_{1234}$ & $\gamma_{13}$ &
$\gamma_{24}$ \\ \hline
$\gamma_{1234}$ & $\gamma_{1234}$ &
$\sI$ & $\gamma_{24}$ & $\gamma_{13}$\\ \hline
$\gamma_{13}$ & $\gamma_{13}$ & $\gamma_{24}$ & $-\sI$ &
$-\gamma_{1234}$\\ \hline
$\gamma_{24}$ & $\gamma_{24}$ & $\gamma_{13}$ & 
$-\gamma_{1234}$ & $-\sI$ \\ \hline
\end{tabular}
}
\;\;\sim\;\;{\renewcommand{\arraystretch}{1.4}
\begin{tabular}{|c||c|c|c|c|}\hline  
    & $\sI$ & $\sW$ & $\sE$ & $\sC$\\ \hline\hline
$\sI$ & $\sI$ & $\sW$ & $\sE$ & $\sC$\\ \hline
$\sW$ & $\sW$ & $\sI$ & $\sC$ & $\sE$\\ \hline
$\sE$ & $\sE$ & $\sC$ & $-\sI$& $-\sW$\\ \hline
$\sC$ & $\sC$ & $\sE$ & $-\sW$& $-\sI$\\ \hline
\end{tabular}.
}
\end{equation}
However, in this representation we cannot directly to identify 
$\sW=\gamma_1\gamma_2\gamma_3\gamma_4$ with the space inversion $P$, since
in this case the Dirac equation $(i\gamma_4\frac{\partial}{\partial x_4}-
i\boldsymbol{\gamma}\frac{\partial}{\partial\bx}-m)\psi(x_4,\bx)=0$ to be not
invariant with respect to $P$. On the other hand, for the canonical basis
(\ref{e20}) there exists a standard representation $P=\gamma_4,\,T=\gamma_1
\gamma_3$ \cite{BLP89}. The multiplication table of a group $\{1,P,T,PT\}\sim
\{I,\gamma_4,\gamma_2\gamma_3,\gamma_4\gamma_1\gamma_3\}$ has a form 
\begin{equation}\label{e31}{\renewcommand{\arraystretch}{1.4}
\begin{tabular}{|c||c|c|c|c|}\hline
   & $\sI$ & $\gamma_4$ & $\gamma_{13}$ & $\gamma_{413}$ 
\\ \hline\hline
$\sI$ & $\sI$& $\gamma_4$ & $\gamma_{13}$ & $\gamma_{413}$ 
\\ \hline
$\gamma_4$ & $\gamma_4$ & $\sI$ & $\gamma_{413}$ & $\gamma_{13}$
\\ \hline
$\gamma_{13}$ & $\gamma_{13}$ & $\gamma_{413}$ &
$-\sI$ & $-\gamma_4$ \\ \hline
$\gamma_{413}$ & $\gamma_{413}$ & $\gamma_{13}$
& $-\gamma_4$ & $-\sI$ \\ \hline
\end{tabular}
}
\;\;\sim\;\;{\renewcommand{\arraystretch}{1.4}
\begin{tabular}{|c||c|c|c|c|}\hline
    & $1$ & $P$ & $T$ & $PT$\\ \hline\hline
$1$ & $1$ & $P$ & $T$ & $PT$\\ \hline
$P$ & $P$ & $1$ & $PT$& $T$\\ \hline
$T$ & $T$ & $PT$& $-1$& $-P$\\ \hline
$PT$& $PT$& $T$ & $-P$& $-1$\\ \hline
\end{tabular}.
}
\end{equation}
It is easy to see that the tables (\ref{e30}) and (\ref{e31}) are equivalent,
therefore, we have an isomorphism $\{\sI,\sW,\sE,\sC\}\simeq\{1,P,T,PT\}$. Besides,
each of these groups is isomorphic to the group $\dZ_4$.
\begin{theorem}[{\rm\cite{Var99}}]\label{taut}
Let $\bsA=\{\sI,\,\sW,\,\sE,\,\sC\}$ be the automorphism group of the algebra
$\C_{p+q}$ $(p+q=2m)$, where 
$\sW=\cE_1\cE_2\cdots\cE_m\cE_{m+1}\cE_{m+2}\cdots\cE_{p+q}$,
and $\sE=\cE_1\cE_2\cdots\cE_m$, $\sC=\cE_{m+1}\cE_{m+2}\cdots\cE_{p+q}$ if
$m\equiv 1\pmod{2}$, and $\sE=\cE_{m+1}\cE_{m+2}\cdots\cE_{p+q}$, 
$\sC=\cE_1\cE_2\cdots
\cE_m$ if $m\equiv 0\pmod{2}$. Let $\bsA_-$ and $\bsA_+$ be the automorphism 
groups, in which the all elements correspondingly commute
$(m\equiv 0\pmod{2})$ and anticommute $(m\equiv 1\pmod{2})$.
Then over the field $\F=\C$ there are only two non--isomorphic groups:
$\bA_-\simeq\dZ_2\otimes\dZ_2$ for the signature $(+,\,+,\,+)$ if
$n\equiv 0,1\pmod{4}$ and
$\bA_+\simeq Q_4/\dZ_2$ for the signature
 $(-,\,-,\,-)$ if
$n\equiv 2,3\pmod{4}$.
\end{theorem}
\begin{proof} 1) The first complex type $n\equiv 0\pmod{2}$.\\
Over the field $\F=\C$ we can always to suppose $\sW^2=1$. Further, let us
find now the matrix $\sE$ of the antiautomorphism $\cA\rightarrow
\widetilde{\cA}$ at any $n=2m$, and elucidate the conditions at which the
matrix $\sE$ commutes with $\sW$, and also define a square of the matrix
$\sE$. Follows to \cite{Rash} let us introduce along with the algebra $\C_{p+q}$
an auxiliary algebra $\C_m$ with basis elements
\[
1,\;\varepsilon_\alpha,\;\varepsilon_{\alpha_1\alpha_2}\;(\alpha_1<\alpha_2),\;
\varepsilon_{\alpha_1\alpha_2\alpha_3}\;(\alpha_1<\alpha_2<\alpha_3),\;
\ldots\;\varepsilon_{12\ldots m}.
\]
In so doing, linear operators $\hat{\cE}_i$ acting in the space
$\C^m$ associated with the algebra $\C_m$, are defined by a following rule
\begin{eqnarray}
&&\hat{\cE}_j\phantom{m+}\;:\;\Lambda\longrightarrow\Lambda\varepsilon_j,\nonumber\\
&&\hat{\cE}_{m+j}\;:\;\Lambda^1\longrightarrow-i\varepsilon_j\Lambda^1,\;
\Lambda^0\longrightarrow i\varepsilon_j\Lambda^0,\label{e32}
\end{eqnarray}
where $\Lambda$ is a general element of the auxiliary algebra $\C_m$,
 $\Lambda^1$ and
$\Lambda^0$ are correspondingly odd and even parts of $\Lambda$,
$\varepsilon_j$ are units of the auxiliary algebra, $j=1,2,\ldots m$. 
Analogously,
in the case of matrix representations of $\cl_{p,q}$ we have
\begin{eqnarray}
&&\hat{\cE}_j\phantom{m+}\;:\;\Lambda\longrightarrow\Lambda\beta_j\varepsilon_j,
\nonumber \\
&&\hat{\cE}_{m+j}\;:\;\Lambda^1\longrightarrow-\varepsilon_j\beta_{m+j}
\Lambda^1,\;
\Lambda^0\longrightarrow\varepsilon_j\beta_{m+j}\Lambda^0,\label{e33}
\end{eqnarray}
where $\beta_i$ are arbitrary complex numbers.
It is easy to verify that transposition of the matrices of so defined
operators gives
\begin{equation}\label{e34}
\cE^T_j=\cE_j,\quad \cE^T_{m+j}=-\cE_{m+j}.
\end{equation}
Further, for the antiautomorphism $\cA\rightarrow\widetilde{\cA}:\;
\widetilde{\sA}=
\sE\sA^T\sE^{-1}$, since in this case $\e_i\rightarrow\e_i$, it is sufficient
to select the matrix $\sE$ so that
\[
\sE\cE^T_i\sE^{-1}=\cE_i,
\]
or taking into account (\ref{e34})
\begin{equation}\label{e35}
\sE\cE^T_j\sE^{-1}=\cE_j,\quad \sE\cE^T_{m+j}\sE^{-1}=-\cE_{m+j}. 
\end{equation}
Therefore, if $m$ is odd, then the matrix $\sE$ has a form
\begin{equation}\label{e36}
\sE=\cE_1\cE_2\ldots \cE_m,
\end{equation}
since in this case a product $\cE_1\cE_2\ldots \cE_m$ commutes
with all elements $\cE_j\;(j=1,\ldots m)$ and anticommutes with all
elements $\cE_{m+j}$.
Analogously, if $m$ is even, then
\begin{equation}\label{e37}
\sE=\cE_{m+1}\cE_{m+2}\ldots \cE_{p+q}.
\end{equation}
As required according to (\ref{e35}) in this case a product (\ref{e37}) 
commutes with $\cE_j$ and anticommutes with
$\cE_{m+j}$.

Let us consider now the conditions at which the matrix $\sE$ commutes or
anticommutes with $\sW$. Let $\sE=\cE_1\cE_2\ldots \cE_m$, where $m$ is odd, 
since
$\sW=\cE_1\ldots \cE_m\cE_{m+1}\ldots \cE_{p+q}$, then
\begin{eqnarray}
\cE_1\ldots \cE_m\cE_1\ldots \cE_m\cE_{m+1}\ldots \cE_{p+q}
&=&(-1)^{\frac{m(m-1)}{2}}
\sigma_1\sigma_2\ldots\sigma_m\cE_{m+1}\ldots \cE_{p+q},\nonumber\\
\cE_1\ldots \cE_m\cE_{m+1}\ldots \cE_{p+q}\cE_1\ldots \cE_m&=&
(-1)^{\frac{m(3m-1)}{2}}
\sigma_1\sigma_2\ldots\sigma_m\cE_{m+1}\ldots \cE_{p+q},\nonumber
\end{eqnarray}
where $\sigma_i$ are the functions of the form (\ref{e2}). It is easy to see
that in this case the elements 
$\sW$ and $\sE$ are always anticommute. Indeed, a comparison
$\frac{m(3m-1)}{2}\equiv\frac{m(m-1)}{2}\pmod{2}$ is equivalent to
$m^2\equiv 0,1\pmod{2}$, and since $m$ is odd, then we have always
$m^2\equiv 1\pmod{2}$. At $m$ is even and $\sE=\cE_{m+1}\ldots 
\cE_{p+q}$ it is easy to see that the matrices $\sW$ and $\sE$ 
are always commute $(m\equiv 0\pmod{2})$.
It is obvious that
over the field $\F=\C$ we can suppose $\cE^2_{1\ldots m}=\cE^2_{m+1\ldots
p+q}=\sI$.

Let us find now the matrix $\sC$ of the antiautomorphism $\cA\rightarrow
\widetilde{\cA^\star}$: $\widetilde{\sA^\star}=\sC\sA^T\sC^{-1}$. Since
in this case $\e_i\rightarrow-\e_i$, then it is sufficient to select the
matrix $\sC$ so that
\[
\sC\cE^T_i\sC^{-1}=-\cE_i,
\]
or taking into account (\ref{e34})
\begin{equation}\label{e38}
\sC\cE^T_j\sC^{-1}=-\cE_j,\quad \sC\cE^T_{m+j}\sC^{-1}=\cE_{m+j},
\end{equation}
where $j=1,\ldots m$.
In comparison with (\ref{e35}) it is easy to see that in
(\ref{e38}) the matrices $\cE_j$ and $\cE_{m+j}$ are changed by the roles.
Therefore, if $m$ is odd, then
\begin{equation}\label{e39}
\sC=\cE_{m+1}\cE_{m+2}\ldots \cE_{p+q},
\end{equation}
and if $m$ is even, then
\begin{equation}\label{e40}
\sC=\cE_1\cE_2\ldots \cE_m.
\end{equation}  
Permutation conditions of the matrices $\sC$ and $\sW$ are analogous to the
permutation conditions of $\sE$ with $\sW$, that is, the matrix $\sC$ of the form
(\ref{e39})
always anticommutes with $\sW$ ($m\equiv 1\pmod{2}$), and the matrix $\sC$ of the
form
(\ref{e40}) always commutes with $\sW$ ($m\equiv 0\pmod{2}$). 
Obviously, over the field $\F=\C$ we can suppose $\sC^2=\sI$.

Finally, let us find permutation conditions of the matrices $\sE$ and $\sC$.
First of all, at $m$ is odd $\sE=\cE_1\ldots \cE_m$,
$\sC=\cE_{m+1}\ldots \cE_{p+q}$, alternatively, at $m$ is even 
$\sE=\cE_{m+1}\ldots
\cE_{p+q}$, $\sC=\cE_1\ldots \cE_m$. Therefore,
\[
\cE_1\ldots \cE_m\cE_{m+1}\ldots \cE_{p+q}=(-1)^{m^2}\cE_{m+1}\ldots \cE_{p+q}
\cE_1\ldots
\cE_m,
\]
that is, the matrices $\sE$ and $\sC$ commute at $m\equiv 0\pmod{2}$ and anticommute
at $m\equiv 1\pmod{2}$.

Now we have all the necessary conditions for the definition and
classification of isomorphisms between finite groups and automorphism groups
of Clifford algebras. 
Over the field $\F=\C$ we can suppose 
$\sW^2=\sE^2=\sC^2=\sI$. At $m\equiv 0\pmod{2}$ we have only one signature
$(+,\,+,\,+)$ and an isomorphism $\bsA_-\simeq
\dZ_2\otimes\dZ_2$ if $p+q\equiv 0,4\pmod{8}$, since over the field $\C$ the
signatures
$(+,\,-,\,-),\;(-,\,+,\,-)$ and $(-,\,-\,-)$ are isomorphic to $(+,\,+,\,+)$.
Correspondingly, at $m\equiv 1\pmod{2}$ we have an isomorphism
$\bsA_+\simeq Q_4/\dZ_2$ for the signature $(-,\,-,\,-)$ if
$p+q\equiv 2,6\pmod{8}$. It should be noted that the signatures
$(+,\,+,\,+)$ and
$(-,\,-,\,-)$ are non--isomorphic, since there exists no a group $\bsA$ with the
signature $(+,\,+,\,+)$ in which all the elements anticommute, and also
there exists no a group $\bsA$ with $(-,\,-,\,-)$ in which all the elements
commute. Thus, over the field $\C$ we have only two
non--isomorphic automorphism groups: $\bsA_-\simeq\dZ_2\otimes\dZ_2,\;
\bsA_+\simeq Q_4/\dZ_2$.\\[0.2cm]
2) The second complex type $n\equiv 1\pmod{2}$.\\
For the odd--dimensional complex Clifford algebra $\C_n$
($n\equiv 1\pmod{2}$) there exists a decomposition into a direct sum of the
two even--dimensional subalgebras \cite{Rash}:
$\C_n\simeq\C^+_n\oplus\C^+_n\simeq\C_{n-1}\oplus\C_{n-1}$.
If $n=2m+1$, then in the decompositions we have even--dimensional algebras
$\C_{2m}$ which, as known, admit the Abelian group
$\sAut_-(\C_{2m})\simeq\dZ_2\otimes\dZ_2$ at $m\equiv 0,2\pmod{4}$ and
non--Abelian group $\sAut_+(\C_{2m})\simeq Q_4/\dZ_2$ at
$m\equiv 1,3\pmod{4}$. Therefore, the odd--dimensional subalgebras
$\C_n$ admit the signature $(+,\,+,\,+)$ at $n\equiv 1\pmod{4}$ and the
signature $(-,\,-,\,-)$ at $n\equiv 3\pmod{4}$.
\end{proof}

The following Theorem is a direct consequence of the previous Theorem.
Here we establish a relation between signatures $(a,b,c)$ of the
D\c{a}browski groups and dimensionality $n$ of the complex space $\C^n$
associated with the algebra $\C_n$.
\begin{theorem}[{\rm\cite{Var99}}]\label{t10}
Let $\pin^{a,b,c}(p,q)$ be a double covering of the complex orthogonal group
$O(n,\C)$ of the space $\C^n$ associated with the complex algebra
$\C_n$.
Squares of the
symbols $a,b,c\in
\{-,+\}$ are correspond to the squares of the elements of the finite group 
$\bsA=\{\sI,\sW,\sE,\sC\}:\;a=\sW^2,\,b=\sE^2,\,c=\sC^2$, where $\sW,\sE$
 and $\sC$
are correspondingly the matrices of the fundamental automorphisms $\cA\rightarrow
\cA^\star,\,\cA\rightarrow\widetilde{\cA}$ and $\cA\rightarrow
\widetilde{\cA^\star}$ of $\C_{n}$. Then over the field
$\F=\C$ for the algebra $\C_n$ there exist 
two non--isomorphic double coverings of the group
$O(n,\C)$:\\
1) Non--Cliffordian groups
\[
\pin^{+,+,+}(n,\C)\simeq\frac{(\spin_0(n,\C)\odot\dZ_2\otimes\dZ_2\otimes\dZ_2)}
{\dZ_2},
\]
if $n\equiv 0\pmod{4}$ and
\[
\pin^{+,+,+}(n,\C)\simeq\pin^{+,+,+}(n-1,\C)\bigcup\e_{12\ldots n}
\pin^{+,+,+}(n-1,\C),
\]
if $n\equiv 1\pmod{4}$.\\
2) Cliffordian groups
\[
\pin^{-,-,-}(n,\C)\simeq\frac{(\spin_0(n,\C)\odot Q_4)}{\dZ_2},
\]
if $n\equiv 2\pmod{4}$ and
\[
\pin^{-,-,-}(n,\C)\simeq\pin^{-,-,-}(n-1,\C)\bigcup\e_{12\ldots n}
\pin^{-,-,-}(n-1,\C),
\]
if $n\equiv 3\pmod{4}$.
\end{theorem}

\section{Discrete symmetries over the field $\F=\R$}
A consideration of the discrete symmetries over the field of real numbers
is a much more complicated problem. First of all, in contrast to the field
of complex numbers over the field $\F=\R$ there exist eight different types
of the Clifford algebras and five division rings, which, in virtue of the
Wedderburn--Artin Theorem, impose hard restrictions on existence and choice
of the matrix representations for the fundamental automorphisms.
\begin{theorem}\label{tautr}
Let $\cl_{p,q}$ be a Clifford algebra over a field $\F=\R$ and let
$\sAut(\cl_{p,q})=\{\sI,\sW,\sE,\sC\}$ be a group of fundamental
automorphisms of the algebra $\cl_{p,q}$. Then for eight types of the 
algebras $\cl_{p,q}$ there exist, depending upon a division ring structure
of $\cl_{p,q}$, following isomorphisms between finite groups and groups
$\sAut(\cl_{p,q})$ with different signatures
$(a,b,c)$, where $a,b,c\in\{-,+\}$:\\[0.2cm]
1) $\K\simeq\R$, types $p-q\equiv 0,2\pmod{8}$.\\
If $\sE=\cE_{p+1}\cE_{p+2}\cdots\cE_{p+q}$ and $\sC=\cE_1\cE_2\cdots\cE_p$,
then Abelian groups $\sAut_-(\cl_{p,q})\simeq\dZ_2\otimes\dZ_2$
with the signature $(+,+,+)$ and $\sAut_-(\cl_{p,q})\simeq\dZ_4$ with the
signature
$(+,-,-)$ exist at $p,q\equiv 0\pmod{4}$ and $p,q\equiv 2\pmod{4}$, 
respectively,
for the type $p-q\equiv 0\pmod{8}$, and also Abelian groups
$\sAut_-(\cl_{p,q})\simeq\dZ_4$ with the signature $(-,-,+)$ and
$\sAut_-(\cl_{p,q})
\simeq\dZ_4$ with the signature $(-,+,-)$ exist 
at $p\equiv 0\pmod{4},\,
q\equiv 2\pmod{4}$ and $p\equiv 2\pmod{4},\,q\equiv 0\pmod{4}$ for the type
$p-q\equiv 2\pmod{8}$, respectively.\\
If $\sE=\cE_1\cE_2\cdots\cE_p$ and $\sC=\cE_{p+1}\cE_{p+2}\cdots\cE_{p+q}$,
then non--Abelian groups $\sAut_+(\cl_{p,q})\simeq D_4/\dZ_2$ with the
signature $(+,-,+)$ and $\sAut_+(\cl_{p,q})\simeq D_4/\dZ_2$ with the
signature
$(+,+,-)$ exist at $p,q\equiv 3\pmod{4}$ and $p,q\equiv 1\pmod{4}$, 
respectively,
for the type $p-q\equiv 0\pmod{8}$, and also non--Abelian groups
$\sAut_+(\cl_{p,q})\simeq Q_4/\dZ_2$ with $(-,-,-)$ and 
$\sAut_+(\cl_{p,q})\simeq
D_4/\dZ_2$ with $(-,+,+)$ exist at $p\equiv 3\pmod{4},\,q\equiv 1
\pmod{4}$ and $p\equiv 1\pmod{4},\,q\equiv 3\pmod{4}$ for the type
$p-q\equiv 2\pmod{8}$, respectively.\\[0.2cm]
2) $\K\simeq\BH$, types $p-q\equiv 4,6\pmod{8}$.\\
If $\sE=\cE_{j_1}\cE_{j_2}\cdots\cE_{j_k}$ is a product of $k$
skewsymmetric matrices (among which $l$ matrices have a square $+\sI$
and $t$ matrices have a square $-\sI$)
and $\sC=\cE_{i_1}\cE_{i_2}\cdots\cE_{i_{p+q-k}}$ is a product of $p+q-k$
symmetric matrices (among which $h$ matrices have a square $+\sI$ and
$g$ have a square $-\sI$),
then at $k\equiv 0\pmod{2}$ for the type $p-q\equiv 4\pmod{8}$ there exist
Abelian groups $\sAut_-(\cl_{p,q})\simeq\dZ_2\otimes\dZ_2$ with $(+,+,+)$
and $\sAut_-(\cl_{p,q})\simeq\dZ_4$ with $(+,-,-)$ if
$l-t,\,h-g\equiv 0,1,4,5\pmod{8}$ and
$l-t,\,h-g\equiv 2,3,6,7\pmod{8}$, respectively. And also at
$k\equiv 0\pmod{2}$ for the type $p-q\equiv 6\pmod{8}$ there exist
$\sAut_-(\cl_{p,q})\simeq\dZ_4$ with $(-,+,-)$ and 
$\sAut_-(\cl_{p,q})\simeq\dZ_4$
with $(-,-,+)$ if $l-t\equiv 0,1,4,5\pmod{8},\,
h-g\equiv 2,3,6,7\pmod{8}$ and $l-t\equiv 2,3,6,7\pmod{8},\,
h-g\equiv 0,1,4,5\pmod{8}$,respectively.\\
Inversely, if $\sE=\cE_{i_1}\cE_{i_2}\cdots\cE_{i_{p+q-k}}$ is a product of
$p+q-k$ symmetric matrices and 
$\sC=\cE_{j_1}\cE_{j_2}\cdots\cE_{j_k}$ is a product of $k$ skewsymmetric
matrices, then at $k\equiv 1\pmod{2}$
for the type $p-q\equiv 4\pmod{8}$ there exist non--Abelian groups
$\sAut_+(\cl_{p,q})
\simeq D_4/\dZ_2$ with $(+,-,+)$ and $\sAut_+(\cl_{p,q})\simeq D_4/\dZ_2$ with
$(+,+,-)$ if $h-g\equiv 2,3,6,7\pmod{8},\,l-t\equiv
0,1,4,5\pmod{8}$ and $h-g\equiv 0,1,4,5\pmod{8},\,l-t\equiv 2,3,6,7\pmod{8}$,
respectively.
And also at $k\equiv 1\pmod{2}$ for the type $p-q\equiv 6\pmod{8}$ there exist
$\sAut_+(\cl_{p,q})\simeq Q_4/\dZ_2$ with $(-,-,-)$ and $\sAut_+(\cl_{p,q})
\simeq D_4/\dZ_2$ with $(-,+,+)$ if $h-g,
\,l-t\equiv 2,3,6,7\pmod{8}$ and $h-g,\,l-t\equiv 0,1,4,5
\pmod{8}$, respectively.\\[0.2cm]
3) $\K\simeq\R\oplus\R,\,\K\simeq\BH\oplus\BH$, types $p-q\equiv 1,5\pmod{8}$.\\
For the algebras $\cl_{0,q}$ of the types $p-q\equiv 1,5\pmod{8}$ there exist
Abelian automorphism groups with the signatures
$(-,-,+)$, $(-,+,-)$ and non--Abelian automorphism groups with the signatures
$(-,-,-)$, $(-,+,+)$. Correspondingly, for the algebras $\cl_{p,0}$ of the
types $p-q\equiv 1,5\pmod{8}$ there exist Abelian groups with
$(+,+,+)$, $(+,-,-)$ and non--Abelian groups with $(+,-,+)$,
$(+,+,-)$. In general case for $\cl_{p,q}$, the types $p-q\equiv 1,5\pmod{8}$
admit all eight automorphism groups.\\[0.2cm]
4) $\K=\C$, types $p-q\equiv 3,7\pmod{8}$.\\
The types $p-q\equiv 3,7\pmod{8}$ admit the Abelian group $\sAut_-(\cl_{p,q})
\simeq\dZ_2\otimes\dZ_2$ with the signature $(+,+,+)$ if $p\equiv 0\pmod{2}$ and
$q\equiv 1\pmod{2}$, and also non--Abelian group 
$\sAut_+(\cl_{p,q})\simeq
Q_4/\dZ_2$ with the signature $(-,-,-)$ if $p\equiv 1\pmod{2}$ and
$q\equiv 0\pmod{2}$. 
\end{theorem} 
\begin{proof} Before we proceed to prove this Theorem, let us consider in
more details a matrix (spinor) representations of the antiautomorphisms
$\cA\rightarrow\widetilde{\cA}$ and $\cA\rightarrow\widetilde{\cA^{\star}}$.
According to Wedderburn--Artin Theorem
the antiautomorphism $\cA\rightarrow\widetilde{\cA}$ corresponds to
an antiautomorphism of the full matrix algebra $\M_{2^m}(\K)$:
$\sA\rightarrow\sA^{\sT}$, in virtue of the well--known relation 
$(\sA\sB)^{\sT}=
\sB^{\sT}\sA^{\sT}$, where $\sT$ is a symbol of transposition. On the other hand,
in the matrix representation of the elements $\cA\in\cl_{p,q}$, for the
antiautomorphism
$\cA\rightarrow\widetilde{\cA}$ we have $\sA\rightarrow\widetilde{\sA}$.
A composition of the two antiautomorphisms, $\sA^{\sT}\rightarrow\sA\rightarrow
\widetilde{\sA}$, gives an automorphism  $\sA^{\sT}\rightarrow\widetilde{\sA}$,
which is an internal automorphism of the algebra $\M_{2^m}(\K)$:
\begin{equation}\label{com}
\widetilde{\sA}=\sE\sA^{\sT}\sE^{-1},
\end{equation}
where $\sE$ is a matrix, by means of which the antiautomorphism $\cA\rightarrow
\widetilde{\cA}$ is expressed in the matrix representation of the
algebra $\cl_{p,q}$.
Under action of the antiautomorphism $\cA\rightarrow\widetilde{\cA}$
the units of $\cl_{p,q}$ remain unaltered, $\e_i\rightarrow\e_i$; therefore
in the matrix representation, we must demand $\cE_i\rightarrow\cE_i$,
where $\cE_i=\gamma(\e_i)$ also.
Therefore, for the definition of the matrix $\sE$
in accordance with (\ref{com}) we have
\begin{equation}\label{com1}
\cE_i\longrightarrow\cE_i=\sE\cE^{\sT}\sE^{-1}.
\end{equation}
Or, let $\{\cE_{\alpha_i}\}$ be a set consisting of symmetric matrices
($\cE^{\sT}_{\alpha_i}=\cE_{\alpha_i}$) and let $\{\cE_{\beta_j}\}$ be a set
consisting of skewsymmetric matrices ($\cE^{\sT}_{\beta_j}=-\cE_{\beta_j}$).
Then the transformation (\ref{com1}) may be rewritten in the following form:
\[
\cE_{\alpha_i}\longrightarrow\cE_{\alpha_i}=\sE\cE_{\alpha_i}\sE^{-1},\quad
\cE_{\beta_j}\longrightarrow\cE_{\beta_j}=-\sE\cE_{\beta_j}\sE^{-1}.
\]
Whence
\begin{equation}\label{commut}
\cE_{\alpha_i}\sE=\sE\cE_{\alpha_i},\quad \cE_{\beta_j}\sE=-\sE\cE_{\beta_j}.
\end{equation}
Thus, the matrix $\sE$ of the antiautomorphism $\cA\rightarrow\widetilde{\cA}$
commutes with a symmetric part of the spinbasis of the algebra $\cl_{p,q}$ and
anticommutes with a skewsymmetric part. An explicit form of the matrix $\sE$
in dependence on the type of the algebras $\cl_{p,q}$ will be found later, but
first let us define a general form of $\sE$, that is, let us show that for
the form of $\sE$ there are only two possibilities: 1) $\sE$ is a product of
symmetric matrices or 2) $\sE$ is a product of skewsymmetric matrices.
Let us prove this assertion another way: Let
$\sE=\cE_{\alpha_1}\cE_{\alpha_2}\cdots\cE_{\alpha_s}\cE_{\beta_1}\cE_{\beta_2}
\cdots\cE_{\beta_k}$ be a product of $s$ symmetric and $k$
skewsymmetric matrices. At this point $1< s+k\leq p+q$. The permutation
condition of the matrix $\sE$ with the symmetric basis matrices
$\cE_{\alpha_i}$ have a form
\begin{eqnarray}
\cE_{\alpha_i}\sE&=&
(-1)^{i-1}\sigma(\alpha_i)\cE_{\alpha_1}\cdots\cE_{\alpha_{i-1}}
\cE_{\alpha_{i+1}}\cdots\cE_{\alpha_s}\cE_{\beta_1}\cdots
\cE_{\beta_k},\nonumber\\
\sE\cE_{\alpha_i}&=&
(-1)^{k+s-i}\sigma(\alpha_i)\cE_{\alpha_1}\cdots\cE_{\alpha_{i-1}}
\cE_{\alpha_{i+1}}\cdots\cE_{\alpha_s}\cE_{\beta_1}\cdots
\cE_{\beta_k}.\label{commut1}
\end{eqnarray}
From here we obtain a comparison $k+s-i\equiv i-1\pmod{2}$, that is, at
$k+s\equiv 0\pmod{2}$ $\sE$ and $\cE_{\alpha_i}$ anticommute and at
$k+s\equiv 1\pmod{2}$ commute. Analogously, for the skewsymmetric part
we have
\begin{eqnarray}
\cE_{\beta_j}\sE&=&
(-1)^{s+j-1}\sigma(\beta_j)\cE_{\alpha_1}\cdots\cE_{\alpha_s}
\cE_{\beta_1}\cdots\cE_{\beta_{j-1}}\cE_{\beta_{j+1}}\cdots
\cE_{\beta_k},\nonumber\\
\sE\cE_{\beta_j}&=&
(-1)^{k-j}\sigma(\beta_i)\cE_{\alpha_1}\cdots\cE_{\alpha_s}
\cE_{\beta_1}\cdots\cE_{\beta_{j-1}}\cE_{\beta_{j+1}}\cdots
\cE_{\beta_k}.\label{commut2}
\end{eqnarray}
From the comparison $k-s\equiv 2j-1\pmod{2}$, it
follows that at $k-s\equiv 0\pmod{2}$,
$\sE$ and $\cE_{\beta_j}$ anticommute and at  $k-s\equiv 1\pmod{2}$
commute. Let $k+s=p+q$, then from (\ref{commut1}) we see that
at $p+q\equiv 0\pmod{2}$ $\sE$ and $\cE_{\alpha_i}$ anticommute, which is
inconsistent with (\ref{commut}). The case $p+q\equiv 1\pmod{2}$ is excluded,
since a dimensionality of $\cl_{p,q}$ is even (in the case of odd
dimensionality the algebra $\cl_{p+1,q}\,(\cl_{p,q+1})$ is isomorphic to
$\End_{\K\oplus\hat{\K}}(I_{p,q}\oplus\hat{I}_{p,q})\simeq\M_{2^m}(\K)\oplus
\M_{2^m}(\K)$, where $m=\frac{p+q}{2}$). Let suppose now that
$k+s<p+q$, that is, let us eliminate from the product $\sE$ one symmetric matrix,
then $k+s\equiv 1\pmod{2}$ and in virtue of (\ref{commut1}) the
matrices $\cE_{\alpha_i}$
that belong to $\sE$ commute with $\sE$, but the matrix that does not
belong to $\sE$ anticommutes with $\sE$. Thus, we came to a contradiction
with (\ref{commut}). It is obvious that elimination of two, three,
or more symmetric matrices from $\sE$ gives an analogous situation.
Now, Let us eliminate from
$\sE$ one skewsymmetric matrix, then $k+s\equiv 1\pmod{2}$ and
in virtue of (\ref{commut1}) $\sE$ and all $\cE_{\alpha_i}$ commute with each
other. Further, in virtue of (\ref{commut2}) the matrices $\cE_{\beta_j}$
that belong to the product $\sE$ commute with $\sE$, whereas the
the skewsymmetric matrix that does not belong to $\sE$ anticommute
with $\sE$. Therefore we came again to a contradiction with
(\ref{commut}). We come to an analogous situation if we eliminate two, three, or
more skewsymmetric matrices. Thus, the product
$\sE$ does not contain simultaneously symmetric and skewsymmetric matrices.
Hence it follows that the matrix of the antiautomorphism
$\cA\rightarrow\widetilde{\cA}$ is a product of only symmetric or only
skewsymmetric matrices.

Further, the matrix representations of the antiautomorphism
$\cA\rightarrow\widetilde{\cA^\star}$: 
$\widetilde{\sA^\star}=\sC\sA^{\sT}\sC^{-1}$ is defined in a similar manner.
First of all, since under action of the antiautomorphism $\widetilde{\star}$
we have $\e_i\rightarrow -\e_i$, in the matrix representation we must
demand $\cE_i\rightarrow -\cE_i$ also, or
\begin{equation}\label{com2}
\cE_i\longrightarrow -\cE_i=\sC\cE^{\sT}\sC^{-1}.
\end{equation}
Taking into account the symmetric $\{\cE_{\alpha_i}\}$ and the
skewsymmetric $\{\cE_{\beta_j}\}$ parts of the spinbasis we can write
the transformation (\ref{com2}) in the form
\[
\cE_{\alpha_i}\longrightarrow -\cE_{\alpha_i}=\sC\cE_{\alpha_i}\sC^{-1},\quad
\cE_{\beta_j}\longrightarrow\cE_{\beta_j}=\sC\cE_{\beta_j}\sC^{-1}.
\]
Hence it follows
\begin{equation}\label{commut3}
\sC\cE_{\alpha_i}=-\cE_{\alpha_i}\sC,\quad
\cE_{\beta_j}\sC=\sC\cE_{\beta_j}.
\end{equation}
Thus, in contrast with (\ref{commut}) the matrix $\sC$ of the antiautomorphism
$\widetilde{\star}$ anticommutes with the symmetric part of the spinbasis
of the algebra $\cl_{p,q}$ and commutes with the skewsymmetric part of the
same spinbasis. Further, in virtue of (\ref{auto17}) a matrix representation
of the automorphism $\star$ is defined as follows
\begin{equation}\label{W}
\sA^\star=\sW\sA\sW^{-1},
\end{equation}
where $\sW$ is a matrix representation of the volume element $\omega$.
The antiautomorphism $\cA\rightarrow\widetilde{\cA^\star}$, in turn, is
the composition of the antiautomorphism $\cA\rightarrow\widetilde{\cA}$ with
the automorphism $\cA\rightarrow\cA^\star$; therefore, from (\ref{com}) and
(\ref{W}) it follows (recall that the order of the composition of the
transformations (\ref{com}) and (\ref{W}) is not important, since
$\widetilde{\cA^\star}=(\widetilde{\cA})^\star=\widetilde{(\cA^\star)}$):
$\widetilde{\sA^\star}=\sW\sE\sA^{\sT}\sE^{-1}\sW^{-1}=\sE(\sW\sA\sW^{-1})^{\sT}
\sE^{-1}$, or
\begin{equation}\label{C}
\widetilde{\sA^\star}=(\sW\sE)\sA^{\sT}(\sW\sE)^{-1}=
(\sE\sW)\sA^{\sT}(\sE\sW)^{-1},
\end{equation}
since $\sW^{-1}=\sW^{\sT}$. Therefore, $\sC=\sE\sW$ or $\sC=\sW\sE$.
By this reason a general form of the matrix $\sC$ is similar to the form
of the matrix $\sE$, that is, $\sC$ is a product of symmetric or skewsymmetric
matrices only.

Let us consider in sequence definitions and permutation conditions of matrices
of the fundamental automorphisms (which are the elements of the groups
$\sAut(\cl_{p,q})$) for all eight types of the algebras
$\cl_{p,q}$, depending upon the division ring structure.\\[0.3cm]
1) The type $p-q\equiv 0\pmod{8}$, $\K\simeq\R$.\\
In this case according to Wedderburn--Artin Theorem there is an isomorphism
$\cl_{p,q}\simeq\M_{2^m}(\R)$, where $m=\frac{p+q}{2}$. First, let consider
a case $p=q=m$. In the full matrix algebra
$\M_{2^m}(\R)$, in accordance with the signature of the algebra $\cl_{p,q}$ 
a choice of the symmetric and skewsymmetric matrices $\cE_i=\gamma(\e_i)$
is hardly fixed.
\begin{equation}\label{A1}
\cE^{\sT}_i=\left\{\begin{array}{rl}
\phantom{-}\cE_i, & \mbox{if $1\leq i\leq m$};\\
-\cE_i, & \mbox{if $m+1\leq i\leq 2m$},
\end{array}\right.
\end{equation}
That is, at this point the matrices of the first and second half of the basis
have a square $+\sI$ and $-\sI$, respectively.
Such a form of the basis (\ref{A1}) is explained by the following
reason: Over the field $\R$ there exist only symmetric matrices with a
square $+\sI$, and there exist no symmetric matrices with a square
$-\sI$. Inversely, skewsymmetric matrices over the field $\R$
only have the square $-\sI$.
Therefore, in this case the matrix of the antiautomorphism
$\cA\rightarrow\widetilde{\cA}$ is a product of $m$ symmetric matrices,
$\sE=\cE_1\cE_2\cdots\cE_m$, or is a product of $m$ skewsymmetric matrices,
$\sE=\cE_{m+1}\cE_{m+2}\cdots\cE_{2m}$. In accordance with (\ref{commut})
let us find permutation conditions of the matrix $\sE$ 
with the basis matrices $\cE_i$.
If $\sE=\cE_1\cE_2\cdots\cE_m$, and $\cE_i$ belong to the first half of the
basis (\ref{A1}), $1\leq i\leq m$, then
\begin{eqnarray}
\sE\cE_i&=&(-1)^{m-i}\cE_1\cE_2\cdots\cE_{i-1}\cE_{i+1}\cdots\cE_m,\nonumber\\
\cE_i\sE&=&(-1)^{i-1}\cE_1\cE_2\cdots\cE_{i-1}\cE_{i+1}\cdots\cE_m.\label{A2}
\end{eqnarray}
Therefore, we have a comparison $m-i\equiv i-1\pmod{2}$, whence
$m\equiv 2i-1\pmod{2}$. Thus, the matrix $\sE$ anticommutes
at $m\equiv 0\pmod{2}$ and commutes at $m\equiv 1\pmod{2}$
with the basis matrices $\cE_i$.
Further, let $\sE=\cE_1\cE_2\cdots\cE_m$, and  $\cE_i$ belong to the
second half of the basis $m+1\leq i\leq 2m$, then
\begin{equation}\label{A3}
\sE\cE_i=(-1)^m\cE_i\sE.
\end{equation}
Therefore at $m\equiv 0\pmod{2}$, $\sE$ commutes and at
$m\equiv 1\pmod{2}$ anticommutes with the matrices of the second half of the
basis.

Let now $\sE=\cE_{m+1}\cE_{m+2}\cdots\cE_{2m}$ be a product of
$m$ skewsymmetric matrices, then
\begin{equation}\label{A4}
\sE\cE_i=(-1)^m\cE_i\sE \quad (1\leq i\leq m)
\end{equation}
and
\begin{equation}\label{A4'}
\begin{array}{ccc}
\sE\cE_i&=&-(-1)^{m-i}\cE_{m+1}\cE_{m+2}\cdots\cE_{i-1}\cE_{i+1}\cdots\cE_{2m},\\
\cE_i\sE&=&-(-1)^{i-1}\cE_{m+1}\cE_{m+2}\cdots\cE_{i-1}\cE_{i+1}\cdots\cE_{2m},
\end{array}\quad (m+1\leq i\leq 2m)\end{equation}
that is, at $m\equiv 0\pmod{2}$ $\sE$ commutes with the matrices of the first
half of the basis (\ref{A1}) and anticommutes with the matrices of the second
half of (\ref{A1}).
At $m\equiv 1\pmod{2}$ $\sE$ anticommutes and commutes with the first and
the second half of the basis (\ref{A1}), respectively.

Let us find permutation conditions of the matrix $\sE$ with a matrix $\sW$
of the volume element (a matrix of the automorphism $\star$). Let
$\sE=\cE_1\cE_2\cdots\cE_m$, then
\begin{eqnarray}
\sE\sW&=&\cE_1\cE_2\cdots\cE_{m}\cE_1\cE_2\cdots\cE_{2m}
=(-1)^{\frac{m(m-1)}{2}}\cE_{m+1}\cE_{m+2}
\cdots\cE_{2m},\nonumber\\
\sW\sE&=&\cE_1\cE_2\cdots\cE_{2m}\cE_1\cE_2\cdots\cE_m=
(-1)^{\frac{m(3m-1)}{2}}\cE_{m+1}\cE_{m+2}\cdots\cE_{2m}.\label{A5}
\end{eqnarray}
Whence $\frac{m(3m-1)}{2}\equiv\frac{m(m-1)}{2}\pmod{2}$ and, therefore
at $m\equiv 0\pmod{2}$, $\sE$ and $\sW$ commute, and at $m\equiv 1\pmod{2}$
anticommute. It is easy to verify that analogous conditions take place if
$\sE=\cE_{m+1}\cE_{m+2}\cdots\cE_{2m}$ is the product of 
skewsymmetric matrices.

Since $\sC=\sE\sW$, then a matrix of the antiautomorphism
$\widetilde{\star}$ has a form $\sC=\cE_{m+1}\cE_{m+2}\cdots\cE_{2m}$ if
$\sE=\cE_1\cE_2\cdots\cE_m$ and correspondingly, $\sC=\cE_1\cE_2\cdots\cE_m$ if
$\sE=\cE_{m+1}\cE_{m+2}\cdots\cE_{2m}$. Therefore, permutation conditions
of the matrices $\sC$ and $\sW$ would be the same as that of
$\sE$ and $\sW$, that is, $\sC$ and $\sW$ commute if $m\equiv 0\pmod{2}$ and
anticommute if $m\equiv 1\pmod{2}$. It is easy to see that permutation
conditions of the matrix $\sC$ with the basis matrices $\cE_i$ are coincide
with (\ref{A2})--(\ref{A4'}).

Out of dependence on the choice of the matrices $\sE$ and $\sC$,
the permutation conditions between them in any of the two cases considered 
previously
are defined by the following relation
\begin{equation}\label{A6}
\sE\sC=(-1)^{m^2}\sC\sE,
\end{equation}
that is, the matrices $\sE$ and $\sC$ commute if $m\equiv 0\pmod{2}$ and
anticommute if $m\equiv 1\pmod{2}$.

Now, let us consider squares of the elements of the automorphism groups 
$\sAut(\cl_{p,q})$, $p-q\equiv 0\pmod{8}$, and $p=q=m$. For the matrices of
the automorphisms
$\widetilde{\phantom{cc}}$ and $\widetilde{\star}$ we have the following two
possibilities:\\
a) $\sE=\cE_1\cE_2\cdots\cE_m$, $\sC=\cE_{m+1}\cE_{m+2}\cdots\cE_{2m}$.
\begin{equation}\label{A7}
\sE^2=\left\{\begin{array}{rl}
+\sI, & \mbox{if $m\equiv 0,1\pmod{4}$},\\
-\sI, & \mbox{if $m\equiv 2,3\pmod{4}$};
\end{array}\right.\quad
\sC^2=\left\{\begin{array}{rl}
+\sI, & \mbox{if $m\equiv 0,3\pmod{4}$},\\
-\sI, & \mbox{if $m\equiv 1,2\pmod{4}$}.
\end{array}\right.
\end{equation}
b) $\sE=\cE_{m+1}\cE_{m+2}\cdots\cE_{2m}$, $\sC=\cE_1\cE_2\cdots\cE_m$.
\begin{equation}\label{A8}
\sE^2=\left\{\begin{array}{rl}
+\sI, & \mbox{if $m\equiv 0,3\pmod{4}$},\\
-\sI, & \mbox{if $m\equiv 1,2\pmod{4}$};
\end{array}\right.\quad
\sC^2=\left\{\begin{array}{rl}
+\sI, & \mbox{if $m\equiv 0,1\pmod{4}$},\\
-\sI, & \mbox{if $m\equiv 2,3\pmod{4}$}.
\end{array}\right.
\end{equation}
In virtue of (\ref{e3}), for the matrix of the automorphism $\star$ we have
always $\sW^2=+\sI$.

Now, we are in a position to define automorphism groups
for the type $p-q\equiv 0\pmod{8}$. First of all, let us consider Abelian groups.
In accordance with (\ref{A5}) and
(\ref{A6}), the automorphism group is Abelian if $m\equiv 0\pmod{2}$
($\sW,\sE$ and $\sC$ commute with each other). In virtue of (\ref{commut}) and
(\ref{A1}) the matrix $\sE$ should be commuted with the first (symmetric)
half and anticommuted with the second (skewsymmetric) half of the basis
(\ref{A1}). From (\ref{A2})--(\ref{A4'}) it is easy to see that this
condition is satisfied only if $\sE=\cE_{m+1}\cE_{m+2}\cdots\cE_{2m}$ and
$m\equiv 0\pmod{2}$. Correspondingly, in accordance with (\ref{commut3}) the
matrix $\sC$ should be anticommuted with the symmetric half of the basis
(\ref{A1}) and commuted with the skewsymmetric half of the same basis.
It is obvious that this condition is satisfied only if $\sC=\cE_1\cE_2\cdots
\cE_m$.
Therefore, when $m=p=q$ in accordance with
(\ref{A8}), there exists an Abelian group $\sAut_-(\cl_{p,q})\simeq
\dZ_2\otimes\dZ_2$ with the signature $(+,+,+)$ if $p,q\equiv 0\pmod{4}$,
and $\sAut_-(\cl_{p,q})\simeq\dZ_4$ 
with the signature $(+,-,-)$
if $p,q\equiv 2\pmod{4}$. Further, in accordance with (\ref{A5}) and (\ref{A6}),
the automorphism group is non--Abelian if $m\equiv 1\pmod{2}$.
In this case, from (\ref{A2})--(\ref{A4'}) it follows that the matrix $\sE$
commutes with the symmetric half and anticommutes with the skewsymmetric half
of the basis (\ref{A1}) if and only if $\sE=\cE_1\cE_2\cdots\cE_m$
is a product of $m$ symmetric matrices, $m\equiv 1\pmod{8}$. In its turn,
the matrix $\sC$ anticommutes with the symmetric half and commutes with the
skewsymmetric half of the basis (\ref{A1}) if and only if $\sC=\cE_{m+1}
\cE_{m+2}\cdots\cE_{2m}$.
Therefore,
in accordance with (\ref{A7}), there exist non--Abelian groups
$\sAut_+(\cl_{p,q})\simeq D_4/\dZ_2$
with the signature $(+,-,+)$ if $p,q\equiv 3\pmod{4}$, and 
$\sAut_+(\cl_{p,q})\simeq D_4/\dZ_2$ with the signature $(+,+,-)$ if
$p,q\equiv 1\pmod{4}$.

In addition to the previously considered case $p=q$, the type
$p-q\equiv 0\pmod{8}$ also admits two particular cases in relation with the
algebras 
$\cl_{p,0}$ and $\cl_{0,q}$. In these cases, a spinbasis is defined as follows
\[
\begin{array}{ccc}
\cE^{\sT}_i&=&\phantom{-}\cE_i\quad\mbox{for the algebras $\cl_{8t,0}$},\\
\cE^{\sT}_i&=&-\cE_i\quad\mbox{for the algebras $\cl_{0,8t}$}.
\end{array}\quad t=1,2,\ldots
\]
that is, a spinbasis of the algebra $\cl_{8t,0}$ consists of only symmetric
matrices, and that of $\cl_{0,8t}$ consists of only
skewsymmetric matrices. According to (\ref{commut}), for the algebra
$\cl_{p,0}$ the matrix $\sE$ should commute with all
$\cE_i$. It is obvious that we cannot take the matrix
$\sE$ of the form $\cE_1\cE_2\cdots\cE_s$, where $1<s<p$, since at
$s\equiv 0\pmod{2}$ $\sE$ and $\cE_i$ anticommute, which contradicts with
(\ref{commut}), and at $s\equiv 1\pmod{2}$ $\sE$ and $\cE_i$ that belong to
$\sE$ commute with each other, whereas $\cE_i$ that do not belong
to $\sE$
anticommute with $\sE$, which again conradicts with (\ref{commut}).
The case $s=p$ is also excluded, since $p$ is even. Therefore, only one
possibility remains, that is, the matrix $\sE$ is proportional to the
unit matrix, $\sE\sim\sI$. At this point, from (\ref{C}) it follows that
$\sC\sim\cE_1\cE_2\cdots\cE_p$ and we see that the conditions (\ref{commut3})
are satisfied.
Thus, the matrices
$\sE\sim\sI$, $\sC=\sE\sW$ and $\sW$ of the fundamental automorphisms
$\cA\rightarrow\widetilde{\cA},\,\cA\rightarrow\widetilde{\cA^\star}$ and
$\cA\rightarrow\cA^\star$ of the algebra $\cl_{p,0}$ ($p\equiv 0\pmod{8}$)
from an Abelian group $\sAut_-(\cl_{p,0})\simeq\dZ_2\otimes
\dZ_2$. Further, for the algebras $\cl_{0,8t}$, in accordance with (\ref{commut})
the matrix $\sE$ should anticommute with all $\cE_i$. It is easy to
see that we cannot take the matrix $\sE$ of the form $\cE_1\cE_2\cdots\cE_k$, 
where
$1<k<q$, since at $k\equiv 0\pmod{2}$ the matrix $\sE$ and the matrices $\cE_i$
that belong to $\sE$ anticommute with each other, whereas
$\cE_i$ that do not belong to $\sE$ commute with $\sE$, which
contradicts with
(\ref{commut}). Inversely, if $k\equiv 1\pmod{2}$, $\sE$ and $\cE_i$
that belong to $\sE$ commute, but $\sE$ and $\cE_i$ that do not 
belong to $\sE$ anticommute,
which also contradicts with (\ref{commut}). It is obvious that in this case 
$\sE\sim\sI$ is excluded; therefore, $\sE\sim\cE_1\cE_2\cdots\cE_q$.
In this case, according to (\ref{commut3}) the matrix $\sC$ is proportional
to the unit matrix.
Thus, the matrices $\sE\sim\cE_1\cE_2\cdots\cE_q$, $\sC=\sE\sW\sim\sI$ and
$\sW$ of the automorphisms $\cA\rightarrow\widetilde{\cA},\,
\cA\rightarrow\widetilde{\cA^\star}$ and $\cA\rightarrow\cA^\star$ of the
algebra $\cl_{0,q}$ ($q\equiv 0\pmod{8}$) from the group
$\sAut_-(\cl_{0,q})\simeq\dZ_2\otimes\dZ_2$.\\[0.3cm]
2) The type $p-q\equiv 2\pmod{8}$, $\K\simeq\R$.\\
In virtue of the isomorphism $\cl_{p,q}\simeq\M_{2^{\frac{p+q}{2}}}(\R)$ for
the type $p-q\equiv 2\pmod{8}$ in accordance with the signature of the
algebra $\cl_{p,q}$, we have the following basis
\begin{equation}\label{B1}
\cE^{\sT}_i=\left\{\begin{array}{rl}
\phantom{-}\cE_i, & \mbox{if $1\leq i\leq p$},\\
-\cE_i, & \mbox{if $p+1\leq i\leq p+q$}.
\end{array}\right.
\end{equation}
Therefore, in this case the matrix of the antiautomorphism 
$\widetilde{\phantom{cc}}$ is a product of
$p$ symmetric matrices, $\sE=\cE_1\cE_2\cdots\cE_p$
or is a product of $q$ skewsymmetric matrices, $\sE=\cE_{p+1}\cE_{p+2}
\cdots\cE_{p+q}$. Let us find permutation conditions of the matrix $\sE$ with the
basis matrices $\cE_i$. Let $\sE=\cE_1\cE_2\cdots\cE_p$, then
\begin{equation}\label{B2}
\begin{array}{ccc}
\sE\cE_i&=&(-1)^{p-i}\cE_1\cE_2\cdots\cE_{i-1}\cE_{i+1}\cdots\cE_p,\\
\cE_i\sE&=&(-1)^{i-1}\cE_1\cE_2\cdots\cE_{i-1}\cE_{i+1}\cdots\cE_p
\end{array}\quad(1\leq i\leq p)
\end{equation}
and
\begin{equation}\label{B3}
\sE\cE_i=(-1)^p\cE_i\sE, \quad (p+1\leq i\leq p+q)
\end{equation}
that is, at $p\equiv 0\pmod{2}$ the matrix $\sE$  anticommutes with the 
symmetric and commutes with the skewsymmetric part of the basis
(\ref{B1}). Correspondingly, at $p\equiv 1\pmod{2}$ $\sE$ commutes with the
symmetric and anticommutes with the skewsymmetric part of the basis (\ref{B1}).

Analogously, let $\sE=\cE_{p+1}\cE_{p+2}\cdots\cE_{p+q}$, then
\begin{equation}\label{B4}
\sE\cE_i=(-1)^q\cE_i\sE\quad (1\leq i\leq p)
\end{equation}
and
\begin{equation}\label{B5}
\begin{array}{ccc}
\sE\cE_i&=&-(-1)^{q-i}\cE_{p+1}\cE_{p+2}\cdots\cE_{i-1}\cE_{i+1}\cdots
\cE_{p+q};\\
\cE_i\sE&=&-(-1)^{i-1}\cE_{p+1}\cE_{p+2}\cdots\cE_{i-1}\cE_{i+1}\cdots
\cE_{p+q},
\end{array}\quad(p+1\leq i\leq p+q)
\end{equation}
that is, at $q\equiv 0\pmod{2}$ the matrix $\sE$ commutes with the symmetric
and anticommutes with the skewsymmetric part of the basis (\ref{B1}).
Correspondingly, at $q\equiv 1\pmod{2}$ $\sE$ anticommutes with the symmetric
and commutes with the skewsymmetric part of (\ref{B1}).

Further, permutation conditions of the matrices $\sE=\cE_1\cE_2
\cdots\cE_p$ and $\sW$ are defined by the following relations:
\begin{eqnarray}
\sE\sW&=&\cE_1\cE_2\cdots\cE_p\cE_1\cE_2\cdots\cE_{p+q}=(-1)^{\frac{p(p-1)}{2}}
\cE_{p+1}\cE_{p+2}\cdots\cE_{p+q},\nonumber\\
\sW\sE&=&\cE_1\cE_2\cdots\cE_{p+q}\cE_1\cE_2\cdots\cE_p=
(-1)^{\frac{p(p-1)}{2}+pq}\cE_{p+1}\cE_{p+2}\cdots\cE_{p+q}.\label{B6}
\end{eqnarray}
From a comparison $\frac{p(p-1)}{2}+pq\equiv\frac{p(p-1)}{2}\pmod{2}$ it follows
that the matrices $\sE$ and $\sW$ commute with each other if
$pq\equiv 0\pmod{2}$ and anticommute if
$pq\equiv 1\pmod{2}$. If we take $\sE=\cE_{p+1}\cE_{p+2}\cdots
\cE_{p+q}$, then the relations
\begin{eqnarray}
\sE\sW&=&\cE_{p+1}\cE_{p+2}\cdots\cE_{p+q}\cE_1\cE_2\cdots\cE_{p+q}=
(-1)^{\frac{q(q+1)}{2}+pq}\cE_1\cE_2\cdots\cE_p,\nonumber\\
\sW\sE&=&\cE_1\cE_2\cdots\cE_{p+q}\cE_{p+1}\cE_{p+2}\cdots\cE_{p+q}=
(-1)^{\frac{q(q+1)}{2}}\cE_1\cE_2\cdots\cE_p\label{B7}
\end{eqnarray}
give analogous permutation conditions for $\sE$ and $\sW$
($pq\equiv 0,1\pmod{2}$). It is obvious that permutation conditions of
$\sC$ (the matrix of the antiautomorphism $\widetilde{\star}$) with the
basis matrices $\cE_i$ and
with $\sW$ 
are analogous to the conditions (\ref{B2})--(\ref{B5}) and
(\ref{B6})--(\ref{B7}), respectively.

Out of dependence on the choice of the matrices $\sE$ and $\sC$,
permutation conditions between them are defined by a relation
\begin{equation}\label{B8}
\sE\sC=(-1)^{pq}\sC\sE,
\end{equation}
that is, $\sE$ and $\sC$ commute if $pq\equiv 0\pmod{2}$ and anticommute
if $pq\equiv 1\pmod{2}$.

For the squares of the automorphisms $\widetilde{\phantom{cc}}$ and
$\widetilde{\star}$ we have following two possibilities:\\[0.2cm]
a) $\sE=\cE_1\cE_2\cdots\cE_p$, $\sC=\cE_{p+1}\cE_{p+2}\cdots\cE_{p+q}$.
\begin{equation}\label{B9}
\sE^2=\left\{\begin{array}{rl}
+\sI, & \mbox{if $p\equiv 0,1\pmod{4}$};\\
-\sI, & \mbox{if $p\equiv 2,3\pmod{4}$},
\end{array}\right.\quad
\sC^2=\left\{\begin{array}{rl}
+\sI, & \mbox{if $q\equiv 0,3\pmod{4}$};\\
-\sI, & \mbox{if $q\equiv 1,2\pmod{4}$}.
\end{array}\right.
\end{equation}
b) $\sE=\cE_{p+1}\cE_{p+2}\cdots\cE_{p+q}$, $\sC=\cE_1\cE_2\cdots\cE_p$.
\begin{equation}\label{B10}
\sE^2=\left\{\begin{array}{rl}
+\sI, & \mbox{if $q\equiv 0,3\pmod{4}$};\\
-\sI, & \mbox{if $q\equiv 1,2\pmod{4}$},
\end{array}\right.\quad
\sC^2=\left\{\begin{array}{rl}
+\sI, & \mbox{if $p\equiv 0,1\pmod{4}$};\\
-\sI, & \mbox{if $p\equiv 2,3\pmod{4}$}.
\end{array}\right.
\end{equation}
For the type $p-q\equiv 2\pmod{8}$ in virtue of (\ref{e3}) a square of the
matrix $\sW$ is always equal to $-\sI$.

Now, let us consider automorphism groups for the type
$p-q\equiv 2\pmod{8}$. In accordance with (\ref{B6})--(\ref{B8}), the
automorphism group $\sAut(\cl_{p,q})$ is Abelian if $pq\equiv 0\pmod{2}$.
Further, in virtue of (\ref{commut}) and (\ref{B1}) the matrix of the
antiautomorphism
$\widetilde{\phantom{cc}}$ should commute with the symmetric part of the
basis (\ref{B1}) and anticommute with the skewsymmetric part of the same
basis. From (\ref{B2})--(\ref{B5}), it is easy to see that
this condition is satisfied at $pq\equiv 0\pmod{2}$ if and only if
$\sE=\cE_{p+1}\cE_{p+2}\cdots
\cE_{p+q}$ is a product of $q$ skewsymmetric matrices (recall that for the
type $p-q\equiv 2\pmod{8}$, the numbers $p$ and $q$ are both even or both
odd). Correspondingly, in accordance with (\ref{commut3}), the matrix $\sC$
should anticommute with the symmetric part of the basis (\ref{B1}) and
commute with skewsymmetric part of the same basis. It is obvious that this
requirement is satisfied if and only if $\sC=\cE_1\cE_2\cdots\cE_p$ is a
product of $p$ symmetric matrices.
Thus, in accordance with (\ref{B10}), there exist Abelian groups
$\sAut_-(\cl_{p,q})\simeq\dZ_4$ with the signature $(-,-,+)$ if
$p\equiv 0\pmod{4}$ and $q\equiv 2\pmod{4}$, and 
with the signature $(-,+,-)$ if
$p\equiv 2\pmod{4}$ and $q\equiv 0\pmod{4}$. Further, according to
(\ref{B6})--(\ref{B8}), the automorphism group is non--Abelian if
$pq\equiv 1\pmod{2}$. In this case, from (\ref{B2})--(\ref{B5}) it follows that
the matrix of the antiautomorphism $\widetilde{\phantom{cc}}$ commutes with
the symmetric part of the basis (\ref{B1}) and anticommutes with the
skewsymmetric part if and only if $\sE=\cE_1\cE_2\cdots\cE_p$ is a product
of $p$ symmetric matrices. In its turn, the matrix $\sC$ anticommutes with
the symmetric part of the basis (\ref{B1}) and commutes with the skewsymmetric
part of the same basis if and only if $\sC=\cE_{p+1}\cE_{p+2}\cdots\cE_{p+q}$.
Therefore in accordance with
(\ref{B9}), there exists a non--Abelian group 
$\sAut_+(\cl_{p,q})\simeq Q_4/\dZ_2$ with the signature
$(-,-,-)$ if $p\equiv 3\pmod{4}$ and $q\equiv 1\pmod{4}$ and 
$\sAut_+(\cl_{p,q})\simeq D_4/\dZ_2$ with the
signature
$(-,+,+)$ if $p\equiv 1\pmod{4}$ and $q\equiv 3\pmod{4}$.\\[0.3cm]
3) The type $p-q\equiv 6\pmod{8}$, $\K\simeq\BH$.\\
First of all, over the ring $\K\simeq\BH$ there exists no fixed basis of
the form (\ref{A1}) or (\ref{B1}) for the matrices $\cE_i$. In general,
a number of the skewsymmetric matrices does not coincide with a number of
matrices with the negative square ($\cE^2_j=-\sI$) as it takes place for the
types $p-q\equiv 0,2\pmod{8}$. Thus, the matrix 
$\sE$ is a product of skewsymmetric matrices $\cE_j$, among which there are
matrices with positive and negative squares, or $\sE$ is a product of
symmetric matrices
$\cE_i$, among which also there are matrices with '$+$' and '$-$' squares.
Let $k$ be a number of the skewsymmetric matrices $\cE_j$ of a spinbasis of
the algebra
$\cl_{p,q}$, $0\leq k\leq p+q$. Among the matrices $\cE_j$, $l$ have 
'$+$'-square
and $t$ matrices have '$-$'-square. Let $0<k<p+q$ and let
$\sE=\cE_{j_1}\cE_{j_2}\cdots\cE_{j_k}$ be a matrix of the antiautomorphism
$\cA\rightarrow\widetilde{\cA}$, then permutation conditions of the matrix
$\sE$ with the matrices $\cE_{i_r}$ of the symmetric part
($0<r\leq p+q-k$) and with the matrices $\cE_{j_u}$ of the skewsymmetric part
($0<u\leq k$) have the respective form
\begin{equation}\label{C1}
\sE\cE_{i_r}=(-1)^k\cE_{i_r}\sE\quad (0<r\leq p+q-k),
\end{equation}
\begin{equation}\label{C2}
\begin{array}{ccc}
\sE\cE_{j_u}&=&(-1)^{k-u}\sigma(j_u)\cE_{j_1}\cE_{j_2}\cdots\cE_{j_{u-1}}
\cE_{j_{u+1}}\cdots\cE_{j_k},\\
\cE_{j_u}\sE&=&(-1)^{u-1}\sigma(j_u)\cE_{j_1}\cE_{j_2}\cdots\cE_{j_{u-1}}
\cE_{j_{u+1}}\cdots\cE_{j_k},
\end{array}\quad (0<u\leq k)
\end{equation}
that is, at $k\equiv 0\pmod{2}$ the matrix $\sE$ commutes with the symmetric part
and anticommutes with the skewsymmetric part of the spinbasis. Correspondingly,
at $k\equiv 1\pmod{2}$, $\sE$ anticommutes with the symmetric and commutes
with the skewsymmetric part. Further, let $\sE=\cE_{i_1}\cE_{i_2}
\cdots\cE_{i_{p+q-k}}$ be a product of the symmetric matrices, then
\begin{equation}\label{C3}
\begin{array}{ccc}
\sE\cE_{i_r}&=&(-1)^{p+q-k}\sigma(i_r)\cE_{i_1}\cE_{i_2}\cdots\cE_{i_{r-1}}
\cE_{i_{r+1}}\cdots\cE_{i_{p+q-k}},\\
\cE_{i_r}\sE&=&(-1)^{r-1}\sigma(i_r)\cE_{i_1}\cE_{i_2}\cdots\cE_{i_{r-1}}
\cE_{i_{r+1}}\cdots\cE_{i_{p+q-k}},
\end{array}\quad (0<r\leq p+q-k)
\end{equation}
\begin{equation}\label{C4}
\sE\cE_{j_u}=(-1)^{p+q-k}\cE_{j_u}\sE,\quad (0<u\leq k)
\end{equation}
that is, at $p+q-k\equiv 0\pmod{2}$ the matrix $\sE$ anticommutes with the
symmetric part and commutes with the skewsymmetric part of the spinbasis.
Correspondingly, at $p+q-k\equiv 1\pmod{2}$ $\sE$ commutes with the symmetric
part and anticommutes with the skewsymmetric part. It is easy to see that
permutations conditions of the matrix $\sC$ with the basis matrices $\cE_i$
are coincide with (\ref{C1})--(\ref{C4}).

For the permutation conditions of the matrices $\sW=\cE_{i_1}\cE_{i_2}\cdots
\cE_{i_{p+q-k}}\cE_{j_1}\cE_{j_2}\cdots\cE_{j_k}$, $\sE=\cE_{j_1}\cE_{j_2}
\cdots\cE_{j_k}$, and $\sC=\cE_{i_1}\cE_{i_2}\cdots\cE_{i_{p+q-k}}$ we have
\begin{eqnarray}
\sE\sW&=&(-1)^{\frac{k(k-1)}{2}+t+k(p+q-k)}\cE_{i_1}\cE_{i_2}\cdots
\cE_{i_{p+q-k}},\nonumber\\
\sW\sE&=&(-1)^{\frac{k(k-1)}{2}+t}\cE_{i_1}\cE_{i_2}\cdots\cE_{i_{p+q-k}}.
\label{C5}
\end{eqnarray}
\begin{equation}\label{C6}
\sE\sC=(-1)^{k(p+q-k}\sC\sE.
\end{equation}
Hence it follows that the matrices $\sW,\,\sE$, and $\sC$ commute at
$k(p+q-k)\equiv 0\pmod{2}$ and anticommute at $k(p+q-k)\equiv 1\pmod{2}$.
It is easy to verify that permutation conditions for the matrices
$\sE=\cE_{i_1}\cE_{i_2}\cdots\cE_{i_{p+q-k}}$, $\sC=\cE_{j_1}\cE_{j_2}\cdots
\cE_{j_k}$ would be the same.

In accordance with (\ref{commut}), (\ref{commut3}),
(\ref{C1})--(\ref{C4}), and also with
(\ref{C5})--(\ref{C6}), the Abelian automorphism groups for the type
$p-q\equiv 6\pmod{8}$ exist only if $\sE=\cE_{j_1}\cE_{j_2}\cdots
\cE_{j_k}$ and $\sC=\cE_{i_1}\cE_{i_2}\cdots\cE_{i_{p+q-k}}$, 
$k\equiv 0\pmod{2}$. Let
$l$ and $t$ be the quantities of the matrices in the
product $\cE_{j_1}\cE_{j_2}\cdots
\cE_{j_k}$, which have '$+$' and '$-$'-squares, respectively,
and also let $h$ and $g$ be the quantities of the matrices with the same
meaning in the product $\cE_{i_1}\cE_{i_2}
\cdots\cE_{i_{p+q-k}}$. Then, the group $\sAut_-(\cl_{p,q})\simeq\dZ_4$
with the signature $(-,+,-)$ exists if $l-t\equiv 0,1,4,5\pmod{8}$ and
$h-g\equiv 2,3,6,7\pmod{8}$ (recall that for the type $p-q\equiv 6\pmod{8}$
we have 
$\sW^2=-\sI$), and also, the group $\sAut_-(\cl_{p,q})\simeq\dZ_4$ with the
signature
$(-,-,+)$ exists if $l-t\equiv 2,3,6,7\pmod{8}$ and $h-g\equiv 0,1,4,5
\pmod{8}$. Further, from (\ref{commut}), (\ref{commut3}),
and (\ref{C1})--(\ref{C6}), it follows
that the non--Abelian automorphism groups exist only if
$\sE=\cE_{i_1}\cE_{i_2}\cdots
\cE_{i_{p+q-k}}$ and $\sC=\cE_{j_1}\cE_{j_2}\cdots\cE_{j_k}$,
$k\equiv 1\pmod{2}$. At this point the group
$\sAut_+(\cl_{p,q})\simeq Q_4/\dZ_2$ with the signature $(-,-,-)$
exists if $h-g\equiv 2,3,6,7\pmod{8}$ and $l-t\equiv 2,3,6,7\pmod{8}$.
Correspondingly, the group $\sAut_+(\cl_{p,q})\simeq D_4/\dZ_2$ with the
signature
$(-,+,+)$ exists if $h-g\equiv 0,1,4,5\pmod{8}$ and $l-t\equiv 0,1,4,5
\pmod{8}$. In absence of the skewsymmetric matrices, $k=0$, the spinbasis
of $\cl_{p,q}$ contains only symmetric matrices. In this case, from
(\ref{commut}), it follows that the matrix of the antiautomorphism $\cA
\rightarrow\widetilde{\cA}$ should commute with all the basis matrices.
It is obvious that this condition is satisfied if and only if $\sE$ is
proportional to the unit matrix. At this point, from (\ref{C}), it follows
that $\sC\sim\cE_1\cE_2\cdots\cE_{p+q}$ and we see that condition
(\ref{commut3}) is satisfied.
Thus, we have the Abelian group
$\sAut_-(\cl_{p,q})\simeq\dZ_4$ with the signature $(-,+,-)$. In other
degenerate case $k=p+q$, the spinbasis of $\cl_{p,q}$ contains only
skewsymmetric matrices; therefore, the matrix $\sE$ should  
anticommute with all the basis matrices. This condition is satisfied
if and only if $\sE\sim\cE_1\cE_2\cdots\cE_{p+q}$. In its turn, the matrix
$\sC$ commutes with all the basis matrices if and only if $\sC\sim\sI$.
It is easy to see
that in this case we have the group $\sAut_-(\cl_{p,q})\simeq\dZ_4$
with the signature $(-,-,+)$.\\[0.3cm]
4) The type $p-q\equiv 4\pmod{8}$, $\K\simeq\BH$.\\
It is obvious that a proof for this type is analogous to the case
$p-q\equiv 6\pmod{8}$, where also $\K\simeq\BH$. For the
type $p-q\equiv 4\pmod{8}$
we have $\sW^2=+\sI$. As well as for the type $p-q\equiv 6\pmod{8}$,
the Abelian groups exist only if $\sE=\cE_{j_1}\cE_{j_2}
\cdots\cE_{j_k}$ and $\sC=\cE_{i_1}\cE_{i_2}\cdots\cE_{i_{p+q-k}}$,
$k\equiv 0\pmod{2}$. At this point the group
$\sAut_-(\cl_{p,q})\simeq\dZ_2\otimes\dZ_2$ with $(+,+,+)$ exists if
$l-t,h-g\equiv 0,1,4,5\pmod{8}$, and also the group $\sAut_-(\cl_{p,q})
\simeq\dZ_4$ with $(+,-,-)$ exists if $l-t,h-g\equiv 2,3,6,7\pmod{8}$.
Correspondingly, the non--Abelian group exist only if $\sE$ is a product
of $k$ skewsymmetric matrices and $\sC$ is a product of $p+q-k$ symmetric
matrices,
$k\equiv 1\pmod{2}$.
The group $\sAut_+(\cl_{p,q})\simeq D_4/\dZ_2$ with $(+,-,+)$
exists if $h-g\equiv 2,3,6,7\pmod{8}$, $l-t\equiv 0,1,4,5\pmod{8}$,
and the group $\sAut_+(\cl_{p,q})\simeq D_4/\dZ_2$ with $(+,+,-)$ exists
if $h-g\equiv 0,1,4,5\pmod{8}$, $l-t\equiv 2,3,6,7\pmod{8}$.
For the type $p-q\equiv 4\pmod{8}$ both the degenerate cases $k=0$ and
$k=p+q$ give rise to the group 
$\sAut_-(\cl_{p,q})\simeq\dZ_2\otimes\dZ_2$.\\[0.3cm]
5) The type $p-q\equiv 1\pmod{8}$, $\K\simeq\R\oplus\R$.\\
In this case a dimensionality  $p+q$ is odd and the algebra $\cl_{p,q}$ is
semi--simple. Over the ring $\K\simeq\R\oplus\R$ the algebras of this type
decompose into a direct sum of two subalgebras with even dimensionality.
At this point there exist two types of decomposition \cite{Rash,Port}:
\begin{eqnarray}
\cl_{p,q}&\simeq&\cl_{p,q-1}\oplus\cl_{p,q-1},\label{D1}\\
\cl_{p,q}&\simeq&\cl_{q,p-1}\oplus\cl_{q,p-1},\label{D1'}
\end{eqnarray}
where each algebra $\cl_{p,q-1}$ ($\cl_{q,p-1}$) is obtained by means
of either of the two central idempotents 
$\frac{1}{2}(1\pm\e_1\e_2\ldots\e_{p+q})$
and isomorphisms
\begin{eqnarray}
\cl^+_{p,q}&\simeq&\cl_{p,q-1},\label{D2'}\\
\cl^+_{p,q}&\simeq&\cl_{q,p-1}.\label{D2}
\end{eqnarray}
In general, the structure of the ring $\K\simeq\R\oplus\R$ in virtue of
the decompositions
(\ref{D1})--(\ref{D1'}) and isomorphisms  (\ref{D2'})--(\ref{D2}) 
admits all eight kinds
of the automorphism groups, since the subalgebras in the direct sums
(\ref{D1})--(\ref{D1'}) have the type $p-q\equiv 2\pmod{8}$ or the type
$p-q\equiv 0\pmod{8}$. More precisely, for the algebras $\cl_{0,q}$ of the type
$p-q\equiv 1\pmod{8}$, the subalgebras in the direct sum (\ref{D1}) have the
type $p-q\equiv 2\pmod{8}$ and only this type; therefore, in accordance with
previously obtained conditions for the type $p-q\equiv 2\pmod{8}$, we have
four and only four kinds of the automorphism groups with the signatures
$(-,-,+),\,(-,+,-)$ and $(-,-,-),\,(-,+,+)$.
Further, for the algebra $\cl_{p,0}$ ($p-q\equiv 1\pmod{8}$) the subalgebras
in the direct sum (\ref{D1'}) have the type $p-q\equiv 0\pmod{8}$; therefore,
in this case there exist four and only four kinds of the automorphism groups
with the signatures $(+,+,+),\,(+,-,-)$ and $(+,-,+),\,(+,+,-)$. In general
case, $\cl_{p,q}$, the type $p-q\equiv 1\pmod{8}$ admits all eight kinds of the
automorphism groups.\\[0.3cm]
6) The type $p-q\equiv 5\pmod{8}$, $\K\simeq\BH\oplus\BH$.\\
In this case the algebra $\cl_{p,q}$ is also semi--simple and, therefore,
we have decompositions of the form (\ref{D1})--(\ref{D1'}). By analogy with
the type $p-q\equiv 1\pmod{8}$, a structure of the double quaternionic ring
$\K\simeq\BH\oplus\BH$ in virtue of the decompositions (\ref{D1})--(\ref{D1'}) 
and isomorphisms (\ref{D2'})--(\ref{D2}) 
is also admits, in a general case, all eight kinds
of the automorphism groups, since the subalgebras in the direct sums
(\ref{D1})--(\ref{D1'}) have the type $p-q\equiv 6\pmod{8}$ or the type
$p-q\equiv 4\pmod{8}$. More precisely, for the algebras $\cl_{0,q}$ of the type
$p-q\equiv 5\pmod{8}$, the subalgebras in the direct sum (\ref{D1}) have the
type $p-q\equiv 6\pmod{8}$ and only this type; therefore, in accordance with
previously obtained results for the quaternionic rings we have four and only
four kinds of the automorphism groups with the signatures $(-,+,-),\,(-,-,+)$ 
and $(-,-,-),\,(-,+,+)$. Analogously, for the algebras $\cl_{p,0}$ 
($p-q\equiv 5\pmod{8}$), the subalgebras in the direct sum (\ref{D1'}) have the
type $p-q\equiv 4\pmod{8}$; therefore, in this case there exist four and only
four kinds of the automorphism groups with the signatures
$(+,+,+),\,(+,-,-)$ and $(+,-,+),\,(+,+,-)$. In general case, $\cl_{p,q}$,
the type $p-q\equiv 5\pmod{8}$ admits all eight kinds of the automorphism
groups.\\[0.3cm]
7) The type $p-q\equiv 3\pmod{8}$, $\K\simeq\C$.\\
For this type a center $\bZ$ of the algebra $\cl_{p,q}$ consists of the unit
and the volume element $\omega=\e_1\e_2\ldots\e_{p+q}$, since $p+q$ is odd
and the element $\omega$ commutes with all the basis elements of the algebra
$\cl_{p,q}$. Moreover,
$\omega^2=-1$, hence it follows that $\bZ\simeq\R\oplus i\R$. Thus, for the
algebras $\cl_{p,q}$ of the type $p-q\equiv 3\pmod{8}$, there exists an
isomorphism
\begin{equation}\label{E1}
\cl_{p,q}\simeq\C_{n-1},
\end{equation}
where $n=p+q$. It is easy to see that the algebra $\C_{n-1}=\C_{2m}$ in 
(\ref{E1})
is a complex algebra with even dimensionality, where $m$ is either even or
odd. More precisely, the number $m$ is even if $p\equiv 0\pmod{2}$ and
$q\equiv 1\pmod{2}$, and odd if $p\equiv 1\pmod{2}$ and
$q\equiv 0\pmod{2}$. In accordance with Theorem \ref{taut} 
at $m\equiv 0\pmod{2}$
the algebra $\C_{2m}$ admits the Abelian group $\sAut_-(\C_{2m})\simeq
\dZ_2\otimes\dZ_2$ with $(+,+,+)$, and at $m\equiv 1\pmod{2}$
the non--Abelian group $\sAut_+(\C_{2m})\simeq Q_4/\dZ_2$ with $(-,-,-)$.
Hence it follows the statement of Theorem for this type.\\[0.3cm]
8) The type $p-q\equiv 7\pmod{8}$, $\K\simeq\C$.\\
It is obvious that for this type the isomorphism (\ref{E1}) also takes
place.
Therefore, the type $p-q\equiv 7\pmod{8}$ admits the group $\sAut_-(\cl_{p,q})
\simeq\dZ_2\otimes\dZ_2$ if $p\equiv 0\pmod{2}$ and $q\equiv 1\pmod{2}$,
and also the group $\sAut_+(\cl_{p,q})\simeq Q_4/\dZ_2$ if $p\equiv 1\pmod{2}$
and $q\equiv 0\pmod{2}$.
\end{proof}\begin{cor}
The matrices $\sE$ and $\sC$ of the antiautomorphisms 
$\cA\rightarrow\widetilde{\cA}$ and $\cA\rightarrow\widetilde{\cA^\star}$
over the field $\F=\R$ satisfy the following conditions
\begin{equation}\label{condt}
\sE^{\sT}=(-1)^{\frac{m(m-1)}{2}}\sE,\quad
\sC^{\sT}=(-1)^{\frac{m(m+1)}{2}}\sC,
\end{equation}
that is, $\sE$ is symmetric if $m\equiv 0,1\pmod{4}$ and skewsymmetric if
$m\equiv 2,3\pmod{4}$. Correspondingly, $\sC$ is symmetric if 
$m\equiv 0,3\pmod{4}$ and skewsymmetric if $m\equiv 1,2\pmod{4}$.
\end{cor}
\begin{proof}
Let us consider first the types with the ring $\K\simeq\R$. As follows
from Theorem \ref{tautr}, the type $p-q\equiv 0\pmod{8}$ admits the
Abelian automorphism groups $(\sE\sC=\sC\sE)$ if $\sE$ is the product of
$q$ skewsymmetric matrices ($q\equiv 0,2\pmod{4}$) and $\sC$ is the product
of $p$ symmetric matrices ($p\equiv 0,2\pmod{4}$). Therefore,
\begin{multline}
\sE^{\sT}=(\cE_{m+1}\cE_{m+2}\cdots\cE_{2m})^{\sT}=\cE^{\sT}_{2m}\cdots
\cE^{\sT}_{m+2}\cE^{\sT}_{m+1}=\\
(-\cE_{2m})\cdots(-\cE_{m+2})(-\cE_{m+1})=\cE_{2m}\cdots\cE_{m+2}\cE_{m+1}=
(-1)^{\frac{q(q-1)}{2}}\sE,\label{T1}
\end{multline}
\begin{equation}
\sC^{\sT}=(\cE_1\cE_2\cdots\cE_m)^{\sT}=\cE^{\sT}_m\cdots\cE^{\sT}_2
\cE^{\sT}_1=\cE_m\cdots\cE_2\cE_1=(-1)^{\frac{p(p-1)}{2}}\sC.\label{T2}
\end{equation}
Further, the type $p-q\equiv 0\pmod{8}$ admits the non--Abelian automorphism
groups ($\sE\sC=-\sC\sE$) if $\sE$ is the product of $p$ symmetric matrices
($p\equiv 1,3\pmod{4}$) and $\sC$ is the product of $q$ skewsymmetric
matrices ($q\equiv 1,3\pmod{4}$). In this case, we have
\begin{equation}
\sE^{\sT}=(\cE_1\cE_2\cdots\cE_m)^{\sT}=\cE^{\sT}_m\cdots\cE^{\sT}_2
\cE^{\sT}_1=\cE_m\cdots\cE_2\cE_1=(-1)^{\frac{p(p-1)}{2}}\sE,\label{T3}
\end{equation}
\begin{multline}
\sC^{\sT}=(\cE_{m+1}\cE_{m+2}\cdots\cE_{2})^{\sT}=\cE^{\sT}_{2m}\cdots
\cE^{\sT}_{m+2}\cE^{\sT}_{m+1}=\\
(-\cE_{2m})\cdots(-\cE_{m+2})(-\cE_{m+1})=-\cE_{2m}\cdots\cE_{m+2}\cE_{m+1}=
-(-1)^{\frac{q(q-1)}{2}}\sC.
\label{T4}
\end{multline}
In the degenerate case $\cl_{p,0}$, $p\equiv 0\pmod{8}$, we have $\sE\sim\sI$
and $\sC\sim\cE_1\cE_2\cdots\cE_p$. Therefore, $\sE$ is always symmetric
and $\sC^{\sT}=(-1)^{\frac{p(p-1)}{2}}\sC$. In other degenerate case
$\cl_{0,q}$, $q\equiv 0\pmod{8}$, we have $\sE\sim\cE_1\cE_2\cdots\cE_q$ and
$\sC\sim\sI$; therefore, $\sE^{\sT}=(-1)^{\frac{q(q-1)}{2}}\sE$ and $\sC$
is always symmetric.

Since for the type $p-q\equiv 0\pmod{8}$ we have $p=q=m$, or $m=p$ and
$m=q$ for the degenerate cases (both degenerate cases correspond to the
Abelian group $\dZ_2\otimes\dZ_2$), it is easy to see that the
formulas (\ref{T1}) and (\ref{T3}) coincide with the first formula of
(\ref{condt}). For the matrix $\sC$, we can unite the formulas (\ref{T2})
and (\ref{T4}) into the formula which coincides with the second formula of
(\ref{condt}). Indeed, the factor $(-1)^{\frac{m(m+1)}{2}}$ does not change
sign in $\sC^{\sT}=(-1)^{\frac{m(m+1)}{2}}\sC$ when $m$ is even and
changes sign when $m$ is odd, which is equivalent to both formulas
(\ref{T2}) and (\ref{T4}).

Further, the following real type $p-q\equiv 2\pmod{8}$ admits the Abelian
automorphism groups if $\sE=\cE_{p+1}\cE_{p+2}\cdots\cE_{p+q}$ and
$\sC=\cE_1\cE_2\cdots\cE_p$, where $p$ and $q\equiv 0,2\pmod{4}$. Therefore,
\begin{multline}
\sE^{\sT}=(\cE_{p+1}\cE_{p+2}\cdots\cE_{p+q})^{\sT}=\cE^{\sT}_{p+q}\cdots
\cE^{\sT}_{p+2}\cE^{\sT}_{p+1}=\\
(-\cE_{p+q})\cdots(-\cE_{p+2})(-\cE_{p+1})=\cE_{p+q}\cdots\cE_{p+2}\cE_{p+1}
=(-1)^{\frac{q(q-1)}{2}}\sE,\label{T5}
\end{multline}
\begin{equation}
\sC^{\sT}=(\cE_1\cE_2\cdots\cE_p)^{\sT}=\cE^{\sT}_p\cdots\cE^{\sT}_2
\cE^{\sT}_1=\cE_p\cdots\cE_2\cE_1=(-1)^{\frac{p(p-1)}{2}}\sC.\label{T6}
\end{equation}
Correspondingly, the type $p-q\equiv 2\pmod{8}$ admits the non--Abelian
automorphism groups if $\sE=\cE_1\cE_2\cdots\cE_p$ and $\sC=\cE_{p+1}\cE_{p+2}
\cdots\cE_{p+q}$, where $p$ and $q\equiv 1,3\pmod{4}$. In this case, we have
\begin{equation}
\sE^{\sT}=(\cE_1\cE_2\cdots\cE_p)^{\sT}=\cE^{\sT}_p\cdots\cE^{\sT}_2
\cE^{\sT}_1=\cE_p\cdots\cE_2\cE_1=(-1)^{\frac{p(p-1)}{2}}\sE,\label{T7}
\end{equation}
\begin{multline}
\sC^{\sT}=(\cE_{p+1}\cE_{p+2}\cdots\cE_{p+q})^{\sT}=\cE^{\sT}_{p+q}\cdots
\cE^{\sT}_{p+2}\cE^{\sT}_{p+1}=\\
(-\cE_{p+q})\cdots(-\cE_{p+2})(-\cE_{p+1})=-\cE_{p+q}\cdots\cE_{p+2}
\cE_{p+1}=-(-1)^{\frac{q(q-1)}{2}}\sC.\label{T8}
\end{multline}
It is easy to see that formulas (\ref{T5})--(\ref{T8}) are similar to the
formulas (\ref{T1})--(\ref{T4}) and, therefore, the conditions (\ref{condt})
hold for the type $p-q\equiv 2\pmod{8}$.

Analogously, the quaternionic types $p-q\equiv 4,6\pmod{8}$ admit the
Abelian automorphism groups if $\sE=\cE_{j_1}\cE_{j_2}\cdots\cE_{j_k}$ and
$\sC=\cE_{i_1}\cE_{i_2}\cdots\cE_{i_{p+q-k}}$, where $k$ and $p+q-k$ are
even (Theorem \ref{tautr}). Transposition of these matrices gives
\begin{multline}
\sE^{\sT}=(\cE_{j_1}\cE_{j_2}\cdots\cE_{j_k})^{\sT}=\cE^{\sT}_{j_k}\cdots
\cE^{\sT}_{j_2}\cE^{\sT}_{j_1}=\\
(-\cE_{j_k})\cdots(-\cE_{j_2})(-\cE_{j_1})=\cE_{j_k}\cdots\cE_{j_2}
\cE_{j_1}=(-1)^{\frac{k(k-1)}{2}}\sE,\label{T9}
\end{multline}
\begin{multline}
\sC^{\sT}=(\cE_{i_1}\cE_{i_2}\cdots\cE_{i_{p+q-k}})^{\sT}=\\
\cE^{\sT}_{i_{p+q-k}}\cdots\cE^{\sT}_{i_2}\cE^{\sT}_{i_1}=\cE_{i_{p+q-k}}\cdots
\cE_{i_2}\cE_{i_1}=(-1)^{\frac{(p+q-k)(p+q-k-1)}{2}}\sC.\label{T10}
\end{multline}
The non--Abelian automorphism groups take place for the types $p-q\equiv
4,6\pmod{8}$ if $\sE=\cE_{i_1}\cE_{i_2}\cdots\cE_{i_{p+q-k}}$ and
$\sC=\cE_{j_1}\cE_{j_2}\cdots\cE_{j_k}$, where $k$ and $p+q-k$ are odd.
In this case we have
\begin{multline}
\sE^{\sT}=(\cE_{i_1}\cE_{i_2}\cdots\cE_{i_{p+q-k}})^{\sT}=\\
\cE^{\sT}_{i_{p+q-k}}\cdots\cE^{\sT}_{i_2}\cE^{\sT}_{i_1}=\cE_{p+q-k}\cdots
\cE_{i_2}\cE_{i_1}=(-1)^{\frac{(p+q-k)(p+q-k-1)}{2}}\sE,\label{T11}
\end{multline}
\begin{multline}
\sC^{\sT}=(\cE_{j_1}\cE_{j_2}\cdots\cE_{j_k})^{\sT}=\cE^{\sT}_{j_k}\cdots
\cE^{\sT}_{j_2}\cE^{\sT}_{j_1}=\\
(-\cE_{j_k})\cdots(-\cE_{j_2})(-\cE_{j_1})=-\cE_{j_k}\cdots\cE_{j_1}
\cE_{j_1}=-(-1)^{\frac{k(k-1)}{2}}\sC.\label{T12}
\end{multline}
As it takes place for these two types considered here, we again come to the
same situation. Therefore, the conditions (\ref{condt}) hold for the
quaternionic types $p-q\equiv 4,6\pmod{8}$.

In virtue of the isomorphism (\ref{E1}) and the Theorem \ref{tautr}, the
matrices $\sE$ and $\sC$ for the types $p-q\equiv 3,7\pmod{8}$ with the
ring $\K\simeq\C$ have the following form: $\sE=\cE_1\cE_2\cdots\cE_m$,
$\sC=\cE_{m+1}\cE_{m+2}\cdots\cE_{2m}$ if $m\equiv 1\pmod{2}$ ($\sE\sC=-
\sC\sE$) and $\sE=\cE_{m+1}\cE_{m+2}\cdots\cE_{2m}$, $\sC=\cE_1\cE_2\cdots
\cE_m$ if $m\equiv 0\pmod{2}$ ($\sE\sC=\sC\sE$). It is obvious that for
these types the conditions (\ref{condt}) hold.

Finally, for the semi--simple types $p-q\equiv 1,5\pmod{8}$ in virtue of the
decompositions (\ref{D1})--(\ref{D1'}) we have the formulas 
(\ref{T1})--(\ref{T4}) or (\ref{T5})--(\ref{T8}) in case of the ring
$\K\simeq\R\oplus\R$ ($p-q\equiv 1\pmod{8}$) and the formulas
(\ref{T9})--(\ref{T12}) in case of the ring $\K\simeq\BH\oplus\BH$
($p-q\equiv 5\pmod{8}$).
\end{proof}

An algebraic structure of the discrete transformations is defined by the
isomorphism $\{\Id,\star,\widetilde{\phantom{cc}},\widetilde{\star}\}\simeq
\{1,P,T,PT\}$ \cite{Var99}. Using (\ref{Pin}) or (\ref{Pinabc}), we can apply
this structure to the double coverings of the orthogonal group $O(p,q)$.
Obviously, in case of the types $p-q\equiv 0,2,4,6\pmod{8}$, it is
established directly. Further, in virtue of the isomorphism (\ref{E1}) for
the types $p-q\equiv 3,7\pmod{8}$, we have
\[
\pin(p,q)\simeq\pin(n-1,\C),
\]
where $n=p+q$. Analogously, for the semi--simple types $p-q\equiv 1,5\pmod{8}$
in virtue of the decompositions (\ref{D1})--(\ref{D1'}) the algebra $\cl_{p,q}$
is isomorphic to a direct sum of two mutually annihilating simple ideals
$\frac{1}{2}(1\pm\omega)\cl_{p,q}$: $\cl_{p,q}\simeq\frac{1}{2}(1+\omega)
\cl_{p,q}\oplus\frac{1}{2}(1-\omega)\cl_{p,q}$, where 
$\omega=\e_{12\cdots p+q}$. At this point, each ideal is isomorphic to
$\cl_{p,q-1}$ or $\cl_{q,p-1}$. Therefore, for the Clifford--Lipschitz
groups of these types we have the following isomorphisms
\begin{eqnarray}
\pin(p,q)&\simeq&\pin(p,q-1)\bigcup\e_{12\ldots p+q}\pin(p,q-1),\nonumber\\
\pin(p,q)&\simeq&\pin(q,p-1)\bigcup\e_{12\ldots p+q}\pin(q,p-1).\nonumber
\end{eqnarray} 
\begin{theorem}\label{tgroupr}
Let $\pin^{a,b,c}(p,q)$ be a double covering of the orthogonal group
$O(p,q)$ of the real space $\R^{p,q}$ associated with the algebra
$\cl_{p,q}$.
The squares of symbols $a,b,c\in
\{-,+\}$ correspond to the squares of the elements of a finite group
$\sAut(\cl_{p,q})=\{\sI,\sW,\sE,\sC\}:\;a=\sW^2,\,b=\sE^2,\,c=\sC^2$, 
where $\sW,\sE$ and $\sC$
are the matrices of the fundamental automorphisms $\cA\rightarrow
\cA^\star,\,\cA\rightarrow\widetilde{\cA}$ and $\cA\rightarrow
\widetilde{\cA^\star}$ of the algebra $\cl_{p,q}$, respectively.
Then over the field $\F=\R$ 
in dependence on a division ring structure of the algebra $\cl_{p,q}$,
there exist eight double coverings of the orthogonal group $O(p,q)$:\\[0.2cm]
1) A non--Cliffordian group
\[
\pin^{+,+,+}(p,q)\simeq\frac{(\spin_0(p,q)\odot\dZ_2\otimes\dZ_2\otimes\dZ_2)}
{\dZ_2},
\]
exists if $\K\simeq\R$ and the numbers $p$ and $q$ form the type 
$p-q\equiv 0\pmod{8}$ and $p,q\equiv 0\pmod{4}$, and also if
$p-q\equiv 4\pmod{8}$ and $\K\simeq\BH$. The algebras $\cl_{p,q}$ with the
rings $\K\simeq\R\oplus\R,\,\K\simeq\BH\oplus\BH$ ($p-q\equiv 1,5\pmod{8}$)
admit the group $\pin^{+,+,+}(p,q)$ if in the direct sums there are
addendums of the type
$p-q\equiv 0\pmod{8}$ or $p-q\equiv 4\pmod{8}$. The types $p-q\equiv 3,7
\pmod{8}$, $\K\simeq\C$ admit a non--Cliffordian group $\pin^{+,+,+}(p+q-1,
\C)$ if $p\equiv 0\pmod{2}$ and $q\equiv 1\pmod{2}$. Further, 
non--Cliffordian groups
\[
\pin^{a,b,c}(p,q)\simeq\frac{(\spin_0(p,q)\odot(\dZ_2\otimes\dZ_4)}{\dZ_2},
\]
with $(a,b,c)=(+,-,-)$ exist if $p-q\equiv 0\pmod{8}$, 
$p,q\equiv 2\pmod{4}$ and $\K\simeq\R$, and also if
$p-q\equiv 4\pmod{8}$ and $\K\simeq\BH$. Non--Cliffordian
groups with the signatures
$(a,b,c)=(-,+,-)$ and $(a,b,c)=(-,-,+)$ exist over the ring
$\K\simeq\R$ ($p-q\equiv 2\pmod{8}$) if $p\equiv
2\pmod{4},\,q\equiv 0\pmod{4}$ and $p\equiv 0\pmod{4},\,q\equiv 2\pmod{4}$,
respectively,
and also these groups exist over the ring $\K\simeq\BH$ if
$p-q\equiv 6\pmod{8}$. 
The algebras $\cl_{p,q}$ with the rings
$\K\simeq\R\oplus\R,\,\K\simeq\BH\oplus\BH$ ($p-q\equiv 1,5\pmod{8}$)
admit the group $\pin^{+,-,-}(p,q)$ if in the direct sums there are addendums
of the type $p-q\equiv 0\pmod{8}$ or $p-q\equiv 4\pmod{8}$, and also admit the
groups $\pin^{-,+,-}(p,q)$ and $\pin^{-,-,+}(p,q)$ if in the direct sums
there are addendums of the type $p-q\equiv 2\pmod{8}$ 
or $p-q\equiv 6\pmod{8}$.\\[0.2cm]
2) A Cliffordian group
\[
\pin^{-,-,-}(p,q)\simeq\frac{(\spin_0(p,q)\odot Q_4)}{\dZ_2},
\]
exists if $\K\simeq\R$ ($p-q\equiv 2\pmod{8}$) and $p\equiv 3\pmod{4},\,
q\equiv 1\pmod{4}$, and also if $p-q\equiv 6\pmod{8}$ and $\K\simeq\BH$.
The algebras $\cl_{p,q}$ with the rings 
$\K\simeq\R\oplus\R,\,\K\simeq\BH\oplus\BH$ ($p-q\equiv 1,5\pmod{8}$)
admit the group $\pin^{-,-,-}(p,q)$ if in the direct sums there are 
addendums of the type
$p-q\equiv 2\pmod{8}$ or $p-q\equiv 6\pmod{8}$. The types $p-q\equiv 3,7
\pmod{8}$, $\K\simeq\C$ admit a Cliffordian group $\pin^{-,-,-}(p+q-1,\C)$,
if $p\equiv 1\pmod{2}$ and $q\equiv 0\pmod{2}$. Further, Cliffordian groups
\[
\pin^{a,b,c}(p,q)\simeq\frac{(\spin_0(p,q)\odot D_4)}{\dZ_2},
\]
with $(a,b,c)=(-,+,+)$ exist if $\K\simeq\R$ ($p-q\equiv 2\pmod{8}$)
and $p\equiv 1\pmod{4},\,q\equiv 3\pmod{4}$,
and also if $p-q\equiv 6\pmod{8}$ and $\K\simeq\BH$. Cliffordian groups with
the signatures
$(a,b,c)=(+,-,+)$ and $(a,b,c)=(+,+,-)$ exist over the ring
$\K\simeq\R$ ($p-q\equiv 0\pmod{8}$) if
$p,q\equiv 3\pmod{4}$ and $p,q\equiv 1\pmod{4}$, respectively,
and also these groups
exist over the ring $\K\simeq\BH$ if $p-q\equiv 4\pmod{8}$.
The algebras $\cl_{p,q}$ with the rings 
$\K\simeq\R\oplus\R,\,\K\simeq\BH\oplus\BH$ ($p-q\equiv 1,5\pmod{8}$)
admit the group $\pin^{-,+,+}(p,q)$ if in the direct sums there are addendums
of the type $p-q\equiv 2\pmod{8}$ or $p-q\equiv 6\pmod{8}$, and also admit the
groups $\pin^{+,-,+}(p,q)$ and $\pin^{+,+,-}(p,q)$ if in the direct sums there
are addendums of the type $p-q\equiv 0\pmod{8}$ or $p-q\equiv 4\pmod{8}$.
\end{theorem}\section{The structure of $\pin(p,q)\not\simeq\pin(q,p)$}
It is easy to see that the definitions (\ref{Pin}) and (\ref{Pinabc}) are
equivalent. Moreover, Salingaros showed \cite{Sal81a,Sal82,Sal84} that there
are isomorphisms $\dZ_2\otimes\dZ_2\simeq\cl_{1,0}$ and $\dZ_4\simeq\cl_{0,1}$.
Further, since $\cl^+_{p,q}\simeq\cl^+_{q,p}$, in accordance with the
definition (\ref{Spin}), it follows that $\spin(p,q)\simeq\spin(q,p)$.
On the other hand, since in a general case $\cl_{p,q}\not\simeq\cl_{q,p}$,
from the definition (\ref{Pin}) it follows that $\pin(p,q)\not\simeq\pin(q,p)$
(or $\pin^{a,b,c}(p,q)\not\simeq\pin^{a,b,c}(q,p)$). In connection with this,
some authors \cite{CDD82,DM88,KT89,CGT95,CGT98,FrTr99} 
used notations $\pin^+\simeq\pin(p,q)$ and
$\pin^-\simeq\pin(q,p)$. In Theorems \ref{tautr} and \ref{tgroupr}, we
have establish a relation between the signatures $(p,q)=(\underbrace{+,+,
\ldots,+}_{p\,\text{times}},\underbrace{-,-,\ldots,-}_{q\,\text{times}})$
of the spaces $\R^{p,q}$ and the signatures $(a,b,c)$ of the automorphism
groups of $\cl_{p,q}$ and corresponding D\c{a}browski groups. This relation
allows to completely define the structure of the inequality $\pin(p,q)\not
\simeq\pin(q,p)$ ($\pin^+\not\simeq\pin^-$). Indeed, from (\ref{Pinabc}) and
(\ref{Spin+}), it follows that $\spin_0(p,q)\simeq\spin_0(q,p)$, therefore,
a nature of the inequality $\pin(p,q)\not\simeq\pin(q,p)$ wholly lies in
the double covering $C^{a,b,c}$ of the discrete subgroup. For example,
in accordance with Theorem \ref{tgroupr} for the type $p-q\equiv 2\pmod{8}$
with the division ring $\K\simeq\R$ there exist the groups $\pin^{a,b,c}(p,q)
\simeq\pin^+$, where double coverings of the discrete subgroup have the form:
1) $C^{-,-,-}\simeq Q_4$ if $p\equiv 3\pmod{4}$ and $q\equiv 1\pmod{4}$;
2) $C^{-,+,+}\simeq D_4$ if $p\equiv 1\pmod{4}$ and $q\equiv 3\pmod{4}$;
3) $C^{-,-,+}\simeq\dZ_2\otimes\dZ_4$ if $p\equiv 0\pmod{4}$ and $q\equiv 2
\pmod{4}$; 4) $C^{-,+,-}\simeq\dZ_2\otimes\dZ_4$ if $p\equiv 2\pmod{4}$ and
$q\equiv 0\pmod{4}$. Whereas the groups with opposite signature, $\pin^-
\simeq\pin^{a,b,c}(q,p)$, have the type $q-p\equiv 6\pmod{8}$ with the
ring $\K\simeq\BH$. In virtue of the more wide ring $\K\simeq\BH$, there
exists a far greater choice of the discrete subgroups for each concrete
kind of $\pin^{a,b,c}(q,p)$. Thus,
\[
\begin{array}{rcl}
\pin^{a,b,c}(p,q)&\not\simeq&\pin^{a,b,c}(q,p)\\
p-q\equiv 2\pmod{8}&&q-p\equiv 6\pmod{8}.
\end{array}
\]
Further, the type $p-q\equiv 1\pmod{8}$ with the ring $\K\simeq\R\oplus\R$
in virtue of Theorems \ref{tautr} and \ref{tgroupr} admits the group
$\pin^{a,b,c}(p,q)\simeq\pin^+$, where the double covering $C^{a,b,c}$
adopts all the eight possible values. Whereas the opposite type
$q-p\equiv 7\pmod{8}$ with the ring $\K\simeq\C$ admits the group
$\pin^{a,b,c}(q,p)\simeq\pin^-$, where for the double covering $C^{a,b,c}$ of
the discrete subgroup there are only two possibilities: 1) $C^{+,+,+}\simeq
\dZ_2\otimes\dZ_2\otimes\dZ_2$ if $p\equiv 0\pmod{2}$ and $q\equiv 1\pmod{2}$;
2) $C^{-,-,-}\simeq Q_4$ if $p\equiv 1\pmod{2}$ and $q\equiv 0\pmod{2}$.
The analogous situation takes place for the two mutually opposite types
$p-q\equiv 3\pmod{8}$ with $\K\simeq\C$ and $q-p\equiv 5\pmod{8}$ with
$\K\simeq\BH\oplus\BH$. Therefore,
\[
\begin{array}{rcl}
\pin^{a,b,c}(p,q)&\not\simeq&\pin^{a,b,c}(q,p)\\
p-q\equiv 1\pmod{8}&&q-p\equiv 7\pmod{8};\\
\pin^{a,b,c}(p,q)&\not\simeq&\pin^{a,b,c}(q,p)\\
p-q\equiv 3\pmod{8}&&q-p\equiv 5\pmod{8}.
\end{array}
\]
It is easy to see that an opposite type to the type $p-q\equiv 0\pmod{8}$
with the ring $\K\simeq\R$ is the same type $q-p\equiv 0\pmod{8}$. Therefore,
in virtue of Theorems \ref{tautr} and \ref{tgroupr} double coverings
$C^{a,b,c}$ for the groups $\pin^{a,b,c}(p,q)\simeq\pin^+$ and $\pin^{a,b,c}
(q,p)\simeq\pin^-$ coincide. The same is the situation for the type
$p-q\equiv 4\pmod{8}$ with $\K\simeq\BH$, which has the opposite type
$q-p\equiv 4\pmod{8}$. Thus,
\[
\begin{array}{rcl}
\pin^{a,b,c}(p,q)&\simeq&\pin^{a,b,c}(q,p)\\
p-q\equiv 0\pmod{8}&&q-p\equiv 0\pmod{8};\\
\pin^{a,b,c}(p,q)&\simeq&\pin^{a,b,c}(q,p)\\
p-q\equiv 4\pmod{8}&&q-p\equiv 4\pmod{8}.
\end{array}
\]
We will call the types $p-q\equiv 0\pmod{8}$ and $p-q\equiv 4\pmod{8}$,
which coincide with their opposite types, {\it neutral types}.\\[0.3cm]
{\bf Example}. Let us consider a structure of the inequality $\pin(3,1)\not
\simeq\pin(1,3)$. The groups $\pin(3,1)$ and $\pin(1,3)$ are two different
double coverings of the general Lorentz group. These groups play an
important role in physics \cite{CWM88,DW90,DG90,DWGK}. 
As follows from (\ref{Pin})
the group $\pin(3,1)$ is completely defined in the framework of the
Majorana algebra $\cl_{3,1}$, which has the type $p-q\equiv 2\pmod{8}$ and
the division ring $\K\simeq\R$. As noted previously, the structure of the
inequality $\pin(p,q)\not\simeq\pin(q,p)$ is defined by the double covering
$C^{a,b,c}$. From Theorems \ref{tautr} and \ref{tgroupr}, it follows that
the algebra $\cl_{3,1}\simeq\M_4(\R)$ admits one and only 
one group $\pin^{-,-,-}(3,1)$, where a double covering of the discrete
subgroup has a form $C^{-,-,-}\simeq Q_4$. Indeed, let us consider a matrix
representation of the units of $\cl_{3,1}$, using the Maple V and the
{\sc CLIFFORD} package developed by Rafa\l Ab\l amowicz 
\cite{Abl96,Abl98,Abl00}. Let $f=\frac{1}{4}(1+\e_1)(1+\e_{34})$ be a
primitive idempotent of the algebra $\cl_{3,1}$ (prestored idempotent for
$\cl_{3,1}$ in {\sc CLIFFORD}), then a following {\sc CLIFFORD} command
sequence gives:\\[0.3cm]
{\tt
>\hs restart:with(Cliff4):with(double):\\[0.1cm]
>\hs dim := 4:\\[0.1cm]
>\hs eval(makealiases(dim)):\\[0.1cm]
>\hs B := linalg(diag(1,1,1,-1)):\\[0.1cm]
>\hs clibasis := cbasis(dim):\\[0.1cm]
>\hs data := clidata(B):\\[0.1cm]
>\hs f := data[4]:\\[0.1cm]
>\hs left\_sbasis := minimalideal(clibasis,f,'left'):\\[0.1cm]
>\hs Kbasis := Kfield(left\_sbasis,f):\\[0.1cm]
>\hs SBgens := left\_sbasis[2]:FBgens := Kbasis[2]:\\[0.1cm]
>\hs K\_basis := spinorKbasis(SBgens,f,FBgens,'left'):\\[0.1cm]
>\hs for i from 1 to 4 do}
\begin{gather}
{\tt E[i] := spinorKrepr(e.i.,K\_basis[1],FBgens,'left') od;}\nonumber\\
\ar
E_1:=\begin{bmatrix}
Id & 0 & 0 & 0\\
0 &-Id & 0 & 0\\
0 & 0 &-Id & 0\\
0 & 0 & 0 & Id
\end{bmatrix},\quad
E_2:=\begin{bmatrix}
0 & Id & 0 & 0\\
Id& 0 & 0 & 0\\
0 & 0 & 0 & Id\\
0 & 0 & Id& 0
\end{bmatrix},\nonumber\\
E_3:=\ar\begin{bmatrix}
0 & 0 & Id & 0\\
0 & 0 & 0 &-Id\\
Id & 0 & 0 & 0\\
0 &-Id & 0 & 0
\end{bmatrix},\quad
E_4:=\begin{bmatrix}
0 & 0 & -Id & 0\\
0 & 0 & 0 & Id\\
Id & 0 & 0 & 0\\
0 & -Id & 0 & 0
\end{bmatrix}.\label{matr}
\end{gather}
It is easy to see that the matrices (\ref{matr}) build up a basis of the
form (\ref{B1}). Since the condition $pq\equiv 1\pmod{2}$ is satisfied for
the algebra $\cl_{3,1}$, the automorphism group $\sAut(\cl_{3,1})$ is
non--Abelian. In accordance with (\ref{commut}), the matrix $\sE$ should 
commute with a symmetric part of the basis (\ref{matr}) and anticommute
with a skewsymmetric part of (\ref{matr}). In this case, as follows from
(\ref{B2})--(\ref{B5}) and (\ref{matr}), the matrix $\sE$ is a product of
$p=3$ symmetric matrices, that is,
\[
\sE=\cE_1\cE_2\cE_3=\ar\begin{pmatrix}
0 & 0 & 0 & -1\\
0 & 0 & -1& 0\\
0 & 1 & 0 & 0\\
1 & 0 & 0 & 0
\end{pmatrix}.
\]
Further, matrices of the automorphisms $\star$ and $\widetilde{\star}$ for
the basis (\ref{matr}) have a form
\[
\sW=\cE_1\cE_2\cE_3\cE_4=\ar\begin{pmatrix}
0 & 1 & 0 & 0\\
-1& 0 & 0 & 0\\
0 & 0 & 0 & 1\\
0 & 0 &-1 & 0
\end{pmatrix},\quad\sC=\sE\sW=\begin{pmatrix}
0 & 0 & 1 & 0\\
0 & 0 & 0 & -1\\
-1 & 0 & 0 & 0\\
0 & 1 & 0 & 0
\end{pmatrix}.
\]
Thus, a group of the fundamental automorphisms of the algebra $\cl_{3,1}$
in the matrix representation is defined by a finite group $\{\sI,\sW,\sE,\sC\}
\sim\{I,\cE_{1234},\cE_{123},\cE_4\}$. The multiplication table of this
group has a form
\begin{equation}\label{tab}
{\renewcommand{\arraystretch}{1.4}
\begin{tabular}{|c||c|c|c|c|}\hline
             & $I$ & $\cE_{1234}$ & $\cE_{123}$ & $\cE_4$ \\ \hline\hline
$I$          & $I$ & $\cE_{1234}$ & $\cE_{123}$ & $\cE_4$ \\ \hline
$\cE_{1234}$ & $\cE_{1234}$ & $-I$ & $\cE_4$ & $-\cE_{123}$ \\ \hline
$\cE_{123}$ & $\cE_{123}$ & $-\cE_4$ & $-I$ & $\cE_{1234}$ \\ \hline
$\cE_4$ & $\cE_4$ & $\cE_{123}$ & $-\cE_{1234}$ & $-I$ \\ \hline
\end{tabular}\;\sim\;
\begin{tabular}{|c||c|c|c|c|}\hline
      & $\sI$ & $\sW$ & $\sE$ & $\sC$ \\ \hline\hline
$\sI$ & $\sI$ & $\sW$ & $\sE$ & $\sC$ \\ \hline
$\sW$ & $\sW$ & $-\sI$& $\sC$ & $-\sE$ \\ \hline
$\sE$ & $\sE$ & $-\sC$& $-\sI$& $\sW$ \\ \hline
$\sC$ & $\sC$ & $\sE$ & $-\sW$& $-\sI$ \\ \hline 
\end{tabular}
}.
\end{equation}
From the table, it follows that $\sAut_+(\cl_{3,1})\simeq\{\sI,\sW,\sE,\sC\}
\simeq Q_4/\dZ_2$ and, therefore, the algebra $\cl_{3,1}$ admits a
Cliffordian group $\pin^{-,-,-}(3,1)$ (Theorem \ref{tgroupr}). It is easy
to verify that the double covering $C^{-,-,-}\simeq Q_4$ is an invariant
fact for the algebra $\cl_{3,1}$, that is, $C^{-,-,-}$ does not depend on the
choice of the matrix representation. Indeed, for each  two commuting
elements of the algebra $\cl_{3,1}$ there exist four different primitive
idempotents that generate four different matrix representations of $\cl_{3,1}$.
The invariability of the previously mentioned fact is easily verified with the
help of a procedure {\tt commutingelements} of the {\sc CLIFFORD} package,
which allows to consider in sequence all the possible primitive idempotents
of the algebra $\cl_{3,1}$ and their corresponding matrix representations.

Now, let us consider discrete subgroups of the double covering $\pin(1,3)$.
The group $\pin(1,3)$, in turn, is completely constructed within the
spacetime algebra $\cl_{1,3}$ that has the opposite (in relation to the
Majorana algebra $\cl_{3,1}$) type $p-q\equiv 6\pmod{8}$ with the division
ring $\K\simeq\BH$. According to Wedderburn--Artin Theorem, in this case
there is an isomorphism $\cl_{1,3}\simeq\M_2(\BH)$. The following
{\sc CLIFFORD} command sequence allows to find matrix representations of the
units of the algebra $\cl_{1,3}$ for a prestored primitive idempotent
$f=\frac{1}{2}(1+\e_{14})$:
\begin{eqnarray}
\text{\tt >}&&\text{\tt restart:with(Cliff4):with(double):}\\
\text{\tt >}&&\text{\tt dim := 4: eval(makealiases(dim)):}\\
\text{\tt >}&&\text{\tt B := linalg(diag(1,-1,-1,-1)):}\\
\text{\tt >}&&\text{\tt clibasis := cbasis(dim):}\\
\text{\tt >}&&\text{\tt data := clidata(B): f := data[4]:}\\
\text{\tt >}&&\text{\tt left\_sbasis := minimalideal(clibasis,f,'left'):}
\label{m6}\\
\text{\tt >}&&\text{\tt Kbasis := Kfield(left\_sbasis,f):}\\
\text{\tt >}&&\text{\tt SBgens := left\_sbasis[2]: FBgens := Kbasis[2]:}\\
\text{\tt >}&&\text{\tt K\_basis := spinorKbasis(SBgens,f,FBgens,'left'):}\\
\text{\tt >}&&\text{\tt for i from 1 to 4 do}
\end{eqnarray}
\begin{gather}
\text{\tt E[i] := spinorKrepr(e.i.,K\_basis[1],FBgens,'left') od;}\label{m10}\\
E_1:=\ar\begin{bmatrix}
0 & Id\\
Id & 0
\end{bmatrix},\quad
E_2:=\begin{bmatrix}
e2 & 0\\
0 & -e2
\end{bmatrix},\quad
E_3:=\begin{bmatrix}
e3 & 0\\
0 & -e3
\end{bmatrix},\quad
E_4:=\begin{bmatrix}
0 & -Id\\
Id & 0
\end{bmatrix}.\label{matr2}
\end{gather}
At this point, the division ring $\K\simeq\BH$ is generated by a set
$\{1,\e_2,\e_3,\e_{23}\}\simeq\{1,\bi,\bj,\bk\}$, where $\bi,\bj,\bk$ are
well-known quaternion units. The basis (\ref{matr2}) contains three
symmetric matrices and one skewsymmetric matrix. Therefore, in accordance
with (\ref{commut}) and (\ref{C1})--(\ref{C6}) the matrix of the
antiautomorphism $\cA\rightarrow\widetilde{\cA}$ is a product of symmetric
matrices of the basis (\ref{matr2}). Thus,
\begin{equation}\label{set}
\sW=\cE_1\cE_2\cE_3\cE_4=\ar\begin{pmatrix}
\bk & 0\\
0 & -\bk
\end{pmatrix},\quad
\sE=\cE_1\cE_2\cE_3=\begin{pmatrix}
0 & \bk\\
\bk & 0
\end{pmatrix},\quad
\sC=\sE\sW=\begin{pmatrix}
0 & 1\\
-1 & 0
\end{pmatrix}.
\end{equation}
It is easy to verify that a set of the matrices (\ref{set}) added by the
unit matrix forms the non--Abelian group $\sAut_+(\cl_{1,3})\simeq
Q_4/\dZ_2$ with a multiplication table of the form (\ref{tab}). Therefore,
the spacetime algebra $\cl_{1,3}$ admits the Cliffordian group
$\pin^{-,-,-}(1,3)$, where a double covering of the discrete subgroup has
the form $C^{-,-,-}\simeq Q_4$. However, as follows from Theorem
\ref{tautr}, in virtue of the more wide ring $\K\simeq\BH$ the group
$\pin^{-,-,-}(1,3)$ does not the only possible for the algebra
$\cl_{1,3}\simeq\M_2(\BH)$. Indeed, looking over all the possible commuting
elements of the algebra $\cl_{1,3}$ we find with the help of the procedure
{\tt commutingelements} that\\
{\tt >\hs L1 := commutingelements(clibasis);}
\begin{equation}\label{m11}
L1:=[e1]
\end{equation}
{\tt >\hs L2 := commutingelements(remove(member,clibasis,L1));}
\begin{equation}\label{m12}
L2:=[e12]
\end{equation}
{\tt >\hs L3 := commutingelements(remove(member,clibasis,[op(L1),op(L2)]));}
\begin{equation}\label{m13}
L3:=[e13]
\end{equation}
{\tt >\hs L4 := commutingelements(remove(member,clibasis,[op(L1),op(L2),\\
\phantom{>\hs}op(L3)]));}
\begin{equation}\label{m14}
L4:=[e14]
\end{equation}
{\tt >\hs L5 := commutingelements(remove(member,clibasis,[op(L1),op(L2),\\
\phantom{>\hs}op(L3),op(L4)]));}
\begin{equation}\label{m15}
L5:=[e234]
\end{equation}
{\tt >\hs f := cmulQ((1/2)$\ast$(Id + e2we3we4);}
\begin{equation}\label{m16}
f:=\frac{1}{2}Id+\frac{1}{2}e234
\end{equation}
{\tt >\hs type(f,primitiveidemp);}
\begin{equation}\label{m17}
true
\end{equation}
It is easy to verify that primitive idempotents $\frac{1}{2}(1\pm\e_1),\,
\frac{1}{2}(1\pm\e_{12}),\,\frac{1}{2}(1\pm\e_{13})$ and $\frac{1}{2}(1\pm
\e_{14})$ constructed by means of the commuting elements $\e_1,\,\e_{12},\,
\e_{13}$ and $\e_{14}$ generate matrix representations that give rise to
the group $\sAut_+(\cl_{1,3})\simeq Q_4/\dZ_2$. However, the situation 
changes for the element $\e_{234}$ and corresponding primitive idempotent
$\frac{1}{2}(1+\e_{234})$ ($\frac{1}{2}(1-\e_{234})$). Indeed, executing
the commands (\ref{m16}) and (\ref{m17}) and subsequently the commands
(\ref{m6})--(\ref{m10}), we find that\\
{\tt >\hs for i from 1 to 4 do}
\begin{gather}
\text{\tt E[i] := spinorKrepr(e.i.,Kbasis[1],FBgens,'left') od;}\nonumber\\
E_1:=\ar\begin{bmatrix}
0 & Id\\
Id & 0
\end{bmatrix},\quad
E_2:=\begin{bmatrix}
e2 & 0\\
0 & -e2
\end{bmatrix},\quad
E_3:=\begin{bmatrix}
e34 & 0\\
0 & -e34
\end{bmatrix},\quad
E_4:=\begin{bmatrix}
e4 & 0\\
0 & -e4
\end{bmatrix},\label{matr3}
\end{gather}
where the division ring $\K\simeq\BH$ is generated by a set $\{1,\e_2,\e_4,
\e_{24}\}\simeq\{1,\bi,\bj,\bk\}$. The basis (\ref{matr3}) consists of 
symmetric matrices only. Therefore, in accordance with (\ref{commut}), the
matrix $\sE$ should commute with all the matrices of the basis
(\ref{matr3}). It is obvious that this condition is satisfied only if $\sE$
is proportional to the unit matrix (recall that any element of the
automorphism group may be multiplied by an arbitrary factor $\eta\in\F$, in
this case $\F=\R$). Further, a set of the matrices $\sW=\cE_1\cE_2\cE_3\cE_4$,
$\sE\sim\sI$, $\sC=\sE\sW$ added by the unit matrix forms a finite group
with a following multiplication table
\[
{\renewcommand{\arraystretch}{1.4}
\begin{tabular}{|c||c|c|c|c|}\hline
      & $\sI$ & $\sW$ & $\sE$ & $\sC$\\ \hline\hline
$\sI$ & $\sI$ & $\sW$ & $\sE$ & $\sC$\\ \hline
$\sW$ & $\sW$ & $-\sI$& $\sC$ & $-\sE$\\ \hline
$\sE$ & $\sE$ & $\sC$ & $\sI$ & $\sW$\\ \hline 
$\sC$ & $\sC$ & $-\sE$& $\sW$ & $-\sI$\\ \hline 
\end{tabular}
}.
\]
As follows from the table, we have in this case the Abelian group
$\sAut_-(\cl_{1,3})\simeq\dZ_4$ with the signature $(-,+,-)$. Thus, the
spacetime algebra $\cl_{1,3}$ admits the group $\pin^{-,+,-}(1,3)$, where
a double covering of the discrete subgroup has the form 
$C^{-,+,-}\simeq\dZ_2\otimes\dZ_4$.

The fulfilled analysis explicitly shows a difference between the two
double coverings $\pin(3,1)$ and $\pin(1,3)$ of the Lorentz group. Since
double coverings of the connected components of both groups $\pin(3,1)$
and $\pin(1,3)$ are isomorphic, $\spin_0(3,1)\simeq\spin_0(1,3)$, the
nature of the difference between them consists in a concrete form and a
number of the double covering $C^{a,b,c}$ of the discrete subgroups. So,
for the Majorana algebra $\cl_{3,1}$, all the existing primitive idempotents
$\frac{1}{4}(1\pm\e_1)(1\pm\e_{34}),\,\frac{1}{4}(1\pm\e_1)(1\pm\e_{24}),\,
\frac{1}{4}(1\pm\e_2)(1\pm\e_{14}),\,\frac{1}{4}(1\pm\e_3)(1\pm\e_{134}),\,
\frac{1}{4}(1\pm\e_{34})(1\pm\e_{234})$ generate 20 matrix representations,
each of which gives rise to the double covering $C^{-,-,-}\simeq Q_4$.
On the other hand, for the spacetime algebra $\cl_{1,3}$ primitive
idempotents $\frac{1}{2}(1\pm\e_{14}),\,\frac{1}{2}(1\pm\e_1),\,
\frac{1}{2}(1\pm\e_{12}),\,\frac{1}{2}(1\pm\e_{13})$ generate 8 matrix
representations with $C^{-,-,-}\simeq Q_4$, whereas remaining two primitive
idempotents $\frac{1}{2}(1\pm\e_{234})$ generate matrix representations
with $C^{-,+,-}\simeq\dZ_2\otimes\dZ_4$.\\[0.3cm]
{\bf Remark}. Physicists commonly use a transition from some given
signature to its opposite (signature change) by means of a replacement
$\cE_i\rightarrow i\cE_i$ (so--called Wick rotation). However, such a
transition is unsatisfactory from a mathematical viewpoint. For example,
we can use the replacement $\cE_i\rightarrow i\cE_i$ for a transition
from the spacetime algebra $\cl_{1,3}\simeq\M_2(\BH)$ to the Majorana
algebra $\cl_{3,1}\simeq\M_4(\R)$ since $i\in\M_2(\BH)$, whereas an inverse
transition $\cl_{3,1}\rightarrow\cl_{1,3}$ can not be performed by the
replacement $\cE_i\rightarrow i\cE_i$, since $i\not\in\M_4(\R)$. The
mathematically correct alternative to the Wick rotation is a
tilt--transformation introduced by Lounesto \cite{Lou93}. The
tilt--transformation is expressed by a map $ab\rightarrow a_+b_++b_+a_-+
b_-a_+-b_-a_-$, where $a_\pm,b_\pm\in\cl^\pm_{p,q}$. The further developing
of the tilt--transformation and its application for a formulation of
physical theories in the spaces with different signatures has been
considered in the recent paper \cite{MPV00}.
\section{Discrete transformations and Brauer--Wall groups}
The algebra $\cl$ is naturally $\dZ_2$--graded. Let
$\cl^+$ (correspondingly $\cl^-$) be a set consisting of all even
(correspondingly odd) 
elements of the algebra $\cl$. The set $\cl^+$ is a subalgebra of
$\cl$. It is obvious that
$\cl=\cl^+\oplus\cl^-$, and also $\cl^+\cl^+
\subset\cl^+,\,\cl^+\cl^-\subset\cl^-,\,
\cl^-\cl^+\subset\cl^-,\,\cl^-\cl^-\subset
\cl^+$. A degree $\deg a$ of the even (correspondingly odd) 
element $a\in\cl$ is equal
to 0 (correspondingly 1). 
Let $\mathfrak{A}$ and $\mathfrak{B}$ be the two associative
$\dZ_2$--graded algebras over the field $\F$; then a multiplication of
homogeneous elements
$\mathfrak{a}^\prime\in\mathfrak{A}$ and $\mathfrak{b}\in\mathfrak{B}$ in a
graded tensor product
$\mathfrak{A}\hat{\otimes}\mathfrak{B}$ is defined as follows: 
$(\mathfrak{a}\otimes \mathfrak{b})(\mathfrak{a}^\prime
\otimes \mathfrak{b}^\prime)=(-1)^{\deg\mathfrak{b}\deg\mathfrak{a}^\prime}
\mathfrak{a}\mathfrak{a}^\prime\otimes\mathfrak{b}\mathfrak{b}^\prime$.
The graded tensor product of the two graded central simple algebras is also
graded central simple 
\cite[Theorem 2]{Wal64}. The Clifford algebra $\cl_{p,q}$ is central simple
if $p-q\not\equiv 1,5\pmod{8}$. It is known that for the Clifford algebra
with odd dimensionality, the isomorphisms are as follows:
$\cl^+_{p,q+1}\simeq\cl_{p,q}$ and $\cl^+_{p+1,q}\simeq\cl_{q,p}$ 
\cite{Rash,Port}. Thus, $\cl^+_{p,q+1}$ and $\cl^+_{p+1,q}$
are central simple algebras. Further, in accordance with Chevalley Theorem
\cite{Che55} for the graded tensor product there is an isomorphism
$\cl_{p,q}\hat{\otimes}\cl_{p^{\p},q^{\p}}\simeq
\cl_{p+p^{\p},q+q^{\p}}$. Two algebras $\cl_{p,q}$ and $\cl_{p^{\p},q^{\p}}$
are said to be of the same class if $p+q^{\p}\equiv p^{\p}+q\pmod{8}$.
The graded central simple Clifford algebras over the field $\F=\R$
form eight similarity classes, which, as it is easy to see, coincide
with the eight types of the algebras $\cl_{p,q}$.
The set of these 8 types (classes) forms a Brauer--Wall group $BW_{\R}$
\cite{Wal64} that is isomorphic to a cyclic group $\dZ_8$. Thus, the
algebra $\cl_{p,q}$ is an element of the Brauer--Wall group, and a group
operation is the graded tensor product $\hat{\otimes}$.
A cyclic structure of the group $BW_{\R}\simeq\dZ_8$ may be represented on
the Trautman diagram (spinorial clock) \cite{BTr87,BT88} (Fig. 1) by means
of a transition $\cl^+_{p,q}\stackrel{h}{\longrightarrow}\cl_{p,q}$ 
(the round on the diagram is realized by an hour--hand). At this point, the type
of the algebra is defined on the diagram by an equality
$q-p=h+8r$, where $h\in\{0,\ldots,7\}$, $r\in\dZ$.

\[
\unitlength=0.5mm
\begin{picture}(100.00,110.00)

\put(97,67){$\C$}\put(105,64){$p-q\equiv 7\!\!\!\!\pmod{8}$}
\put(80,80){1}
\put(75,93.3){$\cdot$}
\put(75.5,93){$\cdot$}
\put(76,92.7){$\cdot$}
\put(76.5,92.4){$\cdot$}
\put(77,92.08){$\cdot$}
\put(77.5,91.76){$\cdot$}
\put(78,91.42){$\cdot$}
\put(78.5,91.08){$\cdot$}
\put(79,90.73){$\cdot$}
\put(79.5,90.37){$\cdot$}
\put(80,90.0){$\cdot$}
\put(80.5,89.62){$\cdot$}
\put(81,89.23){$\cdot$}
\put(81.5,88.83){$\cdot$}
\put(82,88.42){$\cdot$}
\put(82.5,87.99){$\cdot$}
\put(83,87.56){$\cdot$}
\put(83.5,87.12){$\cdot$}
\put(84,86.66){$\cdot$}
\put(84.5,86.19){$\cdot$}
\put(85,85.70){$\cdot$}
\put(85.5,85.21){$\cdot$}
\put(86,84.69){$\cdot$}
\put(86.5,84.17){$\cdot$}
\put(87,83.63){$\cdot$}
\put(87.5,83.07){$\cdot$}
\put(88,82.49){$\cdot$}
\put(88.5,81.9){$\cdot$}
\put(89,81.29){$\cdot$}
\put(89.5,80.65){$\cdot$}
\put(90,80){$\cdot$}
\put(90.5,79.32){$\cdot$}
\put(91,78.62){$\cdot$}
\put(91.5,77.89){$\cdot$}
\put(92,77.13){$\cdot$}
\put(92.5,76.34){$\cdot$}
\put(93,75.51){$\cdot$}
\put(93.5,74.65){$\cdot$}
\put(94,73.74){$\cdot$}
\put(94.5,72.79){$\cdot$}
\put(96.5,73.74){\vector(1,-2){1}}
\put(80,20){3}
\put(97,31){$\BH$}\put(105,28){$p-q\equiv 6\!\!\!\!\pmod{8}$}
\put(75,6.7){$\cdot$}
\put(75.5,7){$\cdot$}
\put(76,7.29){$\cdot$}
\put(76.5,7.6){$\cdot$}
\put(77,7.91){$\cdot$}
\put(77.5,8.24){$\cdot$}
\put(78,8.57){$\cdot$}
\put(78.5,8.91){$\cdot$}
\put(79,9.27){$\cdot$}
\put(79.5,9.63){$\cdot$}
\put(80,10){$\cdot$}
\put(80.5,10.38){$\cdot$}
\put(81,10.77){$\cdot$}
\put(81.5,11.17){$\cdot$}
\put(82,11.58){$\cdot$}
\put(82.5,12.00){$\cdot$}
\put(83,12.44){$\cdot$}
\put(83.5,12.88){$\cdot$}
\put(84,13.34){$\cdot$}
\put(84.5,13.8){$\cdot$}
\put(85,14.29){$\cdot$}
\put(85.5,14.79){$\cdot$}
\put(86,15.3){$\cdot$}
\put(86.5,15.82){$\cdot$}
\put(87,16.37){$\cdot$}
\put(87.5,16.92){$\cdot$}
\put(88,17.5){$\cdot$}
\put(88.5,18.09){$\cdot$}
\put(89,18.71){$\cdot$}
\put(89.5,19.34){$\cdot$}
\put(90,20){$\cdot$}
\put(90.5,20.68){$\cdot$}
\put(91,21.38){$\cdot$}
\put(91.5,22.11){$\cdot$}
\put(92,22.87){$\cdot$}
\put(92.5,23.66){$\cdot$}
\put(93,24.48){$\cdot$}
\put(93.5,25.34){$\cdot$}
\put(94,26.25){$\cdot$}
\put(94.5,27.20){$\cdot$}
\put(95,28.20){$\cdot$}
\put(20,80){7}
\put(25,93.3){$\cdot$}
\put(24.5,93){$\cdot$}
\put(24,92.7){$\cdot$}
\put(23.5,92.49){$\cdot$}
\put(23,92.08){$\cdot$}
\put(22.5,91.75){$\cdot$}
\put(22,91.42){$\cdot$}
\put(21.5,91.08){$\cdot$}
\put(21,90.73){$\cdot$}
\put(20.5,90.37){$\cdot$}
\put(20,90){$\cdot$}
\put(19.5,89.62){$\cdot$}
\put(19,89.23){$\cdot$}
\put(18.5,88.83){$\cdot$}
\put(18,88.42){$\cdot$}
\put(17.5,87.99){$\cdot$}
\put(17,87.56){$\cdot$}
\put(16.5,87.12){$\cdot$}
\put(16,86.66){$\cdot$}
\put(15.5,86.19){$\cdot$}
\put(15,85.70){$\cdot$}
\put(14.5,85.21){$\cdot$}
\put(14,84.69){$\cdot$}
\put(13.5,84.17){$\cdot$}
\put(13,83.63){$\cdot$}
\put(12.5,83.07){$\cdot$}
\put(12,82.49){$\cdot$}
\put(11.5,81.9){$\cdot$}
\put(11,81.29){$\cdot$}
\put(10.5,80.65){$\cdot$}
\put(10,80){$\cdot$}
\put(9.5,79.32){$\cdot$}
\put(9,78.62){$\cdot$}
\put(8.5,77.89){$\cdot$}
\put(8,77.13){$\cdot$}
\put(7.5,76.34){$\cdot$}
\put(7,75.51){$\cdot$}
\put(6.5,74.65){$\cdot$}
\put(6,73.79){$\cdot$}
\put(5.5,72.79){$\cdot$}
\put(5,71.79){$\cdot$}
\put(20,20){5}
\put(25,6.7){$\cdot$}
\put(24.5,7){$\cdot$}
\put(24,7.29){$\cdot$}
\put(23.5,7.6){$\cdot$}
\put(23,7.91){$\cdot$}
\put(22.5,8.24){$\cdot$}
\put(22,8.57){$\cdot$}
\put(21.5,8.91){$\cdot$}
\put(21,9.27){$\cdot$}
\put(20.5,9.63){$\cdot$}
\put(20,10){$\cdot$}
\put(19.5,10.38){$\cdot$}
\put(19,10.77){$\cdot$}
\put(18.5,11.17){$\cdot$}
\put(18,11.58){$\cdot$}
\put(17.5,12){$\cdot$}
\put(17,12.44){$\cdot$}
\put(16.5,12.88){$\cdot$}
\put(16,13.34){$\cdot$}
\put(15.5,13.8){$\cdot$}
\put(15,14.29){$\cdot$}
\put(14.5,14.79){$\cdot$}
\put(14,15.3){$\cdot$}
\put(13.5,15.82){$\cdot$}
\put(13,16.37){$\cdot$}
\put(12.5,16.92){$\cdot$}
\put(12,17.5){$\cdot$}
\put(11.5,18.09){$\cdot$}
\put(11,18.71){$\cdot$}
\put(10.5,19.34){$\cdot$}
\put(10,20){$\cdot$}
\put(9.5,20.68){$\cdot$}
\put(9,21.38){$\cdot$}
\put(8.5,22.11){$\cdot$}
\put(8,22.87){$\cdot$}
\put(7.5,23.66){$\cdot$}
\put(7,24.48){$\cdot$}
\put(6.5,25.34){$\cdot$}
\put(6,26.25){$\cdot$}
\put(5.5,27.20){$\cdot$}
\put(5,28.20){$\cdot$}
\put(13,97){$\R\oplus\R$}\put(-55,105){$p-q\equiv 1\!\!\!\!\pmod{8}$}
\put(50,93){0}
\put(50,100){$\cdot$}
\put(49.5,99.99){$\cdot$}
\put(49,99.98){$\cdot$}
\put(48.5,99.97){$\cdot$}
\put(48,99.96){$\cdot$}
\put(47.5,99.94){$\cdot$}
\put(47,99.91){$\cdot$}
\put(46.5,99.86){$\cdot$}
\put(46,99.84){$\cdot$}
\put(45.5,99.8){$\cdot$}
\put(45,99.75){$\cdot$}
\put(44.5,99.7){$\cdot$}
\put(44,99.64){$\cdot$}
\put(43.5,99.57){$\cdot$}
\put(43,99.51){$\cdot$}
\put(42.5,99.43){$\cdot$}
\put(42,99.35){$\cdot$}
\put(41.5,99.27){$\cdot$}
\put(41,99.18){$\cdot$}
\put(40.5,99.09){$\cdot$}
\put(40,98.99){$\cdot$}
\put(39.5,98.88){$\cdot$}
\put(39,98.77){$\cdot$}
\put(38.5,98.66){$\cdot$}
\put(38,98.54){$\cdot$}
\put(37.5,98.41){$\cdot$}
\put(37,98.28){$\cdot$}
\put(50.5,99.99){$\cdot$}
\put(51,99.98){$\cdot$}
\put(51.5,99.97){$\cdot$}
\put(52,99.96){$\cdot$}
\put(52.5,99.94){$\cdot$}
\put(53,99.91){$\cdot$}
\put(53.5,99.86){$\cdot$}
\put(54,99.84){$\cdot$}
\put(54.5,99.8){$\cdot$}
\put(55,99.75){$\cdot$}
\put(55.5,99.7){$\cdot$}
\put(56,99.64){$\cdot$}
\put(56.5,99.57){$\cdot$}
\put(57,99.51){$\cdot$}
\put(57.5,99.43){$\cdot$}
\put(58,99.35){$\cdot$}
\put(58.5,99.27){$\cdot$}
\put(59,99.18){$\cdot$}
\put(59.5,99.09){$\cdot$}
\put(60,98.99){$\cdot$}
\put(60.5,98.88){$\cdot$}
\put(61,98.77){$\cdot$}
\put(61.5,98.66){$\cdot$}
\put(62,98.54){$\cdot$}
\put(62.5,98.41){$\cdot$}
\put(63,98.28){$\cdot$}
\put(68,97){$\R$}\put(73,105){$p-q\equiv 0\!\!\!\!\pmod{8}$}
\put(50,7){4}
\put(68,2){$\BH\oplus\BH$}\put(90,-4){$p-q\equiv 5\!\!\!\!\pmod{8}$}
\put(50,0){$\cdot$}
\put(50.5,0){$\cdot$}
\put(51,0.01){$\cdot$}
\put(51.5,0.02){$\cdot$}
\put(52,0.04){$\cdot$}
\put(52.5,0.06){$\cdot$}
\put(53,0.09){$\cdot$}
\put(53.5,0.12){$\cdot$}
\put(54,0.16){$\cdot$}
\put(54.5,0.2){$\cdot$}
\put(55,0.25){$\cdot$}
\put(55.5,0.3){$\cdot$}
\put(56,0.36){$\cdot$}
\put(56.5,0.42){$\cdot$}
\put(57,0.49){$\cdot$}
\put(57.5,0.56){$\cdot$}
\put(58,0.64){$\cdot$}
\put(58.5,0.73){$\cdot$}
\put(59,0.82){$\cdot$}
\put(59.5,0.91){$\cdot$}
\put(60,1.01){$\cdot$}
\put(60.5,1.11){$\cdot$}
\put(61,1.22){$\cdot$}
\put(61.5,1.34){$\cdot$}
\put(62,1.46){$\cdot$}
\put(62.5,1.59){$\cdot$}
\put(63,1.72){$\cdot$}
\put(49.5,0){$\cdot$}
\put(49,0.01){$\cdot$}
\put(48.5,0.02){$\cdot$}
\put(48,0.04){$\cdot$}
\put(47.5,0.06){$\cdot$}
\put(47,0.09){$\cdot$}
\put(46.5,0.12){$\cdot$}
\put(46,0.16){$\cdot$}
\put(45.5,0.2){$\cdot$}
\put(45,0.25){$\cdot$}
\put(44.5,0.3){$\cdot$}
\put(44,0.36){$\cdot$}
\put(43.5,0.42){$\cdot$}
\put(43,0.49){$\cdot$}
\put(42.5,0.56){$\cdot$}
\put(42,0.64){$\cdot$}
\put(41.5,0.73){$\cdot$}
\put(41,0.82){$\cdot$}
\put(40.5,0.91){$\cdot$}
\put(40,1.01){$\cdot$}
\put(39.5,1.11){$\cdot$}
\put(39,1.22){$\cdot$}
\put(38.5,1.34){$\cdot$}
\put(38,1.46){$\cdot$}
\put(37.5,1.59){$\cdot$}
\put(37,1.72){$\cdot$}
\put(28,3){$\BH$}\put(-40,-4){$p-q\equiv 4\!\!\!\!\pmod{8}$}
\put(93,50){2}
\put(98.28,63){$\cdot$}
\put(98.41,62.5){$\cdot$}
\put(98.54,62){$\cdot$}
\put(98.66,61.5){$\cdot$}
\put(98.77,61){$\cdot$}
\put(98.88,60.5){$\cdot$}
\put(98.99,60){$\cdot$}
\put(99.09,59.5){$\cdot$}
\put(99.18,59){$\cdot$}
\put(99.27,58.5){$\cdot$}
\put(99.35,58){$\cdot$}
\put(99.43,57.5){$\cdot$}
\put(99.51,57){$\cdot$}
\put(99.57,56.5){$\cdot$}
\put(99.64,56){$\cdot$}
\put(99.7,55.5){$\cdot$}
\put(99.75,55){$\cdot$}
\put(99.8,54.5){$\cdot$}
\put(99.84,54){$\cdot$}
\put(99.86,53.5){$\cdot$}
\put(99.91,53){$\cdot$}
\put(99.94,52.5){$\cdot$}
\put(99.96,52){$\cdot$}
\put(99.97,51.5){$\cdot$}
\put(99.98,51){$\cdot$}
\put(99.99,50.5){$\cdot$}
\put(100,50){$\cdot$}
\put(98.28,37){$\cdot$}
\put(98.41,37.5){$\cdot$}
\put(98.54,38){$\cdot$}
\put(98.66,38.5){$\cdot$}
\put(98.77,39){$\cdot$}
\put(98.88,39.5){$\cdot$}
\put(98.99,40){$\cdot$}
\put(99.09,40.5){$\cdot$}
\put(99.18,41){$\cdot$}
\put(99.27,41.5){$\cdot$}
\put(99.35,42){$\cdot$}
\put(99.43,42.5){$\cdot$}
\put(99.51,43){$\cdot$}
\put(99.57,43.5){$\cdot$}
\put(99.64,44){$\cdot$}
\put(99.7,44.5){$\cdot$}
\put(99.75,45){$\cdot$}
\put(99.8,45.5){$\cdot$}
\put(99.84,46){$\cdot$}
\put(99.86,46.5){$\cdot$}
\put(99.91,47){$\cdot$}
\put(99.94,47.5){$\cdot$}
\put(99.96,48){$\cdot$}
\put(99.97,48.5){$\cdot$}
\put(99.98,49){$\cdot$}
\put(99.99,49.5){$\cdot$}
\put(7,50){6}
\put(1,32){$\C$}\put(-65,29){$p-q\equiv 3\!\!\!\!\pmod{8}$}
\put(1.72,63){$\cdot$}
\put(1.59,62.5){$\cdot$}
\put(1.46,62){$\cdot$}
\put(1.34,61.5){$\cdot$}
\put(1.22,61){$\cdot$}
\put(1.11,60.5){$\cdot$}
\put(1.01,60){$\cdot$}
\put(0.99,59.5){$\cdot$}
\put(0.82,59){$\cdot$}
\put(0.73,58.5){$\cdot$}
\put(0.64,58){$\cdot$}
\put(0.56,57.5){$\cdot$}
\put(0.49,57){$\cdot$}
\put(0.42,56.5){$\cdot$}
\put(0.36,56){$\cdot$}
\put(0.3,55.5){$\cdot$}
\put(0.25,55){$\cdot$}
\put(0.2,54.5){$\cdot$}
\put(0.16,54){$\cdot$}
\put(0.12,53.5){$\cdot$}
\put(0.09,53){$\cdot$}
\put(0.06,52.5){$\cdot$}
\put(0.04,52){$\cdot$}
\put(0.02,51.5){$\cdot$}
\put(0.01,51){$\cdot$}
\put(0,50.5){$\cdot$}
\put(0,50){$\cdot$}
\put(1.72,37){$\cdot$}
\put(1.59,37.5){$\cdot$}
\put(1.46,38){$\cdot$}
\put(1.34,38.5){$\cdot$}
\put(1.22,39){$\cdot$}
\put(1.11,39.5){$\cdot$}
\put(1.01,40){$\cdot$}
\put(0.99,40.5){$\cdot$}
\put(0.82,41){$\cdot$}
\put(0.73,41.5){$\cdot$}
\put(0.64,42){$\cdot$}
\put(0.56,42.5){$\cdot$}
\put(0.49,43){$\cdot$}
\put(0.42,43.5){$\cdot$}
\put(0.36,44){$\cdot$}
\put(0.3,44.5){$\cdot$}
\put(0.25,45){$\cdot$}
\put(0.2,45.5){$\cdot$}
\put(0.16,46){$\cdot$}
\put(0.12,46.5){$\cdot$}
\put(0.09,47){$\cdot$}
\put(0.06,47.5){$\cdot$}
\put(0.04,48){$\cdot$}
\put(0.02,48.5){$\cdot$}
\put(0.01,49){$\cdot$}
\put(0,49.5){$\cdot$}
\put(0.5,67){$\R$}\put(-65,75){$p-q\equiv 2\!\!\!\!\pmod{8}$}
\end{picture}
\]
\vspace{2ex}
\begin{center}
\begin{minipage}{25pc}{\small
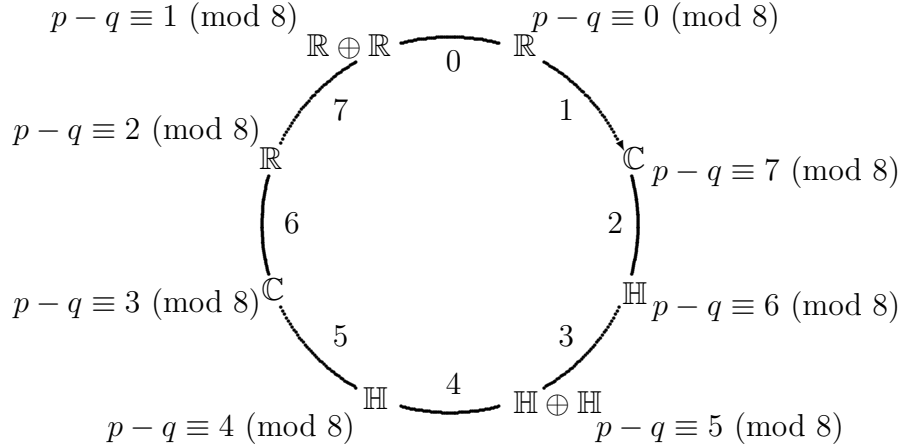
{\bf Fig.1} The Trautman diagram for the Brauer--Wall group
$BW_{\R}\simeq\dZ_8$}
\end{minipage}
\end{center}
\medskip
It is obvious that a group structure over $\cl_{p,q}$, defined by
$BW_{\R}\simeq\dZ_8$, immediately relates with the Atiyah--Bott--Shapiro
periodicity \cite{AtBSh}. In accordance with \cite{AtBSh}, the Clifford
algebra over the field $\F=\R$ is modulo 8 periodic:
$\cl_{p+8,q}\simeq\cl_{p,q}\otimes\cl_{8,0}\,(\cl_{p,q+8}\simeq\cl_{p,q}
\otimes\cl_{0,8})$.

Coming back to Theorem \ref{tautr} we see that for each type of 
algebra $\cl_{p,q}$ there exists some set of the automorphism groups.
If we take into account this relation, then the cyclic structure of a
generalized group $BW^{a,b,c}_{\R}$ would look as follows (Fig. 2).
First of all, the semi--simple algebras $\cl_{p,q}$ with the rings
$\K\simeq\R\oplus\R$ and $\K\simeq\BH\oplus\BH$ ($p-q\equiv 1,5\pmod{8}$)
form an axis of the eighth order, which defines the cyclic group $\dZ_8$.
Further, the neutral types $p-q\equiv 0\pmod{8}$ ($\K\simeq\R$) and
$p-q\equiv 4\pmod{8}$ ($\K\simeq\BH$), which in common admit the
automorphism groups with the signatures $(+,b,c)$, form an axis of the
fourth order corresponding to the cyclic group $\dZ_4$. Analogously, the
two mutually opposite types $p-q\equiv 2\pmod{8}$ ($\K\simeq\R$) and
$p-q\equiv 6\pmod{8}$ ($\K\simeq\BH$), which in common admit the
automorphism groups with the signatures $(-,b,c)$, also form an axis
of the fourth order. Finally, the types $p-q\equiv 3,7\pmod{8}$ ($\K\simeq\C$)
with the $(+,+,+)$ and $(-,-,-)$ automorphism groups form an axis of the
second order. Therefore, $BW^{a,b,c}_{\R}\simeq\dZ_2\otimes(\dZ_4)^2\otimes
\dZ_8$, where $(\dZ_4)^2=\dZ_4\otimes\dZ_4$.

\[
\unitlength=0.5mm
\begin{picture}(100.00,200.00)(0,-50)

\put(97,67){$\C$}\put(105,64){$p-q\equiv 7\!\!\!\!\pmod{8}$}
\put(107,82){$(+,+,+)$}
\put(107,75){$(-,-,-)$}
\put(80,80){1}
\put(75,93.3){$\cdot$}
\put(75.5,93){$\cdot$}
\put(76,92.7){$\cdot$}
\put(76.5,92.4){$\cdot$}
\put(77,92.08){$\cdot$}
\put(77.5,91.76){$\cdot$}
\put(78,91.42){$\cdot$}
\put(78.5,91.08){$\cdot$}
\put(79,90.73){$\cdot$}
\put(79.5,90.37){$\cdot$}
\put(80,90.0){$\cdot$}
\put(80.5,89.62){$\cdot$}
\put(81,89.23){$\cdot$}
\put(81.5,88.83){$\cdot$}
\put(82,88.42){$\cdot$}
\put(82.5,87.99){$\cdot$}
\put(83,87.56){$\cdot$}
\put(83.5,87.12){$\cdot$}
\put(84,86.66){$\cdot$}
\put(84.5,86.19){$\cdot$}
\put(85,85.70){$\cdot$}
\put(85.5,85.21){$\cdot$}
\put(86,84.69){$\cdot$}
\put(86.5,84.17){$\cdot$}
\put(87,83.63){$\cdot$}
\put(87.5,83.07){$\cdot$}
\put(88,82.49){$\cdot$}
\put(88.5,81.9){$\cdot$}
\put(89,81.29){$\cdot$}
\put(89.5,80.65){$\cdot$}
\put(90,80){$\cdot$}
\put(90.5,79.32){$\cdot$}
\put(91,78.62){$\cdot$}
\put(91.5,77.89){$\cdot$}
\put(92,77.13){$\cdot$}
\put(92.5,76.34){$\cdot$}
\put(93,75.51){$\cdot$}
\put(93.5,74.65){$\cdot$}
\put(94,73.74){$\cdot$}
\put(94.5,72.79){$\cdot$}
\put(96.5,73.74){\vector(1,-2){1}}
\put(80,20){3}
\put(97,31){$\BH$}\put(105,38){$p-q\equiv 6\!\!\!\!\pmod{8}$}
\put(112,30){$(-,+,-)$}
\put(112,23){$(-,-,+)$}
\put(112,16){$(-,-,-)$}
\put(112,9){$(-,+,+)$}
\put(75,6.7){$\cdot$}
\put(75.5,7){$\cdot$}
\put(76,7.29){$\cdot$}
\put(76.5,7.6){$\cdot$}
\put(77,7.91){$\cdot$}
\put(77.5,8.24){$\cdot$}
\put(78,8.57){$\cdot$}
\put(78.5,8.91){$\cdot$}
\put(79,9.27){$\cdot$}
\put(79.5,9.63){$\cdot$}
\put(80,10){$\cdot$}
\put(80.5,10.38){$\cdot$}
\put(81,10.77){$\cdot$}
\put(81.5,11.17){$\cdot$}
\put(82,11.58){$\cdot$}
\put(82.5,12.00){$\cdot$}
\put(83,12.44){$\cdot$}
\put(83.5,12.88){$\cdot$}
\put(84,13.34){$\cdot$}
\put(84.5,13.8){$\cdot$}
\put(85,14.29){$\cdot$}
\put(85.5,14.79){$\cdot$}
\put(86,15.3){$\cdot$}
\put(86.5,15.82){$\cdot$}
\put(87,16.37){$\cdot$}
\put(87.5,16.92){$\cdot$}
\put(88,17.5){$\cdot$}
\put(88.5,18.09){$\cdot$}
\put(89,18.71){$\cdot$}
\put(89.5,19.34){$\cdot$}
\put(90,20){$\cdot$}
\put(90.5,20.68){$\cdot$}
\put(91,21.38){$\cdot$}
\put(91.5,22.11){$\cdot$}
\put(92,22.87){$\cdot$}
\put(92.5,23.66){$\cdot$}
\put(93,24.48){$\cdot$}
\put(93.5,25.34){$\cdot$}
\put(94,26.25){$\cdot$}
\put(94.5,27.20){$\cdot$}
\put(95,28.20){$\cdot$}
\put(20,80){7}
\put(25,93.3){$\cdot$}
\put(24.5,93){$\cdot$}
\put(24,92.7){$\cdot$}
\put(23.5,92.49){$\cdot$}
\put(23,92.08){$\cdot$}
\put(22.5,91.75){$\cdot$}
\put(22,91.42){$\cdot$}
\put(21.5,91.08){$\cdot$}
\put(21,90.73){$\cdot$}
\put(20.5,90.37){$\cdot$}
\put(20,90){$\cdot$}
\put(19.5,89.62){$\cdot$}
\put(19,89.23){$\cdot$}
\put(18.5,88.83){$\cdot$}
\put(18,88.42){$\cdot$}
\put(17.5,87.99){$\cdot$}
\put(17,87.56){$\cdot$}
\put(16.5,87.12){$\cdot$}
\put(16,86.66){$\cdot$}
\put(15.5,86.19){$\cdot$}
\put(15,85.70){$\cdot$}
\put(14.5,85.21){$\cdot$}
\put(14,84.69){$\cdot$}
\put(13.5,84.17){$\cdot$}
\put(13,83.63){$\cdot$}
\put(12.5,83.07){$\cdot$}
\put(12,82.49){$\cdot$}
\put(11.5,81.9){$\cdot$}
\put(11,81.29){$\cdot$}
\put(10.5,80.65){$\cdot$}
\put(10,80){$\cdot$}
\put(9.5,79.32){$\cdot$}
\put(9,78.62){$\cdot$}
\put(8.5,77.89){$\cdot$}
\put(8,77.13){$\cdot$}
\put(7.5,76.34){$\cdot$}
\put(7,75.51){$\cdot$}
\put(6.5,74.65){$\cdot$}
\put(6,73.79){$\cdot$}
\put(5.5,72.79){$\cdot$}
\put(5,71.79){$\cdot$}
\put(20,20){5}
\put(25,6.7){$\cdot$}
\put(24.5,7){$\cdot$}
\put(24,7.29){$\cdot$}
\put(23.5,7.6){$\cdot$}
\put(23,7.91){$\cdot$}
\put(22.5,8.24){$\cdot$}
\put(22,8.57){$\cdot$}
\put(21.5,8.91){$\cdot$}
\put(21,9.27){$\cdot$}
\put(20.5,9.63){$\cdot$}
\put(20,10){$\cdot$}
\put(19.5,10.38){$\cdot$}
\put(19,10.77){$\cdot$}
\put(18.5,11.17){$\cdot$}
\put(18,11.58){$\cdot$}
\put(17.5,12){$\cdot$}
\put(17,12.44){$\cdot$}
\put(16.5,12.88){$\cdot$}
\put(16,13.34){$\cdot$}
\put(15.5,13.8){$\cdot$}
\put(15,14.29){$\cdot$}
\put(14.5,14.79){$\cdot$}
\put(14,15.3){$\cdot$}
\put(13.5,15.82){$\cdot$}
\put(13,16.37){$\cdot$}
\put(12.5,16.92){$\cdot$}
\put(12,17.5){$\cdot$}
\put(11.5,18.09){$\cdot$}
\put(11,18.71){$\cdot$}
\put(10.5,19.34){$\cdot$}
\put(10,20){$\cdot$}
\put(9.5,20.68){$\cdot$}
\put(9,21.38){$\cdot$}
\put(8.5,22.11){$\cdot$}
\put(8,22.87){$\cdot$}
\put(7.5,23.66){$\cdot$}
\put(7,24.48){$\cdot$}
\put(6.5,25.34){$\cdot$}
\put(6,26.25){$\cdot$}
\put(5.5,27.20){$\cdot$}
\put(5,28.20){$\cdot$}
\put(13,97){$\R\oplus\R$}\put(-55,87){$p-q\equiv 1\!\!\!\!\pmod{8}$}
\put(-23,95){$(+,+,-)$}
\put(-23,102){$(+,-,+)$}
\put(-23,109){$(+,-,-)$}
\put(-23,116){$(+,+,+)$}
\put(-23,123){$(-,+,+)$}
\put(-23,130){$(-,-,-)$}
\put(-23,137){$(-,+,-)$}
\put(-23,144){$(-,-,+)$}
\put(50,93){0}
\put(50,100){$\cdot$}
\put(49.5,99.99){$\cdot$}
\put(49,99.98){$\cdot$}
\put(48.5,99.97){$\cdot$}
\put(48,99.96){$\cdot$}
\put(47.5,99.94){$\cdot$}
\put(47,99.91){$\cdot$}
\put(46.5,99.86){$\cdot$}
\put(46,99.84){$\cdot$}
\put(45.5,99.8){$\cdot$}
\put(45,99.75){$\cdot$}
\put(44.5,99.7){$\cdot$}
\put(44,99.64){$\cdot$}
\put(43.5,99.57){$\cdot$}
\put(43,99.51){$\cdot$}
\put(42.5,99.43){$\cdot$}
\put(42,99.35){$\cdot$}
\put(41.5,99.27){$\cdot$}
\put(41,99.18){$\cdot$}
\put(40.5,99.09){$\cdot$}
\put(40,98.99){$\cdot$}
\put(39.5,98.88){$\cdot$}
\put(39,98.77){$\cdot$}
\put(38.5,98.66){$\cdot$}
\put(38,98.54){$\cdot$}
\put(37.5,98.41){$\cdot$}
\put(37,98.28){$\cdot$}
\put(50.5,99.99){$\cdot$}
\put(51,99.98){$\cdot$}
\put(51.5,99.97){$\cdot$}
\put(52,99.96){$\cdot$}
\put(52.5,99.94){$\cdot$}
\put(53,99.91){$\cdot$}
\put(53.5,99.86){$\cdot$}
\put(54,99.84){$\cdot$}
\put(54.5,99.8){$\cdot$}
\put(55,99.75){$\cdot$}
\put(55.5,99.7){$\cdot$}
\put(56,99.64){$\cdot$}
\put(56.5,99.57){$\cdot$}
\put(57,99.51){$\cdot$}
\put(57.5,99.43){$\cdot$}
\put(58,99.35){$\cdot$}
\put(58.5,99.27){$\cdot$}
\put(59,99.18){$\cdot$}
\put(59.5,99.09){$\cdot$}
\put(60,98.99){$\cdot$}
\put(60.5,98.88){$\cdot$}
\put(61,98.77){$\cdot$}
\put(61.5,98.66){$\cdot$}
\put(62,98.54){$\cdot$}
\put(62.5,98.41){$\cdot$}
\put(63,98.28){$\cdot$}
\put(68,97){$\R$}\put(75,100){$p-q\equiv 0\!\!\!\!\pmod{8}$}
\put(85,109){$(+,+,-)$}
\put(85,116){$(+,-,+)$}
\put(85,123){$(+,-,-)$}
\put(85,130){$(+,+,+)$}
\put(50,7){4}
\put(68,2){$\BH\oplus\BH$}\put(90,-4){$p-q\equiv 5\!\!\!\!\pmod{8}$}
\put(95,-13){$(-,+,-)$}
\put(95,-20){$(-,-,+)$}
\put(95,-27){$(-,-,-)$}
\put(95,-34){$(-,+,+)$}
\put(95,-41){$(+,+,+)$}
\put(95,-48){$(+,-,-)$}
\put(95,-55){$(+,-,+)$}
\put(95,-62){$(+,+,-)$}
\put(50,0){$\cdot$}
\put(50.5,0){$\cdot$}
\put(51,0.01){$\cdot$}
\put(51.5,0.02){$\cdot$}
\put(52,0.04){$\cdot$}
\put(52.5,0.06){$\cdot$}
\put(53,0.09){$\cdot$}
\put(53.5,0.12){$\cdot$}
\put(54,0.16){$\cdot$}
\put(54.5,0.2){$\cdot$}
\put(55,0.25){$\cdot$}
\put(55.5,0.3){$\cdot$}
\put(56,0.36){$\cdot$}
\put(56.5,0.42){$\cdot$}
\put(57,0.49){$\cdot$}
\put(57.5,0.56){$\cdot$}
\put(58,0.64){$\cdot$}
\put(58.5,0.73){$\cdot$}
\put(59,0.82){$\cdot$}
\put(59.5,0.91){$\cdot$}
\put(60,1.01){$\cdot$}
\put(60.5,1.11){$\cdot$}
\put(61,1.22){$\cdot$}
\put(61.5,1.34){$\cdot$}
\put(62,1.46){$\cdot$}
\put(62.5,1.59){$\cdot$}
\put(63,1.72){$\cdot$}
\put(49.5,0){$\cdot$}
\put(49,0.01){$\cdot$}
\put(48.5,0.02){$\cdot$}
\put(48,0.04){$\cdot$}
\put(47.5,0.06){$\cdot$}
\put(47,0.09){$\cdot$}
\put(46.5,0.12){$\cdot$}
\put(46,0.16){$\cdot$}
\put(45.5,0.2){$\cdot$}
\put(45,0.25){$\cdot$}
\put(44.5,0.3){$\cdot$}
\put(44,0.36){$\cdot$}
\put(43.5,0.42){$\cdot$}
\put(43,0.49){$\cdot$}
\put(42.5,0.56){$\cdot$}
\put(42,0.64){$\cdot$}
\put(41.5,0.73){$\cdot$}
\put(41,0.82){$\cdot$}
\put(40.5,0.91){$\cdot$}
\put(40,1.01){$\cdot$}
\put(39.5,1.11){$\cdot$}
\put(39,1.22){$\cdot$}
\put(38.5,1.34){$\cdot$}
\put(38,1.46){$\cdot$}
\put(37.5,1.59){$\cdot$}
\put(37,1.72){$\cdot$}
\put(28,3){$\BH$}\put(-40,-4){$p-q\equiv 4\!\!\!\!\pmod{8}$}
\put(-20,-13){$(+,+,+)$}
\put(-20,-20){$(+,-,-)$}
\put(-20,-27){$(+,-,+)$}
\put(-20,-34){$(+,+,-)$}
\put(93,50){2}
\put(98.28,63){$\cdot$}
\put(98.41,62.5){$\cdot$}
\put(98.54,62){$\cdot$}
\put(98.66,61.5){$\cdot$}
\put(98.77,61){$\cdot$}
\put(98.88,60.5){$\cdot$}
\put(98.99,60){$\cdot$}
\put(99.09,59.5){$\cdot$}
\put(99.18,59){$\cdot$}
\put(99.27,58.5){$\cdot$}
\put(99.35,58){$\cdot$}
\put(99.43,57.5){$\cdot$}
\put(99.51,57){$\cdot$}
\put(99.57,56.5){$\cdot$}
\put(99.64,56){$\cdot$}
\put(99.7,55.5){$\cdot$}
\put(99.75,55){$\cdot$}
\put(99.8,54.5){$\cdot$}
\put(99.84,54){$\cdot$}
\put(99.86,53.5){$\cdot$}
\put(99.91,53){$\cdot$}
\put(99.94,52.5){$\cdot$}
\put(99.96,52){$\cdot$}
\put(99.97,51.5){$\cdot$}
\put(99.98,51){$\cdot$}
\put(99.99,50.5){$\cdot$}
\put(100,50){$\cdot$}
\put(98.28,37){$\cdot$}
\put(98.41,37.5){$\cdot$}
\put(98.54,38){$\cdot$}
\put(98.66,38.5){$\cdot$}
\put(98.77,39){$\cdot$}
\put(98.88,39.5){$\cdot$}
\put(98.99,40){$\cdot$}
\put(99.09,40.5){$\cdot$}
\put(99.18,41){$\cdot$}
\put(99.27,41.5){$\cdot$}
\put(99.35,42){$\cdot$}
\put(99.43,42.5){$\cdot$}
\put(99.51,43){$\cdot$}
\put(99.57,43.5){$\cdot$}
\put(99.64,44){$\cdot$}
\put(99.7,44.5){$\cdot$}
\put(99.75,45){$\cdot$}
\put(99.8,45.5){$\cdot$}
\put(99.84,46){$\cdot$}
\put(99.86,46.5){$\cdot$}
\put(99.91,47){$\cdot$}
\put(99.94,47.5){$\cdot$}
\put(99.96,48){$\cdot$}
\put(99.97,48.5){$\cdot$}
\put(99.98,49){$\cdot$}
\put(99.99,49.5){$\cdot$}
\put(7,50){6}
\put(1,32){$\C$}\put(-65,29){$p-q\equiv 3\!\!\!\!\pmod{8}$}
\put(-35,20){$(-,-,-)$}
\put(-35,13){$(+,+,+)$}
\put(1.72,63){$\cdot$}
\put(1.59,62.5){$\cdot$}
\put(1.46,62){$\cdot$}
\put(1.34,61.5){$\cdot$}
\put(1.22,61){$\cdot$}
\put(1.11,60.5){$\cdot$}
\put(1.01,60){$\cdot$}
\put(0.99,59.5){$\cdot$}
\put(0.82,59){$\cdot$}
\put(0.73,58.5){$\cdot$}
\put(0.64,58){$\cdot$}
\put(0.56,57.5){$\cdot$}
\put(0.49,57){$\cdot$}
\put(0.42,56.5){$\cdot$}
\put(0.36,56){$\cdot$}
\put(0.3,55.5){$\cdot$}
\put(0.25,55){$\cdot$}
\put(0.2,54.5){$\cdot$}
\put(0.16,54){$\cdot$}
\put(0.12,53.5){$\cdot$}
\put(0.09,53){$\cdot$}
\put(0.06,52.5){$\cdot$}
\put(0.04,52){$\cdot$}
\put(0.02,51.5){$\cdot$}
\put(0.01,51){$\cdot$}
\put(0,50.5){$\cdot$}
\put(0,50){$\cdot$}
\put(1.72,37){$\cdot$}
\put(1.59,37.5){$\cdot$}
\put(1.46,38){$\cdot$}
\put(1.34,38.5){$\cdot$}
\put(1.22,39){$\cdot$}
\put(1.11,39.5){$\cdot$}
\put(1.01,40){$\cdot$}
\put(0.99,40.5){$\cdot$}
\put(0.82,41){$\cdot$}
\put(0.73,41.5){$\cdot$}
\put(0.64,42){$\cdot$}
\put(0.56,42.5){$\cdot$}
\put(0.49,43){$\cdot$}
\put(0.42,43.5){$\cdot$}
\put(0.36,44){$\cdot$}
\put(0.3,44.5){$\cdot$}
\put(0.25,45){$\cdot$}
\put(0.2,45.5){$\cdot$}
\put(0.16,46){$\cdot$}
\put(0.12,46.5){$\cdot$}
\put(0.09,47){$\cdot$}
\put(0.06,47.5){$\cdot$}
\put(0.04,48){$\cdot$}
\put(0.02,48.5){$\cdot$}
\put(0.01,49){$\cdot$}
\put(0,49.5){$\cdot$}
\put(0.5,67){$\R$}\put(-65,75){$p-q\equiv 2\!\!\!\!\pmod{8}$}
\put(-45,65){$(-,-,-)$}
\put(-45,58){$(-,+,+)$}
\put(-45,51){$(-,-,+)$}
\put(-45,44){$(-,+,-)$}
\put(50,50){\line(5,2){60}}
\put(50,50){\line(2,5){30}}
\put(50,50){\line(-2,-5){30}}
\put(50,50){\line(-2,5){30}}
\put(50,50){\line(2,-5){30}}
\put(50,50){\line(-5,-2){60}}
\put(50,50){\line(5,-2){60}}
\put(50,50){\line(-5,2){60}}
\end{picture}
\]

\vspace{2ex}
\begin{center}
\begin{minipage}{25pc}{\small
{\bf Fig.2} The cyclic structure of the generalized group 
$BW^{a,b,c}_{\R}$.}
\end{minipage}
\end{center}
\medskip
Further, over the field $\F=\C$, there exist two types of the complex
Clifford algebras: $\C_n$ and $\C_{n+1}\simeq\C_n\oplus\C_n$. Therefore,
a Brauer--Wall group $BW_{\C}$ acting on the set of 
these two types is isomorphic
to the cyclic group $\dZ_2$. The cyclic structure of the group
$BW_{\C}\simeq\dZ_2$ may be represented on the following Trautman diagram
(Fig. 3) by means of a transition $\C^+_n\stackrel{h}{\longrightarrow}\C_n$
(the round on the diagram is realized by an hour--hand). At this point, the
type of the algebra on the diagram is defined by an equality
$n=h+2r$, where $h\in\{0,1\}$, $r\in\dZ$.

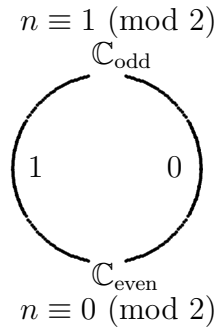
\[
\unitlength=0.5mm
\begin{picture}(50.00,60.00)
\put(5,25){1}
\put(42,25){0}
\put(22,-4){$\C_{{\rm even}}$}
\put(22,55){$\C_{{\rm odd}}$}
\put(3,-13){$n\equiv 0\!\!\!\!\pmod{2}$}
\put(3,64){$n\equiv 1\!\!\!\!\pmod{2}$}
\put(20,49.49){$\cdot$}
\put(19.5,49.39){$\cdot$}
\put(19,49.27){$\cdot$}
\put(18.5,49.14){$\cdot$}
\put(18,49){$\cdot$}
\put(17.5,48.85){$\cdot$}
\put(17,48.68){$\cdot$}
\put(16.5,48.51){$\cdot$}
\put(16,48.32){$\cdot$}
\put(15.5,48.12){$\cdot$}
\put(15,47.91){$\cdot$}
\put(14.5,47.69){$\cdot$}
\put(14,47.45){$\cdot$}
\put(13.5,47.2){$\cdot$}
\put(13,46.93){$\cdot$}
\put(12.5,46.65){$\cdot$}
\put(12,46.35){$\cdot$}
\put(11.5,46.04){$\cdot$}
\put(11,45.71){$\cdot$}
\put(10.5,45.36){$\cdot$}
\put(10,45){$\cdot$}
\put(9.5,44.61){$\cdot$}
\put(9,44.21){$\cdot$}
\put(8.5,43.78){$\cdot$}
\put(8,43.33){$\cdot$}
\put(7.5,42.85){$\cdot$}
\put(7,42.35){$\cdot$}
\put(6.5,41.81){$\cdot$}
\put(6,41.25){$\cdot$}
\put(5.5,40.64){$\cdot$}
\put(5,40){$\cdot$}
\put(4.5,39.3){$\cdot$}
\put(4,38.56){$\cdot$}
\put(3.5,37.76){$\cdot$}
\put(3,36.87){$\cdot$}
\put(2.5,35.89){$\cdot$}
\put(2,34.79){$\cdot$}
\put(1.5,33.53){$\cdot$}
\put(1,32){$\cdot$}
\put(0.5,29.97){$\cdot$}
\put(30,49.49){$\cdot$}
\put(30.5,49.39){$\cdot$}
\put(31,49.27){$\cdot$}
\put(31.5,49.14){$\cdot$}
\put(32,49){$\cdot$}
\put(32.5,48.85){$\cdot$}
\put(33,48.68){$\cdot$}
\put(33.5,48.51){$\cdot$}
\put(34,48.32){$\cdot$}
\put(34.5,48.12){$\cdot$}
\put(35,47.91){$\cdot$}
\put(35.5,47.69){$\cdot$}
\put(36,47.45){$\cdot$}
\put(36.5,47.2){$\cdot$}
\put(37,46.93){$\cdot$}
\put(37.5,46.65){$\cdot$}
\put(38,46.35){$\cdot$}
\put(38.5,46.04){$\cdot$}
\put(39,45.71){$\cdot$}
\put(39.5,45.36){$\cdot$}
\put(40,45){$\cdot$}
\put(40.5,44.61){$\cdot$}
\put(41,44.21){$\cdot$}
\put(41.5,43.78){$\cdot$}
\put(42,43.33){$\cdot$}
\put(42.5,42.85){$\cdot$}
\put(43,42.35){$\cdot$}
\put(43.5,41.81){$\cdot$}
\put(44,41.25){$\cdot$}
\put(44.5,40.64){$\cdot$}
\put(45,40){$\cdot$}
\put(45.5,39.3){$\cdot$}
\put(46,38.56){$\cdot$}
\put(46.5,37.76){$\cdot$}
\put(47,36.87){$\cdot$}
\put(47.5,35.89){$\cdot$}
\put(48,34.79){$\cdot$}
\put(48.5,33.53){$\cdot$}
\put(49,32){$\cdot$}
\put(49.5,29.97){$\cdot$}
\put(0,25){$\cdot$}
\put(0,24.5){$\cdot$}
\put(0.02,24){$\cdot$}
\put(0.04,23.5){$\cdot$}
\put(0.08,23){$\cdot$}
\put(0.12,22.5){$\cdot$}
\put(0.18,22){$\cdot$}
\put(0.25,21.5){$\cdot$}
\put(0.32,21){$\cdot$}
\put(0.4,20.5){$\cdot$}
\put(0.5,20){$\cdot$}
\put(0.61,19.5){$\cdot$}
\put(0.73,19){$\cdot$}
\put(0.85,18.5){$\cdot$}
\put(1,18){$\cdot$}
\put(1.15,17.5){$\cdot$}
\put(1.31,17){$\cdot$}
\put(1.49,16.5){$\cdot$}
\put(1.68,16){$\cdot$}
\put(1.88,15.5){$\cdot$}
\put(2.09,15){$\cdot$}
\put(2.31,14.5){$\cdot$}
\put(2.55,14){$\cdot$}
\put(2.8,13.5){$\cdot$}
\put(3.06,13){$\cdot$}
\put(0,25.5){$\cdot$}
\put(0.02,26){$\cdot$}
\put(0.04,26.5){$\cdot$}
\put(0.08,27){$\cdot$}
\put(0.12,27.5){$\cdot$}
\put(0.18,28){$\cdot$}
\put(0.25,28.5){$\cdot$}
\put(0.32,29){$\cdot$}
\put(0.4,29.5){$\cdot$}
\put(0.5,30){$\cdot$}
\put(0.61,30.5){$\cdot$}
\put(0.73,31){$\cdot$}
\put(0.85,31.5){$\cdot$}
\put(1,32){$\cdot$}
\put(1.15,32.5){$\cdot$}
\put(1.31,33){$\cdot$}
\put(1.49,33.5){$\cdot$}
\put(1.68,34){$\cdot$}
\put(1.88,34.5){$\cdot$}
\put(2.09,35){$\cdot$}
\put(2.31,35.5){$\cdot$}
\put(2.55,36){$\cdot$}
\put(2.8,36.5){$\cdot$}
\put(3.06,37){$\cdot$}
\put(50,25){$\cdot$}
\put(49.99,24.5){$\cdot$}
\put(49.98,24){$\cdot$}
\put(49.95,23.5){$\cdot$}
\put(49.92,23){$\cdot$}
\put(49.87,22.5){$\cdot$}
\put(49.82,22){$\cdot$}
\put(49.75,21.5){$\cdot$}
\put(49.68,21){$\cdot$}
\put(49.51,20.5){$\cdot$}
\put(49.49,20){$\cdot$}
\put(49.39,19.5){$\cdot$}
\put(49.27,19){$\cdot$}
\put(49.14,18.5){$\cdot$}
\put(49,18){$\cdot$}
\put(48.85,17.5){$\cdot$}
\put(48.69,17){$\cdot$}
\put(48.51,16.5){$\cdot$}
\put(48.32,16){$\cdot$}
\put(48.12,15.5){$\cdot$}
\put(47.91,15){$\cdot$}
\put(47.69,14.5){$\cdot$}
\put(47.45,14){$\cdot$}
\put(47.2,13.5){$\cdot$}
\put(46.93,13){$\cdot$}
\put(50,25){$\cdot$}
\put(49.99,25.5){$\cdot$}
\put(49.98,26){$\cdot$}
\put(49.95,26.5){$\cdot$}
\put(49.92,27){$\cdot$}
\put(49.87,27.5){$\cdot$}
\put(49.82,28){$\cdot$}
\put(49.75,28.5){$\cdot$}
\put(49.68,29){$\cdot$}
\put(49.51,29.5){$\cdot$}
\put(49.49,30){$\cdot$}
\put(49.39,30.5){$\cdot$}
\put(49.27,31){$\cdot$}
\put(49.14,31.5){$\cdot$}
\put(49,32){$\cdot$}
\put(48.85,32.5){$\cdot$}
\put(48.69,33){$\cdot$}
\put(48.51,33.5){$\cdot$}
\put(48.32,34){$\cdot$}
\put(48.12,34.5){$\cdot$}
\put(47.91,35){$\cdot$}
\put(47.69,35.5){$\cdot$}
\put(47.45,36){$\cdot$}
\put(47.2,36.5){$\cdot$}
\put(46.93,37){$\cdot$}
\put(20,0.5){$\cdot$}
\put(19.5,0.61){$\cdot$}
\put(19,0.73){$\cdot$}
\put(18.5,0.86){$\cdot$}
\put(18,1){$\cdot$}
\put(17.5,1.15){$\cdot$}
\put(17,1.31){$\cdot$}
\put(16.5,1.49){$\cdot$}
\put(16,1.68){$\cdot$}
\put(15.5,1.87){$\cdot$}
\put(15,2.09){$\cdot$}
\put(14.5,2.31){$\cdot$}
\put(14,2.55){$\cdot$}
\put(13.5,2.8){$\cdot$}
\put(13,3.06){$\cdot$}
\put(12.5,3.35){$\cdot$}
\put(12,3.64){$\cdot$}
\put(11.5,3.96){$\cdot$}
\put(11,4.29){$\cdot$}
\put(10.5,4.63){$\cdot$}
\put(10,5){$\cdot$}
\put(9.5,5.38){$\cdot$}
\put(9,5.79){$\cdot$}
\put(8.5,6.22){$\cdot$}
\put(8,6.67){$\cdot$}
\put(7.5,7.15){$\cdot$}
\put(7,7.65){$\cdot$}
\put(6.5,8.18){$\cdot$}
\put(6,8.75){$\cdot$}
\put(5.5,9.35){$\cdot$}
\put(5,10){$\cdot$}
\put(4.5,10.69){$\cdot$}
\put(4,11.43){$\cdot$}
\put(3.5,12.24){$\cdot$}
\put(3,13.12){$\cdot$}
\put(2.5,14.10){$\cdot$}
\put(2,15.20){$\cdot$}
\put(1.5,16.47){$\cdot$}
\put(1,18){$\cdot$}
\put(0.5,20.02){$\cdot$}
\put(30,0.5){$\cdot$}
\put(30.5,0.61){$\cdot$}
\put(31,0.73){$\cdot$}
\put(31.5,0.86){$\cdot$}
\put(32,1){$\cdot$}
\put(32.5,1.15){$\cdot$}
\put(33,1.31){$\cdot$}
\put(33.5,1.49){$\cdot$}
\put(34,1.68){$\cdot$}
\put(34.5,1.87){$\cdot$}
\put(35,2.09){$\cdot$}
\put(35.5,2.31){$\cdot$}
\put(36,2.55){$\cdot$}
\put(36.5,2.8){$\cdot$}
\put(37,3.06){$\cdot$}
\put(37.5,3.35){$\cdot$}
\put(38,3.64){$\cdot$}
\put(38.5,3.96){$\cdot$}
\put(39,4.29){$\cdot$}
\put(39.5,4.63){$\cdot$}
\put(40,5){$\cdot$}
\put(40.5,5.38){$\cdot$}
\put(41,5.79){$\cdot$}
\put(41.5,6.22){$\cdot$}
\put(42,6.67){$\cdot$}
\put(42.5,7.15){$\cdot$}
\put(43,7.65){$\cdot$}
\put(43.5,8.18){$\cdot$}
\put(44,8.75){$\cdot$}
\put(44.5,9.35){$\cdot$}
\put(45,10){$\cdot$}
\put(45.5,10.69){$\cdot$}
\put(46,11.43){$\cdot$}
\put(46.5,12.24){$\cdot$}
\put(47,13.12){$\cdot$}
\put(47.5,14.10){$\cdot$}
\put(48,15.20){$\cdot$}
\put(48.5,16.47){$\cdot$}
\put(49,18){$\cdot$}
\put(49.5,20.02){$\cdot$}

\end{picture}
\]
\vspace{2ex}
\begin{center}
\begin{minipage}{25pc}{\small
{\bf Fig.3} The Trautman diagram for the Brauer--Wall group
$BW_{\C}\simeq\dZ_2$.}
\end{minipage}
\end{center}
\medskip 
It is obvious that a group structure over $\C_n$, defined by the group
$BW_{\C}\simeq\dZ_2$, immediately relates with a modulo 2 periodicity of the
complex Clifford algebras \cite{AtBSh,Kar79}: $\C_{n+2}\simeq\C_n\otimes\C_2$.

From Theorem \ref{taut}, it follows that the algebra $\C_{2m}\simeq
\M_{2^m}(\C)$ admits the automorphism group $\sAut_-(\C_{2m})\simeq
\dZ_2\otimes\dZ_2$ with the signature $(+,+,+)$ if $m$ is even, and
respectively the group $\sAut_+(\C_{2m})\simeq Q_4/\dZ_2$ with the signature
$(-,-,-)$ if $m$ is odd. In connection with this, the second complex type
$\C_{2m+1}\simeq\C_{2m}\oplus\C_{2m}$ also admits both the previously
mentioned automorphism groups. Therefore, if we take into account this
relation, the cyclic structure of a generalized group $BW^{a,b,c}_{\C}$
would look as follows (Fig. 4). Both complex types $n\equiv 0\pmod{2}$
and $n\equiv 1\pmod{2}$ form an axis of the second order, therefore,
$BW^{a,b,c}_{\C}\simeq\dZ_2$.

\[
\unitlength=0.5mm
\begin{picture}(50.00,80.00)(0,-5)
\put(5,25){1}
\put(42,25){0}
\put(22,-4){$\C_{{\rm even}}$}
\put(22,55){$\C_{{\rm odd}}$}
\put(28,-30){$n\equiv 0\!\!\!\!\pmod{2}$}
\put(28,-13){$(+,+,+)$}
\put(28,-20){$(-,-,-)$}
\put(-31,79){$n\equiv 1\!\!\!\!\pmod{2}$}
\put(-11,70){$(-,-,-)$}
\put(-11,63){$(+,+,+)$}
\put(20,49.49){$\cdot$}
\put(19.5,49.39){$\cdot$}
\put(19,49.27){$\cdot$}
\put(18.5,49.14){$\cdot$}
\put(18,49){$\cdot$}
\put(17.5,48.85){$\cdot$}
\put(17,48.68){$\cdot$}
\put(16.5,48.51){$\cdot$}
\put(16,48.32){$\cdot$}
\put(15.5,48.12){$\cdot$}
\put(15,47.91){$\cdot$}
\put(14.5,47.69){$\cdot$}
\put(14,47.45){$\cdot$}
\put(13.5,47.2){$\cdot$}
\put(13,46.93){$\cdot$}
\put(12.5,46.65){$\cdot$}
\put(12,46.35){$\cdot$}
\put(11.5,46.04){$\cdot$}
\put(11,45.71){$\cdot$}
\put(10.5,45.36){$\cdot$}
\put(10,45){$\cdot$}
\put(9.5,44.61){$\cdot$}
\put(9,44.21){$\cdot$}
\put(8.5,43.78){$\cdot$}
\put(8,43.33){$\cdot$}
\put(7.5,42.85){$\cdot$}
\put(7,42.35){$\cdot$}
\put(6.5,41.81){$\cdot$}
\put(6,41.25){$\cdot$}
\put(5.5,40.64){$\cdot$}
\put(5,40){$\cdot$}
\put(4.5,39.3){$\cdot$}
\put(4,38.56){$\cdot$}
\put(3.5,37.76){$\cdot$}
\put(3,36.87){$\cdot$}
\put(2.5,35.89){$\cdot$}
\put(2,34.79){$\cdot$}
\put(1.5,33.53){$\cdot$}
\put(1,32){$\cdot$}
\put(0.5,29.97){$\cdot$}
\put(30,49.49){$\cdot$}
\put(30.5,49.39){$\cdot$}
\put(31,49.27){$\cdot$}
\put(31.5,49.14){$\cdot$}
\put(32,49){$\cdot$}
\put(32.5,48.85){$\cdot$}
\put(33,48.68){$\cdot$}
\put(33.5,48.51){$\cdot$}
\put(34,48.32){$\cdot$}
\put(34.5,48.12){$\cdot$}
\put(35,47.91){$\cdot$}
\put(35.5,47.69){$\cdot$}
\put(36,47.45){$\cdot$}
\put(36.5,47.2){$\cdot$}
\put(37,46.93){$\cdot$}
\put(37.5,46.65){$\cdot$}
\put(38,46.35){$\cdot$}
\put(38.5,46.04){$\cdot$}
\put(39,45.71){$\cdot$}
\put(39.5,45.36){$\cdot$}
\put(40,45){$\cdot$}
\put(40.5,44.61){$\cdot$}
\put(41,44.21){$\cdot$}
\put(41.5,43.78){$\cdot$}
\put(42,43.33){$\cdot$}
\put(42.5,42.85){$\cdot$}
\put(43,42.35){$\cdot$}
\put(43.5,41.81){$\cdot$}
\put(44,41.25){$\cdot$}
\put(44.5,40.64){$\cdot$}
\put(45,40){$\cdot$}
\put(45.5,39.3){$\cdot$}
\put(46,38.56){$\cdot$}
\put(46.5,37.76){$\cdot$}
\put(47,36.87){$\cdot$}
\put(47.5,35.89){$\cdot$}
\put(48,34.79){$\cdot$}
\put(48.5,33.53){$\cdot$}
\put(49,32){$\cdot$}
\put(49.5,29.97){$\cdot$}
\put(0,25){$\cdot$}
\put(0,24.5){$\cdot$}
\put(0.02,24){$\cdot$}
\put(0.04,23.5){$\cdot$}
\put(0.08,23){$\cdot$}
\put(0.12,22.5){$\cdot$}
\put(0.18,22){$\cdot$}
\put(0.25,21.5){$\cdot$}
\put(0.32,21){$\cdot$}
\put(0.4,20.5){$\cdot$}
\put(0.5,20){$\cdot$}
\put(0.61,19.5){$\cdot$}
\put(0.73,19){$\cdot$}
\put(0.85,18.5){$\cdot$}
\put(1,18){$\cdot$}
\put(1.15,17.5){$\cdot$}
\put(1.31,17){$\cdot$}
\put(1.49,16.5){$\cdot$}
\put(1.68,16){$\cdot$}
\put(1.88,15.5){$\cdot$}
\put(2.09,15){$\cdot$}
\put(2.31,14.5){$\cdot$}
\put(2.55,14){$\cdot$}
\put(2.8,13.5){$\cdot$}
\put(3.06,13){$\cdot$}
\put(0,25.5){$\cdot$}
\put(0.02,26){$\cdot$}
\put(0.04,26.5){$\cdot$}
\put(0.08,27){$\cdot$}
\put(0.12,27.5){$\cdot$}
\put(0.18,28){$\cdot$}
\put(0.25,28.5){$\cdot$}
\put(0.32,29){$\cdot$}
\put(0.4,29.5){$\cdot$}
\put(0.5,30){$\cdot$}
\put(0.61,30.5){$\cdot$}
\put(0.73,31){$\cdot$}
\put(0.85,31.5){$\cdot$}
\put(1,32){$\cdot$}
\put(1.15,32.5){$\cdot$}
\put(1.31,33){$\cdot$}
\put(1.49,33.5){$\cdot$}
\put(1.68,34){$\cdot$}
\put(1.88,34.5){$\cdot$}
\put(2.09,35){$\cdot$}
\put(2.31,35.5){$\cdot$}
\put(2.55,36){$\cdot$}
\put(2.8,36.5){$\cdot$}
\put(3.06,37){$\cdot$}
\put(50,25){$\cdot$}
\put(49.99,24.5){$\cdot$}
\put(49.98,24){$\cdot$}
\put(49.95,23.5){$\cdot$}
\put(49.92,23){$\cdot$}
\put(49.87,22.5){$\cdot$}
\put(49.82,22){$\cdot$}
\put(49.75,21.5){$\cdot$}
\put(49.68,21){$\cdot$}
\put(49.51,20.5){$\cdot$}
\put(49.49,20){$\cdot$}
\put(49.39,19.5){$\cdot$}
\put(49.27,19){$\cdot$}
\put(49.14,18.5){$\cdot$}
\put(49,18){$\cdot$}
\put(48.85,17.5){$\cdot$}
\put(48.69,17){$\cdot$}
\put(48.51,16.5){$\cdot$}
\put(48.32,16){$\cdot$}
\put(48.12,15.5){$\cdot$}
\put(47.91,15){$\cdot$}
\put(47.69,14.5){$\cdot$}
\put(47.45,14){$\cdot$}
\put(47.2,13.5){$\cdot$}
\put(46.93,13){$\cdot$}
\put(50,25){$\cdot$}
\put(49.99,25.5){$\cdot$}
\put(49.98,26){$\cdot$}
\put(49.95,26.5){$\cdot$}
\put(49.92,27){$\cdot$}
\put(49.87,27.5){$\cdot$}
\put(49.82,28){$\cdot$}
\put(49.75,28.5){$\cdot$}
\put(49.68,29){$\cdot$}
\put(49.51,29.5){$\cdot$}
\put(49.49,30){$\cdot$}
\put(49.39,30.5){$\cdot$}
\put(49.27,31){$\cdot$}
\put(49.14,31.5){$\cdot$}
\put(49,32){$\cdot$}
\put(48.85,32.5){$\cdot$}
\put(48.69,33){$\cdot$}
\put(48.51,33.5){$\cdot$}
\put(48.32,34){$\cdot$}
\put(48.12,34.5){$\cdot$}
\put(47.91,35){$\cdot$}
\put(47.69,35.5){$\cdot$}
\put(47.45,36){$\cdot$}
\put(47.2,36.5){$\cdot$}
\put(46.93,37){$\cdot$}
\put(20,0.5){$\cdot$}
\put(19.5,0.61){$\cdot$}
\put(19,0.73){$\cdot$}
\put(18.5,0.86){$\cdot$}
\put(18,1){$\cdot$}
\put(17.5,1.15){$\cdot$}
\put(17,1.31){$\cdot$}
\put(16.5,1.49){$\cdot$}
\put(16,1.68){$\cdot$}
\put(15.5,1.87){$\cdot$}
\put(15,2.09){$\cdot$}
\put(14.5,2.31){$\cdot$}
\put(14,2.55){$\cdot$}
\put(13.5,2.8){$\cdot$}
\put(13,3.06){$\cdot$}
\put(12.5,3.35){$\cdot$}
\put(12,3.64){$\cdot$}
\put(11.5,3.96){$\cdot$}
\put(11,4.29){$\cdot$}
\put(10.5,4.63){$\cdot$}
\put(10,5){$\cdot$}
\put(9.5,5.38){$\cdot$}
\put(9,5.79){$\cdot$}
\put(8.5,6.22){$\cdot$}
\put(8,6.67){$\cdot$}
\put(7.5,7.15){$\cdot$}
\put(7,7.65){$\cdot$}
\put(6.5,8.18){$\cdot$}
\put(6,8.75){$\cdot$}
\put(5.5,9.35){$\cdot$}
\put(5,10){$\cdot$}
\put(4.5,10.69){$\cdot$}
\put(4,11.43){$\cdot$}
\put(3.5,12.24){$\cdot$}
\put(3,13.12){$\cdot$}
\put(2.5,14.10){$\cdot$}
\put(2,15.20){$\cdot$}
\put(1.5,16.47){$\cdot$}
\put(1,18){$\cdot$}
\put(0.5,20.02){$\cdot$}
\put(30,0.5){$\cdot$}
\put(30.5,0.61){$\cdot$}
\put(31,0.73){$\cdot$}
\put(31.5,0.86){$\cdot$}
\put(32,1){$\cdot$}
\put(32.5,1.15){$\cdot$}
\put(33,1.31){$\cdot$}
\put(33.5,1.49){$\cdot$}
\put(34,1.68){$\cdot$}
\put(34.5,1.87){$\cdot$}
\put(35,2.09){$\cdot$}
\put(35.5,2.31){$\cdot$}
\put(36,2.55){$\cdot$}
\put(36.5,2.8){$\cdot$}
\put(37,3.06){$\cdot$}
\put(37.5,3.35){$\cdot$}
\put(38,3.64){$\cdot$}
\put(38.5,3.96){$\cdot$}
\put(39,4.29){$\cdot$}
\put(39.5,4.63){$\cdot$}
\put(40,5){$\cdot$}
\put(40.5,5.38){$\cdot$}
\put(41,5.79){$\cdot$}
\put(41.5,6.22){$\cdot$}
\put(42,6.67){$\cdot$}
\put(42.5,7.15){$\cdot$}
\put(43,7.65){$\cdot$}
\put(43.5,8.18){$\cdot$}
\put(44,8.75){$\cdot$}
\put(44.5,9.35){$\cdot$}
\put(45,10){$\cdot$}
\put(45.5,10.69){$\cdot$}
\put(46,11.43){$\cdot$}
\put(46.5,12.24){$\cdot$}
\put(47,13.12){$\cdot$}
\put(47.5,14.10){$\cdot$}
\put(48,15.20){$\cdot$}
\put(48.5,16.47){$\cdot$}
\put(49,18){$\cdot$}
\put(49.5,20.02){$\cdot$}
\put(25,25){\line(0,1){50}}
\put(25,25){\line(0,-1){50}}
\end{picture}
\]
\vspace{2ex}
\begin{center}
\begin{minipage}{25pc}{\small
{\bf Fig.4} The cyclic structure of the generalized group
$BW^{a,b,c}_{\C}$.}
\end{minipage}
\end{center}
\chapter{Homomorphism $\epsilon:\,\C_{n+1}\rightarrow\C_n$ 
and quotient Clifford algebras}
\setcounter{theorem}{0}
\setcounter{cor}{0}
\setcounter{opr}{0}
\setcounter{lem}{0}
\setcounter{prop}{0}
\setcounter{axiom}{0}
\setcounter{equation}{0}
\markboth{CHAPTER \hspace{0.15cm}\thechapter.\hspace{0.2cm}HOMOMORPHISM
$\epsilon:\,\C_{n+1}\rightarrow\C_n$}
{\thesection.\hspace{0.2cm}ODD--DIMENSIONAL CLIFFORD ALGEBRAS}
\section{Fundamental automorphisms of odd--dimensional Clifford algebras}
\markright{\thesection.\hspace{0.2cm}ODD--DIMENSIONAL CLIFFORD ALGEBRAS}
Let us consider now fundamental automorhisms of the algebras $\cl_{p,q}$ and
$\C_{p+q}$, where $p-q\equiv 1,3,5,7\pmod{8}$, $p+q=2m+1$. As known
\cite{Rash,Port}, the odd--dimensional algebras
$\cl_{p,q}$ and $\C_n$ are isomorphic to direct sums of the two algebras
with even dimensionality if correspondingly
$p-q\equiv 1,5\pmod{8}$
and $p+q\equiv 1,3,5,7\pmod{8}$. Therefore, matrix representations of the
algebras $\cl_{p,q}$, $\C_{p+q}$ $(p+q=2m+1)$ are isomorphic to direct sums
of the full matrix algebras $\M_{2^m}(\K)\oplus\M_{2^m}(\K)$, 
where $\K\simeq\R,\,\K\simeq\BH,\K\simeq\C$. On the other hand, there exists
an homomorphic mapping of the algebras
$\cl_{p,q}$ and $\C_n$ onto one matrix algebra $\M_{2^m}(\K)$
with preservation of addition, multiplication and multiplication by the
number operations. Besides, in case of the field
$\F=\R$ and $p-q\equiv 3,7\pmod{8}$ the algebra $\cl_{p,q}$ is isomorphic to
to a full matrix algebra $\M_{2^m}(\C)$, therefore, representations of the
fundamental automorphisms of this algebra can be realized by means of
$\M_{2^m}(\C)$.
\begin{theorem}
If $p+q=2m+1$, then there exist following homomorphisms:\\
1) $\F=\R$
\[
\epsilon:\;\cl_{p,q}\longrightarrow\M_{2^m}(\K),
\]
where $\K\simeq\R$ if $p-q\equiv 1\pmod{8}$ and $\K\simeq\BH$ if
$p-q\equiv 5\pmod{8}$.\\[0.2cm]
2) $\F=\C$
\[
\epsilon:\;\C_{p+q}\longrightarrow\M_{2^m}(\C),\quad\text{if}\quad
p-q\equiv 1,3,5,7\pmod{8}.
\]
\end{theorem}
\begin{proof} Let us start the proof with a more general case $\F=\C$.
In accordance with (\ref{e4}) the element $\omega$ belongs to a center of the
algebra $\C_n$ and, therefore, this element commutes with all the basis
elements of $\C_n$. Further, we see that the basis vectors
$\{e_1,e_2,\ldots,e_n\}$ generate a subspace
$\C^n\subset\C^{n+1}$. Thus, the algebra $\C_n$ in $\C^n$ is a subalgebra of
$\C_{n+1}$ and consists of the elements which do not contain the element
$\e_{n+1}$. The each element
$\cA\in\C_{n+1}$ can be decomposed as follows
\[
\cA=\cA^1+\cA^0,
\]
where the set $\cA^0$ contains all the elements with $\e_{n+1}$, and
$\cA^1$ does not contain it, therefore, $\cA^1\in\C_n$. At the multiplication
$\cA^0$ by $\epsilon\omega$ the elements $\e_{n+1}$ are mutually annihilate. 
Hence it follows that $\epsilon\omega\cA^0\in\C_n$.
Denoting $\cA^2=\epsilon\omega\cA^0$ and taking into account that
$(\epsilon\omega)^2=1$ we obtain
\[
\cA=\cA^1+\varepsilon\omega\cA^2,
\]
where $\cA^1,\,\cA^2\in\C_n$. Let us define now an homomorphism
$\epsilon:\;\C_{n+1}
\rightarrow\C_n$, acting on the following law
\begin{equation}\label{3.7.1}
\epsilon:\;\cA^1+\varepsilon\omega\cA^2\longrightarrow\cA^1+\cA^2.
\end{equation}
It is obvious that the operations of addition, multiplication and
multiplication by the number are preserved. Indeed, if we take
\[
\cA=\cA^1+\varepsilon\omega\cA^2,\quad\cB=\cB^1+\cB^2,
\]
then in virtue of
$(\varepsilon\omega)^2=1$ and commutativity of
$\omega$ with the all elements we have for the multiplication operation:
\begin{multline}
\cA\cB=(\cA^1\cB^1+\cA^2\cB^2)+\varepsilon\omega(\cA^1\cB^2+\cA^2\cB^1)
\stackrel{\epsilon}{\longrightarrow}\\
(\cA^1\cB^1+\cA^2\cB^2)+(\cA^1\cB^2+\cA^2\cB^1)=(\cA^1+\cA^2)(\cB^1+\cB^2),
\nonumber
\end{multline}
that is, an image of the product is equal to the product of factor images
in the same order. In particular case at
$\cA=\varepsilon\omega$ we have $\cA^1=0$ and
$\cA^2=1$, therefore,
\[
\varepsilon\omega\longrightarrow 1.
\]
Thus, a kernel of the homomorphism $\epsilon$ consists of the all elements
of the form
$\cA^1-\varepsilon\omega\cA^1$, which, as it is easy to see, under action
of $\epsilon$ are mapped into zero. It is clear that $\Ker\epsilon=\{
\cA^1-\varepsilon\omega\cA^1\}$ is a subalgebra of $\C_{n+1}$.
Moreover, the kernel of $\epsilon$ is a bilateral ideal of the algebra
$\C_{n+1}$. Therefore, the algebra ${}^{\epsilon}\C_n$ obtaining in the
result of the mapping $\epsilon:\;\C_{n+1}\rightarrow\C_n$ is 
{\it a quotient algebra on the ideal} $\Ker\epsilon=\{\cA^1-\varepsilon
\omega\cA^1\}$:
\[
{}^{\epsilon}\C_n\simeq\C_{n+1}/\Ker\epsilon.
\] 
Since the algebra $\C_n$ at $n=2m$ is isomorphically mapped onto the full
matrix algebra $\M_{2^m}(\C)$, then in virtue of
$\epsilon:\;\C_{n+1}\rightarrow\C_n\subset\C_{n+1}$ we obtain an
homomorphic mapping of the algebra
$\C_{n+1}$ onto the matrix algebra
$\M_{2^m}(\C)$.

An homomorphism $\epsilon:\;\cl_{p,q}\rightarrow
\M_{2^m}(\K)$ is proved analogously. In this case a quotient algebra
has the form
\[
{}^{\epsilon}\cl_{p-1,q}\simeq\cl_{p,q}/\Ker\epsilon,
\]
or
\[
{}^{\epsilon}\cl_{p,q-1}\simeq\cl_{p,q}/\Ker\epsilon,
\]
where $\Ker\epsilon=\{\cA^1-\omega\cA^1\}$, since in accordance with (\ref{e3})
at $p-q\equiv 1,5\pmod{8}$ we have always $\omega^2=1$ and, therefore,
$\varepsilon=1$.\end{proof}

Let us consider now the automorphisms
$\cA\rightarrow\widetilde{\cA},\;
\cA\rightarrow\cA^\star$ and $\cA\rightarrow\widetilde{\cA^\star}$ defined
in $\C_{n+1}$. We will examine the form of the fundamental automorphisms
defined in $\C_{n+1}$ after the homomorpic mapping
$\epsilon:\;\C_{n+1}\rightarrow\C_n\subset\C_{n+1}$. First of all, for the
antiautomorphism
$\cA\rightarrow\widetilde{\cA}$ it is necessary that elements
$\cA,\cB,\ldots\in\C_{n+1}$, which are mapped into one and the same element
$\cD\in\C_n$ (a kernel of the homomorphism $\epsilon$ if $\cD=0$), after the
transformation $\cA\rightarrow\widetilde{\cA}$ are must converted to the
elements $\widetilde{\cA},\widetilde{\cB},\ldots\in\C_{n+1}$, which are also
mapped into one and the same element $\widetilde{\cD}\in\C_n$. Otherwise,
the transformation $\cA\rightarrow\widetilde{\cA}$ is not transferred from
$\C_{n+1}$ in $\C_n$ as an unambiguous transformation. In particular, it is
necessary that $\widetilde{\varepsilon\omega}=\varepsilon\omega$, since
1 and the element $\varepsilon\omega$ under action of $\epsilon$ are equally
mapped into the unit, then $\widetilde{1}$ and $\widetilde{\varepsilon\omega}$
are also must be mapped into one and the same element in $\C_n$, but
$\widetilde{1}\rightarrow 1$, and $\widetilde{\varepsilon\omega}=\pm
\varepsilon\omega\rightarrow\pm 1$ (in virtue of (\ref{auto19})).
Therefore, we must assume that
\begin{equation}\label{3.7.2}
\widetilde{\varepsilon\omega}=\varepsilon\omega.
\end{equation}
The condition (\ref{3.7.2}) is sufficient for the transfer of the
antiautomorphism
$\cA\rightarrow\widetilde{\cA}$ from $\C_{n+1}$ to $\C_n$. Indeed,
in this case at $\cA\rightarrow\widetilde{\cA}$ we have
\[
\cA^1-\cA^1\varepsilon\omega\longrightarrow\widetilde{\cA^1}-
\widetilde{\varepsilon\omega}\widetilde{\cA^1}=\widetilde{\cA^1}-
\varepsilon\omega\widetilde{\cA^1}.
\]
Therefore, elements of the form $\cA^1-\cA^1\varepsilon\omega$ consisting of
the kernel of the homomorphism $\epsilon$ are converted at the transformation
$\cA\rightarrow\widetilde{\cA}$ into elements of the same form.

Analogous conditions take place for other fundamental automorphisms. However,
for the automorphism $\cA\rightarrow\cA^\star$ the condition
$(\varepsilon\omega)^\star=\varepsilon\omega$ is never fulfilled,
since $\omega$ is odd and in accordance with (\ref{auto16})
we have
\begin{equation}\label{3.7.2'}
\omega^\star=-\omega.
\end{equation}
Therefore, the automorphism $\cA\rightarrow\cA^\star$ is never transferred
from $\C_{n+1}$ to $\C_n$.

Let us return to the antiautomorpism $\cA\rightarrow\widetilde{\cA}$ 
and consider in more detail conditions which sufficient for the transfer of
this transformation
from $\C_{n+1}$ to $\C_n$. At first, in dependence on condition $(\varepsilon
\omega)^2=1$ and square of the element $\omega=e_{12\ldots n+1}$ the factor
$\varepsilon$ has the following values
\begin{equation}\label{3.7.3}
\varepsilon=\begin{cases}
1,& \text{if $p-q\equiv 1,5\!\!\!\pmod{8}$};\\
i,& \text{if $p-q\equiv 3,7\!\!\!\pmod{8}$}.
\end{cases}
\end{equation}
Further, in accordance with the formula (\ref{auto19}) 
for the transformation $\omega
\rightarrow\widetilde{\omega}$ we obtain
\begin{equation}\label{3.7.4}
\widetilde{\omega}=\begin{cases}
\phantom{-}\omega,& \text{if $p-q\equiv 1,5\!\!\!\pmod{8}$};\\
-\omega,& \text{if $p-q\equiv 3,7\!\!\!\pmod{8}$}.
\end{cases}
\end{equation}
Therefore, for the algebras over the field $\F=\R$ 
the antiautomorphism $\cA\rightarrow
\widetilde{\cA}$ is transferred at the mappings $\cl_{p,q}\rightarrow
\cl_{p,q-1},\;\cl_{p,q}\rightarrow\cl_{q,p-1}$, where $p-q\equiv 1,5\pmod{8}$.
Over the field $\F=\C$ the antiautomorphism $\cA\rightarrow\widetilde{\cA}$
is transferred in any case, since the algebras $\C_{n+1}$ with signatures
$p-q\equiv 3,7\pmod{8}$ and $p-q\equiv 1,5\pmod{8}$ are isomorphic. 
At this point, the condition (\ref{3.7.2}) in accordance 
with (\ref{3.7.3}), (\ref{3.7.4})
takes a following form:
\begin{eqnarray}
\phantom{i}\widetilde{\omega}&=&\omega\phantom{i},\quad\text{if
$p-q\equiv 1,5\!\!\!\pmod{8}$};\nonumber\\
\widetilde{i\omega}&=&i\omega,\quad\text{if $p-q\equiv 3,7\!\!\!\pmod{8}$}.
\nonumber
\end{eqnarray}\begin{sloppypar}\noindent
Besides, each of these equalities satisfies to condition
$(\varepsilon\omega)^2=1$.\end{sloppypar}

Let us consider now the antiautomorphism $\cA\rightarrow\widetilde{\cA^\star}$.
It is obvious that for the transfer of $\cA\rightarrow\widetilde{\cA^\star}$
from $\C_{n+1}$ to $\C_n$ it is sufficient that
\begin{equation}\label{3.7.5}
\widetilde{(\varepsilon\omega)^\star}=\varepsilon\omega.
\end{equation}
It is easy to see that in virtue of (\ref{3.7.2'}) and second equality from
(\ref{3.7.4}) the mapping $\C_{p+q}\rightarrow\C_{p+q-1}$ satisfies to this
condition at $p-q\equiv 3,7\pmod{8}$, since in this case
\[
\widetilde{(\varepsilon\omega)^\star}=\varepsilon\widetilde{\omega^\star}=
-\varepsilon\omega^\star=\varepsilon\omega.
\]
Hence it follows that for the algebras over the field
$\F=\R$ the antiautomorphism
$\cA\rightarrow\widetilde{\cA^\star}$ at the mappings $\cl_{p,q}
\rightarrow\cl_{p,q-1},\;\cl_{p,q}\rightarrow\cl_{q,p-1}$ 
($p-q\equiv 1,5\pmod{8}$)
is never transferred.

Summarizing the results obtained previously we come to the following
\begin{theorem}[{\rm\cite{Var99}}]\label{thomo}
1) If $\F=\C$ and $\C_{p+q}\simeq\C_{p+q-1}\oplus\C_{p+q-1}$ is the Clifford
algebra over $\C$, where $p-q\equiv 1,3,5,7\pmod{8}$, then at the homomorphic
mapping $\epsilon:\;\C_{p+q}\rightarrow\C_{p+q-1}$ the antiautomorphism
$\cA\rightarrow\widetilde{\cA}$ is transferred onto the quotient algebra
${}^\epsilon\C_{p+q-1}$ in any case, the automorphism $\cA\rightarrow
\cA^\star$ is never transferred, and the antiautomorphism $\cA\rightarrow
\widetilde{\cA^\star}$ is transferred in the case $p-q\equiv 3,7\pmod{8}$.\\
2) If $\F=R$ and $\cl_{p,q}\simeq\cl_{p,q-1}\oplus\cl_{p,q-1},\;\cl_{p,q}
\simeq\cl_{q,p-1}\oplus\cl_{q,p-1}$ are the Clifford algebras over $\R$, where
$p-q\equiv 1,5\pmod{8}$, then at the homomorphic mappings $\epsilon:\;\cl_{p,q}
\rightarrow\cl_{p,q-1}$ and $\epsilon:\;\cl_{p,q}\rightarrow\cl_{q,p-1}$
the antiautomorphism $\cA\rightarrow\widetilde{\cA}$ is transferred
onto the quotient algebras ${}^\epsilon\cl_{p,q-1}$ and ${}^\epsilon
\cl_{q,p-1}$ in any case, and the automorphism
$\cA\rightarrow\cA^\star$ and antiautomorpism
$\cA\rightarrow\widetilde{\cA^\star}$ are never transferred.
\end{theorem} 

According to Theorem \ref{thomo} in case of odd--dimensional spaces
$\R^{p,q}$ and $\C^{p+q}$ the algebra homomorphisms
$\cl_{p,q}\rightarrow
\cl_{p,q-1},\,\cl_{p,q}\rightarrow\cl_{q,p-1}$ and $\C_{p+q}\rightarrow
\C_{p+q-1}$ induce group homomorphisms $\pin(p,q)\rightarrow\pin(p,q-1),\,
\pin(p,q)\rightarrow\pin(q,p-1)$, $\pin(p+q,\C)\rightarrow\pin(p+q-1,\C)$.
\begin{theorem}[{\rm\cite{Var99}}]\label{todd}\begin{sloppypar}
1) If $\F=\R$ and $\pin^{a,b,c}(p,q)\simeq\pin^{a,b,c}(p,q-1)\cup\omega
\pin^{a,b,c}(p,q-1),\,\pin^{a,b,c}(p,q)\simeq\pin^{a,b,c}(q,p-1)\cup\omega
\pin^{a,b,c}(q,p-1)$ are D\c{a}browski groups over the field $\R$, 
where $p-q\equiv 1,5
\pmod{8}$, then in the results of homomorphic mappings $\pin^{a,b,c}(p,q)
\rightarrow\pin^{a,b,c}(p,q-1)$ and $\pin^{a,b,c}(p,q)\rightarrow
\pin^{a,b,c}(q,p-1)$ take place the following quotient groups:\end{sloppypar}
\begin{eqnarray}
\pin^b(p,q-1)&\simeq&\frac{(\spin_0(p,q-1)\odot\dZ_2\otimes\dZ_2)}{\dZ_2},
\nonumber\\
\pin^b(q,p-1)&\simeq&\frac{(\spin_0(q,p-1)\odot\dZ_2\otimes\dZ_2)}{\dZ_2}.
\nonumber
\end{eqnarray}
2) If $\F=\C$ and $\pin^{a,b,c}(p+q,\C)\simeq\pin^{a,b,c}(p+q-1,\C)\cup
\pin^{a,b,c}(p+q-1,\C)$ are D\c{a}browski groups over the field $\C$, 
where $p+q\equiv
1,3,5,7\pmod{8}$, then in the result of the homomorphic mapping $\pin^{a,b,c}
(p+q,\C)\rightarrow\pin^{a,b,c}(p+q-1,\C)$ take place the following
quotient groups:
\[
\pin^b(p+q-1,\C)\simeq\frac{(\spin_0(p+q-1,\C)\odot\dZ_2\otimes\dZ_2)}{\dZ_2},
\]
if $p+q\equiv 1,5\pmod{8}$ and
\[
\pin^{b,c}(p+q-1,\C),
\]
if $p+q\equiv 3,7\pmod{8}$, at this point the automorphisms corresponded
to discrete transformations of the space
$\C^{p+q-1}$, associated with the quotient algebra
${}^\epsilon\C_{p+q-1}$, do not form a group.
\end{theorem}
\begin{proof} Indeed, according to Theorem \ref{thomo} over the field 
$\F=\R$ at the homomorphic mappings
$\cl_{p,q}\rightarrow\cl_{p,q-1}$ and
$\cl_{p,q}\rightarrow\cl_{q,p-1}$ from all the fundamental automorphisms
only the antiautomorphism $\cA\rightarrow\widetilde{\cA}$ is transferred onto
the quotient algebras ${}^\epsilon\cl_{p,q-1}$ and
${}^\epsilon\cl_{q,p-1}$. Further, in virtue of Proposition  
\ref{prop1} the antiautomorphism $\cA\rightarrow\widetilde{\cA}$ corresponds
the time reversal $T$. Therefore, discrete groups of the spaces
$\R^{p,q-1}$ and $\R^{q,p-1},$ associated with the quotient groups
${}^\epsilon\cl_{p,q-1}$ and
${}^\epsilon\cl_{q,p-1}$, are defined by two--element group $\{1,T\}\sim
\{\sI,\sE\}\simeq\dZ_2$, where $\{\sI,\sE\}$ is an automorphism group of
${}^\epsilon\cl_{p,q-1},\,{}^\epsilon\cl_{q,p-1}$, $\sE$ is the matrix of
$\cA\rightarrow\widetilde{\cA}$. Thus, at
$p-q\equiv 1,5\pmod{8}$ the homomorphic mappings
$\pin^{a,b,c}(p,q)\rightarrow\pin^b(p,q-1)$ and $\pin^{a,b,c}(p,q)\rightarrow
\pin^b(q,p-1)$ take place, where $\pin^b(p,q-1),\,\pin^b(q,p-1)$ 
are the quotient groups,
$b=T^2=\sE^2$. At this point, the double covering $C^b$ of the discrete
group is isomorphic to $\dZ_2\otimes\dZ_2$.
\begin{sloppypar}
Analogously, over the field $\F=\C$ at $p+q\equiv 1,5\pmod{8}$ we have the
quotient group $\pin^b(p+q-1,\C)$. Further, according to Theorem \ref{thomo}
at $p+q\equiv 3,7\pmod{8}$ in the result of the homomorphic mapping
$\C_{p+q}\rightarrow\C_{p+q-1}$ the transformations
$\cA\rightarrow\widetilde{\cA}$ and $\cA\rightarrow\widetilde{\cA^\star}$
are transferred onto ${}^\epsilon\C_{p+q-1}$. Therefore, a set of the
discrete transformations of the space
$\C^{p+q-1}$ associated with ${}^\epsilon\C_{p+q-1}$ is defined by a
trhee--element set $\{1,T,PT\}\sim\{\sI,\sE,\sC\}$, where
$\{\sI,\sE,\sC\}$ is the automorphism set of ${}^\epsilon\C_{p+q-1}$,
$\sE$ and $\sC$ are the matrices of $\cA\rightarrow
\widetilde{\cA}$ and $\cA\rightarrow\widetilde{\cA^\star}$, correspondingly.
It is easy to see that the set
$\{1,T,PT\}\sim\{\sI,\sE,\sC\}$ does not form a group. Thus, the homomorphism
$\pin^{a,b,c}(p+q,\C)\rightarrow\pin^{b,c}(p+q-1,\C)$ takes place
at $p+q\equiv 3,7\pmod{8}$, where
$\pin^{b,c}(p+q-1,\C)$ is the quotient group,
$b=T^2=\sE^2,\,c=(PT)^2=\sC^2$.\end{sloppypar}
\end{proof}
\section{Discrete symmetries of many body system}
In conclusion of this chapter let us consider the many body system described
by the tensor product $\cl_{3,0}\otimes\cdots\otimes\cl_{3,0}\simeq
\C_2\otimes\cdots\otimes\C_2$ \cite{Hol88,DLG93}. We will find an explicit
form of the symmetry in dependence from the number of particles belonged the
system. With this end in view let us recall some basic facts concerning
tensor decompositions of the Clifford algebras.
Let $\cl(V,Q)$ be the Clifford algebra over the field $\F=\R$, where
$V$ is a vector space endowed with quadratic form
$Q=x^2_1+\ldots+x^2_p-\ldots-x^2_{p+q}$. If $p+q$ is even and $\omega^2=1$, 
then
$\cl(V,Q)$ is called {\it positive} and correspondingly
{\it negative} if $\omega^2=-1$, that is, $\cl_{p,q}>0$ if 
$p-q\equiv 0,4\pmod{8}$ and $\cl_{p,q}<0$ if $p-q\equiv 2,6\pmod{8}$.
\begin{theorem}[{\rm Karoubi \cite{Kar79}}]
1) If $\cl(V,Q)>0$ and $\dim V$ is even, then
\[
\cl(V\oplus V^{\p},Q\oplus Q^{\p})\simeq\cl(V,Q)\otimes\cl(V^{\p},Q^{\p}).
\]
2) If $\cl(V,Q)<0$ and $\dim V$ is even, then
\[
\cl(V\oplus V^{\p},Q\oplus Q^{\p})\simeq\cl(V,Q)\otimes\cl(V^{\p},-Q^{\p}).
\]
\end{theorem}
\begin{sloppypar}\noindent
Over the field $\F=\C$ the Clifford algebra is always positive
(if
$\omega^2=\e^2_{12\ldots p+q}=-1$, then we can suppose $\omega=
i\e_{12\ldots p+q}$). Thus, using Karoubi Theorem we find that
\end{sloppypar}
\begin{equation}\label{3}
\underbrace{\C_2\otimes\cdots\otimes\C_2}_{m\,\text{times}}\simeq\C_{2m}.
\end{equation}
Therefore, the tensor product in (\ref{3}) is isomorphic 
to $\C_{2m}$\footnote{According to \cite{Hol88,DLG93,SLD99} the system
consisting of $m$ partices is described by this product.}.
\begin{theorem}[{\rm\cite{Var03}}]\begin{sloppypar}\noindent
Let $\cl_{3,0}\otimes\cdots\otimes\cl_{3,0}\simeq\C_2\otimes\cdots\otimes
\C_2\simeq\C_{2m}$ be the algebra describing the system composed of $m$
identical particles ($m=\frac{p+q}{2}$) and let $\{1,P,T,PT\}\simeq
\{\sI,\sW,\sE,\sC\}$ be a group of discrete transformations of the a
$\C^{2m}$ associated with the algebra $\C_{2m}$. Then\\
1) If the particle number $m$ is even, then the symmetry of the system
is described by the non--Cliffordian group\end{sloppypar}
\[
\pin^{+,+,+}(p+q,\C)\simeq\frac{(\spin_0(p+q,\C)\odot\dZ_2\otimes\dZ_2\otimes
\dZ_2)}{\dZ_2},
\]
at $p+q\equiv 0,1\pmod{4}$.\\
2) If the particle number $m$ is odd
($p+q\equiv 2,3\pmod{4}$), then the symmetry of the system is described
by the Cliffordian group
\[
\pin^{-,-,-}(p+q,\C)\simeq\frac{(\spin_0(p+q,\C)\odot Q_4)}{\dZ_2}.
\]
\end{theorem}
\chapter{Lorentz group}
\setcounter{theorem}{0}
\setcounter{cor}{0}
\setcounter{opr}{0}
\setcounter{lem}{0}
\setcounter{axiom}{0}
\setcounter{equation}{0}
\section{Finite-dimensional representations of the Lorentz group and
complex Clifford algebras}
It is well--known \cite{RF,ED79,BR77} that representations of the Lorentz
group play a fundamental role in the quantum field theory. Physical fields
are defined in terms of finite--dimensional irreducible representations
of the Lorentz group $O(1,3)\simeq O(3,1)$
(correspondingly, Poincar\'{e} group $O(1,3)\odot T(4)$, where
$T(4)$ is a subgroup of four--dimensional translations). It should be
noted that in accordance with \cite{Nai58} any finite--dimensional 
irreducible representation of the proper Lorentz group 
$\fG_+=O_0(1,3)\simeq O_0(3,1)\simeq SL(2;\C)/\dZ_2$ is
equivalent to some spinor representation. Moreover, spinor representations
exhaust in essence all the finite--dimensional representations of the
group $\fG_+$. This fact we will widely use below.

Let us consider in brief the basic facts concerning the theory of spinor
representations of the Lorentz group. The initial point of this theory
is a correspondence between transformations of the proper Lorentz group
and complex matrices of the second order. Indeed, follows to
\cite{GMS} let us compare the Hermitian matrix of the second order
\begin{equation}\label{6.1}
X=\ar\begin{pmatrix}
x_0+x_3 & x_1-ix_2\\
x_1+ix_2 & x_0-x_3
\end{pmatrix}
\end{equation}
to the vector $v$ of the Minkowski space--time $\R^{1,3}$ with coordinates
$x_0,x_1,x_2,x_3$. At this point $\det X=x^2_0-x^2_1-x^2_2-x^2_3=S^2(x)$.
The correspondence between matrices $X$ and vectors $v$ is one--to--one and
linear. Any linear transformation $X^\prime=aXa^\ast$ in a space of the
matrices $X$ may be considered as a linear transformation $g_a$ in
$\R^{1,3}$, where $a$ is a complex matrix of the second order with
$\det a=1$. The correspondence $a\sim g_a$ possesses following properties:
1) $\ar\begin{pmatrix}
1 & 0 \\ 0 & 1\end{pmatrix}\sim e$ (identity element); 2)
$g_{a_1}g_{a_2}=g_{a_1a_2}$ (composition); 3) two different matrices
$a_1$ and $a_2$ correspond to one and the same transformation $g_{a_1}=g_{a_2}$
only in the case $a_1=-a_2$. Since every complex matrix is defined by
eight real numbers, then from the requirement $\det a=1$ it follow two
conditions $\re\det a=1$ and $\im\det a=0$. These conditions leave six
independent parameters, that coincides with parameter number of the
proper Lorentz group.

Further, a set of all complex matrices of the second order forms a full
matrix algebra $\M_2(\C)$ that is isomorphic to a biquaternion algebra
$\C_2$. In its turn, Pauli matrices
\begin{equation}\label{6.2}
\sigma_0=\ar\begin{pmatrix} 
1 & 0 \\
0 & 1
\end{pmatrix},\quad
\sigma_1=\begin{pmatrix}
0 & 1\\
1 & 0
\end{pmatrix},\quad
\sigma_2=\begin{pmatrix}
0 & -i\\
i & 0
\end{pmatrix},\quad
\sigma_3=\begin{pmatrix}
1 & 0\\
0 &-1
\end{pmatrix}.
\end{equation}
form one from a great number of isomorphic spinbasis of the algebra $\C_2$
(by this reason in physics the algebra $\C_2\simeq\cl^+_{1,3}\simeq\cl_{3,0}$ 
is called Pauli algebra). Using the basis (\ref{6.2}) we can write the
matrix (\ref{6.1}) in the form
\begin{equation}\label{6.3}
X=x^\mu\sigma_\mu.
\end{equation}
The Hermitian matrix (\ref{6.3}) is correspond to a spintensor
$(1,1)$ $X^{\lambda\dot{\nu}}$ with following coordinates
\begin{eqnarray}
x^0=+(1/\sqrt{2})(\xi^1\xi^{\dot{1}}+\xi^2\xi^{\dot{2}}),&&
x^1=+(1/\sqrt{2})(\xi^1\xi^{\dot{2}}+\xi^2\xi^{\dot{1}}),\nonumber\\
x^2=-(i/\sqrt{2})(\xi^1\xi^{\dot{2}}-\xi^2\xi^{\dot{1}}),&&
x^3=+(1/\sqrt{2})(\xi^1\xi^{\dot{1}}-\xi^2\xi^{\dot{2}}),\label{6.4}
\end{eqnarray}
where $\xi^\mu$ and $\xi^{\dot{\mu}}$ are correspondingly coordinates of
spinors and cospinors of spinspaces $\dS_2$ and $\dot{\dS}_2$. Linear
transformations of `vectors' (spinors and cospinors) of the spinspaces
$\dS_2$ and $\dot{\dS}_2$ have the form
\begin{equation}\label{6.5}\ar
\begin{array}{ccc}
\begin{array}{ccc}
{}^\prime\xi^1&=&\alpha\xi^1+\beta\xi^2,\\
{}^\prime\xi^2&=&\gamma\xi^1+\delta\xi^2,
\end{array} & \phantom{ccc} &
\begin{array}{ccc}
{}^\prime\xi^{\dot{1}}&=&\dot{\alpha}\xi^{\dot{1}}+\dot{\beta}\xi^{\dot{2}},\\
{}^\prime\xi^{\dot{2}}&=&\dot{\gamma}\xi^{\dot{1}}+\dot{\delta}\xi^{\dot{2}},
\end{array}\\
\sigma=\ar\begin{pmatrix}
\alpha & \beta\\
\gamma & \delta
\end{pmatrix} & \phantom{ccc} &
\dot{\sigma}=\begin{pmatrix}
\dot{\alpha} & \dot{\beta}\\
\dot{\gamma} & \dot{\delta}
\end{pmatrix}.
\end{array}
\end{equation}
Transformations (\ref{6.5}) form the group $SL(2;\C)$, since
$\sigma\in\M_2(\C)$ and
\[
SL(2;\C)=\left\{\ar\begin{pmatrix} \alpha & \beta\\ \gamma &\delta\end{pmatrix}
\in\C_2:\;\det\begin{pmatrix} \alpha & \beta \\ \gamma & \delta\end{pmatrix}=1
\right\}\simeq\spin_+(1,3).
\]
The expressions (\ref{6.4}) and (\ref{6.5}) compose a base of the 2--spinor
van der Waerden formalism \cite{Wa29,Rum36}, in which the spaces
$\dS_2$ and $\dot{\dS}_2$ are called correspondingly spaces of
{\it undotted and dotted spinors}. The each of the spaces
$\dS_2$ and $\dot{\dS}_2$ is homeomorphic to an extended complex plane
$\C\cup\infty$ representing an absolute (the set of infinitely distant points)
of a Lobatchevskii space $S^{1,2}$. At this point a group of fractional
linear transformations of the plane $\C\cup\infty$ is isomorphic to a motion
group of $S^{1,2}$ \cite{Roz55}. Besides, in accordance with
\cite{Kot27} the Lobatchevskii space $S^{1,2}$ is an absolute of the
Minkowski world $\R^{1,3}$ and, therefore, the group of fractional linear
transformations of the plane $\C\cup\infty$ (motion group of
$S^{1,2}$) twice covers a `rotation group' of the space--time $\R^{1,3}$,
that is the proper Lorentz group.
\begin{theorem}\label{tprod}
Let $\C_2$ be a biquaternion algebra and let $\sigma_i$ be a canonical spinor
representations (Pauli matrices) of the units of $\C_2$, then
$2k$ tensor products of the $k$ matrices $\sigma_i$ form a basis of the
full matrix algebra $\M_{2^k}(\C)$, which is a spinor representation of 
a complex Clifford algebra $\C_{2k}$. The set containing $2k+1$ tensor
products of the $k$ matrices $\sigma_i$ is homomorphically mapped onto a
set consisting of the same $2k$ tensor products and forming a basis
of the spinor representation of a quotient algebra 
${}^\epsilon\C_{2k}$.
\end{theorem} 
\begin{proof} 
As a basis of the spinor representation of the algebra
$\C_2$ we take the Pauli matrices (\ref{6.2}). This choice is explained by
physical applications only (from mathematical viewpoint the choice of the
spinbasis for $\C_2$ is not important).
Let us compose now $2k$ $2^k$--dimensional matrices:
\begin{equation}\label{6.6}
{\renewcommand{\arraystretch}{1.2}
\begin{array}{lcl}
\cE_{1}&=&\sigma_{1}\otimes\sigma_{0}\otimes\cdots\otimes\sigma_{0}\otimes 
\sigma_{0}\otimes\sigma_{0},\\
\cE_{2}&=&\sigma_{3}\otimes\sigma_{1}\otimes\sigma_{0}\otimes\cdots\otimes 
\sigma_{0}\otimes\sigma_{0},\\
\cE_{3}&=&\sigma_{3}\otimes\sigma_{3}\otimes\sigma_{1}\otimes\sigma_{0}
\otimes\cdots\otimes\sigma_{0},\\
\hdotsfor[2]{3}\\
\cE_{k}&=&\sigma_{3}\otimes\sigma_{3}\otimes\cdots\otimes\sigma_{3}
\otimes\sigma_{1},\\
\cE_{k+1}&=&\sigma_{2}\otimes\sigma_{0}\otimes\cdots\otimes\sigma_{0},\\
\cE_{k+2}&=&\sigma_{3}\otimes\sigma_{2}\otimes\sigma_{0}\otimes\cdots\otimes 
\sigma_{0},\\
\hdotsfor[2]{3}\\
\cE_{2k}&=&\sigma_{3}\otimes\sigma_{3}\otimes\cdots\otimes\sigma_{3}
\otimes\sigma_{2}.
\end{array}}\end{equation}
Since $\sigma^2_i=\sigma_0$, then for a square of any matrix from the set
(\ref{6.6}) we have
\begin{equation}\label{6.7}
\cE^2_i=\sigma_0\otimes\sigma_0\otimes\cdots\otimes\sigma_0,\quad
i=1,2,\ldots,2k,
\end{equation}
where the product $\sigma_0\otimes\sigma_0\otimes\cdots\otimes\sigma_0$ equals
to $2^k$--dimensional unit matrix. Further,
\begin{equation}\label{6.8}
\cE_{i}\cE_{j}=-\cE_{j}\cE_{i},\quad  i<j;\;\;i,j=1,2,\ldots 2k.
\end{equation}
Indeed, when $i=1$ and $j=3$ we obtain
\begin{eqnarray}
\cE_{1}\cE_{3}&=&\sigma_{1}\sigma_{3}\otimes\sigma_{3}\otimes\sigma_{1}
\otimes\sigma_{0}\otimes\cdots\otimes\sigma_{0},\nonumber \\
\cE_{3}\cE_{1}&=&\sigma_{3}\sigma_{1}\otimes\sigma_{3}\otimes\sigma_{1}
\otimes\sigma_{0}\otimes\cdots\otimes\sigma_{0}.\nonumber
\end{eqnarray}
Thus, the equalities (\ref{6.7}) and (\ref{6.8}) show that the matrices of
the set (\ref{6.6}) satisfy the multiplication law of the Clifford algebra.
Moreover, let us show that a set of the matrices
$\cE^{\alpha_1}_1\cE^{\alpha_2}_2\ldots\cE^{\alpha_{2k}}_{2k}$, where
each of the indices $\alpha_1,\alpha_2,\ldots,\alpha_{2k}$ takes either of
the two values 0 or 1, consists of $2^{2k}$ matrices. At this point
these matrices form a basis of the full
$2^k$--dimensional matrix algebra (spinor representation of $\C_{2k}$).
Indeed, in virtue of $i\sigma_1\sigma_2=\sigma_3$ from (\ref{6.6}) it follows
\begin{gather}
\cN_{j}=\cE_{j}\cE_{k+j}=\sigma_{0}\otimes\sigma_{0}\otimes\cdots\otimes 
\sigma_{3}\otimes\sigma_{0}\otimes\cdots\otimes\sigma_{0},\label{6.9}\\
j=1,2,\ldots k,\nonumber
\end{gather}
here the matrix $\sigma_3$ occurs in the $j$--th position. Further, since
the tensor product
$\sigma_{0}\otimes\cdots\otimes\sigma_{0}$ of the unit matrices of the
second order is also unit matrix $\cE_0$ of the $2^k$--order, then we can
write
\begin{eqnarray}
Z^{++}_{j}&=&\frac{1}{2}(\cE_{0}-\cN_{j})=\sigma_{0}\otimes\sigma_{0}
\otimes\cdots
\otimes Q^{++}\otimes\sigma_{0}\otimes\cdots\otimes\sigma_{0},\nonumber \\
Z^{--}_{j}&=&\frac{1}{2}(\cE_{0}+\cN_{j})=\sigma_{0}\otimes\sigma_{0}
\otimes\cdots
\otimes Q^{--}\otimes\sigma_{0}\otimes\cdots\otimes\sigma_{0},\nonumber
\end{eqnarray}
where the matrices $Q^{++}$ and $Q^{--}$ occur in the $j$--th position
and have the following form
\[
\ar\begin{pmatrix}
1 & 0 \\
0 & 0
\end{pmatrix},\quad
\begin{pmatrix}
0 & 0 \\
0 & 1
\end{pmatrix}.
\]
From (\ref{6.9}) it follows
\[
\cN_{1}\cN_{2}\ldots\cN_{j}=\sigma_{3}\otimes\sigma_{3}\otimes\cdots\otimes
\sigma_{3}\otimes\sigma_{0}\otimes\cdots\otimes\sigma_{0},
\]
where on the left side we have $j$ matrices $\sigma_3$. In virtue of this
equality we obtain
\begin{eqnarray}
\cL_{j}&=&\cN_{1}\ldots\cN_{j-1}\cE_{j}=
\sigma_{0}\otimes\sigma_{0}\otimes\cdots\otimes
\sigma_{1}\otimes\sigma_{0}\otimes\cdots\otimes\sigma_{0},\nonumber \\
\cL_{k+j}&=&\cN_{1}\ldots\cN_{j-1}\cE_{k+j}=
\sigma_{0}\otimes\sigma_{0}\otimes\cdots\otimes
\sigma_{2}\otimes\sigma_{0}\otimes\cdots\otimes\sigma_{0}.\nonumber
\end{eqnarray}
here the matrices $\sigma_1$ and $\sigma_2$ occur in the $j$--th position,
$j=1,2,\ldots,k$. Thus,
\begin{eqnarray}
Z^{+-}_{j}&=&\frac{1}{2}(\cL_{j}+i\cL_{k+j})=
\sigma_{0}\otimes\sigma_{0}\otimes\cdots
\otimes Q^{+-}\otimes\sigma_{0}\otimes\cdots\otimes\sigma_{0},\nonumber \\
Z^{-+}_{j}&=&\frac{1}{2}(\cL_{j}-i\cL_{k+j})=
\sigma_{0}\otimes\sigma_{0}\otimes\cdots
\otimes Q^{-+}\otimes\sigma_{0}\otimes\cdots\otimes\sigma_{0},\nonumber
\end{eqnarray}
where the matrices $Q^{+-}$ and $Q^{-+}$ also occur in the $j$--th position
and have correspondingly the following form
\[
\ar\begin{pmatrix}
0 & 1 \\
0 & 0
\end{pmatrix},\quad
\begin{pmatrix}
0 & 0 \\
1 & 0
\end{pmatrix}.
\]
Therefore, a matrix
\begin{gather}
\prod^{k}_{j}(Z^{s_{j}r_{j}}_{j})=Q^{s_{1}r_{1}}\otimes
Q^{s_{2}r_{2}}\otimes\cdots\otimes Q^{s_{k}r_{k}},\label{6.10}\\
s_{j}=\pm,\;r_{j}=\pm\nonumber
\end{gather}
has unit elements at the intersection of
$(s_1,s_2,\ldots,s_k)$--th row and $(r_1,r_2,\ldots,r_k)$--th column,
the other elements are equal to zero. In virtue of (\ref{6.7}) and (\ref{6.8}) 
each of the matrices $\cL_j,\,\cL_{k+j}$ and, therefore, each of the matrices
$Z^{s_jr_j}_j$, is represented by a linear combination of the matrices
$\cE^{\alpha_1}_1\cE^{\alpha_2}_2\cdots\cE^{\alpha_{2k}}_{2k}$. Hence it
follows that the matrices $\prod^k_j(Z^{s_jr_j}_j)$ and, therefore, all the
$2^k$--dimensional matrices, are represented by such the linear combinations.
Thus, $2k$ matrices $\cE_1,\ldots,\cE_{2k}$ generate a group consisting of
the products $\pm\cE^{\alpha_1}_1\cE^{\alpha_2}_2\cdots
\cE^{\alpha_{2k}}_{2k}$, and an enveloped algebra of this group is a full
$2^k$--dimensional matrix algebra.

The following part of this Theorem tells that the full matrix algebra, forming
by the tensor products (\ref{6.6}), is a spinor representation of the algebra
$\C_{2k}$. Let us prove this part on the several examples. First of all,
in accordance with (\ref{6.6}) tensor products of the $k=2$ order are
\begin{gather}
\ar\cE_{1}=\sigma_{1}\otimes\sigma_{0}=
\begin{pmatrix}
0 & \phantom{-}1 \\
1 & 0
\end{pmatrix}\otimes
\begin{pmatrix}
1 & 0 \\
0 & \phantom{-}1
\end{pmatrix}=
\begin{pmatrix}
0 & 0 & 1 & 0 \\
0 & 0 & 0 & 1 \\
1 & 0 & 0 & 0 \\
0 & 1 & 0 & 0
\end{pmatrix},\nonumber\\
\cE_{2}=\sigma_{3}\otimes\sigma_{1}=\ar
\begin{pmatrix}
1 & 0 \\
0 & -1
\end{pmatrix}\otimes
\begin{pmatrix}
0 & 1 \\
1 & 0
\end{pmatrix}=
\begin{pmatrix}
0 & 1 & 0 & 0 \\
1 & 0 & 0 & 0 \\
0 & 0 & 0 & -1 \\
0 & 0 & -1 & 0
\end{pmatrix},\nonumber\\
\cE_{3}=\sigma_{2}\otimes\sigma_{0}=\ar
\begin{pmatrix}
0 & -i \\
i & 0
\end{pmatrix}\otimes
\begin{pmatrix}
1 & 0 \\
0 & 1
\end{pmatrix}=
\begin{pmatrix}
0 & 0 & -i & 0 \\
0 & 0 & 0 & -i \\
i & 0 & 0 & 0 \\
0 & i & 0 & 0
\end{pmatrix},\nonumber\\
\cE_{4}=\sigma_{3}\otimes\sigma_{2}=\ar
\begin{pmatrix}
1 & 0 \\
0 & -1
\end{pmatrix}\otimes
\begin{pmatrix}
0 & -i \\
i & 0
\end{pmatrix}=
\begin{pmatrix}
0 & -i & 0 & 0 \\
i & 0 & 0 & 0 \\
0 & 0 & 0 & i \\
0 & 0 & -i & 0
\end{pmatrix}.\label{6.11}
\end{gather}
It is easy to see that $\cE^2_i=\cE_0$ and $\cE_i\cE_j=-\cE_j\cE_i$
$(i,j=1,\ldots,4)$. The set of the matrices $\cE^{\alpha_1}_1\cE^{\alpha_2}_2
\cdots\cE^{\alpha_4}_4$, where each of the indices $\alpha_1,\alpha_2,
\alpha_3,\alpha_4$ takes either of the two values 0 or 1, consists of
$2^4=16$ matrices. At this point these matrices form a basis of the full
4--dimensional matrix algebra. Thus, we can define a one--to--one
correspondence between sixteen matrices $\cE^{\alpha_1}_1\cE^{\alpha_2}_2\cdots
\cE^{\alpha_4}_4$ and sixteen basis elements $\e_{i_1}\e_{i_2}\ldots\e_{i_k}$
of the Dirac algebra $\C_4$. Therefore, the matrices (\ref{6.11}) form a
basis of the spinor representation of $\C_4$. Moreover, from (\ref{6.11}) it
follows that $\C_4\simeq\C_2\otimes\C_2$, that is, {\it the Dirac algebra
is a tensor product of the two Pauli algebras}.

Analogously, when $k=3$ from (\ref{6.6}) we obtain following products
(matrices of the eighth order):
\begin{gather}
\cE_1=\sigma_1\otimes\sigma_0\otimes\sigma_0,\quad
\cE_2=\sigma_3\otimes\sigma_1\otimes\sigma_0,\quad
\cE_3=\sigma_3\otimes\sigma_3\otimes\sigma_1,\nonumber\\
\cE_4=\sigma_2\otimes\sigma_0\otimes\sigma_0,\quad
\cE_5=\sigma_3\otimes\sigma_2\otimes\sigma_0,\quad
\cE_6=\sigma_3\otimes\sigma_3\otimes\sigma_2,\nonumber
\end{gather}
The set of the matrices
$\cE^{\alpha_1}_1\cE^{\alpha_2}_2\cdots\cE^{\alpha_6}_6$, consisting of
$2^6=64$ matrices, forms a basis of the full 8--dimensional matrix algebra,
which is isomorphic to a spinor representation of the algebra
$\C_6\simeq\C_2\otimes\C_2\otimes\C_2$. Generalizing we obtain that $2k$
tensor products of the $k$ Pauli matrices form a basis of the spinor
representation of the complex Clifford algebra
\begin{equation}\label{6.11'}
\C_{2k}\simeq
\underbrace{\C_2\otimes\C_2\otimes\cdots\otimes\C_2}_{k\;\text{times}}.
\end{equation}

Let us consider now a case of odd dimensions. When $n=2k+1$ we add to the
set of the tensor products (\ref{6.6}) a matrix
\begin{equation}\label{6.12}
\cE_{2k+1}=\underbrace{\sigma_{3}\otimes\sigma_{3}\otimes\cdots\otimes\sigma_{3}
}_{k\;\text{times}},
\end{equation}
which satisfying the conditions
\begin{gather}
\cE^{2}_{2k+1}=\cE_{0},\quad \cE_{2k+1}\cE_{i}=-\cE_{i}\cE_{2k+1},\nonumber\\
i=1,2,\ldots, k.\nonumber
\end{gather}
It is obvious that a product
$\cE_{1}\cE_{2}\cdots\cE_{2k}\cE_{2k+1}$ commutes with all the products of the
form $\cE^{\alpha_1}_1\cE^{\alpha_2}_2\cdots\cE^{\alpha_{2k}}_{2k}$.
Further, let
\[
\varepsilon=\begin{cases}
1,& \text{if $k\equiv 0\!\!\!\pmod{2}$},\\
i,& \text{if $k\equiv 1\!\!\!\pmod{2}$}.
\end{cases}
\]
Then a product
\[
\sU=\varepsilon\cE_{1}\cE_{2}\cdots\cE_{2k}\cE_{2k+1}
\]
satisfies the condition $\sU^{2}=\cE_{0}$. Let $\sP$ be a set of all
$2^{2k+1}$ matrices $\cE^{\alpha_1}_1\cE^{\alpha_2}_2\cdots
\cE^{\alpha_{2k}}_{2k}\cE^{\alpha_{2k+1}}_{2k+1}$, where $\alpha_j$ equals to 0
or 1, $j=1,2,\ldots,2k+1$. Let us divide the set $\sP$ into two subsets
by the following manner:
\begin{equation}\label{6.13}
\sP=\sP^{1}+\sP^{0},
\end{equation}
where the subset $\sP^0$ contains products with the matrix $\cE_{2k+1}$,
and $\sP^1$ contains products without the matrix $\cE_{2k+1}$. Therefore,
products
$\cE^{\alpha_1}\cE^{\alpha_2}_2\cdots\cE^{\alpha_{2k}}_{2k}\subset\sP^1$
form a full $2^k$--dimensional matrix algebra. Further, when we multiply
the matrices from the subset $\sP^0$ by the matrix $\sU$ the
factors $\cE_{2k+1}$ are mutually annihilate. Thus, matrices of the set
$\sU\sP^0$ are also belong to the $2^k$--dimensional matrix algebra.
Let us denote $\sU\sP^0$ via $\sP^2$. Taking into account that
$\sU^2=\cE_0$ we obtain $\sP^0=\sU\sP^2$ and
\[
\sP=\sP^1+\sP^2,
\]
where $\sP^1,\sP^2\in\M_{2^k}$. Let
\begin{equation}\label{6.14}
\chi\;:\;\;\sP^{1}+\sU\sP^{2}\longrightarrow
\sP^{1}+\sP^{2}
\end{equation}
be an homomorphic mapping of the set (\ref{6.13}) containing the matrix
$\cE_{2k+1}$ onto the set $\sP^1+\sP^2$ which does not contain this matrix.
The mapping $\chi$ preserves addition and multiplication operations.
Indeed, let $\sP=\sP^1+\sU\sP^2$ and $\sQ=\sQ^1+\sU\sQ^2$, then
\[
\sP+\sQ=\sP^1+\sU\sP^2+\sQ^1+\sU\sQ^2\longrightarrow
\sP^1+\sP^2+\sQ^1+\sQ^2,
\]
\begin{multline}
\sP\sQ=(\sP^1+\sU\sP^2)(\sQ^1+\sU\sQ^2)=(\sP^1\sQ^1+\sP^2\sQ^2)+
\sU(\sP^1\sQ^2+\sP^2\sQ^1)\longrightarrow\\
(\sP^1\sQ^1+\sP^2\sQ^2)+(\sP^1\sQ^2+\sP^2\sQ^1)=(\sP^1+\sP^2)(\sQ^1+\sQ^2),
\end{multline}
that is, an image of the products equals to the product of factor images in
the same order. In particular, when $\sP=\sU$ we obtain
$\sP^1=0,\,\sP^2=\cE_0$ and
\begin{equation}\label{6.15}
\sU\longrightarrow\cE_0.
\end{equation}\begin{sloppypar}\noindent
In such a way, at the mapping $\chi$ all the matrices of the form
$\cE^{\alpha_1}_1\cE^{\alpha_2}_2\cdots\cE^{\alpha_{2k}}_{2k}-
\sU\cE^{\alpha_1}_1\cE^{\alpha_2}_2\cdots\cE^{\alpha_{2k}}_{2k}$
are mapped into zero. Therefore, a kernel of the homomorphism $\chi$ is
defined by an expression $\Ker\chi=\left\{\sP^1-\sU\sP^1\right\}$. We obtain
the homomorphic mapping of the set of all the $2^{2k+1}$ matrices
$\cE^{\alpha_1}_1\cE^{\alpha_2}_2\cdots\cE^{\alpha_{2k}}_{2k}
\cE^{\alpha_{2k+1}}_{2k+1}$ onto the full matrix algebra $\M_{2^k}$.
In the result of this mapping we have a quotient algebra
${}^\chi\M_{2^k}\simeq\sP/\Ker\chi$. As noted previously,
$2k$ tensor products of the $k$ Pauli matrices (or other $k$ matrices
defining spinor representation of the biquaternion algebra $\C_2$)
form a basis of the spinor representation of the algebra $\C_{2k}$. 
It is easy to see that there exists one--to--one correspondence 
between $2^{2k+1}$ matrices of the set $\sP$ and 
basis elements of the odd--dimensional
Clifford algebra $\C_{2k+1}$. It is well--known \cite{Rash,Port} that
$\C_{2k+1}$ is isomorphic to a direct sum of two even--dimensional subalgebras:
$\C_{2k+1}\simeq\C_{2k}\oplus\C_{2k}$. Moreover, there exists an
homomorphic mapping $\epsilon:\;\C_{2k+1}\rightarrow\C_{2k}$, in the result
of which we have a quotient algebra
${}^\epsilon\C_{2k}\simeq\C_{2k+1}/\Ker\epsilon$, where
$\Ker\epsilon=\left\{\cA^1-\varepsilon\omega\cA^1\right\}$ is a kernel of the
homomorphism $\epsilon$, $\cA^1$ is an arbitrary element of the subalgebra
$\C_{2k}$, $\omega$ is a volume element of the algebra $\C_{2k+1}$.
It is easy to see that the homomorphisms $\epsilon$ and $\chi$ have a
similar structure. Thus, hence it immediately follows an isomorphism
${}^\epsilon\C_{2k}\simeq{}^\chi\M_{2^k}$. Therefore, a basis of the matrix
quotient algebra ${}^\chi\M_{2^k}$ is also a basis of the spinor
representation of the Clifford quotient algebra ${}^\epsilon\C_{2k}$, that
proves the latter assertion of Theorem.\end{sloppypar}
\end{proof}
Let us consider now spintensor representations of the proper
Lorentz group $O_0(1,3)\simeq SL(2;\C)/\dZ_2\simeq\spin_+(1,3)/\dZ_2$ and their
relations with the complex Clifford algebras. From each complex Clifford
algebra $\C_n=\C\otimes\cl_{p,q}\;
(n=p+q)$ we obtain a spinspace $\dS_{2^{n/2}}$, which is a complexification
of the minimal left ideal of the algebra
$\cl_{p,q}$: $\dS_{2^{n/2}}=\C\otimes I_{p,q}=\C\otimes\cl_{p,q}
e_{pq}$, where $e_{pq}$ is a primitive idempotent of the algebra $\cl_{p,q}$.
Further, a spinspace corresponding the Pauli algebra $\C_2$ has a form
$\dS_2=\C\otimes I_{2,0}=\C\otimes\cl_{2,0}e_{20}$ or
$\dS_2=\C\otimes I_{1,1}=\C\otimes\cl_{1,1}e_{11}(\C\otimes I_{0,2}=
\C\otimes\cl_{0,2}e_{02})$. Therefore, the tensor product
(\ref{6.11'}) of the $k$ algebras $\C_2$ induces a tensor product of the $k$
spinspaces $\dS_2$:
\[
\dS_2\otimes\dS_2\otimes\cdots\otimes\dS_2=\dS_{2^k}.
\]
Vectors of the spinspace $\dS_{2^k}$ (or elements of the minimal left
ideal of $\C_{2k}$) are spintensors of the following form
\begin{equation}\label{6.16}
\xi^{\alpha_1\alpha_2\cdots\alpha_k}=\sum
\xi^{\alpha_1}\otimes\xi^{\alpha_2}\otimes\cdots\otimes\xi^{\alpha_k},
\end{equation}
where summation is produced on all the index collections
$(\alpha_1\ldots\alpha_k)$, $\alpha_i=1,2$. In virtue of (\ref{6.5}) for
each spinor $\xi^{\alpha_i}$ from (\ref{6.16}) we have a transformation rule
${}^\prime\xi^{\alpha^\prime_i}=
\sigma^{\alpha^\prime_i}_{\alpha_i}\xi^{\alpha_i}$. Therefore, in general
case we obtain
\begin{equation}\label{6.17}
{}^\prime\xi^{\alpha^\prime_1\alpha^\prime_2\cdots\alpha^\prime_k}=\sum
\sigma^{\alpha^\prime_1}_{\alpha_1}\sigma^{\alpha^\prime_2}_{\alpha_2}\cdots
\sigma^{\alpha^\prime_k}_{\alpha_k}\xi^{\alpha_1\alpha_2\cdots\alpha_k}.
\end{equation}
A representation (\ref{6.17}) is called
{\it undotted spintensor representation of the proper Lorentz group of the
rank $k$}.

Further, let $\overset{\ast}{\C}_2$ be a biquaternion algebra, the
coefficients of which are complex conjugate. Let us show that the algebra
$\overset{\ast}{\C}_2$ is obtained from $\C_2$ under action of the
automorphism $\cA\rightarrow\cA^\star$ or antiautomorphism
$\cA\rightarrow\widetilde{\cA}$. Indeed, in virtue of an isomorphism
$\C_2\simeq\cl_{3,0}$ a general element
\[
\cA=a^0\e_0+\sum^3_{i=1}a^i\e_i+\sum^3_{i=1}\sum^3_{j=1}a^{ij}\e_{ij}+
a^{123}\e_{123}
\]
of the algebra $\cl_{3,0}$ can be written in the form
\begin{equation}\label{6.17'}
\cA=(a^0+\omega a^{123})\e_0+(a^1+\omega a^{23})\e_1+(a^2+\omega a^{31})\e_2
+(a^3+\omega a^{12})\e_3,
\end{equation}
where $\omega=\e_{123}$. Since $\omega$ belongs to a center of the algebra
$\cl_{3,0}$ (commutes with all the basis elements) and $\omega^2=-1$, then
we can to suppose $\omega\equiv i$. The action of the automorphism $\star$
on the homogeneous element $\cA$ of a degree $k$ is defined by a formula
$\cA^\star=(-1)^k\cA$. In accordance with this the action of the
automorphism $\cA\rightarrow\cA^\star$, where $\cA$ is the element
(\ref{6.17'}), has a form
\begin{equation}\label{In1}
\cA\longrightarrow\cA^\star=-(a^0-\omega a^{123})\e_0-(a^1-\omega a^{23})\e_1
-(a^2-\omega a^{31})\e_2-(a^3-\omega a^{12})\e_3.
\end{equation}
Therefore, $\star:\,\C_2\rightarrow -\overset{\ast}{\C}_2$. Correspondingly,
the action of the antiautomorphism $\cA\rightarrow\widetilde{\cA}$ on the
homogeneous element $\cA$ of a degree $k$ is defined by a formula
$\widetilde{\cA}=(-1)^{\frac{k(k-1)}{2}}\cA$. Thus, for the element
(\ref{6.17'}) we obtain
\begin{equation}\label{In2}
\cA\longrightarrow\widetilde{\cA}=(a^0-\omega a^{123})\e_0+
(a^1-\omega a^{23})\e_1+(a^2-\omega a^{31})\e_2+(a^3-\omega a^{12})\e_3,
\end{equation}
that is, $\widetilde{\phantom{cc}}:\,\C_2\rightarrow
\overset{\ast}{\C}_2$.
This allows to define an algebraic analog of the Wigner's representation
doubling: $\C_2\oplus\overset{\ast}{\C}_2$. 
Further, from (\ref{6.17'})
it follows that $\cA=\cA_1+\omega\cA_2=(a^0\e_0+a^1\e_1+a^2\e_2+a^3\e_3)+
\omega(a^{123}\e_0+a^{23}\e_1+a^{31}\e_2+a^{12}\e_3)$. In general case,
by virtue of an isomorphism $\C_{2k}\simeq\cl_{p,q}$, where $\cl_{p,q}$ is a
real Clifford algebra with a division ring $\K\simeq\C$, $p-q\equiv 3,7
\pmod{8}$, we have for a general element of $\cl_{p,q}$ an expression
$\cA=\cA_1+\omega\cA_2$, here $\omega^2=\e^2_{12\ldots p+q}=-1$ and, therefore,
$\omega\equiv i$. Thus, from $\C_{2k}$ under action of the automorphism
$\cA\rightarrow\cA^\star$ we obtain a general algebraic doubling
\begin{equation}\label{D}
\C_{2k}\oplus\overset{\ast}{\C}_{2k}.
\end{equation}

Correspondingly, a tensor product
$\overset{\ast}{\C}_2\otimes\overset{\ast}{\C}_2\otimes\cdots\otimes
\overset{\ast}{\C}_2\simeq\overset{\ast}{\C}_{2r}$ of $r$ algebras
$\overset{\ast}{\C}_2$ induces a tensor product of $r$ spinspaces
$\dot{\dS}_2$:
\[
\dot{\dS}_2\otimes\dot{\dS}_2\otimes\cdots\otimes\dot{\dS}_2=\dot{\dS}_{2^r}.
\]
Te vectors of the spinspace $\dot{\dS}_{2^r}$ have the form
\begin{equation}\label{6.18}
\xi^{\dot{\alpha}_1\dot{\alpha}_2\cdots\dot{\alpha}_r}=\sum
\xi^{\dot{\alpha}_1}\otimes\xi^{\dot{\alpha}_2}\otimes\cdots\otimes
\xi^{\dot{\alpha}_r},
\end{equation}
where the each cospinor $\xi^{\dot{\alpha}_i}$ from (\ref{6.18}) in virtue of
(\ref{6.5}) is transformed by the rule ${}^\prime\xi^{\dot{\alpha}^\prime_i}=
\sigma^{\dot{\alpha}^\prime_i}_{\dot{\alpha}_i}\xi^{\dot{\alpha}_i}$.
Therefore,
\begin{equation}\label{6.19}
{}^\prime\xi^{\dot{\alpha}^\prime_1\dot{\alpha}^\prime_2\cdots
\dot{\alpha}^\prime_r}=\sum\sigma^{\dot{\alpha}^\prime_1}_{\dot{\alpha}_1}
\sigma^{\dot{\alpha}^\prime_2}_{\dot{\alpha}_2}\cdots
\sigma^{\dot{\alpha}^\prime_r}_{\dot{\alpha}_r}
\xi^{\dot{\alpha}_1\dot{\alpha}_2\cdots\dot{\alpha}_r}.
\end{equation}\begin{sloppypar}\noindent
A representation (\ref{6.19}) is called
{\it a dotted spintensor representation of the proper Lorentz group of the
rank $r$}.\end{sloppypar}

In general case we have a tensor product of $k$ algebras $\C_2$ and
$r$ algebras $\overset{\ast}{\C}_2$:
\[
\C_2\otimes\C_2\otimes\cdots\otimes\C_2\otimes
\overset{\ast}{\C}_2\otimes
\overset{\ast}{\C}_2\otimes\cdots\otimes\overset{\ast}{\C}_2\simeq
\C_{2k}\otimes\overset{\ast}{\C}_{2r},
\]
which induces a spinspace
\[
\dS_2\otimes\dS_2\otimes\cdots\otimes\dS_2\otimes\dot{\dS}_2\otimes
\dot{\dS}_2\otimes\cdots\otimes\dot{\dS}_2=\dS_{2^{k+r}}
\]
with the vectors
\begin{equation}\label{6.20'}
\xi^{\alpha_1\alpha_2\cdots\alpha_k\dot{\alpha}_1\dot{\alpha}_2\cdots
\dot{\alpha}_r}=\sum
\xi^{\alpha_1}\otimes\xi^{\alpha_2}\otimes\cdots\otimes\xi^{\alpha_k}\otimes
\xi^{\dot{\alpha}_1}\otimes\xi^{\dot{\alpha}_2}\otimes\cdots\otimes
\xi^{\dot{\alpha}_r}.
\end{equation}
In this case we have a natural unification of the representations
(\ref{6.17}) and (\ref{6.19}):
\begin{equation}\label{6.20}
{}^\prime\xi^{\alpha^\prime_1\alpha^\prime_2\cdots\alpha^\prime_k
\dot{\alpha}^\prime_1\dot{\alpha}^\prime_2\cdots\dot{\alpha}^\prime_r}=\sum
\sigma^{\alpha^\prime_1}_{\alpha_1}\sigma^{\alpha^\prime_2}_{\alpha_2}\cdots
\sigma^{\alpha^\prime_k}_{\alpha_k}\sigma^{\dot{\alpha}^\prime_1}_{
\dot{\alpha}_1}\sigma^{\dot{\alpha}^\prime_2}_{\dot{\alpha}_2}\cdots
\sigma^{\dot{\alpha}^\prime_r}_{\dot{\alpha}_r}
\xi^{\alpha_1\alpha_2\cdots\alpha_k\dot{\alpha}_1\dot{\alpha}_2\cdots
\dot{\alpha}_r}.
\end{equation}
So, a representation (\ref{6.20}) is called
{\it a spintensor representation of the proper Lorentz group of the
rank $(k,r)$}.

In general case, the representations defining by the formulas
(\ref{6.17}), (\ref{6.19}) and
(\ref{6.20}), are reducible, that is there exists possibility of decomposition 
of the initial spinspace $\dS_{2^{k+r}}$ (correspondingly, spinspaces
$\dS_{2^k}$ and $\dS_{2^r}$) into a direct sum of invariant (with respect to
transformations of the group $\fG_+$) spinspaces
$\dS_{2^{\nu_1}}\oplus\dS_{2^{\nu_2}}\oplus\cdots\oplus\dS_{2^{\nu_s}}$,
where $\nu_1+\nu_2+\ldots+\nu_s=k+r$.
\begin{sloppypar}
Further, an important notion of the {\it physical field} is closely related
with finite--dimensional representations of the proper Lorentz
group $\fG_+$.
In accordance with Wigner interpretation
\cite{Wig39}, an elementary particle is described by some irreducible
finite--dimensional representation of the Poincar\'{e} group. The double
covering of the proper Poincar\'{e} group is isomorphic
to a semidirect product $SL(2;\C)\odot T(4)$, or
$\spin_+(1,3)\odot T(4)$, where $T(4)$ is the subgroup of four--dimensional
translations. Let $\psi(x)$ be a physical field, then at the transformations
$(a,\Lambda)$ of the proper Poincar\'{e} group the field $\psi(x)$ 
is transformed by a following rule\end{sloppypar}
\begin{equation}\label{6.21}
\psi^\prime_mu(x)=\sum_\nu\fC_{\mu\nu}(\sigma)\psi_\nu
(\Lambda^{-1}(x-a)),
\end{equation}\begin{sloppypar}\noindent
where $a\in T(4)$, $\sigma\in\fG_+$, $\Lambda$ is a Lorentz transformation,
and $\fC_{\mu\nu}$ is a representation of the group $\fG_+$ in the space
$\dS_{2^{k+r}}$. Since the group $T(4)$ is Abelian, then all its
representations are one--dimensional. Thus, all the finite--dimensional
representations of the proper Poincar\'{e} group in essence
are equivalent to the representations $\fC$ of the group
$\fG_+$. If the representation $\fC$ is reducible, then the space
$\dS_{2^{k+r}}$ is decomposed into a direct sum of irreducible subspaces, that is,
it is possible to choose in $\dS_{2^{k+r}}$ such a basis, in which all the
matrices $\fC_{\mu\nu}$ take a block--diagonal form. Then the field $\psi(x)$ 
is reduced to some number of the fields corresponding to obtained irreducible
representations of the group $\fG_+$, each of which is transformed
independently from the other, and the field $\psi(x)$ in this case is a
collection of the fields with more simple structure. It is obvious that these
more simple fields correspond to irreducible representations $\fC$. 
As known \cite{Nai58,GMS,RF}, a system of irreducible finite--dimensional
representations of the group $\fG_+$ is realized in the space
$\Sym_{(k,r)}\subset\dS_{2^{k+r}}$ of symmetric spintensors. The
dimensionality of $\Sym_{(k,r)}$ is equal to $(k+1)(r+1)$. 
A representation of the group
$\fG_+$ by such spintensors is irreducible and denoted by the symbol
$\fC^{j,j^\prime}$, where $2j=k,\;2j^\prime=r$, numbers $j$ and
$j^\prime$ defining the spin are integer or half--integer. Then the field
$\psi(x)$ transforming by the formula (\ref{6.21}) is, in general case, a field
of the type $(j,j^\prime)$. In such a way, all the physical fields are
reduced to the fields of this type, the mathematical structure of which
requires a knowledge of representation matrices $\fC^{j,j^\prime}_{\mu\nu}$. 
As a rule, in physics there are two basic types of the fields:\\[0.3cm]
1) The field of type $(j,0)$. The structure of this field 
(or the field $(0,j)$) is described by the representation
$\fC^{j,0}$ ($\fC^{0,j^\prime}$), which is realized in the
space $\Sym_{(k,0)}\subset\dS_{2^k}$
($\Sym_{(0,r)}\subset\dS_{2^r}$). At this point in accordance with
Theorem \ref{tprod} the algebra
$\C_{2k}\simeq\C_2\otimes\C_2\otimes\cdots\otimes\C_2$
(correspondingly, $\overset{\ast}{\C}_2\simeq\overset{\ast}{\C}_2\otimes
\overset{\ast}{\C}_2\otimes\cdots\otimes\overset{\ast}{\C}_2$)
is associated with the field of type $(j,0)$ (correspondingly, $(0,j^\prime)$)
The trivial case $j=0$ corresponds to {\it a Pauli--Weisskopf
field} describing the scalar particles. In other particular case, when
$j=j^\prime=1/2$ we have {\it a Weyl field} describing the neutrino.
At this point the antineutrino is described by a fundamental representation
$\fC^{1/2,0}=\sigma$ of the group $\fG_+$ and the algebra
$\C_2$ related with this representation (Theorem \ref{tprod}). Correspondingly,
the neutrino is described by a conjugated representation $\fC^{0,1/2}$
and the algebra $\overset{\ast}{\C}_2$. In relation with this, it is hardly
too much to say that the neutrino field is a more fundamental physical field,
that is a kind of the basic building block, from which other physical fields
built by means of direct sum or tensor product.\\[0.3cm]
2) The field of type $(j,0)\oplus(0,j)$. The structure of this field admits
a space inversion and, therefore, in accordance with a Wigner's doubling
\cite{Wig64} is described by a representation $\fC^{j,0}\oplus\fC^{0,j}$
of the group $\fG_+$. This representation is realized in the space
$\Sym_{(k,k)}\subset
\dS_{2^{2k}}$. In accordance with (\ref{D})
the Clifford algebra related with this representation is a direct sum
$\C_{2k}\oplus\overset{\ast}{\C}_{2k}\simeq
\C_2\otimes\C_2\otimes\cdots\otimes\C_2\oplus
\overset{\ast}{\C}_2\otimes
\overset{\ast}{\C}_2\otimes\cdots\otimes\overset{\ast}{\C}_2$.
In the simplest case $j=1/2$ we have {\it bispinor (electron--positron)
Dirac field} $(1/2,0)\oplus(0,1/2)$ with the algebra $\C_2\oplus$
\raisebox{0.21ex}
{$\overset{\ast}{\C}_2$}. It should be noted that the Dirac algebra
$\C_4$ considered as a tensor product $\C_2\otimes\C_2$ 
(or
$\C_2\otimes\overset{\ast}{\C}_2$) 
in accordance with (\ref{6.16})
(or (\ref{6.20'})) gives rise to spintensors $\xi^{\alpha_1\alpha_2}$
(or $\xi^{\alpha_1\dot{\alpha}_1}$), but it contradicts with the usual
definition of the Dirac bispinor as a pair 
$(\xi^{\alpha_1},\xi^{\dot{\alpha}_1})$. Therefore, the Clifford algebra
associated with the Dirac field is 
$\C_2\oplus\overset{\ast}{\C}_2$, and
a spinspace of this sum in virtue of unique decomposition
$\dS_2\oplus\dot{\dS}_2=\dS_4$ ($\dS_4$ is a spinspace of $\C_4$) allows to
define $\gamma$--matrices in the Weyl basis.
The case $j=1$ corresponds to 
{\it Maxwell fields} $(1,0)$ and $(0,1)$ with the algebras
$\C_2\otimes\C_2$ and $\overset{\ast}{\C}_2\otimes\overset{\ast}{\C}_2$.
At this point the electromagnetic field is defined by complex linear
combinations $\bF=\bE-i\bH,\;\overset{\ast}{\bF}=\bE+i\bH$ (Helmholtz
representation). Besides, the algebra related with Maxwell field is a tensor
product of the two algebras $\C_2$ describing the neutrino fields.
In this connection it is of interest to recall {\it a neutrino theory of
light} was proposed by de Broglie and Jordan \cite{Bro32,Jor35}.
In the de Broglie--Jordan neutrino theory of light electromagnetic field
is constructed from the two neutrino fields (for more details and related
papers see \cite{Dvo99}). Traditionally, physicists attempt to describe
electromagnetic field in the framework of $(1,0)\oplus(0,1)$ representation
(see old works \cite{LU31,Rum36,RF} and recent developments based on the
Joos--Weinberg formalism \cite{Joo62,Wein} and its relation with a
Bargmann--Wightman--Wigner type quantum field theory \cite{AJG93,Dvo97}).
However, Weinberg's equations (or Weinberg--like equations) for electromagnetic
field obtained within the subspace $\Sym_{(k,r)}$ with dimension
$2(2j+1)$ have acausal (tachionic) solutions \cite{AE92}. 
Electromagnetic field in terms of a quotient representation
$(1,0)\cup (0,1)$ in the full representation space $\dS_{2^{k+r}}$ will be
considered in separate paper.
\end{sloppypar}

In this connection it should be noted two important circumstances related
with irreducible representations of the group $\fG_+$ and complex Clifford
algebras associated with these representations. The first circumstance relates
with the Wigner interpretation of an elementary particle. Namely, a relation
between finite--dimensional representations of the proper Lorentz group
and comlex Clifford algebras (Theorem
\ref{tprod}) allows to essentially extend the Wigner interpretation by means
of the use of an extraordinary rich and universal structure of the Clifford
algebras at the study of space--time (and also intrinsic) symmetries of
elementary particles. The second circumstance relates with the spin.
Usually, the Clifford algebra is associated with a half--integer spin
corresponding to fermionic fields, so--called `matter fields', whilst the
fields with an integer spin (bosonic fields) are eliminated from an algebraic
description. However, such a non--symmetric situation is invalid, since
the fields with integer spin have a natural description in terms of 
spintensor representations of the proper Lorentz group with even rank and
algebras $\C_{2k}$ and $\overset{\ast}{\C}_{2k}$ associated with these
representations, where $k$ is even (for example, Maxwell field). In this
connection it should be noted that generalized statistics in terms of
Clifford algebras have been recently proposed by Finkelstein and
collaborators \cite{FG00,BFGS}.

As known, complex Clifford algebras $\C_n$ are modulo 2 periodic
\cite{AtBSh} and, therefore, there exist two types of $\C_n$:
$n\equiv 0\s\pmod{2}$ and $n\equiv 1\s\pmod{2}$. Let us consider these two
types in the form of following series:
\[
\begin{array}{cccccccccc}
\C_2 && \C_4 && \cdots && \C_{2k} && \cdots \\
& \C_3 && \C_5 && \cdots && \C_{2k+1} && \cdots
\end{array}
\]
Let us consider the decomposition $\C_{2k+1}\simeq\C_{2k}\oplus\C_{2k}$
in more details. This decomposition may be represented by a following scheme
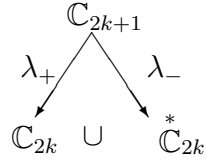
\[
\unitlength=0.5mm
\begin{picture}(70,50)
\put(35,40){\vector(2,-3){15}}
\put(35,40){\vector(-2,-3){15}}
\put(28.25,42){$\C_{2k+1}$}
\put(16,28){$\lambda_{+}$}
\put(49.5,28){$\lambda_{-}$}
\put(13.5,9.20){$\C_{2k}$}
\put(52.75,9){$\stackrel{\ast}{\C}_{2k}$}
\put(32.5,10){$\cup$}
\end{picture}
\]
Here central idempotents
\[
\lambda^+=\frac{1+\varepsilon\e_1\e_2\cdots\e_{2k+1}}{2},\quad
\lambda^-=\frac{1-\varepsilon\e_1\e_2\cdots\e_{2k+1}}{2},
\]
where
\[
\varepsilon=\begin{cases}
1,& \text{if $k\equiv 0\pmod{2}$},\\
i,& \text{if $k\equiv 1\pmod{2}$}
\end{cases}
\]
satisfy the relations $(\lambda^+)^2=\lambda^+$, $(\lambda^-)^2=\lambda^-$,
$\lambda^+\lambda^-=0$. Thus, we have a decomposition of the initial
algebra $\C_{2k+1}$ into a direct sum of two mutually annihilating simple
ideals: $\C_{2k+1}\simeq\frac{1}{2}(1+\varepsilon\omega)\C_{2k+1}\oplus
\frac{1}{2}(1-\varepsilon\omega)\C_{2k+1}$. Each of the ideals
$\lambda^{\pm}\C_{2k+1}$ is isomorphic to the subalgebra 
$\C_{2k}\subset\C_{2k+1}$. In accordance with Chisholm and Farwell \cite{CF97}
the idempotents $\lambda^+$ and $\lambda^-$ can be identified with
helicity projection operators which distinguish left and right handed
spinors. The Chisholm--Farwell notation for $\lambda^\pm$ we will widely
use below.Therefore,
in virtue of the isomorphism $\C_{2k+1}\simeq\C_{2k}\cup\C_{2k}$ and the
homomorphic mapping $\epsilon:\,\C_{2k+1}\rightarrow\C_{2k}$ the second series
(type $n\equiv 1\s\pmod{2}$) is replaced by a sequence of the quotient
algebras ${}^\epsilon\C_{2k}$, that is,
\[
\begin{array}{cccccccccc}
\C_2 && \C_4 && \cdots && \C_{2k} && \cdots \\
& {}^\epsilon\C_2 && {}^\epsilon\C_4 && \cdots && {}^\epsilon\C_{2k} && \cdots
\end{array}
\]
Representations corresponded these two series of $\C_n$ ($n\equiv 0,1\pmod{2}$)
form a full system $\fM=\fM^0\oplus\fM^1$ of finite--dimensional 
representations of the proper Lorentz group $\fG_+$.

All the physical fields used in quantum field theory and related 
representations of the group
$\fG_+$ $\fC^{j,0}\;(\C_{2k})$, $\fC^{j,0}\oplus\fC^{0,j}\;
(\C_{2k}\otimes\overset{\ast}{\C}_{2k}$) are constructed from the upper
series (type $n\equiv 0\s\pmod{2}$). Whilst the lower series
(type $n\equiv 1\s\pmod{2}$) is not considered in physics as yet.
In accordance with Theorem \ref{tprod} we have an isomorphism
${}^\epsilon\C_{2k}\simeq{}^\chi\M_{2^k}$. Therefore, the quotient algebra
${}^\epsilon\C_{2k}$ induces a spinspace
${}^\epsilon\dS_{2^k}$ that is a space of a quotient representation
${}^\chi\fC^{j,0}$ of the group
$\fG_+$. Analogously, a quotient representation ${}^\chi\fC^{0,j^\prime}$ 
is realised in the space ${}^\epsilon\dot{\dS}_{2^r}$ which
induced by the quotient algebra ${}^\epsilon\overset{\ast}{\C}_{2r}$.
In general case, we have a quotient representation ${}^\chi\fC^{j,j^\prime}$ 
defined by a tensor product ${}^\epsilon\C_{2k}\otimes
{}^\epsilon\overset{\ast}{\C}_{2r}$. Thus, {\it the complex type
$n\equiv 1\s\pmod{2}$ corresponds to a full system of irreducible
finite--dimensional quotient representations ${}^\chi\fC^{j,j^\prime}$
of the proper Lorentz group}. Therefore, until now in physics only one half
($n\equiv 0\s\pmod{2}$) of all possible finite--dimensional representations
of the Lorentz group has been used.

Let us consider now a full system of physical fields with different types.
First of all, the field
\begin{equation}\label{F1}
(j,0)=(1/2,0)\otimes(1/2,0)\otimes\cdots\otimes(1/2,0)
\end{equation}
in accordance with Theorem \ref{tprod} is a tensor product of the $k$ fields 
of type $(1/2,0)$, each of which corresponds to the fundamental representation
$\fC^{1/2,0}=\sigma$ of the group $\fG_+$ and the biquaternion
algebra $\C_2$ related with fundamental representation. In its turn, the
field
\begin{equation}\label{F2}
(0,j^\prime)=(0,1/2)\otimes(0,1/2)\otimes\cdots\otimes(0,1/2)
\end{equation}
is a tensor product of the $r$ fundamental fields of the type $(0,1/2)$,
each of which corresponds to a conjugated representation 
$\fC^{0,1/2}=\dot{\sigma}$ and the conjugated algebra
$\overset{\ast}{\C}_2$ obtained in accordance with (\ref{In1})--(\ref{In2})
under action of the automorphism $\cA\rightarrow\cA^\star$ (space inversion),
or under action of the antiautomorphism $\cA\rightarrow\widetilde{\cA}$
(time reversal). The numbers $j$ and $j^\prime$ are integer (bosonic fields) if
in the products (\ref{F1}) and (\ref{F2}) there are $k,r\equiv 0\pmod{2}$
$(1/2,0)$ (or $(0,1/2)$) factors, and the numbers $j$ and $j^\prime$ are
half--integer (fermionic fields) if in the products (\ref{F1})--(\ref{F2})
there are $k,r\equiv 1\pmod{2}$ factors. Further, the field
\begin{equation}\label{F3}
(j,j^\prime)=(1/2,0)\otimes(1/2,0)\otimes\cdots\otimes(1/2,0)\bigotimes
(0,1/2)\otimes(0,1/2)\otimes\cdots\otimes(0,1/2)
\end{equation}
is a tensor product of the fields (\ref{F1}) and (\ref{F2}). As consequence
of the doubling (\ref{D}) we have the field of type $(j,0)\oplus(0,j)$: 
\[
(j,0)\oplus(0,j)=(1/2,0)\otimes(1/2,0)\otimes\cdots\otimes(1/2,0)\bigoplus
(0,1/2)\otimes(0,1/2)\otimes\cdots\otimes(0,1/2)
\]
In general, all the fields (\ref{F1})--(\ref{F3}) describe multiparticle
states. The decompositions of these multiparticle states into single states
provided in the full representation space $\dS_{2^{k+r}}$, where
$\Sym_{(k,r)}$ and $\Sym_{(k,k)}$ with dimensions
$(2j+1)(2j^\prime+1)$ and $2(2j+1)$ are subspaces of $\dS_{2^{k+r}}$
(for example, the Clebsh--Gordan decomposition of two spin 1/2 particles
into singlet and triplet: $(1/2,1/2)=(1/2,0)\otimes(0,1/2)=(0,0)\oplus(1,0)$).
In the papers \cite{Hol88,DLG93,SLD99} a multiparticle state is described in 
the framework of a tensor product 
$\cl_{3,0}\otimes\cl_{3,0}\otimes\cdots\otimes\cl_{3,0}$. It is easy to see
that in virtue of the isomorphism $\cl_{3,0}\simeq\C_2$ the tensor product
of the algebras $\cl_{3,0}$ is isomorphic to the product (\ref{6.11'}).
Therefore, the Holland approach naturally incorporates into a more general
scheme considered here. Finally, for the type $n\equiv 1\pmod{2}$ we have
quotient representations ${}^\chi\fC^{j,j^\prime}$ of the group
$\fG_+$. The physical fields corresponding to the
quotient representations are constructed like the fields (\ref{F1})-(\ref{F3}).
Due to the decomposition
$\C_n\simeq\C_{n-1}\cup\C_{n-1}$ ($n\equiv 1\pmod{2}$) we have a field
\begin{equation}\label{F4}
(j,0)\cup(j,0)=(1/2,0)\otimes(1/2,0)\otimes\cdots\otimes(1/2,0)\bigcup
(1/2,0)\otimes(1/2,0)\otimes\cdots\otimes(1/2,0),
\end{equation}
and also we have fields $(0,j)\cup(0,j)$ and $(j,0)\cup(0,j)$ if the quotient
algebras ${}^\epsilon\C_{n-1}$ admit space inversion or time reversal.
The field
\begin{equation}\label{F5}
(j,0)\cup(0,j)=(1/2,0)\otimes(1/2,0)\otimes\cdots\otimes(1/2,0)\bigcup
(0,1/2)\otimes(0,1/2)\otimes\cdots\otimes(0,1/2)
\end{equation}
is analogous to the field $(j,0)\oplus(0,j)$, but, in general, the field
$(j,0)\cup(0,j)$ has a quantity of violated discrete symmetries.
An explicit form of the quotient representations and their relations with
discrete symmetries will be explored in the following sections.
\section{Discrete symmetries on the representation spaces of the
Lorentz group}
Since all the physical fields are defined in terms of finite--dimensional
representations of the group $\fG_+$, then a construction of the
discrete symmetries 
(space inversion $P$, time reversal $T$ and combination 
$PT$) on the representation spaces of the Lorentz group has a primary
importance.

In general case, according to Theorem \ref{tprod} a space of the
finite--dimensional representation of the group $SL(2;\C)$ is a spinspace
$\dS_{2^{k+r}}$, or a minimal left ideal of the algebra 
$\C_{2k}\otimes\overset{\ast}{\C}_{2r}$. 
Therefore, in the spinor representation the
fundamental automorphisms of the algebra
$\C_{2k}\otimes\overset{\ast}{\C}_{2r}$ 
by virtue of the
isomorphism $\Aut(\cl)\simeq\{1,P,T,PT\}$ induce discrete transformations
on the representation spaces (spinspaces) of the Lorentz group.
\begin{theorem}\label{tinf}
1) The field $\F=\C$. The tensor products 
$\C_2\otimes\C_2\otimes\cdots\otimes\C_2$, 
$\overset{\ast}{\C}_2\otimes\overset{\ast}{\C}_2\otimes\cdots\otimes
\overset{\ast}{\C}_2$, 
$\C_2\otimes\C_2\otimes\cdots\otimes\C_2\otimes
\overset{\ast}{\C}_2\otimes\overset{\ast}{\C}_2\otimes\cdots
\overset{\ast}{\C}_2$ of the Pauli algebra $\C_2$ correspond to
finite--dimensional representations $\fC^{l_0+l_1-1,0}$, $\fC^{0,l_0-l_1+1}$,
$\fC^{l_0+l_1-1,l_0-l_1+1}$ of the proper Lorentz group $\fG_+$, where
$(l_0,l_1)=\left(\frac{k}{2},\frac{k}{2}+1\right)$, $(l_0,l_1)=
\left(-\frac{r}{2},\frac{r}{2}+1\right)$, $(l_0,l_1)=\left(\frac{k-r}{2},
\frac{k+r}{2}+1\right)$, and spinspaces $\dS_{2^k}$, $\dS_{2^r}$,
$\dS_{2^{k+r}}$ are representation spaces of the group $\fG_+$,
$\C_2\leftrightarrow\fC^{1,0}$ 
($\overset{\ast}{\C}_2\leftrightarrow\fC^{0,-1}$) 
is a fundamental
representation of $\fG_+$. Then in a spinor representation of the
fundamental automorphisms of the algebra $\C_n$ for the matrix $\sW$ of the
automorphism $\cA\rightarrow\cA^\star$ (space inversion) and also for the
matrices $\sE$ and $\sC$ of the antiautomorphisms $\cA\rightarrow
\widetilde{\cA}$ (time reversal) and $\cA\rightarrow\widetilde{\cA^\star}$
(full reflection) the following permutation relations with infinitesimal
operators of the group $\fG_+$ take place:
\begin{alignat}{7}
\ld\sW,A_{23}\rd &=\ld\sW,A_{13}\rd &=\ld\sW,A_{12}\rd &=0,\quad
\lf\sW,B_1\rf &=\lf\sW,B_2\rf &=\lf\sW,B_3\rf &=0\label{T0}\\
\ld\sW,H_+\rd &=\ld\sW,H_-\rd &=\ld\sW,H_3\rd &=0,\quad
\lf\sW,F_+\rf &=\lf\sW,F_-\rf &=\lf\sW,F_3\rf &=0.\label{T0'}
\end{alignat}
\begin{alignat}{7}
\ld\sE,A_{23}\rd &=\ld\sE,A_{13}\rd &=\ld\sE,A_{12}\rd &=0,\quad
\lf\sE,B_1\rf &=\lf\sE,B_2\rf &=\lf\sE,B_3\rf &=0,\label{bT1}\\
\ld\sC,A_{23}\rd &=\ld\sC,A_{13}\rd &=\ld\sC,A_{12}\rd &=0,\quad
\ld\sC,B_1\rd &=\ld\sC,B_2\rd &=\ld\sC,B_3\rd &=0,\label{bT2}\\
\ld\sE,H_+\rd &=\ld\sE,H_-\rd &=\ld\sE,H_3\rd &=0,\quad
\lf\sE,F_+\rf &=\lf\sE,F_-\rf &=\lf\sE,F_3\rf &=0,\label{bT3}\\
\ld\sC,H_+\rd &=\ld\sC,H_-\rd &=\ld\sC,H_3\rd &=0,\quad
\ld\sC,F_+\rd &=\ld\sC,F_-\rd &=\ld\sC,F_3\rd &=0.\label{bT4}
\end{alignat}
\begin{alignat}{7}
\ld\sE,A_{23}\rd &=\ld\sE,A_{13}\rd &=\ld\sE,A_{12}\rd &=0,\quad
\ld\sE,B_1\rd &=\ld\sE,B_2\rd &=\ld\sE,B_3\rd &=0,\label{bT5}\\
\ld\sC,A_{23}\rd &=\ld\sC,A_{13}\rd &=\ld\sC,A_{12}\rd &=0,\quad
\lf\sC,B_1\rf &=\lf\sC,B_2\rd &=\lf\sC,B_3\rd &=0,\label{bT6}\\
\ld\sE,H_+\rd &=\ld\sE,H_-\rd &=\ld\sE,H_3\rd &=0,\quad
\ld\sE,F_+\rd &=\ld\sE,F_-\rd &=\ld\sE,F_3\rd &=0,\label{bT7}\\
\ld\sC,H_+\rd &=\ld\sC,H_-\rd &=\ld\sC,H_3\rd &=0,\quad
\lf\sC,F_+\rf &=\lf\sC,F_-\rf &=\lf\sC,F_3\rf &=0.\label{bT8}
\end{alignat}
\begin{alignat}{7}
\ld\sE,A_{23}\rd &=0,\quad\lf\sE,A_{13}\rf &=\lf\sE,A_{12}\rf &=0,\quad
\ld\sE,B_1\rd &=0,\quad\lf\sE,B_2\rf &=\lf\sE,B_3\rf &=0,\label{bT9}\\
\ld\sC,A_{23}\rd &=0,\quad\lf\sC,A_{13}\rf &=\lf\sC,A_{12}\rf &=0,\quad
\lf\sC,B_1\rf &=0,\quad\ld\sC,B_2\rd &=\ld\sC,B_3\rd &=0.\label{bT10}
\end{alignat}
\begin{alignat}{7}
\ld\sE,A_{23}\rd &=0,\quad\lf\sE,A_{13}\rf &=\lf\sE,A_{12}\rf &=0,\quad
\lf\sE,B_1\rf &=0,\quad\ld\sE,B_2\rd &=\ld\sE,B_3\rd &=0,\label{bT11}\\
\ld\sC,A_{23}\rd &=0,\quad\lf\sC,A_{13}\rf &=\lf\sC,A_{12}\rf &=0,\quad
\ld\sC,B_1\rd &=0,\quad\lf\sC,B_2\rf &=\ld\sC,B_3\rf &=0.\label{bT12}
\end{alignat}
\begin{alignat}{7}
\lf\sE,A_{23}\rf &=\lf\sE,A_{13}\rf &=0,\quad\ld\sE,A_{12}\rd &=0,\quad
\ld\sE,B_1\rd &=\ld\sE,B_2\rd &=0,\quad\lf\sE,B_3\rf &=0,\label{bT13}\\
\lf\sC,A_{23}\rf &=\lf\sC,A_{13}\rf &=0,\quad\ld\sC,A_{12}\rd &=0,\quad
\lf\sC,B_1\rf &=\lf\sC,B_2\rf &=0,\quad\ld\sC,B_3\rd &=0,\label{bT14}\\
\lf\sE,H_+\rf &=\lf\sE,H_-\rf &=0,\quad\ld\sE,H_3\rd &=0,\quad
\ld\sE,F_+\rd &=\ld\sE,F_-\rd &=0,\quad\lf\sE,F_3\rf &=0,\label{bT15}\\
\lf\sC,H_+\rf &=\lf\sC,H_-\rf &=0,\quad\ld\sC,H_3\rd &=0,\quad
\lf\sC,F_+\rf &=\lf\sC,F_-\rf &=0,\quad\ld\sC,F_3\rd &=0.\label{bT16}
\end{alignat}
\begin{alignat}{7}
\lf\sE,A_{23}\rf &=0,\;\;\ld\sE,A_{13}\rd &=0,\;\;\lf\sE,A_{12}\rf &=0,\;\;
\ld\sE,B_1\rd &=0,\;\;\lf\sE,B_2\rf &=0,\;\;\ld\sE,B_3\rd &=0,\label{bT17}\\
\lf\sC,A_{23}\rf &=0,\;\;\ld\sC,A_{13}\rd &=0,\;\;\lf\sC,A_{12}\rf &=0,\;\;
\lf\sC,B_1\rf &=0,\;\;\ld\sC,B_2\rd &=0,\;\;\lf\sC,B_3\rf &=0.\label{bT18}
\end{alignat}
\begin{alignat}{7}
\lf\sE,A_{23}\rf &=0,\;\;\ld\sE,A_{13}\rd &=0,\;\;\lf\sE,A_{12}\rf &=0,\;\;
\lf\sE,B_1\rf &=0,\;\;\ld\sE,B_2\rd &=0,\;\;\lf\sE,B_3\rf &=0,\label{bT19}\\
\lf\sC,A_{23}\rf &=0,\;\;\ld\sC,A_{13}\rd &=0,\;\;\lf\sC,A_{12}\rf &=0,\;\;
\ld\sC,B_1\rd &=0,\;\;\lf\sC,B_2\rf &=0,\;\;\ld\sC,B_3\rd &=0.\label{bT20}
\end{alignat}
\begin{alignat}{7}
\lf\sE,A_{23}\rf &=\lf\sE,A_{13}\rf &=0,\quad\ld\sE,A_{12}\rd &=0,\quad
\lf\sE,B_1\rf &=\lf\sE,B_2\rf &=0,\quad\ld\sE,B_3\rd &=0,\label{bT21}\\
\lf\sC,A_{23}\rf &=\lf\sC,A_{13}\rf &=0,\quad\ld\sC,A_{12}\rd &=0,\quad
\ld\sC,B_1\rd &=\ld\sC,B_2\rd &=0,\quad\lf\sC,B_3\rf &=0,\label{bT22}\\
\lf\sE,H_+\rf &=\lf\sE,H_-\rf &=0,\quad\ld\sE,H_3\rd &=0,\quad
\lf\sE,F_+\rf &=\lf\sE,F_-\rf &=0,\quad\ld\sE,F_3\rd &=0,\label{bT23}\\
\lf\sC,H_+\rf &=\lf\sC,H_-\rf &=0,\quad\ld\sC,H_3\rd &=0,\quad
\ld\sC,F_+\rd &=\ld\sC,F_-\rd &=0,\quad\lf\sC,F_3\rf &=0.\label{bT24}
\end{alignat}
where $A_{23},\,A_{13},\,A_{12}$ are infinitesimal operators of a subgroup of
three--dimensional rotations, $B_1,\,B_2,\,B_3$ are infinitesimal operators of
hyperbolic rotations.\\[0.2cm]
2) The field $\F=\R$. The factorization $\cl_{s_i,t_j}\otimes\cl_{s_i,t_j}
\otimes\cdots\otimes\cl_{s_i,t_j}$ of the real Clifford algebra $\cl_{p,q}$
corresponds to a real finite--dimensional representation of the group $\fG_+$
with a pair $(l_0,l_1)=\left(\frac{p+q}{4},0\right)$, that is equivalent to a
representation of the subgroup $SO(3)$ of three--dimensional rotations
($B_1=B_2=B_3=0$). Then there exist two classes of real representations
$\fR^{l_0}_{0,2}$ of the group $\fG_+$ corresponding to the algebras
$\cl_{p,q}$ with a division ring $\K\simeq\R$, $p-q\equiv 0,2\pmod{8}$,
and also there exist two classes of quaternionic representations 
$\fH^{l_0}_{4,6}$ of $\fG_+$ corresponding to the algebras $\cl_{p,q}$
with a ring $\K\simeq\BH$, $p-q\equiv 4,6\pmod{8}$. For the real
representations $\fR^{l_0}_{0,2}$ operators of the discrete subgroup of 
$\fG_+$ defining by the matrices $\sW,\,\sE,\,\sC$ of the fundamental
automorphisms of $\cl_{p,q}$ with $p-q\equiv 0,2\pmod{8}$ are always
commute with all the infinitesimal operators of the representation.
In turn, for the quaternionic representations $\fH^{l_0}_{4,6}$ following
relations hold:
\begin{equation}
\ld\sW,A_{23}\rd=\ld\sW,A_{13}\rd=\ld\sW,A_{12}\rd=0,\quad
\ld\sW,H_+\rd=\ld\sW,H_-\rd=\ld\sW,H_3\rd=0.
\end{equation}
\begin{alignat}{7}
\ld\sE,A_{23}\rd &=\ld\sE,A_{13}\rd &=\ld\sE,A_{12}\rd &=0,\quad
\ld\sC,A_{23}\rd &=\ld\sC,A_{13}\rd &=\ld\sC,A_{12}\rd &=0,\label{TR11}\\
\ld\sE,H_+\rd &=\ld\sE,H_-\rd &=\ld\sE,H_3\rd &=0,\quad
\ld\sC,H_+\rd &=\ld\sC,H_-\rd &=\ld\sC,H_3\rd &=0.
\end{alignat}
\begin{equation}
\ld\sE,A_{23}\rd=0,\quad\lf\sE,A_{13}\rf=\lf\sE,A_{12}\rf=0,\quad
\ld\sC,A_{23}\rd=0,\quad\lf\sC,A_{13}\rf=\lf\sC,A_{12}\rf=0.\label{TR12}
\end{equation}
\begin{alignat}{7}
\lf\sE,A_{23}\rf &=\lf\sE,A_{13}\rf &=0,\quad\ld\sE,A_{12}\rd=0,\quad
\lf\sC,A_{23}\rf &=\lf\sC,A_{13}\rf &=0,\quad\ld\sC,A_{12}\rd=0,\\
\lf\sE,H_+\rf &=\lf\sE,H_-\rf &=0,\quad\ld\sE,H_3\rd=0,\quad
\lf\sC,H_+\rf &=\lf\sC,H_-\rf &=0,\quad\ld\sC,H_3\rd=0.
\end{alignat}
\begin{equation}
\lf\sE,A_{23}\rf=0,\;\;\ld\sE,A_{13}\rd=0,\;\;\lf\sE,A_{12}\rf=0,\;\;
\lf\sC,A_{23}\rf=0,\;\;\ld\sC,A_{13}\rd=0,\;\;\lf\sC,A_{12}\rf=0.
\end{equation}
\end{theorem}
\begin{proof}
1) Complex representations.\\
As noted previously, a full representation space of the finite--dimensional
representation of the proper Lorentz group $\fG_+$ is defined in terms of
the minimal left ideal of the algebra $\C_2\otimes\C_2\otimes\cdots\C_2
\simeq\C_{2k}$. Indeed, in virtue of an isomorphism
\begin{equation}\label{Iso}
\C_{2k}\simeq\cl_{p,q},
\end{equation}
where $\cl_{p,q}$ is a Clifford algebra over the field $\F=\R$ with a
division ring $\K\simeq\C$, $p-q\equiv 3,7\pmod{8}$, we have for the minimal
left ideal of $\C_{2k}$ an expression $\dS=\cl_{p,q}f$, here
\[
f=\frac{1}{2}(1\pm\e_{\alpha_1})\frac{1}{2}(1\pm\e_{\alpha_2})\cdots
\frac{1}{2}(1\pm\e_{\alpha_t})
\]
is a primitive idempotent of the algebra $\cl_{p,q}$ \cite{Lou81}, and
$\e_{\alpha_1},\e_{\alpha_2},\ldots,\e_{\alpha_t}$ are commuting elements
with square 1 of the canonical basis of $\cl_{p,q}$ generating a group of
order $2^t$. The values of $t$ are defined by a formula $t=q-r_{q-p}$,
where $r_i$ are the Radon--Hurwitz numbers \cite{Rad22,Hur23}, values of
which form a cycle of the period 8: $r_{i+8}=r_i+4$. 
The dimension of the minimal left ideal $\dS$ is equal to 
$2^k=2^{\frac{p+q-1}{2}}$. Therefore, for the each finite--dimensional
representation of the group $\fG_+$ we have $2^t$ copies of the spinspace
$\dS_{2^k}$ (full representation space). It should be noted that not all
these copies are equivalent to each other, some of them give rise to
different reflection groups (see \cite{Var00}).

In general, all the finite--dimensional representations of 
group $\fG_+$ in the spinspace $\dS_{2^k}$
are reducible.
Therefore, there exists a decomposition of the spinspace
$\dS_{2^k}\simeq\dS_2\otimes\dS_2\otimes\cdots\dS_2$ into a direct sum of
invariant subspaces $\Sym_{(k_j,0)}$ of symmetric spintensors with
dimensions $(k_j+1)$:
\[
\dS_{2^k}=\Sym_{(k_1,0)}\oplus\Sym_{(k_2,0)}\oplus\cdots\oplus\Sym_{k_s,0)},
\]
where $k_1+k_2+\ldots+k_s=k$, $k_j\in\dZ$. At this point there exists an
orthonormal basis with matrices of the form
\[
\ar\begin{pmatrix}
A^0_t & & & & & & &\\
& A^{1/2}_t & & & & & &\\
& & \ddots & & & & & \\
& & & A^s_t & & & &\\
& & & & B^0_t & & &\\
& & & & & B^{1/2}_t & &\\
& & & & & & \ddots &\\
& & & & & & & B^s_t
\end{pmatrix},
\]
where for the matrices $A^i_1,\,A^j_2,\,A^j_3,\,B^j_1,\,B^j_2,\,B^j_3$
(matrices of the infinitesimal operators of $\fG_+$) in accordance with
Gel'fand--Naimark formulas \cite{GMS,Nai58}
\begin{equation}\label{I1}
A_{23}\xi_{l,m}=-\frac{i}{2}\sqrt{(l+m+1)(l-m)}\xi_{l,m+1}-
\frac{i}{2}\sqrt{(l+m)(l-m+1)}\xi_{l,m-1},
\end{equation}
\begin{equation}\label{I2}
A_{13}\xi_{l,m}=\frac{1}{2}\sqrt{(l+m)(l-m+1)}\xi_{l,m-1}-
\frac{1}{2}\sqrt{(l+m+1)(l-m)}\xi_{l,m+1},
\end{equation}
\begin{equation}\label{I3}
A_{12}\xi_{l,m}=-im\xi_{l,m},
\end{equation}
\begin{multline}\label{I4}
B_1\xi_{l,m}=-\frac{i}{2}C_l\sqrt{(l-m)(l-m-1)}\xi_{l-1,m+1}+
\frac{i}{2}A_l\sqrt{(l-m)(l+m+1)}\xi_{l,m+1}-\\
\frac{i}{2}C_{l+1}\sqrt{(l+m+1)(l+m+2)}\xi_{l+1,m+1}+
\frac{i}{2}C_l\sqrt{(l+m)(l+m-1)}\xi_{l-1,m-1}+\\
\frac{i}{2}A_l\sqrt{(l+m)(l-m+1)}\xi_{l,m-1}+
\frac{i}{2}C_{l+1}\sqrt{(l-m+1)(l-m+2)}\xi_{l+1,m-1},
\end{multline}
\begin{multline}\label{I5}
B_2\xi_{l,m}=-\frac{1}{2}C_l\sqrt{(l+m)(l+m-1)}\xi_{l-1,m-1}-
\frac{1}{2}A_l\sqrt{(l+m)(l-m+1)}\xi_{l,m-1}-\\
\frac{1}{2}C_{l+1}\sqrt{(l-m+1)(l-m+2)}\xi_{l+1,m-1}-
\frac{1}{2}C_l\sqrt{(l-m)(l-m-1)}\xi_{l-1,m+1}+\\
\frac{1}{2}A_l\sqrt{(l-m)(l+m+1)}\xi_{l,m+1}-
\frac{1}{2}C_{l+1}\sqrt{(l+m+1)(l+m+2)}\xi_{l+1,m+1},
\end{multline}
\begin{equation}\label{I6}
B_3\xi_{l,m}=-iC_l\sqrt{l^2-m^2}\xi_{l-1,m}+iA_lm\xi_{l,m}+
iC_{l+1}\sqrt{(l+1)^2-m^2}\xi_{l+1,m},
\end{equation}
\begin{equation}\label{I6'}
A_l=\frac{il_0l_1}{l(l+1)},\quad
C_l=\frac{i}{l}\sqrt{\frac{(l^2-l^2_0)(l^2-l^2_1)}{4l^2-1}}
\end{equation}
\begin{gather}
m=-l,-l+1,\ldots,l-1,l\nonumber\\
l=l_0,l_0+1,\ldots\nonumber
\end{gather}
we have
\begin{equation}\label{bA1}
A^j_{23}=-\frac{i}{2}
\ar\begin{bmatrix}
0 & \alpha_{-l_j+1} & 0 & \dots & 0 & 0\\
\alpha_{-l_j+1} & 0 & \alpha_{-l_j+2} & \dots & 0 & 0\\
0 & \alpha_{-l_j+2} & 0 & \dots & 0 & 0\\
\hdotsfor[2]{6}\\
\hdotsfor[2]{6}\\
0 & 0 & 0 & \dots & 0 & \alpha_{l_j}\\
0 & 0 & 0 & \dots & \alpha_{l_j} & 0
\end{bmatrix}
\end{equation}
\begin{equation}\label{bA2}
A^j_{13}=\frac{1}{2}
\ar\begin{bmatrix}
0 & \alpha_{-l_j+1} & 0 & \dots & 0 & 0\\
-\alpha_{-l_j+1} & 0 & \alpha_{-l_j+2} & \dots & 0 & 0\\
0 & -\alpha_{-l_j+2} & 0 & \dots & 0 & 0\\
\hdotsfor[2]{6}\\
\hdotsfor[2]{6}\\
0 & 0 & 0 & \dots & 0 & \alpha_{l_j}\\
0 & 0 & 0 & \dots & -\alpha_{l_j} & 0
\end{bmatrix}
\end{equation}
\begin{equation}\label{bA3}
A^j_{12}=
\ar\begin{bmatrix}
il_j & 0 & 0 & \dots & 0 & 0\\
0 & i(l_j-1) & 0 & \dots & 0 & 0\\
0 & 0 & i(l_j-2) & \dots & 0 & 0\\
\hdotsfor[2]{6}\\
\hdotsfor[2]{6}\\
0 & 0 & 0 & \dots & -i(l_j-1) & 0\\
0 & 0 & 0 & \dots & 0 & -il_j
\end{bmatrix}
\end{equation}
\begin{equation}\label{bB1}
B^j_1=\frac{i}{2}A_j
\ar\begin{bmatrix}
0 & \alpha_{-l_j+1} & 0 & \dots & 0 & 0\\
\alpha_{-l_j+1} & 0 & \alpha_{-l_j+2} & \dots & 0 & 0\\
0 & \alpha_{-l_j+2} & 0 & \dots & 0 & 0\\
\hdotsfor[2]{6}\\
\hdotsfor[2]{6}\\
0 & 0 & 0 & \dots & 0 & \alpha_{l_j}\\
0 & 0 & 0 & \dots & \alpha_{l_j} & 0
\end{bmatrix}
\end{equation}
\begin{equation}\label{bB2}
B^j_2=\frac{1}{2}A_j
\ar\begin{bmatrix}
0 & -\alpha_{-l_j+1} & 0 & \dots & 0 & 0\\
\alpha_{-l_j+1} & 0 & -\alpha_{-l_j+2} & \dots & 0 & 0\\
0 & \alpha_{-l_j+2} & 0 & \dots & 0 & 0\\
\hdotsfor[2]{6}\\
\hdotsfor[2]{6}\\
0 & 0 & 0 & \dots & 0 & -\alpha_{l_j}\\
0 & 0 & 0 & \dots & \alpha_{l_j} & 0
\end{bmatrix}
\end{equation}
\begin{equation}\label{bB3}
B^j_3=\frac{1}{2}A_j
\ar\begin{bmatrix}
il_j & 0 & 0 & \dots & 0 & 0\\
0 & i(l_j-1) & 0 & \dots & 0 & 0\\
0 & 0 & i(l_j-2) & \dots & 0 & 0\\
\hdotsfor[2]{6}\\
\hdotsfor[2]{6}\\
0 & 0 & 0 & \dots & -i(l_j-1) & 0\\
0 & 0 & 0 & \dots & 0 & -il_j
\end{bmatrix}
\end{equation}
where $\alpha_m=\sqrt{(l_j+m)(l_j-m+1)}$. The formulas (\ref{I1})--(\ref{I6'})
define a finite--dimensional representation of the group $\fG_+$ when
$l^2_1=(l_0+p)^2$, $p$ is some natural number,
$l_0$ is an integer or half--integer number, $l_1$ is an arbitrary
complex number. In the case $l^2_1\neq(l_0+p)^2$ we have an
infinite--dimensional representation of $\fG_+$. We will deal below only
with the finite--dimensional representations, because these representations
are most useful in physics.

The relation between the numbers $l_0$, $l_1$ and the number $k$ of the
factors $\C_2$ in the product $\C_2\otimes\C_2\otimes\cdots\otimes\C_2$ is
given by a following formula
\[
(l_0,l_1)=\left(\frac{k}{2},\frac{k}{2}+1\right),
\]
whence it immediately follows that $k=l_0+l_1-1$. Thus, we have
{\it a complex representation $\fC^{l_0+l_1-1,0}$ of the proper Lorentz
group $\fG_+$ in the spinspace $\dS_{2^k}$}.

Let us calculate now infinitesimal operators of the fundamental representation
$\fC^{1,0}$ of $\fG_+$. The representation $\fC^{1,0}$ is defined by a pair
$(l_0,l_1)=\left(\frac{1}{2},\frac{3}{2}\right)$. In accordance with
(\ref{I1})--(\ref{I6'}) from (\ref{bA1})--(\ref{bB3}) we obtain
\begin{align}
A^{1/2}_{23} &=-\frac{i}{2}\ar\begin{bmatrix}
0 & \alpha_{1/2}\\
\alpha_{1/2} & 0
\end{bmatrix}=-\frac{i}{2}\ar\begin{bmatrix}
0 & 1\\
1 & 0
\end{bmatrix},\label{18}\\[0.2cm]
A^{1/2}_{13} &=\frac{1}{2}\ar\begin{bmatrix}
0 & \alpha_{1/2}\\
-\alpha_{1/2} & 0
\end{bmatrix}=\frac{1}{2}\ar\begin{bmatrix}
0 & 1\\
-1 & 0
\end{bmatrix},\label{19}\\[0.2cm]
A^{1/2}_{12} &=\frac{1}{2}\ar\begin{bmatrix}
i & 0\\
0 & -i
\end{bmatrix},\label{b20}\\[0.2cm]
B^{1/2}_1 &=\frac{i}{2}A_{1/2}\ar\begin{bmatrix}
0 & \alpha_{1/2}\\
\alpha_{1/2} & 0
\end{bmatrix}=-\frac{1}{2}\ar\begin{bmatrix}
0 & 1\\
1 & 0
\end{bmatrix},\label{21}\\[0.2cm]
B^{1/2}_2 &=\frac{1}{2}A_{1/2}\ar\begin{bmatrix}
0 & -\alpha_{1/2}\\
\alpha_{1/2} & 0
\end{bmatrix}=\frac{1}{2}\ar\begin{bmatrix}
0 & -i\\
i & 0
\end{bmatrix},\label{22}\\[0.2cm]
B^{1/2}_3 &=\frac{i}{2}A_{1/2}\ar\begin{bmatrix}
1 & 0\\
0 & -1
\end{bmatrix}=\frac{1}{2}\ar\begin{bmatrix}
-1 & 0\\
0 & 1
\end{bmatrix}.\label{23}
\end{align}
The operators (\ref{18})--(\ref{23}) satisfy the relations
\begin{alignat}{3}
\ld A_{23},A_{13}\rd &=A_{12}, & \quad \ld A_{13},A_{12}\rd &=A_{23}, \quad
\ld A_{12},A_{23}\rd &=A_{13},\nonumber\\
\ld B_1,B_2\rd &=-A_{12}, & \quad \ld B_2,B_3\rd &=A_{23}, \quad
\ld B_3,B_1\rd &=A_{13},\nonumber\\
\ld A_{23},B_1\rd &=0, & \quad \ld A_{13},B_2\rd &=0, \quad
\ld A_{12},B_3\rd &=0,\nonumber\\
\ld A_{23},B_2\rd &=-B_3, & \quad \ld A_{23},B_3\rd &=B_2,&&\nonumber\\
\ld A_{13},B_3\rd &=-B_1, & \quad \ld A_{13},B_1\rd &=B_3,&&\nonumber\\
\ld A_{12},B_1\rd &=B_2, & \quad \ld A_{12},B_2\rd &=-B_1.&&\label{bcommut}
\end{alignat}
From (\ref{21})--(\ref{23}) it is easy to see that there is an equivalence
between infinitesimal operators $B^{1/2}_i$ and Pauli matrices:
\begin{equation}\label{IF1}
B^{1/2}_1=-\frac{1}{2}\sigma_1,\quad B^{1/2}_2=\frac{1}{2}\sigma_2,\quad
B^{1/2}_3=-\frac{1}{2}\sigma_3.
\end{equation}
In its turn, from (\ref{18})--(\ref{b20}) it follows
\begin{equation}\label{IF2}
A^{1/2}_{23}=-\frac{1}{2}\sigma_2\sigma_3,\quad
A^{1/2}_{13}=-\frac{1}{2}\sigma_1\sigma_2,\quad
A^{1/2}_{12}=\frac{1}{2}\sigma_1\sigma_2.
\end{equation}
It is obvious that this equivalence takes place also for high dimensions,
that is, there exists an equivalence between infinitesimal operators
(\ref{bA1})--(\ref{bB3}) and tensor products of the Pauli matrices.
In such a way, let us suppose that 
\begin{alignat}{3}
A^j_{23} &\sim-\frac{1}{2}\cE_a\cE_b, & \quad
A^j_{13} &\sim-\frac{1}{2}\cE_c\cE_b, & \quad
A^j_{12} &\sim\frac{1}{2}\cE_c\cE_a,\label{O1}\\
B^j_1 &\sim-\frac{1}{2}\cE_c, & \quad
B^j_2 &\sim\frac{1}{2}\cE_a, & \quad
B^j_3 &\sim-\frac{1}{2}\cE_b,\label{O2}
\end{alignat}
where $\cE_i$ ($i=a,b,c$) are $k$--dimensional matrices (the tensor products
(\ref{6.6})) and $c<a<b$. It is easy to verify that the operators
(\ref{O1})--(\ref{O2}) satisfy the relations (\ref{bcommut}). Indeed,
for the commutator $\ld A_{23},A_{13}\rd$ we obtain
\begin{multline}
\ld A_{23},A_{13}\rd=A_{23}A_{13}-A_{13}A_{23}=
\frac{1}{4}\cE_a\cE_b\cE_c\cE_b-\\
-\frac{1}{4}\cE_c\cE_b\cE_a\cE_b=-\frac{1}{4}\cE_a\cE_c+
\frac{1}{4}\cE_c\cE_a=\frac{1}{2}\cE_c\cE_a=A_{12}\nonumber
\end{multline}
and so on. Therefore, the operator set (\ref{O1})--(\ref{O2}) isomorphically
defines the set of infinitesimal operators of the group $\fG_+$.

In accordance with Gel'fand--Yaglom approach \cite{GY48} (see also
\cite{GMS,Nai58}) an operation of space inversion $P$ commutes with all the
operators $A_{ik}$ and anticommutes with all the operators $B_i$:
\begin{alignat}{3}
PA_{23}P^{-1} &=A_{23}, & \quad
PA_{13}P^{-1} &=A_{13}, & \quad
PA_{12}P^{-1} &=A_{12}, \nonumber\\
PB_1P^{-1} &=-B_1, & \quad
PB_2P^{-1} &=-B_2, & \quad
PB_3P^{-1} &=-B_3.\label{GY}
\end{alignat}

Let us consider permutation conditions of the operators (\ref{O1})--(\ref{O2})
with the matrix $\sW$ of the automorphism $\cA\rightarrow\cA^\star$
(space inversion). Since $\sW=\cE_1\cE_2\cdots\cE_n$ is a volume element of
$\C_n$, then $\cE_a,\,\cE_b,\,\cE_c\in\sW$, $\cE^2_i=\sI$,
$\sW^2=\sI$ at $n\equiv 0\pmod{4}$ and $\sW^2=-\sI$ at $n\equiv 2\pmod{4}$.
Therefore, for the operator $A_{23}\sim-\frac{1}{2}\cE_a\cE_b$ we obtain
\begin{eqnarray}
A_{23}\sW&=&-(-1)^{a+b-2}\frac{1}{2}\cE_1\cE_2\cdots\cE_{a-1}\cE_{a+1}\cdots
\cE_{b-1}\cE_{b+1}\cdots\cE_n,\nonumber\\
\sW A_{23}&=&-(-1)^{2n-a-b}\frac{1}{2}\cE_1\cE_2\cdots\cE_{a-1}\cE_{a+1}\cdots
\cE_{b-1}\cE_{b+1}\cdots\cE_n,\nonumber
\end{eqnarray}
whence it immediately follows a comparison $a+b-2\equiv 2n-a-b\pmod{2}$ or
$2(a+b)\equiv 2(n+1)\pmod{2}$. Thus, $\sW$ and $A_{23}$ are always commute.
It is easy to verify that analogous conditions take place for the operators
$A_{13},\,A_{12}$ (except the case $n=2$). Further, for 
$B_1\sim-\frac{1}{2}\cE_c$ we obtain
\begin{eqnarray}
B_1\sW&=&-(-1)^{c-1}\frac{1}{2}\cE_1\cE_2\cdots\cE_{c-1}\cE_{c+1}\cdots\cE_n,
\nonumber\\
\sW B_1&=&-(-1)^{n-c}\frac{1}{2}\cE_1\cE_2\cdots\cE_{c-1}\cE_{c+1}\cdots\cE_n,
\nonumber
\end{eqnarray}
that is, $c-1\equiv n-c\pmod{2}$ or $n\equiv 2c-1\pmod{2}$. Therefore, the
matrix $\sW$ always anticommute with $B_1$ (correspondingly with
$B_1,\,B_2$), since $n\equiv 0\pmod{2}$. Thus, in full accordance with
Gel'fand--Yaglom relations (\ref{GY}) we have\footnote{Except the case of the
fundamental representation $\fC^{1,0}$ for which the automorphism group is
$\sAut_+(\C_2)=\{\sI,\sW,\sE,\sC\}=\{\sigma_0,-\sigma_3,\sigma_1,-i\sigma_2\}
\simeq D_4/\dZ_2$. It is easy to verify that the matrix $\sW\sim\sigma_3$
does not satisfy the relations (\ref{bC1}). Therefore, in case of $\fC^{1,0}$
we have an anomalous behaviour of the parity transformation $\sW\sim P$.
This fact will be explained further within quotient representations.}
\begin{alignat}{3}
\sW A_{23}\sW^{-1} &=A_{23}, &\quad
\sW A_{13}\sW^{-1} &=A_{13}, &\quad
\sW A_{12}\sW^{-1} &=A_{12}, \nonumber\\
\sW B_1\sW^{-1} &=-B_1, &\quad
\sW B_2\sW^{-1} &=-B_2, &\quad
\sW B_3\sW^{-1} &=-B_3,\label{bC1}
\end{alignat}
where the matrix $\sW$ of the automorphism $\cA\rightarrow\cA^\star$ is an
element of an Abelian automorphism group $\sAut_-(\C_n)\simeq\dZ_2\otimes\dZ_2$
with the signature $(+,+,+)$ at $n\equiv 0\pmod{4}$ and also a 
non--Abelian automorphism group $\sAut_+(\C_n)\simeq Q_4/\dZ_2$ with the
signature $(-,-,-)$ at $n\equiv 2\pmod{4}$, here $\dZ_2\otimes\dZ_2$ is a
Gauss--Klein group and $Q_4$ is a quaternionic group (see Theorem \ref{taut} in
Chapter 2).

Let us consider now permutation conditions of the operators
(\ref{O1})--(\ref{O2}) with the matrix $\sE$ of the antiautomorphism
$\cA\rightarrow\widetilde{\cA}$ (time reversal). Over the field $\F=\C$
the matrix $\sE$ has two forms (Theorem \ref{taut}):
1) $\sE=\cE_1\cE_2\cdots\cE_m$ at $m\equiv 1\pmod{2}$, the group
$\sAut_+(\C_n)\simeq Q_4/\dZ_2$; 2) $\sE=\cE_{m+1}\cE_{m+2}\cdots\cE_n$ at
$m\equiv 0\pmod{2}$, the group $\sAut_+(\C_n)\simeq\dZ_2\otimes\dZ_2$.
Obviously, in both cases $\sW=\cE_1\cE_2\cdots\cE_m\cE_{m+1}\cE_{m+2}\cdots
\cE_n$, where the matrices $\cE_i$ are symmetric for $1<i\leq m$ and
skewsymmetric for $m<i\leq n$. 

So, let $\sE=\cE_{m+1}\cE_{m+2}\cdots\cE_n$ be a matrix of 
$\cA\rightarrow\widetilde{\cA}$, $m\equiv 0\pmod{2}$. Let us assume that
$\cE_a,\,\cE_b,\,\cE_c\in\sE$, then for the operator $A_{23}\sim
-\frac{1}{2}\cE_a\cE_b$ we obtain
\begin{eqnarray}
A_{23}\sE&=&-(-1)^{b+a-2}\frac{1}{2}\cE_{m+1}\cE_{m+2}\cdots\cE_{a-1}\cE_{a+1}
\cdots\cE_{b-1}\cE_{b+1}\cdots\cE_n,\nonumber\\
\sE A_{23}&=&-(-1)^{2m-a-b}\frac{1}{2}\cE_{m+1}\cE_{m+2}\cdots\cE_{a-1}\cE_{a+1}
\cdots\cE_{b-1}\cE_{b+1}\cdots\cE_n,\label{R3}
\end{eqnarray}
that is, $b+a-2\equiv 2m-a-b\pmod{2}$ and in this case the matrix $\sE$
commutes with $A_{23}$ (correspondingly with $A_{13},\,A_{12}$). For the
operator $B_1$ we have
(analogously for $B_2$ and $B_3$):
\begin{eqnarray}
B_1\sE&=&-(-1)^{c-1}\frac{1}{2}\cE_{m+1}\cE_{m+2}\cdots\cE_{c-1}\cE_{c+1}
\cdots\cE_n,\nonumber\\
\sE B_1&=&-(-1)^{m-c}\frac{1}{2}\cE_{m+1}\cE_{m+2}\cdots\cE_{c-1}\cE_{c+1}
\cdots\cE_n,\label{R4}
\end{eqnarray}
that is, $m\equiv 2c-1\pmod{2}$ and, therefore, the matrix $\sE$ in this case
always anticommutes with $B_i$, since $m\equiv 0\pmod{2}$. Thus,
\begin{alignat}{3}
\sE A_{23}\sE^{-1} &=A_{23}, &\quad
\sE A_{13}\sE^{-1} &=A_{13}, &\quad
\sE A_{12}\sE^{-1} &=A_{12}, \nonumber\\
\sE B_1\sE^{-1} &=-B_1, &\quad
\sE B_2\sE^{-1} &=-B_2, &\quad
\sE B_3\sE^{-1} &=-B_3.\label{bC2}
\end{alignat}
Let us assume now that $\cE_a,\,\cE_b,\,\cE_c\not\in\sE$, then
\begin{equation}\label{R5}
A_{23}\sE=(-1)^{2m}\sE A_{23},
\end{equation}
that is, in this case $\sE$ and $A_{ik}$ are always commute. For the
hyperbolic operators we obtain
\begin{equation}\label{R6}
B_i\sE=(-1)^m\sE B_i\quad(i=1,2,3)
\end{equation}
and since $m\equiv 0\pmod{2}$, then $\sE$ and $B_i$ are also commute.
Therefore,
\begin{equation}\label{bC3}
\ld\sE,A_{23}\rd=0,\;\;\ld\sE,A_{13}\rd=0,\;\;\ld\sE,A_{12}\rd=0,\;\;
\ld\sE,B_1\rd=0,\;\;\ld\sE,B_2\rd=0,\;\;\ld\sE,B_3\rd=0.
\end{equation}
Assume now that $\cE_a,\,\cE_b\in\sE$ and $\cE_c\not\in\sE$. Then in
accordance with (\ref{R3}) and (\ref{R6}) the matrix $\sE$ commutes with
$A_{23}$ and $B_1$, and according to (\ref{R4}) anticommutes with $B_2$ and
$B_3$. For the operator $A_{13}$ we find
\begin{eqnarray}
A_{13}\sE&=&-(-1)^{b-1}\frac{1}{2}\cE_c\cE_{m+1}\cE_{m+2}\cdots\cE_{b-1}
\cE_{b+1}\cdots\cE_n,\nonumber\\
\sE A_{23}&=&-(-1)^{2m-b}\frac{1}{2}\cE_c\cE_{m+1}\cE_{m+2}\cdots\cE_{b-1}
\cE_{b+1}\cdots\cE_n,\label{R7}
\end{eqnarray}
that is, $2m\equiv 2b-1\pmod{2}$ and, therefore, in this case $\sE$ always
anticommutes with $A_{13}$ and correspondingly with the operator $A_{12}$
which has the analogous structure. Thus,
\begin{equation}\label{bC4}
\ld\sE,A_{23}\rd=0,\;\;\lf\sE,A_{13}\rf=0,\;\;\lf\sE,A_{12}\rf=0,\;\;
\ld\sE,B_1\rd=0,\;\;\lf\sE,B_2\rf=0,\;\;\lf\sE,B_3\rf=0.
\end{equation}
If we take $\cE_a\in\sE$, $\cE_b,\,\cE_c\not\in\sE$, then for the operator
$A_{23}$ it follows that
\begin{eqnarray}
A_{23}\sE&=&-(-1)^{a-2}\frac{1}{2}\cE_b\cE_{m+1}\cE_{m+2}\cdots\cE_{a-1}
\cE_{a+1}\cdots\cE_n,\nonumber\\
\sE A_{23}&=&-(-1)^{2m-a-1}\frac{1}{2}\cE_b\cE_{m+1}\cE_{m+2}\cdots\cE_{a-1}
\cE_{a+1}\cdots\cE_n,\label{R8}
\end{eqnarray}
that is, $a-2\equiv 2m-a-1\pmod{2}$ or $2m\equiv 2a-1\pmod{2}$. Therefore,
$\sE$ anticommutes with $A_{23}$. Further, in virtue of (\ref{R5}) $\sE$
commutes with $A_{13}$ and anticommutes with $A_{12}$ in virtue of (\ref{R7}).
Correspondingly, from (\ref{R6}) and (\ref{R4}) it follows that $\sE$
commutes with $B_1,\,B_3$ and anticommutes with $B_2$. Thus,
\begin{equation}\label{bC5}
\lf\sE,A_{23}\rf=0,\;\;\ld\sE,A_{13}\rd=0,\;\;\lf\sE,A_{12}\rf=0,\;\;
\ld\sE,B_1\rd=0,\;\;\lf\sE,B_2\rf=0,\;\;\ld\sE,B_3\rd=0.
\end{equation}
Cyclic permutations of the indices in $\cE_i,\,\cE_j\in\sE$, 
$\cE_k\not\in\sE$ and $\cE_i\in\sE$, $\cE_j,\,\cE_k\not\in\sE$,
$i,j,k=\{a,b,c\}$, give the following relations
\begin{equation}\label{bC6}
\begin{array}{lll}
\lf\sE,A_{23}\rf=0,&\lf\sE,A_{13}\rf=0,&\ld\sE,A_{12}\rd=0,\\
\lf\sE,B_1\rf=0,&\lf\sE,B_2\rf=0,&\ld\sE,B_3\rd=0,
\end{array}\quad\cE_a,\,\cE_c\in\sE,\;\cE_b\not\in\sE.
\end{equation}
\begin{equation}\label{C7}
\begin{array}{lll}
\lf\sE,A_{23}\rf=0,&\ld\sE,A_{13}\rd=0,&\lf\sE,A_{12}\rf=0,\\
\lf\sE,B_1\rf=0,&\ld\sE,B_2\rd=0,&\lf\sE,B_3\rf=0,
\end{array}\quad\cE_b,\,\cE_c\in\sE,\;\cE_a\not\in\sE.
\end{equation}
\begin{equation}\label{C8}
\begin{array}{lll}
\lf\sE,A_{23}\rf=0,&\lf\sE,A_{13}\rf=0,&\ld\sE,A_{12}\rd=0,\\
\ld\sE,B_1\rd=0,&\ld\sE,B_2\rd=0,&\lf\sE,B_3\rf=0,
\end{array}\quad\cE_b\in\sE,\;\cE_a,\,\cE_c\not\in\sE.
\end{equation}
\begin{equation}\label{C9}
\begin{array}{lll}
\ld\sE,A_{23}\rd=0,&\lf\sE,A_{13}\rf=0,&\lf\sE,A_{12}\rf=0,\\
\lf\sE,B_1\rf=0,&\ld\sE,B_2\rd=0,&\ld\sE,B_3\rd=0,
\end{array}\quad\cE_c\in\sE,\;\cE_a,\,\cE_b\not\in\sE.
\end{equation}

Let us consider now the matrix $\sE$ of the group $\sAut_+(\C_n)\simeq
Q_4/\dZ_2$. In this case $\sE=\cE_1\cE_2\cdots\cE_m$, $m\equiv 1\pmod{2}$.
At $\cE_a,\,\cE_b,\,\cE_c\in\sE$ from (\ref{R3}) it follows that
$2(a+b)\equiv 2(m+1)\pmod{2}$, therefore, in this case $\sE$ always commutes
with $A_{ik}$. In its turn, from (\ref{R4}) it follows that $\sE$ always
commutes with $B_i$, since $m\equiv 1\pmod{2}$. Thus, we have the relations
(\ref{bC3}). At $\cE_a,\cE_b,\cE_c\not\in\sE$ (except the cases $n=2$ and
$n=4$) from (\ref{R5}) it follows that $\sE$ always commutes with $A_{ik}$,
and from (\ref{R6}) it follows that $\sE$ always anticommutes with $B_i$.
Therefore, for the case $\cE_a,\cE_b,\cE_c\not\in\sE$ we have the relations
(\ref{bC2}). Analogously, at $\cE_a,\cE_b\in\sE$, $\cE_c\not\in\sE$ in
accordance with (\ref{R3}) and (\ref{R4}) $\sE$ commutes with $A_{23}$ and
$B_1,B_3$, and in accordance with (\ref{R7}) and (\ref{R6}) anticommutes
with $A_{13},A_{12}$ and $B_1$. Therefore, for this case we have the
relations (\ref{C9}). At $\cE_a\in\sE$, $\cE_b,\cE_c\not\in\sE$ from
(\ref{R8}) it follows that $\sE$ anticommutes with $A_{23}$. In virtue of
(\ref{R5}) $\sE$ commutes with $A_{13}$ and anticommutes with $A_{12}$ in
virtue of (\ref{R7}). Correspondingly, from (\ref{R6}) and (\ref{R4}) it
follows that $\sE$ anticommutes with $B_1,B_3$ and commutes with $B_2$.
Therefore, for the case $\cE_a\in\sE$, $\cE_b,\cE_c\not\in\sE$ we have the
relations (\ref{C7}). Further, cyclic permutations of the indices in
$\cE_i,\cE_j\in\sE$, $\cE_k\not\in\sE$ and $\cE_i\in\sE$, 
$\cE_j,\cE_k\not\in\sE$, $i,j,k=\{a,b,c\}$, give for $\cE_a,\cE_c\in\sE$,
$\cE_b\not\in\sE$ the relations (\ref{C8}), for $\cE_b,\cE_c\in\sE$,
$\cE_a\not\in\sE$ the relations (\ref{bC5}), for $\cE_b\in\sE$,
$\cE_a,\cE_c\not\in\sE$ the relations (\ref{bC6}) and for $\cE_c\in\sE$,
$\cE_a,\cE_b\not\in\sE$ the relations (\ref{bC4}).

Let us consider now the permutation conditions of the operators
(\ref{O1})--(\ref{O2}) with the matrix $\sC$ of the antiautomorphism
$\cA\rightarrow\widetilde{\cA^\star}$ (full reflection). Over the field
$\F=\C$ the matrix $\sC$ has two different forms (Theorem \ref{taut}):
1) $\sC=\cE_1\cE_2\cdots\cE_m$ at $m\equiv 0\pmod{2}$, the group
$\sAut_-(\C_n)\simeq\dZ_2\otimes\dZ_2$; 2) $\sC=\cE_{m+1}\cE_{m+2}\cdots\cE_n$
at $m\equiv 1\pmod{2}$, the group $\sAut_+(\C_n)\simeq Q_4/\dZ_2$. So, let
$\sC=\cE_1\cE_2\cdots\cE_m$ be a matrix of 
$\cA\rightarrow\widetilde{\cA^\star}$, $m\equiv 1\pmod{2}$, then by analogy
with the matrix $\sE=\cE_{m+1}\cE_{m+2}\cdots\cE_n$ of $\cA\rightarrow
\widetilde{\cA}$, $m\equiv 0\pmod{2}$, we have for $\sC$ the relations of
the form (\ref{bC2})--(\ref{C9}). In its turn, the matrix 
$\sC=\cE_{m+1}\cE_{m+2}\cdots\cE_n$ is analogous to 
$\sE=\cE_1\cE_2\cdots\cE_m$, $m\equiv 1\pmod{2}$, therefore, in this case
we have also the relations (\ref{bC2})--(\ref{C9}) for $\sC$.

Now we have all possible combinations of permutation relations between the
matrices of infinitesimal operators (\ref{O1})--(\ref{O2}) of the proper
Lorentz group $\fG_+$ and matrices of the fundamental automorphisms of the
complex Clifford algebra $\C_n$ associated with the complex representation
$\fC^{l_0+l_1-1,0}$ of $\fG_+$. It is obvious that the relations (\ref{bC1})
take place for any representation $\fC^{l_0+l_1-1,0}$ of the group
$\fG_+$. Further, if $\fC^{l_0+l_1-1,0}$ with $2(l_0+l_1-1)\equiv 0\pmod{4}$
and if $\sAut_-(\C_n)\simeq\dZ_2\otimes\dZ_2$ with $\sE=\cE_{m+1}\cE_{m+2}
\cdots\cE_n$, $\sC=\cE_1\cE_2\cdots\cE_m$, $m\equiv 0\pmod{2}$, then at
$\cE_a,\cE_b,\cE_c\in\sE$ ($\cE_a,\cE_b,\cE_c\not\in\sC$) we have the
relations (\ref{bC2}) for $\sE$ and the relations of the form (\ref{bC3}) for
$\sC$ and, therefore, we have the relations (\ref{bT1}) and (\ref{bT2}) of the
present Theorem. Correspondingly, at $\cE_a,\cE_b,\cE_c\not\in\sE$
($\cE_a,\cE_b,\cE_c\in\sC$) we have the relations (\ref{bC3}) for $\sE$ and
the relations of the form (\ref{bC2}) for $\sC$, that is, the relations 
(\ref{bT5}) and (\ref{bT6}) of Theorem. At $\cE_a,\cE_b\in\sE$, 
$\cE_c\not\in\sE$ ($\cE_a,\cE_b\not\in\sC$, $\cE_c\in\sC$) we obtain the
relations (\ref{bC4}) for $\sE$ and the relations of the form (\ref{C9}) for
$\sC$ ((\ref{bT9})--(\ref{bT10}) in Theorem) and so on. Analogously, if
$\fC^{l_0+l_1-1,0}$ with $2(l_0+l_1-1)\equiv 2\pmod{4}$ and if
$\sAut_+(\C_n)\simeq Q_4/\dZ_2$ with $\sE=\cE_1\cE_2\cdots\cE_m$,
$\sC=\cE_{m+1}\cE_{m+2}\cdots\cE_n$, $m\equiv 1\pmod{2}$, then at
$\cE_a,\cE_b,\cE_c\in\sE$ ($\cE_a,\cE_b,\cE_c\not\in\sC$) we have the
relations (\ref{bC3}) for $\sE$ and the relations of the form (\ref{bC2})
for $\sC$ (relations (\ref{bT2}) and (\ref{bT1}) in Theorem) and so on.

Further, let us consider the following combinations of $A_{ik}$ and $B_i$
(rising and lowering operators):
\begin{alignat}{3}
H_+ &=iA_{23}-A_{13}, &\quad
H_- &=iA_{23}+A_{13}, &\quad
H_3 &=iA_{13},\nonumber\\
F_+ &=iB_1-B_2, &\quad
F_- &=iB_1+B_2, &\quad
F_3 &=iB_3,\label{I}
\end{alignat}
satisfying in virtue of (\ref{bcommut}) the relations
\begin{gather}
\ld H_+,H_3\rd=-H_+,\;\;\ld H_-,H_3\rd=H_-,\;\;\ld H_+,H_-\rd=2H_3,\nonumber\\
\ld H_+,F_+\rd=\ld H_-,F_-\rd=\ld H_3,F_3\rd=0,\nonumber\\
\ld F_+,F_3\rd=-H_+,\;\;\ld F_-,F_3\rd=H_-,\;\;\ld F_+,F_-\rd=-2H_3,\nonumber\\
\ld H_+,F_3\rd=F_+,\;\;\ld H_-,F_3\rd=-F_-,\nonumber\\
\ld H_-,F_+\rd=-\ld H_+,F_-\rd=2F_3,\nonumber\\
\ld F_+,H_3\rd=-F_+,\;\;\ld F_-,H_3\rd=F_-.\label{bcommut2}
\end{gather}
It is easy to see that for $\sW$ from (\ref{bC1}) and (\ref{I}) we have always
the relations (\ref{T0}). Further, for $\sAut_-(\C_n)\simeq\dZ_2\otimes\dZ_2$
at $\cE_a,\cE_b,\cE_c\in\sE$ from (\ref{bC2}) and (\ref{bC3}) in virtue of
(\ref{I}) it follow the relations (\ref{bT3}) and (\ref{bT4}). Analogously,
at $\cE_a,\cE_b,\cE_c\not\in\sE$ from (\ref{bC3}) and (\ref{bC2}) we obtain
the relations (\ref{bT7}) and (\ref{bT8}). In contrast with this, at
$\cE_a,\cE_b\in\sE$, $\cE_c\not\in\sE$ the combinations (\ref{bC4}) and
(\ref{C9}) do not form permutation relations with operators $H_{+,-,3}$ and
$F_{+,-,3}$, since $\sE$ and $\sC$ commute with $A_{23}$ and anticommute
with $A_{13}$, and $B_1$ commutes with $\sC$ and anticommutes with $\sE$
(inverse relations take place for $B_2$). Other two relations
(\ref{bT21})--(\ref{bT24}) and (\ref{bT13})--(\ref{bT16}) for
$\sAut_-(\C_n)\simeq\dZ_2\otimes\dZ_2$ correspond to $\cE_a,\cE_c\in\sE$,
$\cE_b\not\in\sE$ and $\cE_b\in\sE$, $\cE_a,\cE_c\not\in\sE$. It is easy
to see that in such a way for the group $\sAut_+(\C_n)\simeq Q_4/\dZ_2$ we
obtain the relations (\ref{bT7})--(\ref{bT8}), (\ref{bT3})--(\ref{bT4}),
(\ref{bT21})--(\ref{bT24}) and (\ref{bT13})--(\ref{bT16}) correspondingly
for $\cE_a,\cE_b,\cE_c\in\sE$, $\cE_a,\cE_b,\cE_c\not\in\sE$,
$\cE_a,\cE_c\in\sE$, $\cE_b\not\in\sE$ and $\cE_b\in\sE$, 
$\cE_a,\cE_c\not\in\sE$.

In accordance with \cite{GMS} a representation conjugated to
$\fC^{l_0+l_1-1,0}$ is defined by a pair
\[
(l_0,l_1)=\left(-\frac{r}{2},\,\frac{r}{2}+1\right),
\]
that is, this representation has a form $\fC^{0,l_0-l_1+1}$. In its turn,
a representation conjugated to fundamental representation $\fC^{1,0}$ is
$\fC^{0,-1}$. Let us find infinitesimal operators of the representation
$\fC^{0,-1}$. At $l_0=-1/2$ and $l_1=3/2$ from (\ref{I1})--(\ref{I6'}) and
(\ref{bA1})--(\ref{bB3}) we obtain
\begin{alignat}{3}
A^{-1/2}_{23} &=-\frac{i}{2}\ar\begin{bmatrix}
0 & 1\\
1 & 0
\end{bmatrix}, &\quad
A^{-1/2}_{13} &=\frac{1}{2}\ar\begin{bmatrix}
0 & 1\\
-1 & 0
\end{bmatrix}, &\quad
A^{-1/2}_{12} &=\frac{1}{2}\ar\begin{bmatrix}
i & 0\\
0 &-i
\end{bmatrix},\nonumber\\[0.2cm]
B^{-1/2}_1 &=\frac{1}{2}\ar\begin{bmatrix}
0 & 1\\
1 & 0
\end{bmatrix}, &\quad
B^{-1/2}_2 &=\frac{1}{2}\ar\begin{bmatrix}
0 & i\\
-i & 0
\end{bmatrix}, &\quad
B^{-1/2}_3 &=\frac{i}{2}\ar\begin{bmatrix}
-i & 0\\
0 & i
\end{bmatrix}.\nonumber
\end{alignat}
Or
\begin{alignat}{3}
A^{-1/2}_{23} &=-\frac{1}{2}\sigma_2\sigma_3, &\quad
A^{-1/2}_{13} &=-\frac{1}{2}\sigma_1\sigma_3, &\quad
A^{-1/2}_{12} &=\frac{1}{2}\sigma_1\sigma_2,\nonumber\\
B^{-1/2}_1 &=\frac{1}{2}\sigma_1, &\quad
B^{-1/2}_2 &=-\frac{1}{2}\sigma_2, &\quad
B^{-1/2}_3 &=\frac{1}{2}\sigma_3.\label{Op}
\end{alignat}
It is easy to see that operators (\ref{Op}) differ from 
(\ref{IF1})--(\ref{IF2}) only in the sign at the operators $B_i$ of the
hyperbolic rotations. This result is a direct consequence of the
well--known definition of the group $SL(2;\C)$ as a complexification of the
special unimodular group $SU(2)$ (see \cite{Vil68}). Indeed, the group
$SL(2;\C)$ has six parameters $a_1,a_2,a_3,ia_1,ia_2,ia_3$, where
$a_1,a_2,a_3\in SU(2)$. It is easy to verify that operators (\ref{Op})
satisfy the relations (\ref{bcommut}). Therefore, as in case of the
the representation $\fC^{l_0+l_1-1,0}$ infinitesimal operators of the
conjugated representation $\fC^{0,l_0-l_1+1}$ are defined as follows
\begin{alignat}{3}
A^{j^\prime}_{23} &\sim-\frac{1}{2}\cE_a\cE_b, &\quad
A^{j^\prime}_{13} &\sim-\frac{1}{2}\cE_c\cE_b, &\quad
A^{j^\prime}_{12} &\sim\frac{1}{2}\cE_c\cE_a,\nonumber\\
B^{j^\prime}_1 &\sim\frac{1}{2}\cE_c, &\quad
B^{j^\prime}_2 &\sim-\frac{1}{2}\cE_a, &\quad
B^{j^\prime}_3 &\sim\frac{1}{2}\cE_b,\label{Op2}
\end{alignat}
where $\cE_a,\cE_b,\cE_c$ are tensor products of the form (\ref{6.6}).
It is not difficult to verify that operators (\ref{Op2}) satisfy the
relations (\ref{bcommut}), and their linear combinations $H_{+,-,3}$,
$F_{+,-,3}$ satisfy the relations (\ref{bcommut2}). Since the structure of
the operators (\ref{Op2}) is analogous to the structure of the operators
(\ref{O1})--(\ref{O2}), then all the permutation conditions between the
operators of discrete symmetries and operators (\ref{O1})--(\ref{O2})
of the representation $\fC^{l_0+l_1-1,0}$ are valid also for the conjugated
representation $\fC^{0,l_0-l_1+1}$ and, obviously, for a representation
$\fC^{l_0+l_1-1,l_0-l_1+1}$.\\[0.2cm]
2) Real representations.\\
As known \cite{GMS}, if an irreducible representation of the proper Lorentz
group $\fG_+$ is defined by the pair $(l_0,l_1)$, then a conjugated
representation is also irreducible and defined by a pair $\pm(l_0,-l_1)$.
Hence it follows that the irreducible representation is equivalent to its
conjugated representation only in case when this representation is defined
by a pair $(0,l_1)$ or $(l_0,0)$, that is, either of the two numbers
$l_0$ and $l_1$ is equal to zero. We assume that $l_1=0$. In its turn,
for the complex Clifford algebra $\C_n$
($\overset{\ast}{\C}_n$) associated
with the representation $\fC^{l_0+l_1-1,0}\,(\fC^{0,l_0-l_1+1})$ of the
group $\fG_+$ the equivalence of the representation to its conjugated
representation induces a relation 
$\C_n=\overset{\ast}{\C}_n$, which,
obviously, is fulfilled only in case when the algebra $\C_n=\C\otimes\cl_{p,q}$
is reduced into its real subalgebra $\cl_{p,q}$ ($p+q=n$). Thus, a
restriction of the complex representation $\fC^{l_0+l_1-1,0}$
(or $\fC^{0,l_0-l_1+1}$) of the group $\fG_+$ onto a real representation,
$(l_0,l_1)\rightarrow(l_0,0)$, induces a restriction $\C_n\rightarrow\cl_{p,q}$.
Further, over the field $\F=\R$ at $p+q\equiv 0\pmod{2}$ there are four
types of real algebras $\cl_{p,q}$: two types $p-q\equiv 0,2\pmod{8}$ with
a real division ring $\K\simeq\R$ and two types $p-q\equiv 4,6\pmod{8}$ with a
quaternionic division ring $\K\simeq\BH$. Thus, we have four classes of the
real representations of the group $\fG_+$:
\begin{eqnarray}
\fR^{l_0}_0&\leftrightarrow&\cl_{p,q},\;p-q\equiv 0\pmod{8},\;\K\simeq\R,
\nonumber\\
\fR^{l_0}_2&\leftrightarrow&\cl_{p,q},\;p-q\equiv 2\pmod{8},\;\K\simeq\R,
\nonumber\\
\fH^{l_0}_4&\leftrightarrow&\cl_{p,q},\;p-q\equiv 4\pmod{8},\;\K\simeq\BH,
\nonumber\\
\fH^{l_0}_6&\leftrightarrow&\cl_{p,q},\;p-q\equiv 6\pmod{8},\;\K\simeq\BH,
\label{Ident}
\end{eqnarray}
We will call the representations $\fH^{l_0}_4$ and $\fH^{l_0}_6$ are
{\it quaternionic representations} of the group $\fG_+$. It is not difficult
to see that for the real representations with the pair $(l_0,0)$ all the
coefficients $A_l=il_0l_1/l(l+1)$ are equal to zero, since $l_1=0$.
Therefore, all the infinitesimal operators $B^j_1,B^j_2,B^j_3$ (see
formulas (\ref{bB1})--(\ref{bB3})) of hyperbolic rotations are
also equal to zero. Hence it follows that the restriction
$(l_0,l_1)\rightarrow(l_0,0)$ induces a restriction of the group $\fG_+$
onto its subgroup $SO(3)$ of three--dimensional rotations. Thus, real
representations with the pair $(l_0,0)$ are representations of the subgroup
$SO(3)$. This result directly follows from the complexification of $SU(2)$
which is equivalent to $SL(2;\C)$. Indeed, the parameters $a_1,a_2,a_3$
compose a real part of $SL(2;\C)$ which under complex conjugation remain
unaltered, whereas the parameters $\pm ia_1,\pm ia_2,\pm ia_3$ of the complex
part of $SL(2;\C)$ under complex conjugation are mutually annihilate for
the representations with the pairs $(l_0,l_1)$ and $\pm(l_0,-l_1)$.

Let us find a relation of the number $l_0$ with dimension of the real algebra
$\cl_{p,q}$. If $p+q\equiv 0\pmod{2}$ and $\omega^2=\e^2_{12\ldots p+q}=1$,
then $\cl_{p,q}$ is called {\it positive} ($\cl_{p,q}>0$ at 
$p-q\equiv 0,4\pmod{8}$) and correspondingly {\it negative} if $\omega^2=-1$
($\cl_{p,q}<0$ at $p-q\equiv 2,6\pmod{8}$). Further, in accordance with
Karoubi Theorem \cite[Prop.~3.16]{Kar79} it follows that if $\cl(V,Q)>0$,
and $\dim V$ is even, then $\cl(V\oplus V^\prime,Q\oplus Q^\prime)\simeq
\cl(V,Q)\otimes\cl(V^\prime,Q^\prime)$, and also if $\cl(V,Q)<0$, and
$\dim V$ is even, then $\cl(V\oplus V^\prime, Q\oplus Q^\prime)\simeq
\cl(V,Q)\otimes\cl(V^\prime,-Q^\prime)$, where $V$ is a vector space
associated with $\cl_{p,q}$, $Q$ is a quadratic form of $V$. Using the
Karoubi Theorem we obtain for the algebra $\cl_{p,q}$ a following
factorization
\begin{equation}\label{Ten}
\cl_{p,q}\simeq\underbrace{\cl_{s_i,t_j}\otimes\cl_{s_i,t_j}\otimes\cdots
\otimes\cl_{s_i,t_j}}_{r\;\text{times}}
\end{equation}
where $s_i,t_j\in\{0,1,2\}$. For example, there are two different
factorizations $\cl_{1,1}\otimes\cl_{0,2}$ and $\cl_{1,1}\otimes\cl_{2,0}$
for the spacetime algebra $\cl_{1,3}$ and Majorana algebra $\cl_{3,1}$.
It is obvious that $l_0=r/2$ and $n=2r=p+q=4l_0$, therefore, $l_0=(p+q)/4$.

So, we begin with the representation of the class $\fR^{l_0}_0$. In accordance
with Theorem \ref{tautr} (see Chapter 2) for the algebra $\cl_{p,q}$ of the type
$p-q\equiv 0\pmod{8}$ a matrix of the automorphism $\cA\rightarrow\cA^\star$
has a form $\sW=\cE_1\cE_2\cdots\cE_{p+q}$ and $\sW^2=\sI$. It is obvious
that for $\cE^2_a=\cE^2_b=\cE^2_c=\sI$ permutation conditions of the matrix
$\sW$ with $A_{ik}$ are analogous to (\ref{bC1}), that is, $\sW$ always
commutes with the operators $A_{ik}$ of the subgroup $SO(3)$. It is
sufficient to consider permutation conditions of $\sW$ with the operators
$A_{ik}$ of $SO(3)$ only, since $B_i=0$ for the real representations.
In this case the relations (\ref{bcommut}) take a form
\begin{equation}\label{bcommut3}
\ld A_{23},A_{13}\rd=A_{12}, \quad \ld A_{13},A_{12}\rd=A_{23}, \quad
\ld A_{12},A_{23}\rd=A_{13}.
\end{equation}
Assume now that $\cE^2_a=\cE^2_b=\cE^2_c=-\sI$, then it is easy to verify
that operators
\begin{equation}\label{O3}
A_{23}\sim\frac{1}{2}\cE_a\cE_b,\quad A_{13}\sim\frac{1}{2}\cE_c\cE_b,\quad
A_{12}\sim-\frac{1}{2}\cE_c\cE_a
\end{equation}
satisfy the relations (\ref{bcommut3}) and commute with the matrix $\sW$ of
$\cA\rightarrow\cA^\star$. It is easy to see that at $\cE^2_i=-\sI$,
$\cE^2_j=\cE^2_k=\sI$ and $\cE^2_i=\cE^2_j=-\sI$, $\cE^2_k=\sI$
($i,j,k=\{a,b,c\}$) the operators $A_{ik}$ do not satisfy the relations
(\ref{bcommut3}). Therefore, there exist only two possibilities
$\cE^2_a=\cE^2_b=\cE^2_c=\sI$ and $\cE^2_a=\cE^2_b=\cE^2_c=-\sI$
corresponding to the operators (\ref{O1}) and (\ref{O3}), respectively.

Further, for the type $p-q\equiv 0\pmod{8}$ ($p=q=m$) at
$\sE=\cE_{p+1}\cE_{p+2}\cdots\cE_{p+q}$ and $\sC=\cE_1\cE_2\cdots\cE_p$
there exist Abelian groups $\sAut_-(\cl_{p,q})\simeq\dZ_2\otimes\dZ_2$
with the signature $(+,+,+)$ and $\sAut_-(\cl_{p,q})\simeq\dZ_4$ with
$(+,-,-)$ correspondingly at $p,q\equiv 0\pmod{4}$ and $p,q\equiv 2\pmod{4}$,
and also at $\sE=\cE_1\cE_2\cdots\cE_p$ and $\sC=\cE_{p+1}\cE_{p+2}\cdots
\cE_{p+q}$ there exist non--Abelian groups $\sAut_+(\cl_{p,q})\simeq
D_4/\dZ_2$ with $(+,-,+)$ and $\sAut_+(\cl_{p,q})\simeq D_4/\dZ_2$ with
$(+,+,-)$ correspondingly at $p,q\equiv 3\pmod{4}$ and $p,q\equiv 1\pmod{4}$
(Theorem \ref{tautr}). Besides, for the algebras $\cl_{8t,0}$ of the
type $p-q\equiv 0\pmod{8}$, $t=1,2,\ldots$, the matrices $\sE\sim\sI$,
$\sC\sim\cE_1\cE_2\cdots\cE_p$ and $\sW$ form an Abelian group
$\sAut_-(\cl_{p,0})\simeq\dZ_2\otimes\dZ_2$. Correspondingly, for the
algebras $\cl_{0,8t}$ of the type $p-q\equiv 0\pmod{8}$ the matrices
$\sE\sim\cE_1\cE_2\cdots\cE_q$, $\sC\sim\sI$ and $\sW$ also form the
group $\sAut_-(\cl_{0,q})\simeq\dZ_2\otimes\dZ_2$.

So, let $\sE=\cE_{m+1}\cE_{m+2}\cdots\cE_{2m}$ and $\sC=\cE_1\cE_2\cdots\cE_m$
be matrices of $\cA\rightarrow\widetilde{\cA}$ and
$\cA\rightarrow\widetilde{\cA^\star}$, $p=q=m$, $m\equiv 0\pmod{2}$.
Then for the operators (\ref{O1}) we obtain
\begin{equation}\label{C34}
A_{ik}\sE=(-1)^{2m}\sE A_{ik},
\end{equation}
\begin{eqnarray}
A_{ik}\sC&=&\pm(-1)^{i+j-2}\frac{1}{2}\sigma(i)\sigma(j)\cE_1\cE_2\cdots
\cE_{i-1}\cE_{i+1}\cdots\cE_{j-1}\cE_{j+1}\cdots\cE_m,\nonumber\\
\sC A_{ik}&=&\pm(-1)^{2m-i-j}\frac{1}{2}\sigma(i)\sigma(j)\cE_1\cE_2\cdots
\cE_{i-1}\cE_{i+1}\cdots\cE_{j-1}\cE_{j+1}\cdots\cE_m,\label{C35}
\end{eqnarray}
since $\cE_a,\cE_b,\cE_c\not\in\sE$ and a function $\sigma(n)=\sigma(m-n)$
has a form
\[
\sigma(n)=\begin{cases}
-1 & \text{if $n\leq 0$}\\
+1 & \text{if $n> 0$}
\end{cases}
\]
Analogously, for the operators (\ref{O3}) we find
\begin{eqnarray}
A_{ik}\sE&=&\pm(-1)^{i+j-2}\frac{1}{2}\sigma(i)\sigma(j)\cE_{m+1}\cE_{m+2}
\cdots\cE_{i-1}\cE_{i+1}\cdots\cE_{j-1}\cE_{j+1}\cdots\cE_{2m},\nonumber\\
\sE A_{ik}&=&\pm(-1)^{2m-i-j}\frac{1}{2}\sigma(i)\sigma(j)\cE_{m+1}\cE_{m+2}
\cdots\cE_{i-1}\cE_{i+1}\cdots\cE_{j-1}\cE_{j+1}\cdots\cE_{2m},\label{C36}
\end{eqnarray}
\begin{equation}\label{C37}
A_{ik}\sC=(-1)^{2m}\sC A_{ik}.
\end{equation}
It is easy to see that in both cases the matrices $\sE$ and $\sC$ always
commute with the operators $A_{ik}$. Therefore, for the type 
$p-q\equiv 0\pmod{8}$ at $p,q\equiv 0\pmod{4}$ and $p,q\equiv 2\pmod{4}$ the
elements of the Abelian groups $\sAut_-(\cl_{p,q})\simeq\dZ_2\otimes\dZ_2$
and $\sAut_-(\cl_{p,q})\simeq\dZ_4$ with $(+,-,-)$ are always commute with
infinitesimal operators $A_{ik}$ of $SO(3)$. Correspondingly, for the algebra
$\cl_{8t,0}$ the elements of $\sAut_-(\cl_{8t,0})\simeq\dZ_2\otimes\dZ_2$
are also commute with $A_{ik}$. The analogous statement takes place for
other degenerate case $\sAut_-(\cl_{0,8t})\simeq\dZ_2\otimes\dZ_2$. In the
case of non--Abelian groups $\sAut_+(\cl_{p,q})\simeq D_4/\dZ_2$
(signatures $(+,-,+)$ and $(+,+,-)$) we obtain the same permutation conditions
as (\ref{C34})--(\ref{C37}), that is, in this case the matrices of the
fundamental automorphisms always commute with $A_{ik}$. Thus, for the real
representation of the class $\fR^{l_0}_0$ operators of the discrete subgroup
always commute with all the infinitesimal operators of $SO(3)$.

Further, for the real representation of the class $\fR^{l_0}_2$, type
$p-q\equiv 2\pmod{8}$, at $\sE=\cE_{p+1}\cE_{p+2}\cdots\cE_{p+q}$ and
$\sC=\cE_1\cE_2\cdots\cE_p$ there exist Abelian groups $\sAut_-(\cl_{p,q})
\simeq\dZ_4$ with $(-,-,+)$ and $\sAut_-(\cl_{p,q})\simeq\dZ_4$ with
$(-,+,-)$ correspondingly at $p\equiv 0\pmod{4}$, $q\equiv 2\pmod{4}$ and
$p\equiv 2\pmod{4}$, $q\equiv 0\pmod{4}$, and also at $\sE=\cE_1\cE_2\cdots
\cE_p$ and $\sC=\cE_{p+1}\cE_{p+2}\cdots\cE_{p+q}$ there exist
non--Abelian groups $\sAut_+(\cl_{p,q})\simeq Q_4/\dZ_2$ with $(-,-,-)$ and
$\sAut_+(\cl_{p,q})\simeq D_4/\dZ_2$ with $(-,+,+)$ correspondingly at
$p\equiv 3\pmod{4}$, $q\equiv 1\pmod{4}$ and $p\equiv 1\pmod{4}$,
$q\equiv 3\pmod{4}$ (Theorem \ref{tautr}). So, for the Abelian groups
at $\cE_a,\cE_b,\cE_c\not\in\sE$ (operators (\ref{O1})) we obtain
\begin{equation}\label{C38}
A_{ik}\sE=(-1)^{2q}\sE A_{ik},
\end{equation}
\begin{eqnarray}
A_{ik}\sC&=&\pm(-1)^{i+j-2}\frac{1}{2}\sigma(i)\sigma(j)\cE_1\cE_2\cdots
\cE_{i-1}\cE_{i+1}\cdots\cE_{j-1}\cE_{j+1}\cdots\cE_p,\nonumber\\
\sC A_{ik}&=&\pm(-1)^{2p-i-j}\frac{1}{2}\sigma(i)\sigma(j)\cE_1\cE_2\cdots
\cE_{i-1}\cE_{i+1}\cdots\cE_{j-1}\cE_{j+1}\cdots\cE_p.\label{C39}
\end{eqnarray}
Correspondingly, for the operators (\ref{O3}) ($\cE_a,\cE_b,\cE_c\in\sE$)
we have
\begin{eqnarray}
A_{ik}\sE&=&\pm(-1)^{i+j-2}\frac{1}{2}\sigma(i)\sigma(j)\cE_{p+1}\cE_{p+2}
\cdots\cE_{i-1}\cE_{i+1}\cdots\cE_{j-1}\cE_{j+1}\cdots\cE_{p+q},\nonumber\\
\sE A_{ik}&=&\pm(-1)^{2q-i-j}\frac{1}{2}\sigma(i)\sigma(j)\cE_{p+1}\cE_{p+2}
\cdots\cE_{i-1}\cE_{i+1}\cdots\cE_{j-1}\cE_{j+1}\cdots\cE_{p+q},\label{C40}
\end{eqnarray}
\begin{equation}\label{C41}
A_{ik}\sC=(-1)^{2p}\sC A_{ik}.
\end{equation}
The analogous relations take place for the non--Abelian groups. From
(\ref{C38})--(\ref{C41}) it is easy to see that the matrices $\sE$ and
$\sC$ always commute with $A_{ik}$. Therefore, for the real representation
of the class $\fR^{l_0}_2$ operators of the discrete subgroup always commute
with all the infinitesimal operators of $SO(3)$.

Let us consider now quaternionic representations. Quaternionic representations
of the classes $\fH^{l_0}_4$ and $\fH^{l_0}_6$, types $p-q\equiv 4,6\pmod{8}$,
in virtue of the more wide ring $\K\simeq\BH$ have a more complicated
structure of the reflection groups than in the case of $\K\simeq\R$.
Indeed, if $\sE=\cE_{j_1}\cE_{j_2}\cdots\cE_{j_k}$ is a product of $k$
skewsymmetric matrices (among which $l$ matrices have `$+$'-square and
$t$ matrices have `$-$'-square) and 
$\sC=\cE_{i_1}\cE_{i_2}\cdots\cE_{i_{p+q-k}}$ is a product of $p+q-k$
symmetric matrices (among which there are $h$ `$+$'-squares and $g$ 
`$-$-squares), then at $k\equiv 0\pmod{2}$ for the type $p-q\equiv 4\pmod{8}$
there exist Abelian groups $\sAut_-(\cl_{p,q})\simeq\dZ_2\otimes\dZ_2$ and
$\sAut_-(\cl_{p,q})\simeq\dZ_4$ with $(+,-,-)$ if correspondingly
$l-t,h-g\equiv 0,1,4,5\pmod{8}$ and $l-t,h-g\equiv 2,3,6,7\pmod{8}$, and
also at $k\equiv 1\pmod{2}$ for the type $p-q\equiv 6\pmod{8}$ there exist
$\sAut_-(\cl_{p,q})\simeq\dZ_4$ with $(-,+,-)$ and $\sAut_-(\cl_{p,q})\simeq
\dZ_4$ with $(-,-,+)$ if correspondingly $l-t\equiv 0,1,4,5\pmod{8}$,
$h-g\equiv 2,3,6,7\pmod{8}$ and $l-t\equiv 2,3,6,7\pmod{8}$,
$h-g\equiv 0,1,4,5\pmod{8}$. Inversely, if $\sE=\cE_{i_1}\cE_{i_2}\cdots
\cE_{i_{p+q-k}}$ and $\sC=\cE_{j_1}\cE_{j_2}\cdots\cE_{j_k}$, then
at $k\equiv 1\pmod{8}$ for the type $p-q\equiv 4\pmod{8}$ there exist
non--Abelian groups $\sAut_+(\cl_{p,q})\simeq D_4/\dZ_2$ with $(+,-,+)$ and
$\sAut_+(\cl_{p,q})\simeq D_4/\dZ_2$ with $(+,+,-)$ if correspondingly
$h-g\equiv 2,3,6,7\pmod{8}$, $l-t\equiv 0,1,4,5\pmod{8}$ and
$h-g\equiv 0,1,4,5\pmod{8}$, $l-t\equiv 2,3,6,7\pmod{8}$, and also at
$k\equiv 1\pmod{2}$ for the type $p-q\equiv 6\pmod{8}$ there exist
$\sAut_+(\cl_{p,q})\simeq Q_4/\dZ_2$ with $(-,-,-)$ and
$\sAut_+(\cl_{p,q})\simeq D_4/\dZ_2$ with $(-,+,+)$ if correspondingly
$h-g,l-t\equiv 2,3,6,7\pmod{8}$ and $h-g,l-t\equiv 0,1,4,5\pmod{8}$
(see Theorem \ref{tautr}).

So, let $\sE=\cE_{j_1}\cE_{j_2}\cdots\cE_{j_k}$ and
$\sC=\cE_{i_1}\cE_{i_2}\cdots\cE_{i_{p+q-k}}$ be the matrices of
$\cA\rightarrow\widetilde{\cA}$ and $\cA\rightarrow\widetilde{\cA^\star}$,
$k\equiv 0\pmod{2}$. Assume that $\cE_a,\cE_b,\cE_c\in\sE$, that is, all the
matrices $\cE_i$ in the operators (\ref{O1}) or (\ref{O3}) are
skewsymmetric. Then for the operators (\ref{O1}) and (\ref{O3}) we obtain
\begin{eqnarray}
A_{23}\sE&=&-(-1)^{b+a-2}\frac{1}{2}\sigma(j_a)\sigma(j_b)\cE_{j_1}\cE_{j_2}
\cdots\cE_{j_{a-1}}\cE_{j_{a+1}}\cdots\cE_{j_{b-1}}\cE_{j_{b+1}}\cdots
\cE_{j_k},\nonumber\\
\sE A_{23}&=&-(-1)^{2k-a-b}\frac{1}{2}\sigma(j_a)\sigma(j_b)\cE_{j_1}\cE_{j_2}
\cdots\cE_{j_{a-1}}\cE_{j_{a+1}}\cdots\cE_{j_{b-1}}\cE_{j_{b+1}}\cdots
\cE_{j_k},\label{C42}
\end{eqnarray}
\begin{equation}\label{C43}
A_{23}\sC=(-1)^{2(p+q-k)}\sC A_{23},
\end{equation}
that is, $\sE$ and $\sC$ commute with $A_{23}$ (correspondingly with
$A_{13}$, $A_{12}$). It is easy to see that relations (\ref{C42}) and
(\ref{C43}) are analogous to the relations (\ref{R3}) and (\ref{R5}) for the
field $\F=\C$. Therefore, from (\ref{C42}) and (\ref{C43}) we obtain the
relations (\ref{TR11}) of Theorem. Further, assume that $\cE_a,\cE_b,\cE_c
\not\in\sE$, that is, all the matrices $\cE_i$ in the operators
(\ref{O1}) and (\ref{O3}) are symmetric. Then
\begin{equation}\label{C44}
A_{23}\sE=(-1)^{2k}\sE A_{23},
\end{equation}
\begin{eqnarray}
A_{23}\sC&=&-(-1)^{b+a-2}\frac{1}{2}\sigma(i_a)\sigma(i_b)\cE_{i_1}\cE_{i_2}
\cdots\cE_{i_{a-1}}\cE_{i_{a+1}}\cdots\cE_{i_{b-1}}\cE_{i_{b+1}}\cdots
\cE_{i_{p+q-k}},\nonumber\\
\sC A_{23}&=&-(-1)^{2(p+q-k)-a-b}\frac{1}{2}\sigma(i_a)\sigma(i_b)\cE_{i_1}
\cE_{i_2}\cdots\cE_{i_{a-1}}\cE_{i_{a+1}}\cdots\cE_{i_{b-1}}\cE_{i_{b+1}}
\cdots\cE_{i_{p+q-k}}\label{C45}
\end{eqnarray}
and analogous relations take place for $A_{13},A_{12}$. It is easy to verify
that from (\ref{C44}) and (\ref{C45}) we obtain the same relations
(\ref{TR11}), since (\ref{TR11}) are relations (\ref{bT1}) or (\ref{bT5})
at $B_i=0$. Therefore, over the ring $\K\simeq\BH$ (quasicomplex case)
the elements of Abelian reflection groups of the quaternionic
representations $\fH^{l_0}_{4,6}$ satisfy the relations
(\ref{bT1})--(\ref{bT24}) over the field $\F=\C$ at $B_i=0$. Indeed, at
$\cE_a,\cE_b\in\sE$, $\cE_c\not\in\sE$ and $\cE_c\in\sE$,
$\cE_a,\cE_b\not\in\sE$ we have relations (\ref{TR12}) which are particular
cases of (\ref{bT9}) at $B_i=0$ and so on. It is easy to verify that the
same relations take place for non--Abelian reflection groups at
$\sE=\cE_{i_1}\cE_{i_2}\cdots\cE_{i_{p+q-k}}$ and $\sC=\cE_{j_1}\cE_{j_2}
\cdots\cE_{j_k}$, $k\equiv 1\pmod{2}$.
\end{proof}
{\bf Remark}. Theorem exhausts all possible permutation relations between
transformations $P,\,T,\,PT$ and infinitesimal operators of the group
$\fG_+$. The relations (\ref{T0})--(\ref{T0'}) take place always, that is,
at any $n\equiv 0\pmod{2}$ (except the case $n=2$). In turn, the relations
(\ref{bT1})--(\ref{bT24}) are divided into two classes. The first class
contains relations with operators $H_{+,-,3}$ and $F_{+,-,3}$ (the relations
(\ref{bT1})--(\ref{bT4}), (\ref{bT5})--(\ref{bT8}), (\ref{bT13})--(\ref{bT16}),
(\ref{bT21})--(\ref{bT24})). The second class does not contain the relations
with $H_{+,-,3}$, $F_{+,-,3}$ (the relations (\ref{bT9})--(\ref{bT10}),
(\ref{bT11})--(\ref{bT12}), (\ref{bT17})--(\ref{bT18}), (\ref{bT19})--(\ref{bT20}).
Besides, in accordance with \cite{GMS} for the transformation $T$ there
are only two possibilities $T=P$ and $T=-P$ (both these cases correspond
to relation (\ref{bT1})). However, from other relations it follows that
$T\neq\pm P$, as it should be take place in general case. The exceptional
case $n=2$  corresponds to neutrino field and further it will be explored
in the following sections within quotient representations of the group
$\fG_+$. Permutations relations with respect to symmetric subspaces
$\Sym_{(k,r)}$ can be obtained by similar manner.
\section{Atiyah--Bott--Shapiro periodicity on the Lorentz group}
In accordance with the section 2 the finite--dimensional representations
$\fC$, $\overset{\ast}{\fC}$, 
$\fC\oplus\overset{\ast}{\fC}$ related with
the algebras $\C_{2k}$, $\overset{\ast}{\C}_{2r}$, 
$\C_{2k}\oplus\overset{\ast}{\C}_{2k}$
of the type $n\equiv 0\pmod{2}$ and the quotient
representations ${}^\chi\fC$, ${}^\chi\overset{\ast}{\fC}$,
${}^\chi\fC\cup{}^\chi\fC$
(${}^\chi\fC\oplus{}^\chi\overset{\ast}{\fC}$) 
related with the quotient
algebras ${}^\epsilon\C_{2k}$, ${}^\epsilon\overset{\ast}{\C}_{2r}$,
${}^\epsilon\C_{2k}\cup{}^\epsilon\overset{\ast}{\C}_{2k}$ corresponding to the type
$n\equiv 1\pmod{2}$ form a full system $\fM=\fM^0\oplus\fM^1$ of the
finite--dimensional representations of the proper Lorentz group $\fG_+$.
This extension of the Lorentz group allows to apply Atiyah--Bott--Shapiro
periodicity \cite{AtBSh} on the system $\fM$, and also it allows to define on
$\fM$ some cyclic relations that give rise to interesting applications in
particle physics (in the spirit of Gell-Mann--Ne'eman eightfold way 
\cite{GN64}).

As known, ABS--periodicity based on the
$\dZ_2$--graded structure of the Clifford algebra $\cl$. 
The graded central
simple Clifford algebras over the field $\F=\C$ form two similarity classes,
which, as it is easy to see, coincide with the two types of the algebras
$\C_n$: $n\equiv 0,1\pmod{2}$. The set of these 2 types (classes) forms
a Brauer--Wall group $BW_{\C}$ \cite{Wal64,Lou81} that is isomorphic to a
cyclic group $\dZ_2$. Thus, the algebra $\C_n$ is an element of the
Brauer--Wall group $BW_{\C}\simeq\dZ_2$, and a group operation is the graded
tensor product $\hat{\otimes}$. Coming back to representations of the group
$\fG_+$ we see that in virtue of identifications $\C_n\leftrightarrow\fC$
($n\equiv 0\pmod{2}$) and $\C_n\leftrightarrow\fC\cup\fC$ ($n\equiv 1\pmod{2}$)
a group action of $BW_{\C}\simeq\dZ_2$ can be transferred onto the system
$\fM=\fM^0\oplus\fM^1$. Indeed, a cyclic structure of the group
$BW_{\C}\simeq\dZ_2$ is defined by a transition 
$\C^+_n\overset{h}{\longrightarrow}\C_n$, where the type of the algebra
$\C_n$ is defined by a formula $n=h+2r$, here $h\in\{0,1\}$, $r\in\dZ$
\cite{BTr87,BT88}. Therefore, the action of $BW_{\C}\simeq\dZ_2$ on $\fM$
is defined by a transition 
$\overset{+}{\fC}\overset{h}{\longrightarrow}
\fC$, where $\overset{+}{\fC}\!{}^{l_0+l_1-1,0}\simeq\fC^{l_0+l_1-2,0}$ when
$\fC\in\fM^0$ ($h=1$) and $\overset{+}{\fC}=\left(\fC^{l_0+l_1-1,0}\cup
\fC^{l_0+l_1-1,0}\right)^+\sim{}^\chi\fC^{l_0+l_1-1,0}$ when $\fC\in\fM^1$
($h=0$), $\dim\fC=h+2r$ ($\dim\fC=l_0+l_1-1$ if $\fC\in\fM^0$ and
$\dim\fC=2(l_0+l_1-1)$ if $\fC\in\fM^1$). For example, in virtue of
$\C^+_2\simeq\C_1$ a transition $\C^+_2\rightarrow\C_2$ ($\C_1\rightarrow
\C_2$) induces on the system $\fM$ a transition 
$\overset{+}{\fC}\!{}^{1,0}
\rightarrow\fC^{1,0}$ that in virtue of $\overset{+}{\fC}\!{}^{1,0}\simeq
\fC^{0,0}$ is equivalent to $\fC^{0,0}\rightarrow\fC^{1,0}$ ($\fC^{0,0}$ is
one--dimensional representation of $\fG_+$) and, therefore, $h=1$.
In its turn, a transition $\C^+_3\rightarrow\C_3$ induces on the system
$\fM$ a transition $\left(\fC^{1,0}\cup\fC^{1,0}\right)^+\rightarrow
\fC^{1,0}\cup\fC^{1,0}$ or ${}^\chi\fC^{1,0}\rightarrow\fC^{1,0}\cup
\fC^{1,0}$ and, therefore, $h=0$. In such a way, we see that a cyclic
structure of the group $BW_{\C}\simeq\dZ_2$ induces on the system $\fM$
modulo 2 periodic relations which can be explicitly showed on the
Trautmann--like diagram (spinorial clock \cite{BTr87,BT88}, see also
\cite{Var00} and Chapter 2):
\bigskip\bigskip
\[
\unitlength=0.5mm
\begin{picture}(50.00,50.00)(0,0)
\put(5,25){0}
\put(42,25){1}
\put(22,-4){$\fC$}
\put(17,55){$\fC\cup\fC$}
\put(3,-13){$n\equiv 0\!\!\!\!\pmod{2}$}
\put(3,64){$n\equiv 1\!\!\!\!\pmod{2}$}
\put(20,49.49){$\cdot$}
\put(19.5,49.39){$\cdot$}
\put(19,49.27){$\cdot$}
\put(18.5,49.14){$\cdot$}
\put(18,49){$\cdot$}
\put(17.5,48.85){$\cdot$}
\put(17,48.68){$\cdot$}
\put(16.5,48.51){$\cdot$}
\put(16,48.32){$\cdot$}
\put(15.5,48.12){$\cdot$}
\put(15,47.91){$\cdot$}
\put(14.5,47.69){$\cdot$}
\put(14,47.45){$\cdot$}
\put(13.5,47.2){$\cdot$}
\put(13,46.93){$\cdot$}
\put(12.5,46.65){$\cdot$}
\put(12,46.35){$\cdot$}
\put(11.5,46.04){$\cdot$}
\put(11,45.71){$\cdot$}
\put(10.5,45.36){$\cdot$}
\put(10,45){$\cdot$}
\put(9.5,44.61){$\cdot$}
\put(9,44.21){$\cdot$}
\put(8.5,43.78){$\cdot$}
\put(8,43.33){$\cdot$}
\put(7.5,42.85){$\cdot$}
\put(7,42.35){$\cdot$}
\put(6.5,41.81){$\cdot$}
\put(6,41.25){$\cdot$}
\put(5.5,40.64){$\cdot$}
\put(5,40){$\cdot$}
\put(4.5,39.3){$\cdot$}
\put(4,38.56){$\cdot$}
\put(3.5,37.76){$\cdot$}
\put(3,36.87){$\cdot$}
\put(2.5,35.89){$\cdot$}
\put(2,34.79){$\cdot$}
\put(1.5,33.53){$\cdot$}
\put(1,32){$\cdot$}
\put(0.5,29.97){$\cdot$}
\put(30,49.49){$\cdot$}
\put(30.5,49.39){$\cdot$}
\put(31,49.27){$\cdot$}
\put(31.5,49.14){$\cdot$}
\put(32,49){$\cdot$}
\put(32.5,48.85){$\cdot$}
\put(33,48.68){$\cdot$}
\put(33.5,48.51){$\cdot$}
\put(34,48.32){$\cdot$}
\put(34.5,48.12){$\cdot$}
\put(35,47.91){$\cdot$}
\put(35.5,47.69){$\cdot$}
\put(36,47.45){$\cdot$}
\put(36.5,47.2){$\cdot$}
\put(37,46.93){$\cdot$}
\put(37.5,46.65){$\cdot$}
\put(38,46.35){$\cdot$}
\put(38.5,46.04){$\cdot$}
\put(39,45.71){$\cdot$}
\put(39.5,45.36){$\cdot$}
\put(40,45){$\cdot$}
\put(40.5,44.61){$\cdot$}
\put(41,44.21){$\cdot$}
\put(41.5,43.78){$\cdot$}
\put(42,43.33){$\cdot$}
\put(42.5,42.85){$\cdot$}
\put(43,42.35){$\cdot$}
\put(43.5,41.81){$\cdot$}
\put(44,41.25){$\cdot$}
\put(44.5,40.64){$\cdot$}
\put(45,40){$\cdot$}
\put(45.5,39.3){$\cdot$}
\put(46,38.56){$\cdot$}
\put(46.5,37.76){$\cdot$}
\put(47,36.87){$\cdot$}
\put(47.5,35.89){$\cdot$}
\put(48,34.79){$\cdot$}
\put(48.5,33.53){$\cdot$}
\put(49,32){$\cdot$}
\put(49.5,29.97){$\cdot$}
\put(0,25){$\cdot$}
\put(0,24.5){$\cdot$}
\put(0.02,24){$\cdot$}
\put(0.04,23.5){$\cdot$}
\put(0.08,23){$\cdot$}
\put(0.12,22.5){$\cdot$}
\put(0.18,22){$\cdot$}
\put(0.25,21.5){$\cdot$}
\put(0.32,21){$\cdot$}
\put(0.4,20.5){$\cdot$}
\put(0.5,20){$\cdot$}
\put(0.61,19.5){$\cdot$}
\put(0.73,19){$\cdot$}
\put(0.85,18.5){$\cdot$}
\put(1,18){$\cdot$}
\put(1.15,17.5){$\cdot$}
\put(1.31,17){$\cdot$}
\put(1.49,16.5){$\cdot$}
\put(1.68,16){$\cdot$}
\put(1.88,15.5){$\cdot$}
\put(2.09,15){$\cdot$}
\put(2.31,14.5){$\cdot$}
\put(2.55,14){$\cdot$}
\put(2.8,13.5){$\cdot$}
\put(3.06,13){$\cdot$}
\put(0,25.5){$\cdot$}
\put(0.02,26){$\cdot$}
\put(0.04,26.5){$\cdot$}
\put(0.08,27){$\cdot$}
\put(0.12,27.5){$\cdot$}
\put(0.18,28){$\cdot$}
\put(0.25,28.5){$\cdot$}
\put(0.32,29){$\cdot$}
\put(0.4,29.5){$\cdot$}
\put(0.5,30){$\cdot$}
\put(0.61,30.5){$\cdot$}
\put(0.73,31){$\cdot$}
\put(0.85,31.5){$\cdot$}
\put(1,32){$\cdot$}
\put(1.15,32.5){$\cdot$}
\put(1.31,33){$\cdot$}
\put(1.49,33.5){$\cdot$}
\put(1.68,34){$\cdot$}
\put(1.88,34.5){$\cdot$}
\put(2.09,35){$\cdot$}
\put(2.31,35.5){$\cdot$}
\put(2.55,36){$\cdot$}
\put(2.8,36.5){$\cdot$}
\put(3.06,37){$\cdot$}
\put(50,25){$\cdot$}
\put(49.99,24.5){$\cdot$}
\put(49.98,24){$\cdot$}
\put(49.95,23.5){$\cdot$}
\put(49.92,23){$\cdot$}
\put(49.87,22.5){$\cdot$}
\put(49.82,22){$\cdot$}
\put(49.75,21.5){$\cdot$}
\put(49.68,21){$\cdot$}
\put(49.51,20.5){$\cdot$}
\put(49.49,20){$\cdot$}
\put(49.39,19.5){$\cdot$}
\put(49.27,19){$\cdot$}
\put(49.14,18.5){$\cdot$}
\put(49,18){$\cdot$}
\put(48.85,17.5){$\cdot$}
\put(48.69,17){$\cdot$}
\put(48.51,16.5){$\cdot$}
\put(48.32,16){$\cdot$}
\put(48.12,15.5){$\cdot$}
\put(47.91,15){$\cdot$}
\put(47.69,14.5){$\cdot$}
\put(47.45,14){$\cdot$}
\put(47.2,13.5){$\cdot$}
\put(46.93,13){$\cdot$}
\put(50,25){$\cdot$}
\put(49.99,25.5){$\cdot$}
\put(49.98,26){$\cdot$}
\put(49.95,26.5){$\cdot$}
\put(49.92,27){$\cdot$}
\put(49.87,27.5){$\cdot$}
\put(49.82,28){$\cdot$}
\put(49.75,28.5){$\cdot$}
\put(49.68,29){$\cdot$}
\put(49.51,29.5){$\cdot$}
\put(49.49,30){$\cdot$}
\put(49.39,30.5){$\cdot$}
\put(49.27,31){$\cdot$}
\put(49.14,31.5){$\cdot$}
\put(49,32){$\cdot$}
\put(48.85,32.5){$\cdot$}
\put(48.69,33){$\cdot$}
\put(48.51,33.5){$\cdot$}
\put(48.32,34){$\cdot$}
\put(48.12,34.5){$\cdot$}
\put(47.91,35){$\cdot$}
\put(47.69,35.5){$\cdot$}
\put(47.45,36){$\cdot$}
\put(47.2,36.5){$\cdot$}
\put(46.93,37){$\cdot$}
\put(20,0.5){$\cdot$}
\put(19.5,0.61){$\cdot$}
\put(19,0.73){$\cdot$}
\put(18.5,0.86){$\cdot$}
\put(18,1){$\cdot$}
\put(17.5,1.15){$\cdot$}
\put(17,1.31){$\cdot$}
\put(16.5,1.49){$\cdot$}
\put(16,1.68){$\cdot$}
\put(15.5,1.87){$\cdot$}
\put(15,2.09){$\cdot$}
\put(14.5,2.31){$\cdot$}
\put(14,2.55){$\cdot$}
\put(13.5,2.8){$\cdot$}
\put(13,3.06){$\cdot$}
\put(12.5,3.35){$\cdot$}
\put(12,3.64){$\cdot$}
\put(11.5,3.96){$\cdot$}
\put(11,4.29){$\cdot$}
\put(10.5,4.63){$\cdot$}
\put(10,5){$\cdot$}
\put(9.5,5.38){$\cdot$}
\put(9,5.79){$\cdot$}
\put(8.5,6.22){$\cdot$}
\put(8,6.67){$\cdot$}
\put(7.5,7.15){$\cdot$}
\put(7,7.65){$\cdot$}
\put(6.5,8.18){$\cdot$}
\put(6,8.75){$\cdot$}
\put(5.5,9.35){$\cdot$}
\put(5,10){$\cdot$}
\put(4.5,10.69){$\cdot$}
\put(4,11.43){$\cdot$}
\put(3.5,12.24){$\cdot$}
\put(3,13.12){$\cdot$}
\put(2.5,14.10){$\cdot$}
\put(2,15.20){$\cdot$}
\put(1.5,16.47){$\cdot$}
\put(1,18){$\cdot$}
\put(0.5,20.02){$\cdot$}
\put(30,0.5){$\cdot$}
\put(30.5,0.61){$\cdot$}
\put(31,0.73){$\cdot$}
\put(31.5,0.86){$\cdot$}
\put(32,1){$\cdot$}
\put(32.5,1.15){$\cdot$}
\put(33,1.31){$\cdot$}
\put(33.5,1.49){$\cdot$}
\put(34,1.68){$\cdot$}
\put(34.5,1.87){$\cdot$}
\put(35,2.09){$\cdot$}
\put(35.5,2.31){$\cdot$}
\put(36,2.55){$\cdot$}
\put(36.5,2.8){$\cdot$}
\put(37,3.06){$\cdot$}
\put(37.5,3.35){$\cdot$}
\put(38,3.64){$\cdot$}
\put(38.5,3.96){$\cdot$}
\put(39,4.29){$\cdot$}
\put(39.5,4.63){$\cdot$}
\put(40,5){$\cdot$}
\put(40.5,5.38){$\cdot$}
\put(41,5.79){$\cdot$}
\put(41.5,6.22){$\cdot$}
\put(42,6.67){$\cdot$}
\put(42.5,7.15){$\cdot$}
\put(43,7.65){$\cdot$}
\put(43.5,8.18){$\cdot$}
\put(44,8.75){$\cdot$}
\put(44.5,9.35){$\cdot$}
\put(45,10){$\cdot$}
\put(45.5,10.69){$\cdot$}
\put(46,11.43){$\cdot$}
\put(46.5,12.24){$\cdot$}
\put(47,13.12){$\cdot$}
\put(47.5,14.10){$\cdot$}
\put(48,15.20){$\cdot$}
\put(48.5,16.47){$\cdot$}
\put(49,18){$\cdot$}
\put(49.5,20.02){$\cdot$}

\end{picture}
\]
\vspace{1ex}
\begin{center}
\begin{minipage}{25pc}{\small
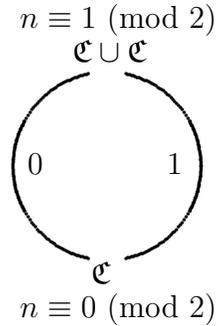
{\bf Fig.1} The action of the Brauer--Wall group 
$BW_{\C}\simeq\dZ_2$ on the full system $\fM=\fM^0\oplus\fM^1$ of complex
finite--dimensional representations $\fC$ of the proper Lorentz group $\fG_+$.} 
\end{minipage}
\end{center}
\bigskip
It is obvious that a group structure over $\C_n$, defined by the group
$BW_{\C}\simeq\dZ_2$, immediately relates with a modulo 2 periodicity of the
complex Clifford algebras \cite{AtBSh,Kar79}: $\C_{n+2}\simeq\C_n\otimes\C_2$.
Therefore, we have the following relations for $\fC$ and
$\overset{\ast}{\fC}$:
\begin{eqnarray}
\fC^{l_0+l_1,0}&\simeq&\fC^{l_0+l_1-1,0}\otimes\fC^{1,0},\nonumber\\
\fC^{0,l_0-l_1}&\simeq&\fC^{0,l_0-l_1+1}\otimes\fC^{0,-1}\nonumber
\end{eqnarray}
and correspondingly for 
$\fC\otimes\overset{\ast}{\fC}$
\[
\fC^{l_0+l_1,l_0-l_1}\simeq\fC^{l_0+l_1-1,l_0-l_1+1}\otimes\fC^{1,-1}.
\]
Thus, the action of $BW_{\C}\simeq\dZ_2$ form a cycle of the period 2 on
the system $\fM$, where the basic 2-period factor is the fundamental
representation $\fC^{1,0}$ ($\fC^{0,-1}$) of the group $\fG_+$. This
cyclic structure is intimately related with de Broglie--Jordan neutrino
theory of light and, moreover, this structure is a natural generalization of
BJ--theory. Indeed, in the simplest case we obtain two relations
$\fC^{2,0}\simeq\fC^{1,0}\otimes\fC^{1,0}$ and $\fC^{0,-2}\simeq
\fC^{0,-1}\otimes\fC^{0,-1}$, which, as it is easy to see, correspond two
helicity states of the photon 
(left-- and right--handed polarization)\footnote{Such a description
corresponds to Helmholtz--Silberstein representation of the electromagnetic
field as the complex linear combinations $\bF=\bE+i\bH$,
$\overset{\ast}{\bF}=\bE-i\bH$ that form a basis of the 
Majorana--Oppenheimer quantum electrodynamics 
\cite{Opp31,RF68,Esp98,Dvo97b} (see also recent development on this
subject based on the Joos--Weinberg and Bargmann--Wigner formalisms
\cite{Dvo97}).}. In such a way, subsequent rotations of the
representation 2-cycle give all other higher spin physical fields and
follows to de Broglie and Jordan this structure should be called as
`{\it neutrino theory of everything}'.

Further, when we consider restriction of the complex representation $\fC$
onto real representations $\fR$ and $\fH$, that corresponds to restriction
of the group $\fG_+$ onto its subgroup $SO(3)$\footnote{Physical feilds
defined within such representations describe neutral particles, or particles
at rest such as atomic nuclei.}, we come to a more high--graded modulo 8
periodicity over the field $\F=\R$. 
Indeed,
the Clifford algebra $\cl_{p,q}$ is central simple
if $p-q\not\equiv 1,5\pmod{8}$. It is known that for the Clifford algebra
with odd dimensionality, the isomorphisms are as follows:
$\cl^+_{p,q+1}\simeq\cl_{p,q}$ and $\cl^+_{p+1,q}\simeq\cl_{q,p}$ 
\cite{Rash,Port}. Thus, $\cl^+_{p,q+1}$ and $\cl^+_{p+1,q}$
are central simple algebras. Further, in accordance with Chevalley Theorem
\cite{Che55} for the graded tensor product there is an isomorphism
$\cl_{p,q}\hat{\otimes}\cl_{p^{\p},q^{\p}}\simeq
\cl_{p+p^{\p},q+q^{\p}}$. Two algebras $\cl_{p,q}$ and $\cl_{p^{\p},q^{\p}}$
are said to be of the same class if $p+q^{\p}\equiv p^{\p}+q\pmod{8}$.
The graded central simple Clifford algebras over the field $\F=\R$
form eight similarity classes, which, as it is easy to see, coincide
with the eight types of the algebras $\cl_{p,q}$.
The set of these 8 types (classes) forms a Brauer--Wall group $BW_{\R}$
\cite{Wal64,Lou81} that is isomorphic to a cyclic group $\dZ_8$. 
Therefore, in virtue of identifications (\ref{Ident}) a group action of
$BW_{\R}\simeq\dZ_8$ can be transferred onto the system $\fM=\fM^+\oplus\fM^-$.
In its turn, a cyclic structure of the group $BW_{\R}\simeq\dZ_8$ is defined by
a transition $\cl^+_{p,q}\overset{h}{\longrightarrow}\cl_{p,q}$, where the
type of the algebra $\cl_{p,q}$ is defined by a formula
$q-p=h+8r$, here $h\in\{0,\ldots,7\}$, $r\in\dZ$ \cite{BTr87,BT88}. Thus,
the action of $BW_{\R}\simeq\dZ_8$ on $\fM$ is defined by a transition
$\overset{+}{\fD}\!{}^{l_0}\overset{h}{\longrightarrow}\fD^{l_0}$, where
$\fD^{\l_0}=\left\{\fR^{l_0}_{0,2},\fH^{l_0}_{4,6},\fC^{l_0}_{3,7},
\fR^{l_0}_{0,2}\cup\fR^{l_0}_{0,2},\fH^{l_0}_{4,6}\cup\fH^{l_0}_{4,6}\right\}$,
and $\overset{+}{\fD}\!{}^{r/2}\simeq\fD^{\frac{r-1}{2}}$ when $\fD\in\fM^+$ and
$\left(\fD^{r/2}\cup\fD^{r/2}\right)^+\simeq{}^\chi\fD^{r/2}$ when
$\fD\in\fM^-$, $r$ is a number of tensor products in (\ref{Ten}). Therefore,
a cyclic structure of the group $BW_{\R}\simeq\dZ_8$ induces on the system
$\fM$ modulo 8 periodic relations which can be explicitly showed on the
following diagram (the round on the diagram is realized by an hour--hand):
\bigskip
\[
\unitlength=0.5mm
\begin{picture}(100.00,110.00)

\put(97,67){$\fC^{l_0}_7$}\put(108,64){$p-q\equiv 7\!\!\!\!\pmod{8}$}
\put(80,80){1}
\put(75,93.3){$\cdot$}
\put(75.5,93){$\cdot$}
\put(76,92.7){$\cdot$}
\put(76.5,92.4){$\cdot$}
\put(77,92.08){$\cdot$}
\put(77.5,91.76){$\cdot$}
\put(78,91.42){$\cdot$}
\put(78.5,91.08){$\cdot$}
\put(79,90.73){$\cdot$}
\put(79.5,90.37){$\cdot$}
\put(80,90.0){$\cdot$}
\put(80.5,89.62){$\cdot$}
\put(81,89.23){$\cdot$}
\put(81.5,88.83){$\cdot$}
\put(82,88.42){$\cdot$}
\put(82.5,87.99){$\cdot$}
\put(83,87.56){$\cdot$}
\put(83.5,87.12){$\cdot$}
\put(84,86.66){$\cdot$}
\put(84.5,86.19){$\cdot$}
\put(85,85.70){$\cdot$}
\put(85.5,85.21){$\cdot$}
\put(86,84.69){$\cdot$}
\put(86.5,84.17){$\cdot$}
\put(87,83.63){$\cdot$}
\put(87.5,83.07){$\cdot$}
\put(88,82.49){$\cdot$}
\put(88.5,81.9){$\cdot$}
\put(89,81.29){$\cdot$}
\put(89.5,80.65){$\cdot$}
\put(90,80){$\cdot$}
\put(90.5,79.32){$\cdot$}
\put(91,78.62){$\cdot$}
\put(91.5,77.89){$\cdot$}
\put(92,77.13){$\cdot$}
\put(92.5,76.34){$\cdot$}
\put(93,75.51){$\cdot$}
\put(93.5,74.65){$\cdot$}
\put(94,73.74){$\cdot$}
\put(94.5,72.79){$\cdot$}
\put(96.5,73.74){\vector(1,-2){1}}
\put(80,20){3}
\put(97,31){$\fH^{l_0}_6$}\put(108,28){$p-q\equiv 6\!\!\!\!\pmod{8}$}
\put(75,6.7){$\cdot$}
\put(75.5,7){$\cdot$}
\put(76,7.29){$\cdot$}
\put(76.5,7.6){$\cdot$}
\put(77,7.91){$\cdot$}
\put(77.5,8.24){$\cdot$}
\put(78,8.57){$\cdot$}
\put(78.5,8.91){$\cdot$}
\put(79,9.27){$\cdot$}
\put(79.5,9.63){$\cdot$}
\put(80,10){$\cdot$}
\put(80.5,10.38){$\cdot$}
\put(81,10.77){$\cdot$}
\put(81.5,11.17){$\cdot$}
\put(82,11.58){$\cdot$}
\put(82.5,12.00){$\cdot$}
\put(83,12.44){$\cdot$}
\put(83.5,12.88){$\cdot$}
\put(84,13.34){$\cdot$}
\put(84.5,13.8){$\cdot$}
\put(85,14.29){$\cdot$}
\put(85.5,14.79){$\cdot$}
\put(86,15.3){$\cdot$}
\put(86.5,15.82){$\cdot$}
\put(87,16.37){$\cdot$}
\put(87.5,16.92){$\cdot$}
\put(88,17.5){$\cdot$}
\put(88.5,18.09){$\cdot$}
\put(89,18.71){$\cdot$}
\put(89.5,19.34){$\cdot$}
\put(90,20){$\cdot$}
\put(90.5,20.68){$\cdot$}
\put(91,21.38){$\cdot$}
\put(91.5,22.11){$\cdot$}
\put(92,22.87){$\cdot$}
\put(92.5,23.66){$\cdot$}
\put(93,24.48){$\cdot$}
\put(93.5,25.34){$\cdot$}
\put(94,26.25){$\cdot$}
\put(94.5,27.20){$\cdot$}
\put(95,28.20){$\cdot$}
\put(20,80){7}
\put(25,93.3){$\cdot$}
\put(24.5,93){$\cdot$}
\put(24,92.7){$\cdot$}
\put(23.5,92.49){$\cdot$}
\put(23,92.08){$\cdot$}
\put(22.5,91.75){$\cdot$}
\put(22,91.42){$\cdot$}
\put(21.5,91.08){$\cdot$}
\put(21,90.73){$\cdot$}
\put(20.5,90.37){$\cdot$}
\put(20,90){$\cdot$}
\put(19.5,89.62){$\cdot$}
\put(19,89.23){$\cdot$}
\put(18.5,88.83){$\cdot$}
\put(18,88.42){$\cdot$}
\put(17.5,87.99){$\cdot$}
\put(17,87.56){$\cdot$}
\put(16.5,87.12){$\cdot$}
\put(16,86.66){$\cdot$}
\put(15.5,86.19){$\cdot$}
\put(15,85.70){$\cdot$}
\put(14.5,85.21){$\cdot$}
\put(14,84.69){$\cdot$}
\put(13.5,84.17){$\cdot$}
\put(13,83.63){$\cdot$}
\put(12.5,83.07){$\cdot$}
\put(12,82.49){$\cdot$}
\put(11.5,81.9){$\cdot$}
\put(11,81.29){$\cdot$}
\put(10.5,80.65){$\cdot$}
\put(10,80){$\cdot$}
\put(9.5,79.32){$\cdot$}
\put(9,78.62){$\cdot$}
\put(8.5,77.89){$\cdot$}
\put(8,77.13){$\cdot$}
\put(7.5,76.34){$\cdot$}
\put(7,75.51){$\cdot$}
\put(6.5,74.65){$\cdot$}
\put(6,73.79){$\cdot$}
\put(5.5,72.79){$\cdot$}
\put(5,71.79){$\cdot$}
\put(20,20){5}
\put(25,6.7){$\cdot$}
\put(24.5,7){$\cdot$}
\put(24,7.29){$\cdot$}
\put(23.5,7.6){$\cdot$}
\put(23,7.91){$\cdot$}
\put(22.5,8.24){$\cdot$}
\put(22,8.57){$\cdot$}
\put(21.5,8.91){$\cdot$}
\put(21,9.27){$\cdot$}
\put(20.5,9.63){$\cdot$}
\put(20,10){$\cdot$}
\put(19.5,10.38){$\cdot$}
\put(19,10.77){$\cdot$}
\put(18.5,11.17){$\cdot$}
\put(18,11.58){$\cdot$}
\put(17.5,12){$\cdot$}
\put(17,12.44){$\cdot$}
\put(16.5,12.88){$\cdot$}
\put(16,13.34){$\cdot$}
\put(15.5,13.8){$\cdot$}
\put(15,14.29){$\cdot$}
\put(14.5,14.79){$\cdot$}
\put(14,15.3){$\cdot$}
\put(13.5,15.82){$\cdot$}
\put(13,16.37){$\cdot$}
\put(12.5,16.92){$\cdot$}
\put(12,17.5){$\cdot$}
\put(11.5,18.09){$\cdot$}
\put(11,18.71){$\cdot$}
\put(10.5,19.34){$\cdot$}
\put(10,20){$\cdot$}
\put(9.5,20.68){$\cdot$}
\put(9,21.38){$\cdot$}
\put(8.5,22.11){$\cdot$}
\put(8,22.87){$\cdot$}
\put(7.5,23.66){$\cdot$}
\put(7,24.48){$\cdot$}
\put(6.5,25.34){$\cdot$}
\put(6,26.25){$\cdot$}
\put(5.5,27.20){$\cdot$}
\put(5,28.20){$\cdot$}
\put(-1,97){$\fR^{l_0}_{0,2}\cup\fR^{l_0}_{0,2}$}
\put(-55,107){$p-q\equiv 1\!\!\!\!\pmod{8}$}
\put(50,93){0}
\put(50,100){$\cdot$}
\put(49.5,99.99){$\cdot$}
\put(49,99.98){$\cdot$}
\put(48.5,99.97){$\cdot$}
\put(48,99.96){$\cdot$}
\put(47.5,99.94){$\cdot$}
\put(47,99.91){$\cdot$}
\put(46.5,99.86){$\cdot$}
\put(46,99.84){$\cdot$}
\put(45.5,99.8){$\cdot$}
\put(45,99.75){$\cdot$}
\put(44.5,99.7){$\cdot$}
\put(44,99.64){$\cdot$}
\put(43.5,99.57){$\cdot$}
\put(43,99.51){$\cdot$}
\put(42.5,99.43){$\cdot$}
\put(42,99.35){$\cdot$}
\put(41.5,99.27){$\cdot$}
\put(41,99.18){$\cdot$}
\put(40.5,99.09){$\cdot$}
\put(40,98.99){$\cdot$}
\put(39.5,98.88){$\cdot$}
\put(39,98.77){$\cdot$}
\put(38.5,98.66){$\cdot$}
\put(38,98.54){$\cdot$}
\put(37.5,98.41){$\cdot$}
\put(37,98.28){$\cdot$}
\put(50.5,99.99){$\cdot$}
\put(51,99.98){$\cdot$}
\put(51.5,99.97){$\cdot$}
\put(52,99.96){$\cdot$}
\put(52.5,99.94){$\cdot$}
\put(53,99.91){$\cdot$}
\put(53.5,99.86){$\cdot$}
\put(54,99.84){$\cdot$}
\put(54.5,99.8){$\cdot$}
\put(55,99.75){$\cdot$}
\put(55.5,99.7){$\cdot$}
\put(56,99.64){$\cdot$}
\put(56.5,99.57){$\cdot$}
\put(57,99.51){$\cdot$}
\put(57.5,99.43){$\cdot$}
\put(58,99.35){$\cdot$}
\put(58.5,99.27){$\cdot$}
\put(59,99.18){$\cdot$}
\put(59.5,99.09){$\cdot$}
\put(60,98.99){$\cdot$}
\put(60.5,98.88){$\cdot$}
\put(61,98.77){$\cdot$}
\put(61.5,98.66){$\cdot$}
\put(62,98.54){$\cdot$}
\put(62.5,98.41){$\cdot$}
\put(63,98.28){$\cdot$}
\put(65,97){$\fR^{l_0}_0$}\put(73,108){$p-q\equiv 0\!\!\!\!\pmod{8}$}
\put(50,7){4}
\put(67,2){$\fH^{l_0}_{4,6}\cup\fH^{l_0}_{4,6}$}
\put(90,-6){$p-q\equiv 5\!\!\!\!\pmod{8}$}
\put(50,0){$\cdot$}
\put(50.5,0){$\cdot$}
\put(51,0.01){$\cdot$}
\put(51.5,0.02){$\cdot$}
\put(52,0.04){$\cdot$}
\put(52.5,0.06){$\cdot$}
\put(53,0.09){$\cdot$}
\put(53.5,0.12){$\cdot$}
\put(54,0.16){$\cdot$}
\put(54.5,0.2){$\cdot$}
\put(55,0.25){$\cdot$}
\put(55.5,0.3){$\cdot$}
\put(56,0.36){$\cdot$}
\put(56.5,0.42){$\cdot$}
\put(57,0.49){$\cdot$}
\put(57.5,0.56){$\cdot$}
\put(58,0.64){$\cdot$}
\put(58.5,0.73){$\cdot$}
\put(59,0.82){$\cdot$}
\put(59.5,0.91){$\cdot$}
\put(60,1.01){$\cdot$}
\put(60.5,1.11){$\cdot$}
\put(61,1.22){$\cdot$}
\put(61.5,1.34){$\cdot$}
\put(62,1.46){$\cdot$}
\put(62.5,1.59){$\cdot$}
\put(63,1.72){$\cdot$}
\put(49.5,0){$\cdot$}
\put(49,0.01){$\cdot$}
\put(48.5,0.02){$\cdot$}
\put(48,0.04){$\cdot$}
\put(47.5,0.06){$\cdot$}
\put(47,0.09){$\cdot$}
\put(46.5,0.12){$\cdot$}
\put(46,0.16){$\cdot$}
\put(45.5,0.2){$\cdot$}
\put(45,0.25){$\cdot$}
\put(44.5,0.3){$\cdot$}
\put(44,0.36){$\cdot$}
\put(43.5,0.42){$\cdot$}
\put(43,0.49){$\cdot$}
\put(42.5,0.56){$\cdot$}
\put(42,0.64){$\cdot$}
\put(41.5,0.73){$\cdot$}
\put(41,0.82){$\cdot$}
\put(40.5,0.91){$\cdot$}
\put(40,1.01){$\cdot$}
\put(39.5,1.11){$\cdot$}
\put(39,1.22){$\cdot$}
\put(38.5,1.34){$\cdot$}
\put(38,1.46){$\cdot$}
\put(37.5,1.59){$\cdot$}
\put(37,1.72){$\cdot$}
\put(28,3){$\fH^{l_0}_4$}\put(-40,-4){$p-q\equiv 4\!\!\!\!\pmod{8}$}
\put(93,50){2}
\put(98.28,63){$\cdot$}
\put(98.41,62.5){$\cdot$}
\put(98.54,62){$\cdot$}
\put(98.66,61.5){$\cdot$}
\put(98.77,61){$\cdot$}
\put(98.88,60.5){$\cdot$}
\put(98.99,60){$\cdot$}
\put(99.09,59.5){$\cdot$}
\put(99.18,59){$\cdot$}
\put(99.27,58.5){$\cdot$}
\put(99.35,58){$\cdot$}
\put(99.43,57.5){$\cdot$}
\put(99.51,57){$\cdot$}
\put(99.57,56.5){$\cdot$}
\put(99.64,56){$\cdot$}
\put(99.7,55.5){$\cdot$}
\put(99.75,55){$\cdot$}
\put(99.8,54.5){$\cdot$}
\put(99.84,54){$\cdot$}
\put(99.86,53.5){$\cdot$}
\put(99.91,53){$\cdot$}
\put(99.94,52.5){$\cdot$}
\put(99.96,52){$\cdot$}
\put(99.97,51.5){$\cdot$}
\put(99.98,51){$\cdot$}
\put(99.99,50.5){$\cdot$}
\put(100,50){$\cdot$}
\put(98.28,37){$\cdot$}
\put(98.41,37.5){$\cdot$}
\put(98.54,38){$\cdot$}
\put(98.66,38.5){$\cdot$}
\put(98.77,39){$\cdot$}
\put(98.88,39.5){$\cdot$}
\put(98.99,40){$\cdot$}
\put(99.09,40.5){$\cdot$}
\put(99.18,41){$\cdot$}
\put(99.27,41.5){$\cdot$}
\put(99.35,42){$\cdot$}
\put(99.43,42.5){$\cdot$}
\put(99.51,43){$\cdot$}
\put(99.57,43.5){$\cdot$}
\put(99.64,44){$\cdot$}
\put(99.7,44.5){$\cdot$}
\put(99.75,45){$\cdot$}
\put(99.8,45.5){$\cdot$}
\put(99.84,46){$\cdot$}
\put(99.86,46.5){$\cdot$}
\put(99.91,47){$\cdot$}
\put(99.94,47.5){$\cdot$}
\put(99.96,48){$\cdot$}
\put(99.97,48.5){$\cdot$}
\put(99.98,49){$\cdot$}
\put(99.99,49.5){$\cdot$}
\put(7,50){6}
\put(1,32){$\fC^{l_0}_3$}\put(-65,29){$p-q\equiv 3\!\!\!\!\pmod{8}$}
\put(1.72,63){$\cdot$}
\put(1.59,62.5){$\cdot$}
\put(1.46,62){$\cdot$}
\put(1.34,61.5){$\cdot$}
\put(1.22,61){$\cdot$}
\put(1.11,60.5){$\cdot$}
\put(1.01,60){$\cdot$}
\put(0.99,59.5){$\cdot$}
\put(0.82,59){$\cdot$}
\put(0.73,58.5){$\cdot$}
\put(0.64,58){$\cdot$}
\put(0.56,57.5){$\cdot$}
\put(0.49,57){$\cdot$}
\put(0.42,56.5){$\cdot$}
\put(0.36,56){$\cdot$}
\put(0.3,55.5){$\cdot$}
\put(0.25,55){$\cdot$}
\put(0.2,54.5){$\cdot$}
\put(0.16,54){$\cdot$}
\put(0.12,53.5){$\cdot$}
\put(0.09,53){$\cdot$}
\put(0.06,52.5){$\cdot$}
\put(0.04,52){$\cdot$}
\put(0.02,51.5){$\cdot$}
\put(0.01,51){$\cdot$}
\put(0,50.5){$\cdot$}
\put(0,50){$\cdot$}
\put(1.72,37){$\cdot$}
\put(1.59,37.5){$\cdot$}
\put(1.46,38){$\cdot$}
\put(1.34,38.5){$\cdot$}
\put(1.22,39){$\cdot$}
\put(1.11,39.5){$\cdot$}
\put(1.01,40){$\cdot$}
\put(0.99,40.5){$\cdot$}
\put(0.82,41){$\cdot$}
\put(0.73,41.5){$\cdot$}
\put(0.64,42){$\cdot$}
\put(0.56,42.5){$\cdot$}
\put(0.49,43){$\cdot$}
\put(0.42,43.5){$\cdot$}
\put(0.36,44){$\cdot$}
\put(0.3,44.5){$\cdot$}
\put(0.25,45){$\cdot$}
\put(0.2,45.5){$\cdot$}
\put(0.16,46){$\cdot$}
\put(0.12,46.5){$\cdot$}
\put(0.09,47){$\cdot$}
\put(0.06,47.5){$\cdot$}
\put(0.04,48){$\cdot$}
\put(0.02,48.5){$\cdot$}
\put(0.01,49){$\cdot$}
\put(0,49.5){$\cdot$}
\put(0.5,67){$\fR^{l_0}_2$}\put(-65,75){$p-q\equiv 2\!\!\!\!\pmod{8}$}
\end{picture}
\]

\vspace{2ex}
\begin{center}
\begin{minipage}{25pc}{\small
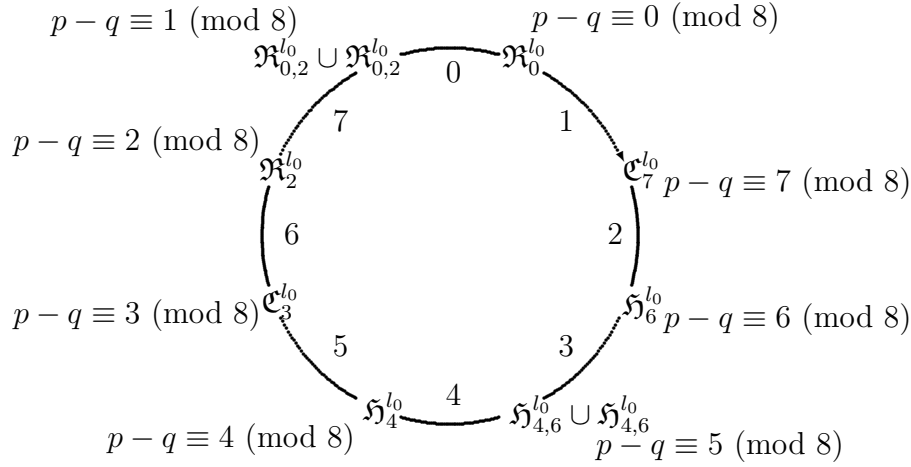
{\bf Fig.2} The action of the Brauer--Wall group $BW_{\R}\simeq\dZ_8$ on
the full system $\fM=\fM^+\oplus\fM^-$ of real representations $\fD$ of the
proper Lorentz group $\fG_+$, $l_0=\frac{p+q}{4}$.}
\end{minipage}
\end{center}
\medskip
Further, it is well--known that a group structure over $\cl_{p,q}$, defined by
$BW_{\R}\simeq\dZ_8$, immediately relates with the Atiyah--Bott--Shapiro
periodicity \cite{AtBSh}. In accordance with \cite{AtBSh}, the Clifford
algebra over the field $\F=\R$ is modulo 8 periodic:
$\cl_{p+8,q}\simeq\cl_{p,q}\otimes\cl_{8,0}\,(\cl_{p,q+8}\simeq\cl_{p,q}
\otimes\cl_{0,8})$. Therefore, we have a following relation
\[
\fD^{l_0+2}\simeq\fD^{l_0}\otimes\fR^2_0,
\]
since $\fR^2_0\leftrightarrow\cl_{8,0}\;(\cl_{0,8})$ and in virtue of
Karoubi Theorem from (\ref{Ten}) it follows that
$\cl_{8,0}\simeq\cl_{2,0}\otimes\cl_{0,2}\otimes\cl_{0,2}\otimes\cl_{2,0}$
($\cl_{0,8}\simeq\cl_{0,2}\otimes\cl_{2,0}\otimes\cl_{2,0}
\otimes\cl_{0,2}$)\footnote{The minimal left ideal of $\cl_{8,0}$ is equal
to $\dS_{16}$ and in virtue of the real ring $\K\simeq\R$ is defined within
the full matrix algebra $\M_{16}(\R)$. At first glance, from the
factorization of $\cl_{8,0}$ it follows that
$\M_2(\R)\otimes\BH\otimes\BH\otimes\M_2(\R)\not\simeq\M_{16}(\R)$, but it
is wrong, since there is an isomorphism $\BH\otimes\BH\simeq\M_4(\R)$
(see Appendix B in \cite{BDGK}).}, therefore, $r=4$, $l_0=r/2=2$.
On the other hand, in terms of minimal left ideal the modulo 8
periodicity looks like
\[
\dS_{n+8}\simeq\dS_n\otimes\dS_{16}.
\]
In virtue of the mapping $\gamma_{8,0}:\,\cl_{8,0}\rightarrow\M_2(\dO)$
\cite{MS96} (see also excellent review \cite{Bae01}) the latter relation
can be written in the form
\[
\dS_{n+8}\simeq\dS_n\otimes\dO^2,
\]
where $\dO$ is {\it an octonion algebra}. Since the algebra $\cl_{8,0}\simeq
\cl_{0,8}$ admits an octonionic representation, then in virtue of the
modulo 8 periodicity the octonionic representations can be defined for all
high dimensions and, therefore, on the system $\fM=\fM^+\oplus\fM^-$ we
have a relation
\[
\fD^{l_0+2}\simeq\fD^{l_0}\otimes\fO,
\]
where $\fO$ is {\it an octonionic representation} of the group $\fG_+$
($\fO\sim\fR^2_0$). Thus, the action of $BW_{\R}\simeq\dZ_8$ form a cycle
of the period 8 on the system $\fM$. This is intimately related with an
octonionic structure. In 1973, G\"{u}naydin and G\"{u}rsey showed that an
automorphism group of the algebra $\dO$ is isomorphic to an exceptional Lie
group $G_2$ that contains $SU(3)$ as a subgroup \cite{GG73}. The
G\"{u}naydin--G\"{u}rsey construction allows to incorporate the quark
phenomenology into a general algebraic framework. Moreover, this
construction allows to define the quark structure on the system $\fM$
within octonionic representations of the proper Lorentz group $\fG_+$.
It is obvious that within such a framework the quark structure cannot be
considered as a fundamental physical structure underlieing of the world
(as it suggested by QCD). This is fully derivative structure firstly
appearred in 8-dimension and further reproduced into high dimensions by the
round of 8-cycle generated by the group $BW_{\R}\simeq\dZ_8$ from 8 ad
infinitum (growth of quark's flavors with increase of energy). One can say
that such a description, included very powerful algebraic tools, opens an
another way of understanding of the Gell-Mann--Ne'emann eightfold way in
particle physics.\section{Pseudoautomorphism $\cA\longrightarrow\overline{\cA}$
and charge conjugation}
As noted previously, an extraction of the minimal left ideal of the complex
algebra $\C_n\simeq\C_2\otimes\C_2\otimes\cdots\otimes\C_2$ induces a space
of the finite--dimensional spintensor representation of the group
$\fG_+$. Besides, the algebra $\C_n$ is associated with a complex vector
space $\C^n$. Let $n=p+q$, then an extraction operation of the real subspace
$\R^{p,q}$ in $\C^n$  forms the foundation of definition of the discrete
transformation known in physics as
{\it a charge conjugation} $C$. Indeed, let
$\{\e_1,\ldots,\e_n\}$ be an orthobasis in the space $\C^n$, $\e^2_i=1$.
Let us remain the first $p$ vectors of this basis unchanged, and other $q$
vectors multiply by the factor $i$. Then the basis
\begin{equation}\label{6.23}
\left\{\e_1,\ldots,\e_p,i\e_{p+1},\ldots,i\e_{p+q}\right\}
\end{equation}
allows to extract the subspace $\R^{p,q}$ in $\C^n$. Namely,
for the vectors $\R^{p,q}$ we take the vectors of
$\C^n$ which decompose on the basis
(\ref{6.23}) with real coefficients. In such a way we obtain a real vector
space $\R^{p,q}$ endowed (in general case) with a non--degenerate
quadratic form
\[
Q(x)=x^2_1+x^2_2+\ldots+x^2_p-x^2_{p+1}-x^2_{p+2}-\ldots-x^2_{p+q},
\]
where $x_1,\ldots,x_{p+q}$ are coordinates of the vector $\bx$ 
in the basis (\ref{6.23}).
It is easy to see that the extraction of
$\R^{p,q}$ in $\C^n$ induces an extraction of
{\it a real subalgebra} $\cl_{p,q}$ in $\C_n$. Therefore, any element
$\cA\in\C_n$ can be unambiguously represented in the form
\[
\cA=\cA_1+i\cA_2,
\]
where $\cA_1,\,\cA_2\in\cl_{p,q}$. The one--to--one mapping
\begin{equation}\label{6.24}
\cA\longrightarrow\overline{\cA}=\cA_1-i\cA_2
\end{equation}
transforms the algebra $\C_n$ into itself with preservation of addition
and multiplication operations for the elements $\cA$; the operation of
multiplication of the element $\cA$ by the number transforms to an operation
of multiplication by the complex conjugate number.
Any mapping of $\C_n$ satisfying these conditions is called
{\it a pseudoautomorphism}. Thus, the extraction of the subspace
$\R^{p,q}$ in the space $\C^n$ induces in the algebra $\C_n$ 
a pseudoautomorphism $\cA\rightarrow\overline{\cA}$ \cite{Rash}.

Let us consider a spinor representation of the pseudoautomorphism
$\cA\rightarrow\overline{\cA}$ of the algebra $\C_n$ when $n\equiv 0\s\pmod{2}$.
In the spinor representation the every element $\cA\in\C_n$ should be
represented by some matrix $\sA$, and the pseudoautomorphism (\ref{6.24})
takes a form of the pseudoautomorphism of the full matrix algebra 
$\M_{2^{n/2}}$:
\[
\sA\longrightarrow\overline{\sA}.
\]\begin{sloppypar}\noindent
On the other hand, a transformation replacing the matrix $\sA$ by the
complex conjugate matrix, $\sA\rightarrow\dot{\sA}$, is also some
pseudoautomorphism of the algebra $\M_{2^{n/2}}$. The composition of the two
pseudoautomorpisms $\dot{\sA}\rightarrow\sA$ and
$\sA\rightarrow\overline{\sA}$, $\dot{\sA}\rightarrow\sA\rightarrow
\overline{\sA}$, is an internal automorphism
$\dot{\sA}\rightarrow\overline{\sA}$ of the full matrix algebra $\M_{2^{n/2}}$:
\end{sloppypar}
\begin{equation}\label{6.25}
\overline{\sA}=\Pi\dot{\sA}\Pi^{-1},
\end{equation}
where $\Pi$ is a matrix of the pseudoautomorphism 
$\cA\rightarrow\overline{\cA}$ in the spinor representation.
The sufficient condition for definition of the pseudoautomorphism
$\cA\rightarrow\overline{\cA}$ is a choice of the matrix
$\Pi$ in such a way that the transformation 
$\sA\rightarrow\Pi\dot{\sA}\Pi^{-1}$ transfers into itself the matrices
$\cE_1,\ldots,\cE_p,i\cE_{p+1},\ldots,i\cE_{p+q}$
(the matrices of the spinbasis of $\cl_{p,q}$), that is,
\begin{equation}\label{6.26}
\cE_i\longrightarrow\cE_i=\Pi\dot{\cE}_i\Pi^{-1}\quad
(i=1,\ldots,p+q).
\end{equation}
\begin{theorem}\label{tpseudo}
Let $\C_n$ be a complex Clifford algebra when $n\equiv 0\s\pmod{2}$
and let $\cl_{p,q}\subset\C_n$ be its subalgebra with a real division ring
$\K\simeq\R$ when $p-q\equiv 0,2\s\pmod{8}$ and quaternionic division ring
$\K\simeq\BH$ when $p-q\equiv 4,6\s\pmod{8}$, $n=p+q$. Then in dependence
on the division ring structure of the real subalgebra $\cl_{p,q}$ the matrix
$\Pi$ of the pseudoautomorphism $\cA\rightarrow\overline{\cA}$ 
has the following form:\\[0.2cm]
1) $\K\simeq\R$, $p-q\equiv 0,2\s\pmod{8}$.\\[0.1cm]
The matrix $\Pi$ for any spinor representation over the ring $\K\simeq\R$
is proportional to the unit matrix.\\[0.2cm]
2) $\K\simeq\BH$, $p-q\equiv 4,6\s\pmod{8}$.\\[0.1cm]
$\Pi=\cE_{\alpha_1}\cE_{\alpha_2}\cdots\cE_{\alpha_a}$ when 
$a\equiv 0\s\pmod{2}$ and
$\Pi=\cE_{\beta_1}\cE_{\beta_2}\cdots\cE_{\beta_b}$ when $b\equiv 1\s\pmod{2}$,
where $a$ complex matrices $\cE_{\alpha_t}$ 
and $b$ real matrices $\cE_{\beta_s}$ form a basis of the spinor
representation of the algebra $\cl_{p,q}$ over the ring $\K\simeq\BH$,
$a+b=p+q,\,0<t\leq a,\,0<s\leq b$. At this point
\begin{eqnarray}
\Pi\dot{\Pi}&=&\phantom{-}\sI\quad\text{if $a,b\equiv 0,1\s\pmod{4}$},
\nonumber\\
\Pi\dot{\Pi}&=&-\sI\quad\text{if $a,b\equiv 2,3\s\pmod{4}$},\nonumber
\end{eqnarray}
where $\sI$ is the unit matrix.
\end{theorem}
\begin{proof}\begin{sloppypar}\noindent
The algebra $\C_n$ ($n\equiv 0\s\pmod{2}$, $n=p+q$) in virtue of
$\C_n=\C\otimes\cl_{p,q}$ and definition of the division ring
$\K\simeq f\cl_{p,q}f$ 
($f$ is a primitive idempotent of the algebra $\cl_{p,q}$)
has four different real subalgebras: $p-q\equiv 0,2\s\pmod{8}$
for the real division ring $\K\simeq\R$ and $p-q\equiv 4,6\s\pmod{8}$ for
the quaternionic division ring $\K\simeq\BH$.\\[0.2cm]
1) $\K\simeq\R$.\\[0.1cm]
Since for the types $p-q\equiv 0,2\s\pmod{8}$ there is an isomorphism
$\cl_{p,q}\simeq\M_{2^{\frac{p+q}{2}}}(\R)$ (Wedderburn--Artin Theorem), then
all the matrices $\cE_i$ of the spinbasis of $\cl_{p,q}$ are real and
$\dot{\cE}_i=\cE_i$. Therefore, in this case the condition (\ref{6.26})
can be written as follows\end{sloppypar}
\[
\cE_i\longrightarrow\cE_i=\Pi\cE_i\Pi^{-1},
\]
whence $\cE_i\Pi=\Pi\cE_i$. Thus, for the algebras $\cl_{p,q}$ of the types
$p-q\equiv 0,2\s\pmod{8}$ the matrix $\Pi$ of the pseudoautomorphism
$\cA\rightarrow\overline{\cA}$ commutes with all the matrices $\cE_i$.
It is easy to see that
$\Pi\sim\sI$.\\[0.2cm]
2) $\K\simeq\BH$.\\[0.1cm]
In turn, for the quaternionic types $p-q\equiv 4,6\s\pmod{8}$ there is an
isomorphism $\cl_{p,q}\simeq\M_{2^{\frac{p+q}{2}}}(\BH)$. Therefore, among
the matrices of the spinbasis of the algebra $\cl_{p,q}$ there are matrices
$\cE_\alpha$ satisfying the condition $\dot{\cE}_\alpha=-\cE_\alpha$. 
Let $a$ be a quantity of the complex matrices, then the spinbasis of $\cl_{p,q}$
is divided into two subsets. The first subset
$\{\dot{\cE}_{\alpha_t}=-\cE_{\alpha_t}\}$ contains complex matrices,
$0<t\leq a$, and the second subset
$\{\dot{\cE}_{\beta_s}=\cE_{\beta_s}\}$ contains real matrices,
$0<s\leq p+q-a$. In accordance with a spinbasis structure of the algebra
$\cl_{p,q}\simeq\M_{2^{\frac{p+q}{2}}}(\BH)$ the condition (\ref{6.26})
can be written as follows
\[
\cE_{\alpha_t}\longrightarrow-\cE_{\alpha_t}=\Pi\cE_{\alpha_t}\Pi^{-1},\quad
\cE_{\beta_s}\longrightarrow\cE_{\beta_s}=\Pi\cE_{\beta_s}\Pi^{-1}.
\]
Whence
\begin{equation}\label{6.27}
\cE_{\alpha_t}\Pi=-\Pi\cE_{\alpha_t},\quad
\cE_{\beta_s}\Pi=\Pi\cE_{\beta_s}.
\end{equation}
Thus, for the quaternionic types $p-q\equiv 4,6\s\pmod{8}$ the matrix
$\Pi$ of the pseudoautomorphism $\cA\rightarrow\overline{\cA}$ anticommutes
with a complex part of the spinbasis of $\cl_{p,q}$ and commutes with
a real part of the same spinbasis. From (\ref{6.27}) it follows that a
structure of the matrix $\Pi$ is analogous to the structure of
the matrices $\sE$ and $\sC$ of the antiautomorphisms
$\cA\rightarrow\widetilde{\cA}$ and
$\cA\rightarrow\widetilde{\cA^\star}$, correspondingly 
(see Theorem \ref{tautr}), that is, the matrix
$\Pi$ of the pseudoautomorphism $\cA\rightarrow\overline{\cA}$ of the algebra
$\C_n$ is a product of only complex matrices, or only real matrices
of the spinbasis of the subalgebra $\cl_{p,q}$.

So, let $0<a<p+q$ and let $\Pi=\cE_{\alpha_1}\cE_{\alpha_2}\cdots
\cE_{\alpha_a}$ be a matrix of $\cA\rightarrow\overline{\cA}$,
then permutation conditions of the matrix $\Pi$ 
with the matrices $\cE_{\beta_s}$
of the real part ($0<s\leq p+q-a$) and with the matrices
$\cE_{\alpha_t}$ of the complex part ($0<t\leq a$) have the form
\begin{equation}\label{6.28}
\Pi\cE_{\beta_s}=(-1)^a\cE_{\beta_s}\Pi,
\end{equation}
\begin{eqnarray}
\Pi\cE_{\alpha_t}&=&(-1)^{a-t}\sigma(\alpha_t)\cE_{\alpha_1}\cE_{\alpha_2}
\cdots\cE_{\alpha_{t-1}}\cE_{\alpha_{t+1}}\cdots\cE_{\alpha_a},\nonumber\\
\cE_{\alpha_t}\Pi&=&(-1)^{t-1}\sigma(\alpha_t)\cE_{\alpha_1}\cE_{\alpha_2}
\cdots\cE_{\alpha_{t-1}}\cE_{\alpha_{t+1}}\cdots\cE_{\alpha_a},\label{6.29}
\end{eqnarray}
that is, when $a\equiv 0\s\pmod{2}$ the matrix $\Pi$ commutes with the real
part and anticommutes with the complex part of the spinbasis of $\cl_{p,q}$.
Correspondingly, when $a\equiv 1\s\pmod{2}$ the matrix $\Pi$ anticommutes
with the real part and commutes with the complex part. Further, let
$\Pi=\cE_{\beta_1}\cE_{\beta_2}\cdots\cE_{\beta_{p+q-a}}$ be a product of the
real matrices, then
\begin{eqnarray}
\Pi\cE_{\beta_s}&=&(-1)^{p+q-a-s}\sigma(\beta_s)\cE_{\beta_1}\cE_{\beta_2}
\cdots\cE_{\beta_{s-1}}\cE_{\beta_{s+1}}\cdots\cE_{\beta_{p+q-a}},\nonumber\\
\cE_{\beta_s}\Pi&=&(-1)^{s-1}\sigma(\beta_s)\cE_{\beta_1}\cE_{\beta_2}
\cdots\cE_{\beta_{s-1}}\cE_{\beta_{s+1}}\cdots\cE_{\beta_{p+q-a}},\label{6.30}
\end{eqnarray}
\begin{equation}\label{6.31}
\Pi\cE_{\alpha_t}=(-1)^{p+q-a}\cE_{\alpha_t}\Pi,
\end{equation}
that is, when $p+q-a\equiv 0\s\pmod{2}$ the matrix $\Pi$ anticommutes with
the real part and commutes with the complex part 
of the spinbasis of $\cl_{p,q}$. Correspondingly, when
$p+q-a\equiv 1\s\pmod{2}$ the matrix $\Pi$ commutes with the real part and
anticommutes with the complex part.

The comparison of the conditions (\ref{6.28})--(\ref{6.29}) 
with the condition (\ref{6.27}) shows that the matrix
$\Pi=\cE_{\alpha_1}\cE_{\alpha_2}\cdots\cE_{\alpha_a}$ exists only at
$a\equiv 0\s\pmod{2}$, that is, $\Pi$ is a product of the complex matrices
$\cE_{\alpha_t}$ of the even number. In its turn, a comparison of
(\ref{6.30})--(\ref{6.31}) with
(\ref{6.27}) shows that the matrix $\Pi=\cE_{\beta_1}\cE_{\beta_2}\cdots
\cE_{\beta_{p+q-a}}$ exists only at $p+q-a\equiv 1\s\pmod{2}$, that is,
$\Pi$ is a product of the real matrices $\cE_{\beta_s}$ of the odd number.

Let us calculate now the product $\Pi\dot{\Pi}$. 
Let $\Pi=\cE_{\beta_1}\cE_{\beta_2}\cdots
\cE_{\beta_{p+q-a}}$ be a product of the $p+q-a$ real matrices.
Since $\dot{\cE}_{\beta_s}=\cE_{\beta_s}$, then
$\dot{\Pi}=\Pi$ and $\Pi\dot{\Pi}=\Pi^2$. Therefore,
\begin{equation}\label{6.32}
\Pi\dot{\Pi}=(\cE_{\beta_1}\cE_{\beta_2}\cdots\cE_{\beta_{p+q-a}})^2=
(-1)^{\frac{(p+q-a)(p+q-a-1)}{2}}\cdot\sI.
\end{equation}
Further, let $\Pi=\cE_{\alpha_1}\cE_{\alpha_2}\cdots\cE_{\alpha_a}$ be a
product of the $a$ complex matrices. Then
$\dot{\cE}_{\alpha_t}=-\cE_{\alpha_t}$ and $\dot{\Pi}=(-1)^a\Pi=\Pi$, since
$a\equiv 0\s\pmod{2}$. Therefore,
\begin{equation}\label{6.33}
\Pi\dot{\Pi}=(\cE_{\alpha_1}\cE_{\alpha_2}\cdots\cE_{\alpha_a})^2=
(-1)^{\frac{a(a-1)}{2}}\cdot\sI.
\end{equation}
Let $p+q-a=b$ be a quantity of the real matrices $\cE_{\beta_s}$ of the
spinbasis of $\cl_{p,q}$, then $p+q=a+b$. Since $p+q$ is always even number
for the quaternionic types $p-q\equiv 4,6\s\pmod{8}$, then $a$ and $b$ 
are simultaneously even or odd numbers. Thus, from (\ref{6.32}) and (\ref{6.33})
it follows
\[
\Pi\dot{\Pi}=\begin{cases}
\phantom{-}\sI,& \text{if $a,b\equiv 0,1\s\!\!\pmod{4}$},\\
-\sI,& \text{if $a,b\equiv 2,3\s\!\!\pmod{4}$},
\end{cases}
\]
which required to be proved.
\end{proof}

In the present form of quantum field theory complex fields correspond
to charged particles. Thus, the extraction of the subalgebra $\cl_{p,q}$ with
the real ring $\K\simeq\R$ in $\C_n$, $p-q\equiv 0,2\s\pmod{8}$,
corresponds to physical fields describing {\it truly neutral particles}
such as photon and neutral mesons ($\pi^0,\,\eta^0,\,\rho^0,\,
\omega^0,\,\varphi^0,\,K^0$). In turn, the subalgebras $\cl_{p,q}$ with the
ring $\K\simeq\BH$, $p-q\equiv 4,6\s\pmod{8}$ correspond to charged or
neutral fields.

As known \cite{GY48b}, the charge conjugation $C$ should be satisfied the
following requirement
\begin{equation}\label{Com}
CI^{ik}=I^{ik}C,
\end{equation}
where $I^{ik}$ are infinitesimal operators of the group $\fG_+$\footnote{The
requirement $CP=PC$ presented also in the Gel'fand--Yaglom work \cite{GY48b}
is superfluous, since the inverse relation $CP=-PC$ is valid in
BWW--type quantum field theories \cite{AJG93}.}. This requirement is necessary
for the definition of the operation $C$ on the representation spaces of
$\fC$. Let us find permutation conditions of the matrix $\Pi$ with $I^{ik}$
defined by the relations (\ref{O1})--(\ref{O2}). It is obvious that in the
case of $\K\simeq\R$ the matrix $\Pi\simeq\sI$ commutes with all the
operators $I^{ik}$ and, therefore, the relations (\ref{Com}) hold. In the
case of $\K\simeq\BH$ and $\Pi=\cE_{\alpha_1}\cE_{\alpha_2}\cdots\cE_{\alpha_a}$
it is easy to verify that at $\cE_a,\cE_b,\cE_c\in\Pi$ (all 
$\cE_a,\cE_b,\cE_c$ are complex matrices) the matrix $\Pi$ commutes with
$A_{ik}$ and anticommutes with $B_i$ (permutation conditions in this case
are analogous to (\ref{R3}) and (\ref{R4})). In its turn, when
$\cE_a,\cE_b,\cE_c\not\in\Pi$ (all $\cE_a,\cE_b,\cE_c$ are real matrices) the
matrix $\Pi$ commutes with all the operators $I^{ik}$. It is easy to see
that all other cases given by the cyclic permutations $\cE_i,\cE_j\in\Pi$,
$\cE_k\not\in\Pi$ and $\cE_i\in\Pi$, $\cE_j,\cE_k\not\in\Pi$
($i,j,k\in\{a,b,c\}$) do not satisfy the relations (\ref{Com}). For example,
at $\cE_a,\cE_b\in\Pi$ and $\cE_c\not\in\Pi$ the matrix $\Pi$ commutes with
$A_{23}$ and $B_1$ and anticommutes with $A_{13},A_{12},B_2,B_3$. Further,
in the case of $\Pi=\cE_{\beta_1}\cE_{\beta_2}\cdots\cE_{\beta_b}$ it is not
difficult to see that only at $\cE_a,\cE_b,\cE_c\in\Pi$ (all
$\cE_a,\cE_b,\cE_c$ are real matrices) the matrix $\Pi$ commutes with all
$I^{ik}$. Therefore, in both cases $\Pi=\cE_{\alpha_1}\cE_{\alpha_2}\cdots
\cE_{\alpha_a}$ and $\Pi=\cE_{\beta_1}\cE_{\beta_2}\cdots\cE_{\beta_b}$ the
relations (\ref{Com}) hold when all the matrices $\cE_a,\cE_b,\cE_c$
belonging to (\ref{O1})--(\ref{O2}) are real.

When we restrict the complex representation $\fC$ (charged particles)
of $\fG_+$ to real representation $\fR$ (truly neutral particles) and
$\fH$ (neutral particles) we see that in this case the charge conjugation
is reduced to an identical transformation $\sI$ for $\fR$ and to a
particle--antiparticle conjugation $C^\prime$ 
for $\fH$. Moreover, as follows from
Theorem \ref{tinf} for the real representations $B_i=0$ and, therefore,
the relations (\ref{Com}) take a form
\begin{equation}\label{Com2}
C^\prime A_{ik}=A_{ik}C^\prime.
\end{equation}
Over the ring $\K\simeq\R$ the relations (\ref{Com2}) hold identically.
It is easy to verify that over the ring $\K\simeq\BH$ for the matrix
$\Pi=\cE_{\alpha_1}\cE_{\alpha_2}\cdots\cE_{\alpha_a}$ the relations
(\ref{Com2}) hold at $\cE_a,\cE_b,\cE_c\in\Pi$ and 
$\cE_a,\cE_b,\cE_c\not\in\Pi$. The same result takes place for the matrix
$\Pi=\cE_{\beta_1}\cE_{\beta_2}\cdots\cE_{\beta_b}$. All other cases given by
the cyclic permutations do not satisfy the relations (\ref{Com2}).
Therefore, in both cases the relations (\ref{Com2}) hold when all the
matrices $\cE_a,\cE_b,\cE_c$ in (\ref{O1}) (correspondingly in (\ref{O3}))
are complex or real.
Let us consider the action of the pseudoautomorphism 
$\cA\rightarrow\overline{\cA}$ on the spinors
(\ref{6.5}) (`vectors' of the fundamental representation of the group
$\fG_+$). The matrix $\Pi$ allows to compare to each spinor
$\xi^\alpha$ its conjugated spinor $\overline{\xi}^\alpha$ by the following
rule
\begin{equation}\label{6.33'}
\overline{\xi}^\alpha=\Pi^\alpha_{\dot{\alpha}}\xi^{\dot{\alpha}},
\end{equation}
here $\xi^{\dot{\alpha}}=(\xi^\alpha)^\cdot$. In accordance with Theorem
\ref{tpseudo} for the matrix $\Pi^\alpha_{\dot{\beta}}$ 
we have $\dot{\Pi}=\Pi^{-1}$ or
$\dot{\Pi}=-\Pi^{-1}$, where $\Pi^{-1}=\Pi^{\dot{\alpha}}_\beta$.
Then a twice conjugated spinor looks like
\[
\overline{\xi}^\alpha=\overline{\Pi^\alpha_{\dot{\beta}}\xi^{\dot{\beta}}}=
\Pi^\alpha_{\dot{\alpha}}(\Pi^\alpha_{\dot{\beta}}\xi^{\dot{\beta}})^\cdot=
\Pi^\alpha_{\dot{\alpha}}(\pm\Pi^{\dot{\alpha}}_\beta)\xi^\beta=
\pm\xi^\alpha.
\]
Therefore, the twice conjugated spinor coincides with the initial spinor
in the case of the real subalgebra of $\C_2$ with the ring
$\K\simeq\R$ (the algebras $\cl_{1,1}$ and $\cl_{2,0}$), and also in the case
of $\K\simeq\BH$ (the algebra $\cl_{0,2}\simeq\BH$) at
$a-b\equiv 0,1\pmod{4}$. Since for the algebra $\cl_{0,2}\simeq\BH$ we have
always $a-b\equiv 0\pmod{4}$, then a property of the reciprocal conjugacy
of the spinors
$\xi^\alpha$ ($\alpha=1,2$) is an invariant fact for the fundamental
representation of the group $\fG_+$ (this property is very important in
physics, since this is an algebraic expression of the requirement
$C^2=1$). Further, since the `vector' (spintensor) of the 
finite--dimensional representation of the group
$\fG_+$ is defined by the tensor product
$\xi^{\alpha_1\alpha_2\cdots\alpha_k}=\sum\xi^{\alpha_1}\otimes
\xi^{\alpha_2}\otimes\cdots\otimes\xi^{\alpha_k}$, then its conjugated
spintensor takes a form
\begin{equation}\label{6.33''}
\overline{\xi}^{\alpha_1\alpha_2\cdots\alpha_k}=
\sum\Pi^{\alpha_1}_{\dot{\alpha}_1}\Pi^{\alpha_2}_{\dot{\alpha}_2}\cdots
\Pi^{\alpha_k}_{\dot{\alpha}_k}\xi^{\dot{\alpha}_1\dot{\alpha}_2\cdots
\dot{\alpha}_k},
\end{equation}\begin{sloppypar}\noindent
It is obvious that the condition of reciprocal conjugacy 
$\overline{\overline{\xi}}\!{}^{\alpha_1\alpha_2\cdots\alpha_k}=
\xi^{\alpha_1\alpha_2
\cdots\alpha_k}$ is also fulfilled for (\ref{6.33''}), since for each matrix
$\Pi^{\alpha_i}_{\dot{\alpha}_i}$ in (\ref{6.33''}) we have
$\dot{\Pi}=\Pi^{-1}$ (all the matrices $\Pi^{\alpha_i}_{\dot{\alpha}_i}$
are defined for the algebra $\C_2$). \end{sloppypar}

Let us define now permutation conditions of the matrix $\Pi$ of the
pseudoautomorphism
$\cA\rightarrow\overline{\cA}$ (charge conjugation) with the matrix $\sW$ 
of the automorphism $\cA\rightarrow\cA^\star$ (space inversion).
First of all, in accordance with Theorem \ref{tpseudo} in the case of
$\cl_{p,q}$ with the real ring $\K\simeq\R$ (types $p-q\equiv 0,2\pmod{8}$)
the matrix $\Pi$ is proportional to the unit matrix and, therefore,
commutes with the matrix $\sW$. In the case of
$\K\simeq\BH$ (types $p-q\equiv 4,6\pmod{8}$) from Theorem
\ref{tpseudo} it follows two possibilities:
$\Pi=\cE_{\alpha_1}\cE_{\alpha_2}\cdots\cE_{\alpha_a}$ is a product of
$a$ complex matrices at $a\equiv 0\pmod{2}$ and
$\Pi=\cE_{\beta_1}\cE_{\beta_2}\cdots\cE_{\beta_b}$ is a product of
$b$ real matrices at $b\equiv 1\pmod{2}$. 
Since $a+b=p+q$, then the matrix $\sW$ can be represented by the product
$\cE_{\alpha_1}\cE_{\alpha_2}\cdots\cE_{\alpha_a}\cE_{\beta_1}\cE_{\beta_2}
\cdots\cE_{\beta_b}$. Then for $\Pi=\cE_{\alpha_1}\cE_{\alpha_2}\cdots
\cE_{\alpha_a}$ we have
\begin{eqnarray}
\Pi\sW&=&(-1)^{\frac{a(a-1)}{2}}\sigma(\alpha_1)\sigma(\alpha_2)\cdots
\sigma(\alpha_a)\cE_{\beta_1}\cE_{\beta_2}\cdots\cE_{\beta_b},\nonumber\\
\sW\Pi&=&(-1)^{\frac{a(a-1)}{2}+ba}\sigma(\alpha_1)\sigma(\alpha_2)\cdots
\sigma(\alpha_a)\cE_{\beta_1}\cE_{\beta_2}\cdots\cE_{\beta_b}.\nonumber
\end{eqnarray}
Hence it follows that at $ab\equiv 0\pmod{2}$ the matrices $\Pi$ and $\sW$
always commute, since $a\equiv 0\pmod{2}$. Taking
$\Pi=\cE_{\beta_1}\cE_{\beta_2}\cdots\cE_{\beta_b}$ we obtain following
conditions:
\begin{eqnarray}
\Pi\sW&=&(-1)^{\frac{b(b-1)}{2}+ab}\sigma(\beta_1)\sigma(\beta_2)\cdots
\sigma(\beta_b)\cE_{\alpha_1}\cE_{\alpha_2}\cdots\cE_{\alpha_a},\nonumber\\
\sW\Pi&=&(-1)^{\frac{b(b-1)}{2}}\sigma(\beta_1)\sigma(\beta_2)\cdots
\sigma(\beta_b)\cE_{\alpha_1}\cE_{\alpha_2}\cdots\cE_{\alpha_a}.\nonumber
\end{eqnarray}
Hence it follows that $ab\equiv 1\pmod{2}$, since in this case 
$b\equiv 1\pmod{2}$,
and $p+q=a+b$ is even number, $a$ is odd number. Therefore, at
$ab\equiv 1\pmod{2}$ the matrices $\Pi$ and $\sW$ always anticommute.

It should be noted one important feature related with the anticommutation of
the matrices
$\Pi$ and $\sW$, $\Pi\sW=-\sW\Pi$, that corresponds to relation
$CP=-PC$. The latter relation holds for Bargmann--Wightmann--Wigner type
quantum field theories in which bosons and antibosons have mutually
opposite intrinsic parities \cite{AJG93}. Thus, in this case
the matrix of the operator $C$ is a product of real matrices of odd number.
\section{Quotient representations of the Lorentz group}
\begin{theorem}\label{tfactor}1) $\F=\C$.
Let $\cA\rightarrow\overline{\cA}$, $\cA\rightarrow\cA^\star$,
$\cA\rightarrow\widetilde{\cA}$ be the automorphisms of the odd--dimensional
complex Clifford algebra $\C_{n+1}$ ($n+1\equiv 1,3\s\pmod{4}$) corresponding
the discrete transformations $C,\,P,\,T$ (charge conjugation, space inversion,
time reversal) and let ${}^\epsilon\C_n$ be a quotient algebra obtained in the
result of the homomorphic mapping $\epsilon:\;\C_{n+1}
\rightarrow\C_n$. Then over the field $\F=\C$ 
in dependence on the structure of
${}^\epsilon\C_n$ all the quotient representations of the
Lorentz group are divided in the following six classes:
\begin{eqnarray}
1)\;{}^\chi\fC^{l_0+l_1-1,0}_{a_1}&:&\{T,\,C\sim I\},\nonumber\\
2)\;{}^\chi\fC^{l_0+l_1-1,0}_{a_2}&:&\{T,\,C\},\nonumber\\
3)\;{}^\chi\fC^{l_0+l_1-1,0}_{b}&:&\{T,\,CP,\,CPT\},\nonumber\\
4)\;{}^\chi\fC^{l_0+l_1-1,0}_{c}&:&\{PT,\,C,\,CPT\},\nonumber\\
5)\;{}^\chi\fC^{l_0+l_1-1,0}_{d_1}&:&\{PT,\,CP\sim IP,\,CT\sim IT\},\nonumber\\
6)\;{}^\chi\fC^{l_0+l_1-1,0}_{d_2}&:&\{PT,\,CP,\,CT\}.\nonumber
\end{eqnarray}
2) $\F=\R$. Real quotient representations are divided into four different
classes:
\begin{eqnarray}
7)\;{}^\chi\fR^{l_0}_{e_1}&:&\{T,\,C\sim I,\,CT\sim IT\},\nonumber\\
8)\;{}^\chi\fR^{l_0}_{e_2}&:&\{T,\,CP\sim IP,\,CPT\sim IPT\},\nonumber\\
9)\;{}^\chi\fH^{l_0}_{f_1}&:&\{T,\,C\sim C^\prime,\,CT\sim C^\prime T \},
\nonumber\\
10)\;{}^\chi\fH^{l_0}_{f_2}&:&\{T,\,CP\sim C^\prime P,\,CPT\sim C^\prime PT\}.
\nonumber
\end{eqnarray}
\end{theorem}
\begin{proof}
1) Complex representations.\\
Before we proceed to find an explicit form of the quotient representations
${}^\chi\fC$ it is necessary to consider in details a
structure of the quotient algebras
${}^\epsilon\C_n$ obtaining in the result of the homomorphic mapping
$\epsilon:\,\C_{n+1}\rightarrow\C_n$. The structure of the quotient algebra
${}^\epsilon\C_n$ depends on the transfer of the automorphisms
$\cA\rightarrow\cA^\star,
\;\cA\rightarrow\widetilde{\cA},\;\cA\rightarrow\widetilde{\cA^\star},\;
\cA\rightarrow\overline{\cA}$ of the algebra $\C_{n+1}$ under action of the
homomorphism $\epsilon$ onto its subalgebra $\C_n$. As noted previously
(see conclusion of Theorem \ref{tprod}), the homomorphisms $\epsilon$ and
$\chi$ have an analogous texture. The action of the homomorphism
$\epsilon$ is defined as follows
\[
\epsilon:\;\cA^1+\varepsilon\omega\cA^2\longrightarrow\cA^1+\cA^2,
\]
where $\cA^1,\,\cA^2\in\C_n$, $\omega=\e_{12\cdots n+1}$, and
\[
\varepsilon=\begin{cases}
1,& \text{if $n+1\equiv 1\s\!\!\pmod{4}$},\\
i,& \text{if $n+1\equiv 3\s\!\!\pmod{4}$};
\end{cases}
\]
so that $(\varepsilon\omega)^2=1$. At this point $\varepsilon\omega\rightarrow 1$
and the quotient algebra has a form
\[
{}^\epsilon\C_n\simeq\C_{n+1}/\Ker\epsilon,
\]
where $\Ker\epsilon=\left\{\cA^1-\varepsilon\omega\cA^1\right\}$ is a kernel
of the homomorphism $\epsilon$. 

For the transfer of the antiautomorphism $\cA\rightarrow\widetilde{\cA}$ 
from $\C_{n+1}$ into $\C_n$ it is necessary that
\begin{equation}\label{6.34}
\widetilde{\varepsilon\omega}=\varepsilon\omega.
\end{equation}
Indeed, since under action of $\epsilon$ the elements
1 and $\varepsilon\omega$ are equally mapped into the unit, then transformed
elements $\widetilde{1}$ and $\widetilde{\varepsilon\omega}$ are also
should be mapped into 1, but $\widetilde{1}=1\rightarrow 1$, and
$\widetilde{\varepsilon\omega}=\pm\varepsilon\omega\rightarrow\pm 1$ in
virtue of $\widetilde{\omega}=(-1)^{\frac{n(n-1)}{2}}\omega$, whence
\begin{equation}\label{6.34'}
\widetilde{\omega}=\begin{cases}
\omega,& \text{if $n+1\equiv 1\s\!\!\pmod{4}$};\\
-\omega,& \text{if $n+1\equiv 3\s\!\!\pmod{4}$}.
\end{cases}
\end{equation}
Therefore, {\it under action of the homomorphism $\epsilon$ the antiautomorphism
$\cA\rightarrow\widetilde{\cA}$ is transferred from $\C_{n+1}$ 
into $\C_n$ only at $n\equiv 0\s\pmod{4}$}.

In its turn, for the transfer of the automorphism $\cA\rightarrow\cA^\star$
it is necessary that $(\varepsilon\omega)^\star=\varepsilon\omega$. However,
since the element $\omega$ is odd and $\omega^\star=(-1)^{n+1}
\omega$, then we have always
\begin{equation}\label{6.35}
\omega^\star=-\omega.
\end{equation}
Thus, {\it the automorphism $\cA\rightarrow\cA^\star$ is never transferred
from $\C_{n+1}$ into $\C_n$}.

Further, for the transfer of the antiautomorphism 
$\cA\rightarrow\widetilde{\cA^\star}$
from $\C_{n+1}$ into $\C_n$ it is necessary that
\begin{equation}\label{6.36}
\widetilde{(\varepsilon\omega)^\star}=\varepsilon\omega.
\end{equation}
It is easy to see that the condition (\ref{6.36}) is satisfied only at
$n+1\equiv 3\s\pmod{4}$, since in this case from the second equality of
(\ref{6.34'}) and (\ref{6.35}) it follows
\begin{equation}\label{6.36'}
\widetilde{(\varepsilon\omega)^\star}=\varepsilon\widetilde{\omega^\star}=
-\varepsilon\omega^\star=\varepsilon\omega.
\end{equation}
Therefore, {\it under action of the homomorphism $\epsilon$ the 
antiautomorphism
$\cA\rightarrow\widetilde{\cA^\star}$ is transferred from $\C_{n+1}$ into $\C_n$
only at $n\equiv 2\s\pmod{4}$}.

Let $n+1=p+q$. Defining in $\C_{n+1}$ the basis $\{\e_1,\ldots,\e_p,i\e_{p+1},
\ldots, i\e_{p+q}\}$ we extract the real subalgebra $\cl_{p,q}$, where at
$p-q\equiv 3,7\s\pmod{8}$ we have a complex division ring
$\K\simeq\C$, and at $p-q\equiv 1\s\pmod{8}$ and $p-q\equiv 5\s\pmod{8}$
correspondingly a double real division ring $\K\simeq\R\oplus\R$ and a double
quaternionic division ring $\K\simeq\BH\oplus\BH$. The product
$\e_1\e_2\cdots\e_pi\e_{p+1}\cdots i\e_{p+q}=i^q\omega\in\C_{n+1}$ sets
a volume element of the real subalgebra $\cl_{p,q}$. At this point we have a
condition $\overline{(i^q\omega)}=i^q\omega$, that is,
$(-i)^q\overline{\omega}=i^q\omega$, whence
\begin{equation}\label{6.37}
\overline{\omega}=(-1)^q\omega.
\end{equation}
When $q$ is even, from (\ref{6.37}) it follows $\overline{\omega}=\omega$ and,
therefore, the pseudoautomorphism $\cA\rightarrow\overline{\cA}$ is transferred
at $q\equiv 0\s\pmod{2}$, and since $p+q$ is odd number, then we have always
$p\equiv 1\s\pmod{2}$. In more detail, at $n+1\equiv 3\s\pmod{4}$ the
pseudoautomorphism $\cA\rightarrow\overline{\cA}$ is transferred from $\C_{n+1}$
into $\C_n$ if the real subalgebra $\cl_{p,q}$ possesses the complex ring
$\K\simeq\C$, $p-q\equiv 3,7\s\pmod{8}$, and is not transferred
($\overline{\omega}=-\omega,\;q\equiv 1\s\pmod{2},\,p\equiv 0\s\pmod{2}$)
in the case of $\cl_{p,q}$ with double rings $\K\simeq\R\oplus\R$ and
$\K\simeq\BH\oplus\BH$, $p-q\equiv 1,5\s\pmod{8}$. In its turn, at
$n+1\equiv 1\s\pmod{4}$ the pseudoautomorphism $\cA\rightarrow\overline{\cA}$
is transferred from $\C_{n+1}$ into $\C_n$ if the subalgebra $\cl_{p,q}$ has
the type $p-q\equiv 1,5\s\pmod{8}$ and is not transferred in the case
of $\cl_{p,q}$ with $p-q\equiv 3,7\s\pmod{8}$. 
Besides, in virtue of (\ref{6.35}) at
$n+1\equiv 3\s\pmod{4}$ with $p-q\equiv 1,5\s\pmod{8}$ 
and at $n+1\equiv 1\s\pmod{4}$ with 
$p-q\equiv 3,7\s\pmod{8}$ a pseudoautomorphism
$\cA\rightarrow\overline{\cA^\star}$ (a composition of the pseudoautomorphism
$\cA\rightarrow\overline{\cA}$ with the automorphism $\cA\rightarrow\cA^\star$)
is transferred from $\C_{n+1}$ into $\C_n$, since
\[
\overline{\varepsilon\omega^\star}=\varepsilon\omega.
\]
Further, in virtue of the second equality of (\ref{6.34'}) 
at $n+1\equiv 3\s\pmod{4}$ with
$p-q\equiv 1,5\s\pmod{8}$  
a pseudoantiautomorphism $\cA\rightarrow\overline{\widetilde{\cA}}$
(a composition of the pseudoautomorphism $\cA\rightarrow\overline{\cA}$ with
the antiautomorphism $\cA\rightarrow\widetilde{\cA}$) is transferred from
$\C_{n+1}$ into $\C_n$, since
\[
\overline{\widetilde{\varepsilon\omega}}=\varepsilon\omega.
\]
Finally, a pseudoantiautomorphism 
$\cA\rightarrow\overline{\widetilde{\cA^\star}}$ (a composition of the
pseudoautomorphism $\cA\rightarrow\overline{\cA}$ with the antiautomorphism
$\cA\rightarrow\widetilde{\cA^\star}$), corresponded to $CPT$--transformation,
is transferred from $\C_{n+1}$ into $\C_n$ at $n+1\equiv 3\pmod{4}$ and
$\cl_{p,q}$ with $p-q\equiv 3,7\pmod{8}$, since in this case in virtue of
(\ref{6.36'}) and (\ref{6.37}) we have
\[
\overline{\widetilde{(\varepsilon\omega^\star)}}=\varepsilon\omega.
\]
Also at $n+1\equiv 1\pmod{4}$ and $q\equiv 1\pmod{2}$ we obtain
\[
\overline{\widetilde{(\varepsilon\omega^\star)}}=-
\widetilde{(\varepsilon\omega^\star)}=-(\varepsilon\omega)^\star=
\varepsilon\omega,
\]
therefore, the transformation $\cA\rightarrow\overline{\widetilde{\cA^\star}}$
is transferred at $n+1\equiv 1\pmod{4}$ and $\cl_{p,q}$ with
$p-q\equiv 3,7\pmod{8}$.

The conditions for the transfer of the fundamental automorphisms of the algebra
$\C_{n+1}$ into its subalgebra $\C_n$ under action of the homomorphism
$\epsilon$ allow to define in evident way an explicit form of the quotient
algebras ${}^\epsilon\C_n$.\\[0.2cm]
1) The quotient algebra ${}^\epsilon\C_n$, $n\equiv 0\s\pmod{4}$.\\[0.1cm]
As noted previously, in the case $n+1\equiv 1\s\pmod{4}$ 
the antiautomorphism $\cA\rightarrow\widetilde{\cA}$ and
pseudoautomorphism $\cA\rightarrow\overline{\cA}$ are transferred from
$\C_{n+1}$ into $\C_n$ if the subalgebra
$\cl_{p,q}\subset\C_{n+1}$ possesses the double rings $\K\simeq\R\oplus\R$,
$\K\simeq\BH\oplus\BH$ ($p-q\equiv 1,5\s\pmod{8}$), and also the
pseudoautomorphism $\cA\rightarrow\overline{\cA^\star}$ and
pseudoantiautomorpism $\cA\rightarrow\overline{\widetilde{\cA^\star}}$
are transferred if
$\cl_{p,q}$ has the complex ring $\K\simeq\C$ ($p-q\equiv 3,7\s\pmod{8}$).
It is easy to see that in dependence on the type of $\cl_{p,q}$ the structure
of the quotient algebras ${}^\epsilon\C_n$ of this type is divided into two
different classes:\\[0.1cm]
{\bf a}) The class of quotient algebras ${}^\epsilon\C_n$ containing the
antiautomorphism
$\cA\rightarrow\widetilde{\cA}$ and pseudoautomorphism $\cA\rightarrow
\overline{\cA}$. It is obvious that in dependence on a division ring
structure of the subalgebra
$\cl_{p,q}\subset\C_{n+1}$ this class is divided into two subclasses:
\begin{description}
\item[$a_1$] ${}^\epsilon\C_n$ with $\cA\rightarrow\widetilde{\cA}$,
$\cA\rightarrow\overline{\cA}$ at $\cl_{p,q}$ with the ring
$\K\simeq\R\oplus\R$, $p-q\equiv 1\s\pmod{8}$.
\item[$a_2$] ${}^\epsilon\C_n$ with $\cA\rightarrow\widetilde{\cA}$,
$\cA\rightarrow\overline{\cA}$ at $\cl_{p,q}$ with the ring
$\K\simeq\BH\oplus\BH$, $p-q\equiv 5\s\pmod{8}$.
\end{description}
{\bf b}) The class of quotient algebras ${}^\epsilon\C_n$ containing the
transformations
$\cA\rightarrow\widetilde{\cA}$, $\cA\rightarrow
\overline{\cA^\star}$, $\cA\rightarrow\overline{\widetilde{\cA^\star}}$ 
if the subalgebra $\cl_{p,q}\subset\C_{n+1}$ has the
complex ring $\K\simeq\C$, $p-q\equiv 3,7\s\pmod{8}$.\\[0.2cm]
2) The quotient algebra ${}^\epsilon\C_n$, $n\equiv 2\s\pmod{4}$.\\[0.1cm]
In the case $n+1\equiv 3\s\pmod{4}$ the antiautomorphism
$\cA\rightarrow\widetilde{\cA^\star}$, pseudoautomorphism
$\cA\rightarrow\overline{\cA}$ and pseudoantiautomorphism
$\cA\rightarrow\overline{\widetilde{\cA^\star}}$
are transferred from $\C_{n+1}$ into $\C_n$
if the subalgebra $\cl_{p,q}\subset\C_{n+1}$
possesses the complex ring $\K\simeq\C$ ($p-q\equiv 3,7\s\pmod{8}$), and also
the pseudoautomorphism $\cA\rightarrow\overline{\cA^\star}$ and
pseudoantiautomorphism $\cA\rightarrow\overline{\widetilde{\cA}}$ are
transferred if $\cl_{p,q}$ has the double rings $\K\simeq\R\oplus\R$,
$\K\simeq\BH\oplus\BH$ ($p-q\equiv 1,5\s\pmod{8}$). In dependence on the type of
$\cl_{p,q}\subset\C_{n+1}$ all the quotient algebras ${}^\epsilon\C_n$ of this
type are divided into following two classes:\\[0.1cm]
{\bf c}) The class of quotient algebras ${}^\epsilon\C_n$ containing the
transformations
$\cA\rightarrow\widetilde{\cA^\star}$,
$\cA\rightarrow\overline{\cA}$, 
$\cA\rightarrow\overline{\widetilde{\cA^\star}}$
if the subalgebra $\cl_{p,q}$ has the ring
$\K\simeq\C$, $p-q\equiv 3,7\s\pmod{8}$.\\[0.1cm]
{\bf d}) The class of quotient algebras ${}^\epsilon\C_n$ containing the
antiautomorphism
$\cA\rightarrow\widetilde{\cA^\star}$, pseudoautomorphism
$\cA\rightarrow\overline{\cA^\star}$ and pseudoautomorphism
$\cA\rightarrow\overline{\widetilde{\cA}}$. At this point, in dependence on the
division ring structure of $\cl_{p,q}$ we have two subclasses
\begin{description}
\item[$d_1$] ${}^\epsilon\C_n$ with $\cA\rightarrow\widetilde{\cA^\star}$,
$\cA\rightarrow\overline{\cA^\star}$ and $\cA\rightarrow
\overline{\widetilde{\cA}}$ at $\cl_{p,q}$ with the ring $\K\simeq\R\oplus\R$,
$p-q\equiv 1\s\pmod{8}$.
\item[$d_2$] ${}^\epsilon\C_n$ with $\cA\rightarrow\widetilde{\cA^\star}$,
$\cA\rightarrow\overline{\cA^\star}$ and $\cA\rightarrow
\overline{\widetilde{\cA}}$ at $\cl_{p,q}$ with the ring $\K\simeq\BH\oplus\BH$,
$p-q\equiv 5\s\pmod{8}$.
\end{description}
Thus, we have 6 different classes of the quotient algebras ${}^\epsilon\C_n$.
Further, in accordance with \cite{Var99} 
the automorphism $\cA\rightarrow\cA^\star$ corresponds to space inversion
$P$, the antiautomorphisms
$\cA\rightarrow\widetilde{\cA}$ and $\cA\rightarrow\widetilde{\cA^\star}$
set correspondingly time reversal $T$ and full reflection
$PT$, and the pseudoautomorphism $\cA\rightarrow\overline{\cA}$ corresponds to
charge conjugation $C$. Taking into account this relation and Theorem
\ref{tprod} we come to classification presented in Theorem for complex
quotient representations.\\[0.2cm]
2) Real representations.\\
Let us define real quotient representations of the group $\fG_+$. First
of all, in the case of types $p-q\equiv 3,7\pmod{8}$ we have the
isomorphism (\ref{Iso}) and, therefore, these representations are equivalent
to complex representations considered in the section 3. Further, when
$p-q\equiv 1,5\pmod{8}$ we have the real algebras $\cl_{p,q}$ with the
rings $\K\simeq\R\oplus\R$, $\K\simeq\BH\oplus\BH$ and, therefore, there
exist homomorphic mappings $\epsilon:\,\cl_{p,q}\rightarrow\cl_{p,q-1}$,
$\epsilon:\,\cl_{p,q}\rightarrow\cl_{q,p-1}$. In this case the quotient
algebra has a form
\[
{}^\epsilon\cl_{p,q-1}\simeq\cl_{p,q}/\Ker\epsilon
\]
or
\[
{}^\epsilon\cl_{q,p-1}\simeq\cl_{p,q}/\Ker\epsilon,
\]
where $\Ker\epsilon=\left\{\cA^1-\omega\cA^1\right\}$ is a kernel of $\epsilon$, 
since in accordance with
\[
\omega^2=\begin{cases}
-1& \text{if $p-q\equiv 2,3,6,7\pmod{8}$},\\
+1& \text{if $p-q\equiv 0,1,4,5\pmod{8}$}
\end{cases}
\]
at $p-q\equiv 1,5\pmod{8}$ we have always $\omega^2=1$ and, therefore,
$\varepsilon=1$. Thus, for the transfer of the antiautomorphism
$\cA\rightarrow\widetilde{\cA}$ from $\cl_{p,q}$ into $\cl_{p,q-1}$
($\cl_{q,p-1}$) it is necessary that
\[
\widetilde{\omega}=\omega
\]
In virtue of the relation $\widetilde{\omega}=(-1)^{\frac{(p+q)(p+q+1)}{2}}
\omega$ we obtain
\begin{equation}\label{Real1}
\widetilde{\omega}=\begin{cases}
+\omega& \text{if $p-q\equiv 1,5\pmod{8}$},\\
-\omega& \text{if $p-q\equiv 3,7\pmod{8}$}.
\end{cases}
\end{equation}
Therefore, for the algebras over the field $\F=\R$ the antiautomorphism
$\cA\rightarrow\widetilde{\cA}$ is transferred at the mappings
$\cl_{p,q}\rightarrow\cl_{p,q-1}$, $\cl_{p,q}\rightarrow\cl_{q,p-1}$, where
$p-q\equiv 1,5\pmod{8}$.

In its turn, for the transfer of the automorphism $\cA\rightarrow\cA^\star$
it is necessary that $\omega^\star=\omega$. However, since the element
$\omega$ is odd and $\omega^\star=(-1)^{p+q}\omega$, then we have always
\begin{equation}\label{Real2}
\omega^\star=-\omega.
\end{equation}
Thus, the automorphism $\cA\rightarrow\cA^\star$ is never transferred from
$\cl_{p,q}$ into $\cl_{p,q-1}$ ($\cl_{q,p-1}$)

Further, for the transfer of the antiautomorphism 
$\cA\rightarrow\widetilde{\cA^\star}$ it is necessary that
\[
\widetilde{\omega^\star}=\omega.
\]
From (\ref{Real1}) and (\ref{Real2}) for the types $p-q\equiv 1,5\pmod{8}$
we obtain
\begin{equation}\label{Real3}
\widetilde{\omega^\star}=\omega^\star=-\omega.
\end{equation}
Therefore, under action of the homomorphism $\epsilon$ the antiautomorphism
$\cA\rightarrow\widetilde{\cA^\star}$ is never transferred from $\cl_{p,q}$
into $\cl_{p,q-1}$ ($\cl_{q,p-1}$).

As noted previously, for the real representations of $\fG_+$ the
pseudoautomorphism $\cA\rightarrow\overline{\cA}$ is reduced into identical
transformation $\sI$ for $\fR^{l_0}_{0,2}$ and to particle--antiparticle
conjugation $C^\prime$ for $\fH^{l_0}_{4,6}$. The volume element $\omega$
of $\cl_{p,q}$ (types $p-q\equiv 1,5\pmod{8}$) can be represented by the
product $\e_1\e_2\cdots\e_p\e^{\p}_{p+1}\e^{\p}_{p+2}\cdots\e^{\p}_{p+q}$,
where $\e^{\p}_{p+j}=i\e_{p+j}$, $\e^2_j=1$, $(\e^{\p}_{p+j})^2=-1$.
Therefore, for the transfer of $\cA\rightarrow\overline{\cA}$ from $\cl_{p,q}$
into $\cl_{p,q-1}$ ($\cl_{q,p-1}$) we have a condition
\[
\overline{\omega}=\omega,
\]
and in accordance with (\ref{6.37}) it follows that the pseudoautomorphism
$\cA\rightarrow\overline{\cA}$ is transferred at $q\equiv 0\pmod{2}$. 
Further, in virtue of the relation (\ref{Real2}) the pseudoautomorphism
$\cA\rightarrow\overline{\cA^\star}$ is transferred at $q\equiv 1\pmod{2}$,
since in this case we have
\[
\overline{\omega^\star}=\omega.
\]
Also from (\ref{Real1}) it follows that the pseudoantiautomorphism
$\cA\rightarrow\overline{\widetilde{\cA}}$ is transferred at
$p-q\equiv 1,5\pmod{8}$ and $q\equiv 0\pmod{2}$, since
\[
\overline{\widetilde{\omega}}=\omega.
\]
Finally, the pseudoantiautomorphism 
$\cA\rightarrow\overline{\widetilde{\cA^\star}}$ ($CPT$--transformation)
in virtue of (\ref{Real3}) and (\ref{6.37}) is transferred from $\cl_{p,q}$
into $\cl_{p,q-1}$ ($\cl_{q,p-1}$) at $p-q\equiv1,5\pmod{8}$ and
$q\equiv 1\pmod{2}$.

Now we are in a position that allows to classify the real quotient algebras
${}^\epsilon\cl_{p,q-1}$ (${}^\epsilon\cl_{q,p-1}$).\\
1) The quotient algebra ${}^\epsilon\cl_{p,q-1}$ (${}^\epsilon\cl_{q,p-1}$,
$p-q\equiv 1\pmod{8}$.\\
In this case the initial algebra $\cl_{p,q}$ has the double real division
ring $\K\simeq\R\oplus\R$ and its subalgebras $\cl_{p,q-1}$ and
$\cl_{q,p-1}$ are of the type $p-q\equiv 0\pmod{8}$ or $p-q\equiv 2\pmod{8}$
with the ring $\K\simeq\R$. Therefore, in accordance with Theorem 
\ref{tpseudo} for all such quotient algebras the pseudoautomorphism
$\cA\rightarrow\overline{\cA}$ is equivalent to the identical transformation
$\sI$. The antiautomorphism $\cA\rightarrow\widetilde{\cA}$ in this case
is transferred into $\cl_{p,q-1}$ ($\cl_{q,p-1}$) at any $p-q\equiv 1\pmod{8}$.
Further, in dependence on the number $q$ we have two different classes of
the quotient algebras of this type:
\begin{description}
\item[$e_1$] ${}^\epsilon\cl_{p,q-1}$ (${}^\epsilon\cl_{q,p-1}$) with
$\cA\rightarrow\widetilde{\cA}$, $\cA\rightarrow\overline{\cA}$,
$\cA\rightarrow\overline{\widetilde{\cA}}$, $p-q\equiv 1\pmod{8}$,
$q\equiv 0\pmod{2}$.
\item[$e_2$] ${}^\epsilon\cl_{p,q-1}$ (${}^\epsilon\cl_{q,p-1}$) with
$\cA\rightarrow\widetilde{\cA}$, $\cA\rightarrow\overline{\cA^\star}$,
$\cA\rightarrow\overline{\widetilde{\cA^\star}}$, $p-q\equiv 1\pmod{8}$,
$q\equiv 1\pmod{2}$.
\end{description}
2) The quotient algebras ${}^\epsilon\cl_{p,q-1}$ (${}^\epsilon\cl_{q,p-1}$),
$p-q\equiv 5\pmod{8}$.\\
In this case the initial algebra $\cl_{p,q}$ has the double quaternionic
division ring $\K\simeq\BH\oplus\BH$ and its subalgebras $\cl_{p,q-1}$ and
$\cl_{q,p-1}$ are of the type $p-q\equiv 4\pmod{8}$ or $p-q\equiv 6\pmod{8}$
with the ring $\K\simeq\BH$. Therefore, in this case the pseudoautomorphism
$\cA\rightarrow\overline{\cA}$ is equivalent to the particle--antiparticle
conjugation $C^\prime$. As in the previous case the antiautomorphism
$\cA\rightarrow\widetilde{\cA}$ is transferred at any $p-q\equiv 5\pmod{8}$.
For this type in dependence on the number $q$ there are two different classes:
\begin{description}
\item[$f_1$] ${}^\epsilon\cl_{p,q-1}$ (${}^\epsilon\cl_{q,p-1}$) with
$\cA\rightarrow\widetilde{\cA}$, $\cA\rightarrow\overline{\cA}$,
$\cA\rightarrow\overline{\widetilde{\cA}}$, $p-q\equiv 5\pmod{8}$,
$q\equiv 0\pmod{2}$.
\item[$f_2$] ${}^\epsilon\cl_{p,q-1}$ (${}^\epsilon\cl_{q,p-1}$) with
$\cA\rightarrow\widetilde{\cA}$, $\cA\rightarrow\overline{\cA^\star}$,
$\cA\rightarrow\overline{\widetilde{\cA^\star}}$, $p-q\equiv 5\pmod{8}$,
$q\equiv 1\pmod{2}$.
\end{description}
\end{proof}
\section{Algebraic construction of physical fields}
Many years ago Bogoliubov and Shirkov \cite{BS93} pointed out that among
all physical fields the electromagnetic field (beyond all shadow of doubt
the main physical field) is quantized with the most difficulty. In the
standard Gupta--Bleuler approach an unobservable magnitude 
(electromagnetic four--potential $\bA$)
is quantized. At this point, the four--potential has four degrees of
freedom, but in nature there are only two degrees of freedom for a
photon field (left and right handed polarizations). Besides, the electromagnetic
four--potential is transformed within $(1/2,1/2)$--representaion of the
homogeneous Lorentz group and, therefore, in accordance with 
a well-known Weinberg Theorem \cite{Wein} the field described by $\bA$ has 
a null helicity, that also contradicts with experience. Moreover,
at the present time electromagnetic field is understood as a
`gauge field' that gives rise to a peculiar opposition with other physical
fields called by this reason as `matter fields'.

With the aim of overcoming this unnatural opposition all the physical
fields should be considered on an equal footing. In this section we present
an algebraic construction of the most fundamental physical fields such as
neutrino field, electron--positron field and 
electromagnetic field (see also \cite{Var022}).
Our consideration based mainly on the relation between Clifford algebras and
Lorentz group considered in the previous sections.
In \cite{Var01} all the Clifford algebras are understood as
`algebraic coverings' of finite--dimensional representations of the proper
Lorentz group $\fG_+$. In the previous sections
(see also \cite{Var01}) it has been shown that there is a
following classification:\\[0.2cm]
{\bf I}. Complex representations.
\begin{description}
\item[1)] Representations $\fC^{l_0+l_1-1,0}\leftrightarrow\C_n$ with
the field $(j,0)$, where $j=\frac{l_0+l_1-1}{2}$.
\item[2)] Representations 
$\fC^{0,l^\prime_0-l^\prime_1+1}\leftrightarrow\overset{\ast}{\C}_n$ 
with the field
$(0,j^\prime)$, where $j^\prime=\frac{l^\prime_0-l^\prime_1+1}{2}$.
\item[3)] Representations 
$\fC^{l_0+l_1-1,l^\prime_0-l^\prime_1+1}\leftrightarrow
\C_n\otimes\overset{\ast}{\C}_n$
with the field $(j,j^\prime)$.
\item[4)] Representations
$\fC^{l_0+l_1-1,0}\oplus\fC^{0,l_0-l_1+1}\leftrightarrow
\C_n\oplus\overset{\ast}{\C}_n$ with the field $(j,0)\oplus(0,j)$, 
$j=j^\prime$.
\item[5)] Quotient representations
${}^\chi\fC^{l_0+l_1-1,0}\cup{}^\chi\fC^{0,l_0-l_1+1}\leftrightarrow
{}^\epsilon\C_n\cup{}^\epsilon\overset{\ast}{\C}_n$ with the field
$(j,0)\cup(0,j)$.
\end{description}
{\bf II}. Real representations.
\begin{description}
\item[6)] Real representations $\fR^{l_0}_{0,2}\leftrightarrow\cl_{p,q}$,
$p-q\equiv 0,2\pmod{8}$, with the field $[j]$, where $j=\frac{l_0}{2}$,
$l_0=\frac{p+q}{4}$.
\item[7)] Quaternionic representations
$\fH^{l_0}_{4,6}\leftrightarrow\cl_{p,q}$, $p-q\equiv 4,6\pmod{8}$,
with the field $[j]$.
\item[8)] Quotient representations
${}^\chi\fD^{l_0}\cup{}^\chi\fD^{l_0}\leftrightarrow
{}^\epsilon\cl_{p,q}\cup{}^\epsilon\cl_{p,q}$ with the field
$[j]\cup[j]$, where ${}^\chi\fD^{l_0}=\{{}^\chi\fR^{l_0}_{0,2},
{}^\chi\fH^{l_0}_{4,6}\}$.
\end{description} 
Here the numbers $l_0$ and $l_1$ define the finite--dimensional
representation in the Gel'fand--Naimark representation theory of the Lorentz
group \cite{GMS,Nai58}. In its turn, quotient representations correspond to
the type $n\equiv 1\pmod{2}$ of $\C_n$ (or to the types
$p-q\equiv 1,5\pmod{8}$ for the real representations). Over the field
$\F=\C$ these representations obtained in the result of the following
decomposition
\[
\unitlength=0.5mm
\begin{picture}(70,50)
\put(35,40){\vector(2,-3){15}}
\put(35,40){\vector(-2,-3){15}}
\put(28.25,42){$\C_{2k+1}$}
\put(16,28){$\lambda_{+}$}
\put(49.5,28){$\lambda_{-}$}
\put(13.5,9.20){$\C_{2k}$}
\put(52.75,9){$\stackrel{\ast}{\C}_{2k}$}
\put(32.5,10){$\cup$}
\end{picture}
\]
Here central idempotents
\[
\lambda^+=\frac{1+\varepsilon\e_1\e_2\cdots\e_{2k+1}}{2},\quad
\lambda^-=\frac{1-\varepsilon\e_1\e_2\cdots\e_{2k+1}}{2},
\]
where
\[
\varepsilon=\begin{cases}
1,& \text{if $k\equiv 0\pmod{2}$},\\
i,& \text{if $k\equiv 1\pmod{2}$}
\end{cases}
\]
satisfy the relations $(\lambda^+)^2=\lambda^+$, $(\lambda^-)^2=\lambda^-$,
$\lambda^+\lambda^-=0$. 

Analysing the quotient representations of the group $\fG_+$ presented in
Theorem \ref{tfactor}, we see that only a repsesentation of the class
$c$ at $j=(l_0+l_1-1)/2=1/2$ is adequate for description of the neutrino
field. This representation admits full reflection
$PT$, charge conjugation $C$ and $CPT$--transformation
(space inversion $P$ is not defined). In contrast with this, the first three
classes $a_1,a_2$ and $b$ are unsuitable for description of neutrino,
since in this case $j$ is an integer number, $n\equiv 0\pmod{4}$ (bosonic
fields). In turn, the classes $d_1$ and $d_2$ admit 
$CT$--transformation that in accordance with $CPT$--Theorem is equivalent to
space inversion $P$, which, as known, is a forbidden operation for the
neutrino field. So, we have an homomorphic mapping
$\epsilon:\,\C_3\rightarrow\C_2$, where $\C_3$ is a simplest Clifford algebra
of the type $n+1\equiv 3\s\pmod{4}$. In accordance with Theorem
\ref{tfactor} under action of the homomorphism $\epsilon:\,\C_3
\rightarrow\C_2$ the transformations
$\cA\rightarrow\widetilde{\cA^\star}$, $\cA\rightarrow\overline{\cA}$ and
$\cA\rightarrow\overline{\widetilde{\cA^\star}}$ are transferred from $\C_3$
into $\C_2$. At this point, the real subalgebra $\cl_{3,0}\subset\C_3$ has
the complex ring $\K\simeq\C$, $p-q\equiv 3\s\pmod{8}$, and, therefore, the
matrix $\Pi$ of the pseudoautomorphism $\cA\rightarrow\overline{\cA}$ is not
unit, that according to Theorem \ref{tpseudo} corresponds to charged or
{\it neutral} fields.

The first simplest case of the decomposition 
$\C_{2k}\cup\overset{\ast}{\C}_{2k}$
is presented by the algebra
$\C_3$ related with the neutrino field. Indeed,
\[
\unitlength=0.5mm
\begin{picture}(70,50)
\put(35,40){\vector(2,-3){15}}
\put(35,40){\vector(-2,-3){15}}
\put(32.25,42){$\C_{3}$}
\put(16,28){$\lambda_{+}$}
\put(49.5,28){$\lambda_{-}$}
\put(13.5,9.20){$\C_{2}$}
\put(52.75,9){$\stackrel{\ast}{\C}_{2}$}
\put(32.5,10){$\cup$}
\end{picture}
\]
here central idempotents
\begin{equation}\label{Cent}
\lambda_{-}=\frac{1-i\e_1\e_2\e_3}{2},\quad
\lambda_{+}=\frac{1+i\e_1\e_2\e_3}{2} 
\end{equation}
in accordance with \cite{CF97}
can be identified with helicity projection operators. In such a way, we have
two helicity states describing by the quotient algebras 
${}^\epsilon\C_2$ and ${}^\epsilon\overset{\ast}{\C}_2$, and a full
neutrino--antineutrino algebra is ${}^\epsilon\C_2\cup{}^\epsilon
\overset{\ast}{\C}_2$ (cf. electron--positron algebra
$\C_2\oplus\overset{\ast}{\C}_2$). 

Let $\varphi\in\C_3$ be an algebraic spinor of the form
(sometimes called operator spinor, see \cite{FRO90a})
\begin{equation}\label{Neut1}
\varphi=a^0+a^1\e_1+a^2\e_2+a^3\e_3+a^{12}\e_1\e_2+a^{13}\e_1\e_3+
a^{23}\e_2\e_3+a^{123}\e_1\e_2\e_3.
\end{equation}
Then it is easy to verify that spinors
\begin{equation}\label{Neut2}
\varphi^+=\lambda_+\varphi=\frac{1}{2}(1+i\e_1\e_2\e_3)\varphi,\quad
\varphi^-=\lambda_-\varphi=\frac{1}{2}(1-i\e_1\e_2\e_3)\varphi
\end{equation}
are mutually orthogonal, $\varphi^+\varphi^-=0$, since 
$\lambda_+\lambda_-=0$, and also $\varphi^+\in\C_2$,
$\varphi^-\in\overset{\ast}{\C}_2$. Further, it is obvious that a spinspace
of the algebra ${}^\epsilon\C_2\cup{}^\epsilon\overset{\ast}{\C}_2$ is
$\dS_2\cup\dot{\dS}_2$. It should be noted here that structures of the
spinspaces $\dS_2\cup\dot{\dS}_2$ and $\dS_2\oplus\dot{\dS}_2$
are different. Indeed\footnote{See also \cite{Abl98}.},
\[
\dS_2\cup\dot{\dS}_2=\ar\begin{pmatrix}
\left[00,\dot{0}\dot{0}\right] & \left[01,\dot{0}\dot{1}\right]\\
\left[10,\dot{1}\dot{0}\right] & \left[11,\dot{1}\dot{1}\right]
\end{pmatrix},\quad
\dS_2\oplus\dot{\dS}_2=\begin{pmatrix}
00 & 01 & & \\
10 & 11 & & \\
   &    &\dot{0}\dot{0} & \dot{0}\dot{1}\\
   &    &\dot{1}\dot{0} & \dot{1}\dot{1}
\end{pmatrix}.
\]
Under action of the pseudoautomorphism $\cA\rightarrow\overline{\cA}$
(charge conjugation $C$, see \cite{Var01}) spinspace $\dS_2\cup\dot{\dS}_2$
take a form
\[
\dot{\dS}_2\cup\dS_2=\ar\begin{pmatrix}
\left[\dot{0}\dot{0},00\right] & \left[\dot{0}\dot{1},01\right]\\
\left[\dot{1}\dot{0},10\right] & \left[\dot{1}\dot{1},11\right]
\end{pmatrix}.
\]
Since spinor representations of the quotient algebras ${}^\epsilon\C_2$ and
${}^\epsilon\overset{\ast}{\C}_2$ are defined in terms of Pauli matrices
$\sigma_i$, then the algebraic spinors $\varphi^+\in{}^\epsilon\C_2$ and
$\varphi^-\in{}^\epsilon\overset{\ast}{\C}_2$ correspond to spinors
$\xi^{\alpha_i}\in\dS_2$ and $\xi^{\dot{\alpha}_i}\in\dot{\dS}_2$
($i=0,1$). Hence we have Weyl equations
\begin{equation}\label{Weyl}
\left(\frac{\partial}{\partial x^0}-\boldsymbol{\sigma}
\frac{\partial}{\partial\bx}\right)\xi^{\alpha}=0,\quad
\left(\frac{\partial}{\partial x^0}+\boldsymbol{\sigma}
\frac{\partial}{\partial\bx}\right)\xi^{\dot{\alpha}}=0.
\end{equation}
Therefore, two--component Weyl theory can be formulated naturally within
quotient representation ${}^\chi\fC^{1,0}_c\cup{}^\chi\fC^{0,-1}_c$ of the
group $\fG_+$. Further, in virtue of an isomorphism 
$\C_2\simeq\cl_{3,0}\simeq\cl^+_{1,3}$ ($\cl_{1,3}$ is the space--time
algebra) the spinor field of the quotient representation ${}^\chi\fC^{0,-1}_c$
(${}^\chi\fC^{1,0}_c$) can be expressed via the Dirac--Hestenes spinor
field $\phi(x)\in\cl_{3,0}$ \cite{Hest66,Hest67,Lou93}. 
Indeed, the Dirac--Hestenes spinor is represented by the
biquaternion number (\ref{2e3}) (see Chapter 1),
or using $\gamma$--matrix basis (\ref{Gamma})
we can write $\phi(x)$ in the matrix form
\begin{equation}\label{174}
\ar\phi=\begin{pmatrix}
\phi_1 & -\phi^\ast_2 & \phi_3 & \phi^\ast_4 \\
\phi_2 & \phi^\ast_1 & \phi_4 & -\phi^\ast_3\\
\phi_3 & \phi^\ast_4 & \phi_1 & -\phi^\ast_2\\
\phi_4 & -\phi^\ast_3 & \phi_2 & \phi^\ast_1
\end{pmatrix},
\end{equation}
where
\[
\phi_1=a^0-ia^{12},\quad
\phi_2=a^{13}-ia^{23},\quad
\phi_3=a^{03}-ia^{0123},\quad
\phi_4=a^{01}+ia^{02}.
\]
From (\ref{Neut1})--(\ref{Neut2}) and (\ref{2e3}) it is easy to see that
spinors $\varphi^+$ and $\varphi^-$ are algebraically equivalent to the
spinor $\phi\in\C_2\simeq\cl_{3,0}$. Further, since $\phi\in\cl^+_{1,3}$,
then actions of the antiautomorphisms $\cA\rightarrow\widetilde{\cA}$ and
$\cA\rightarrow\widetilde{\cA^\star}$ on the field $\phi$ are equivalent.
On the other hand, in accordance with Feynman--Stueckelberg interpretation,
time reversal for the chiral field is equivalent to charge conjugation
(particles reversed in time are antiparticles). Thus, for the field
$\phi\in{}^\chi\fC^{0,-1}_c$ we have $C\sim T$ and, therefore, this field
is $CP$--invariant.

As known, the spinor (\ref{2e3}) (or (\ref{174})) satisfies the Dirac--Hestenes
equation
\begin{equation}\label{DH}
\partial\phi\gamma_2\gamma_1-\frac{mc}{\hbar}\phi\gamma_0=0,
\end{equation}
where $\partial=\gamma^\mu\frac{\partial}{\partial x^\mu}$ is the Dirac
operator. Let us show that a massless Dirac--Hestenes equation
\begin{equation}\label{DHM}
\partial\phi\gamma_2\gamma_1=0
\end{equation}
describes the neutrino field. Indeed, the matrix 
$\gamma_0\gamma_1\gamma_2\gamma_3$ commutes with all the elements
of the biquaternion (\ref{2e3}) and, therefore, 
$\gamma_0\gamma_1\gamma_2\gamma_3$ is equivalent
to the volume element $\omega=\e_1\e_2\e_3$ of the biquaternion algebra
$\cl_{3,0}$. In such a way, we see that idempotents
\[
P_+=\frac{1+\gamma_5}{2},\quad P_-=\frac{1-\gamma_5}{2}
\]
cover the central idempotents (\ref{Cent}), where 
$\gamma_5=-i\gamma_0\gamma_1\gamma_2\gamma_3$.
Further, from (\ref{DHM}) we
obtain
\[
P_\pm\gamma^\mu\frac{\partial}{\partial x^\mu}\phi\gamma_2\gamma_1=
\gamma^\mu P_mp\frac{\partial}{\partial x^\mu}\phi\gamma_2\gamma_1=0,
\]
that is, there are two separated equations for 
$\phi^\pm=P_\pm\phi\gamma_2\gamma_1$:
\begin{equation}\label{DHM2}
\gamma^\mu\frac{\partial}{\partial x^\mu}\phi^\pm=0,
\end{equation}
where
\[
\phi^\pm=\frac{1}{2}(1\pm\gamma_5)\phi\gamma_2\gamma_1=\frac{i}{2}
\ar\begin{pmatrix}
\phi_1\mp\phi_3 & \phi^\ast_2\pm\phi^\ast_4 & \phi_3\mp\phi_1 &
-\phi^\ast_4\mp\phi^\ast_2\\
\phi_2\mp\phi_4 & -\phi^\ast_1\mp\phi^\ast_3 & \phi_4\mp\phi_2 &
\phi^\ast_3\pm\phi^\ast_1\\
\mp\phi_1+\phi_2 & \mp\phi^\ast_2-\phi^\ast_4 & \mp\phi_3+\phi_1 &
\pm\phi^\ast_4+\phi^\ast_2\\
\mp\phi_2+\phi_4 & \pm\phi^\ast_1+\phi^\ast_3 & \mp\phi_4+\phi_2 &
\mp\phi^\ast_3-\phi^\ast_1
\end{pmatrix}
\]
Therefore, each of the functions $\phi^+$ and $\phi^-$ contains only four
independent components and in the split form we have
\[
\phi^+=\ar\begin{pmatrix}
\psi_1 & \psi_2 & \psi_3 & \psi_4\\
-\psi_1 & -\psi_2 & -\psi_3 & -\psi_4
\end{pmatrix},\quad
\phi^-=\begin{pmatrix}
\psi_5 & \psi_6 & \psi_7 & \psi_8\\
\psi_5 & \psi_6 & \psi_7 & \psi_8
\end{pmatrix},
\]
where
\begin{gather}\ar
\psi_1=\frac{i}{2}\begin{pmatrix}
\phi_1-\phi_3\\
\phi_2-\phi_4
\end{pmatrix},\;\;
\psi_2=\frac{i}{2}\begin{pmatrix}
\phi^\ast_2+\phi^\ast_4\\
-\phi^\ast_1-\phi^\ast_3
\end{pmatrix},\;\;
\psi_3=\frac{i}{2}\begin{pmatrix}
\phi_3-\phi_1\\
\phi_4-\phi_2
\end{pmatrix},\;\;
\psi_4=\frac{i}{2}\begin{pmatrix}
-\phi^\ast_4-\phi^\ast_2\\
\phi^\ast_3+\phi^\ast_1
\end{pmatrix},\nonumber\\
\psi_5=\frac{i}{2}\ar\begin{pmatrix}
\phi_1+\phi_3\\
\phi_2+\phi_4
\end{pmatrix},\;\;
\psi_6=\frac{i}{2}\begin{pmatrix}
\phi^\ast_2-\phi^\ast_4\\
-\phi^\ast_1+\phi^\ast_3
\end{pmatrix},\;\;
\psi_7=\frac{i}{2}\begin{pmatrix}
\phi_3+\phi_1\\
\phi_4+\phi_2
\end{pmatrix},\;\;
\psi_8=\frac{i}{2}\begin{pmatrix}
-\phi^\ast_4+\phi^\ast_2\\
\phi^\ast_3-\phi^\ast_1
\end{pmatrix}.\nonumber
\end{gather}
Thus, in the $\gamma$--matrix basis we obtain from (\ref{DHM2})
\[
\left(\frac{\partial}{\partial x^0}-\boldsymbol{\sigma}
\frac{\partial}{\partial\bx}\right)\psi_i=0,\quad
\left(\frac{\partial}{\partial x^0}+\boldsymbol{\sigma}
\frac{\partial}{\partial\bx}\right)\psi_{i+4}=0,\quad(i=1,2,3,4)
\]
These equations are equivalent to Weyl equations (\ref{Weyl})
for neutrino field and, therefore, by analogy 
with the Dirac--Hestenes equations for $m\neq 0$ it
should be called {\it Weyl--Hestenes equations for neutrino field}.

The Dirac electron--positron field
$(1/2,0)\oplus(0,1/2)$ corresponds to the algebra $\C_2\oplus
\overset{\ast}{\C}_2$. It should be noted that the Dirac algebra
$\C_4$ considered as a tensor product $\C_2\otimes\C_2$ 
(or $\C_2\otimes\overset{\ast}{\C}_2$) 
gives rise to spintensors $\xi^{\alpha_1\alpha_2}$
(or $\xi^{\alpha_1\dot{\alpha}_1}$), but it contradicts with the usual
definition of the Dirac bispinor as a pair 
$(\xi^{\alpha_1},\xi^{\dot{\alpha}_1})$. Therefore, the Clifford algebra
associated with the Dirac field is $\C_2\oplus\overset{\ast}{\C}_2$, and
a spinspace of this sum in virtue of unique decomposition
$\dS_2\oplus\dot{\dS}_2=\dS_4$ ($\dS_4$ is a spinspace of $\C_4$) allows to
define $\gamma$--matrices in the Weyl basis.

In common with other massless fields (such as the neutrino field
$(1/2,0)\cup(0,1/2)$) the Maxwell electromagnetic field is also described
within the quotient representations of the Lorentz group. In accordance
with Theorem \ref{tfactor} in \cite{Var01} 
the photon field can be described by a
quotient representation of the class 
${}^\chi\fC^{2,0}_{a_1}\cup{}^\chi\fC^{0,-2}_{a_1}$. This representation
admits time reversal $T$ and an identical charge conjugation $C\sim\sI$
that corresponds to truly neutral particles 
(see Theorem \ref{tpseudo}).
The quotient algebra ${}^\epsilon\C_4\cup{}^\epsilon\overset{\ast}{\C}_4$
associated with the Maxwell field $(1,0)\cup(0,1)$ is obtained in the 
result of an homomorphic mapping $\epsilon:\,C_5\rightarrow\C_4$. Indeed,
for the algebra $\C_5$ we have a decomposition
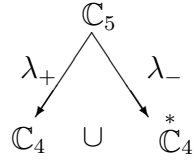
\[
\unitlength=0.5mm
\begin{picture}(70,50)
\put(35,40){\vector(2,-3){15}}
\put(35,40){\vector(-2,-3){15}}
\put(32.25,42){$\C_{5}$}
\put(16,28){$\lambda_{+}$}
\put(49.5,28){$\lambda_{-}$}
\put(13.5,9.20){$\C_{4}$}
\put(52.75,9){$\stackrel{\ast}{\C}_{4}$}
\put(32.5,10){$\cup$}
\end{picture}
\]
where the central idempotents
\[
\lambda_+=\frac{1+\e_1\e_2\e_3\e_4\e_5}{2},\quad
\lambda_-=\frac{1-\e_1\e_2\e_3\e_4\e_5}{2}
\]
correspond to the helicity projection operators of the Maxwell field.
As known, for the photon there are two helicity states: left and right
handed polarizations.

Let $\varphi\in\C_5$ be an algebraic spinor of the form
\[
\varphi=a^0+\sum^5_{i=1}a^i\e_i+\sum^5_{i,j=1}a^{ij}\e_i\e_j+
\sum^5_{i,j,k=1}a^{ijk}\e_i\e_j\e_k+
\sum^5_{i,j,k,l=1}a^{ijkl}\e_i\e_j\e_k\e_l+a^{12345}\e_1\e_2\e_3\e_4\e_5,
\]
then the spinors 
\[
\varphi^+=\lambda_+\varphi=\frac{1}{2}(1+\e_1\e_2\e_3\e_4\e_5)\varphi,\quad
\varphi^-=\lambda_-\varphi=\frac{1}{2}(1-\e_1\e_2\e_3\e_4\e_5)\varphi
\]
are mutually orthogonal, $\varphi^+\varphi^-=0$, and $\varphi^+\in\C_4$,
$\varphi^-\in\overset{\ast}{\C}_4$. The spinspace of the algebra
${}^\epsilon\C_4\cup{}^\epsilon\overset{\ast}{\C}_4$ has a form
\[
{\renewcommand{\arraystretch}{1.4}
\dS_4\cup\dot{\dS}_4=\begin{pmatrix}
\left[0000,\dot{0}\dot{0}\dot{0}\dot{0}\right] &
\left[0001,\dot{0}\dot{0}\dot{0}\dot{1}\right] &
\left[0010,\dot{0}\dot{0}\dot{1}\dot{0}\right] &
\left[0011,\dot{0}\dot{0}\dot{1}\dot{1}\right] \\
\left[0100,\dot{0}\dot{1}\dot{0}\dot{0}\right] &
\left[0101,\dot{0}\dot{1}\dot{0}\dot{1}\right] &
\left[0110,\dot{0}\dot{1}\dot{1}\dot{0}\right] &
\left[0111,\dot{0}\dot{1}\dot{1}\dot{1}\right] \\
\left[1000,\dot{1}\dot{0}\dot{0}\dot{0}\right] &
\left[1001,\dot{1}\dot{0}\dot{0}\dot{1}\right] &
\left[1010,\dot{1}\dot{0}\dot{1}\dot{0}\right] &
\left[1011,\dot{1}\dot{0}\dot{1}\dot{1}\right] \\
\left[1100,\dot{1}\dot{1}\dot{0}\dot{0}\right] &
\left[1101,\dot{1}\dot{1}\dot{0}\dot{1}\right] &
\left[1110,\dot{1}\dot{1}\dot{1}\dot{0}\right] &
\left[1111,\dot{1}\dot{1}\dot{1}\dot{1}\right]
\end{pmatrix}}
\]

Let us consider now an explicit construction of the Maxwell field
$(1,0)\cup(0,1)$ within the quotient algebra
${}^\epsilon\C_4\cup{}^\epsilon\overset{\ast}{\C}_4$. First of all,
let us define a spinor representation of the field
$(1,0)\cup(0,1)$. As a rule, a spinor field of the quotient algebra
${}^\epsilon\C_4\simeq\cl_{4,1}$ is appearred at the extraction of the
minimal left ideal \cite{FRO90a,FRO90b,RSVL} (see also Chapter 1):
\begin{equation}\label{left}
I_{4,1}=\cl_{4,1}e_{41}\simeq\cl^+_{1,3}e_{13}\frac{1}{2}(1+i\gamma_1\gamma_2)=
\ar\begin{pmatrix}
\psi_1 & 0 & 0 & 0\\
\psi_2 & 0 & 0 & 0\\
\psi_3 & 0 & 0 & 0\\
\psi_4 & 0 & 0 & 0
\end{pmatrix},
\end{equation}
where $e_{13}\frac{1}{2}(1+\gamma_0)$ and $e_{41}=\frac{1}{2}(1+\gamma_0)
\frac{1}{2}(1+i\gamma_1\gamma_2)$ are primitive idempotents of the
algebras $\cl_{1,3}$ and $\C_4$. Further, since $\cl^+_{1,3}\simeq\cl_{3,0}
\simeq\C_2$, then the spinor field $\psi$ in (\ref{left}) can be expressed
via the Dirac--Hestenes spinor field $\phi\in\cl_{3,0}\simeq\C_2$. Let
\[
\nabla=\partial^{0}\e_{0}+\partial^{1}\e_{1}+\partial^{2}\e_{2}+
\partial^{3}\e_{3},\quad
A=A^{0}\e_{0}+A^{1}\e_{1}+A^{2}\e_{2}+A^{3}\e_{3}
\]
be linear elements of the algebra $\cl_{3,0}$, where $A_i$ are the components
of the electromagnetic four--potential. Then
\begin{multline}\label{e10}
\nabla A=(\partial^{0}\e_{0}+\partial^{1}\e_{1}+
\partial^{2}\e_{2}+\partial^{3}\e_{3})(A^{0}\e_{0}+A^{1}\e_{1}+A^{2}\e_{2}+
A^{3}\e_{3})=\\
(\underbrace{\partial^{0}A^{0}+\partial^{1}A^{1}+\partial^{2}A^{2}+
\partial^{3}A^{3}}_{E^{0}})\e_{0}+(\underbrace{\partial^{0}A^{1}+
\partial^{1}A^{0}}_{E^{1}})\e_{0}\e_{1}+\\
(\underbrace{\partial^{0}A^{2}+\partial^{2}A^{0}}_{E^{2}})\e_{0}\e_{2}+
(\underbrace{\partial^{0}A^{3}+\partial^{3}A^{0}}_{E^{3}})\e_{0}\e_{3}+
(\underbrace{\partial^{2}A^{3}-\partial^{3}A^{2}}_{H^{1}})\e_{2}\e_{3}+\\
(\underbrace{\partial^{3}A^{1}-\partial^{1}A^{3}}_{H^{2}})\e_{3}\e_{1}+
(\underbrace{\partial^{1}A^{2}-\partial^{2}A^{1}}_{H^{3}})\e_{1}\e_{2}.
\end{multline}
The scalar part $E_{0}\equiv 0$, since the first bracket in (\ref{e10}) is
a Lorentz
condition $\partial^{0}A^{0}+\mbox{div}{\bf A}=0$. It is easy to see that 
other brackets are components of electric and magnetic fields:
$-E^{i}=-(\partial^{i}A^{0}+\partial^{0}A^{i})$, $H^{i}=
(\mbox{curl\bf A})^{i}$.

Since $\omega=\e_{123}$ belongs to a center of $\cl_{3,0}$, then
$$\omega \e_{1}=\e_{1}\omega=\e_{2}\e_{3},\quad\omega \e_{2}=\e_{2}
\omega=\e_{3}\e_{1},
\quad\omega \e_{3}=\e_{3}\omega=\e_{1}\e_{2}.$$

In accordance with these relations we can write (\ref{e10}) as follows
\begin{equation}\label{e11}
\nabla A=(E^{1}+\omega H^{1})\e_{1}+(E^{2}+\omega H^{2})
\e_{2}+(E^{3}+\omega H^{3})\e_{3}
\end{equation}
Further, let us compose the product 
$\nabla\bF$, where
$\bF$ is an 
expression
of the type (\ref{e10}):
\begin{multline}
\nabla\bF=\mbox{div\bf E}\e_{0}-((\mbox{curl\bf H})^{1}-
\partial^{0}E_{1})\e_{1}-((\mbox{curl\bf H})^{2}-\partial^{0}E^{2})\e_{2}-\\
-((\mbox{curl\bf H})^{3}-\partial^{0}E^{3})\e_{3}+((\mbox{curl\bf E})^{1}+
\partial^{0}H^{1})\e_{2}\e_{3}+((\mbox{curl\bf E})^{2}+\partial^{0}H^{2})
\e_{3}\e_{1}+\\
((\mbox{curl\bf E})^{3}+\partial^{0}H^{3})\e_{1}\e_{2}+\mbox{div\bf H}
\e_{1}\e_{2}\e_{3}.
\end{multline}
It is easy to see that the first coefficient of the product 
$\nabla\bF$
is a left part of the equation $\mbox{div\bf E}=\varrho$. The following three
coefficients compose a left part of the equation $\mbox{curl\bf H}-
\partial^{0}\mbox{\bf E}=j$,
other coefficients compose the equations 
$\mbox{curl\bf E}+\partial^{0}\mbox{\bf H}=0$
and $\mbox{div\bf H}=0$, respectively.

Further, since the element $\gamma_5=\gamma_0\gamma_1\gamma_2\gamma_3$
commutes with all other elements of the biquaternion (\ref{2e3}) and
$\gamma^2_5=-1$, then we can rewrite (\ref{2e3}) as follows
\[
\phi=(a^0+\gamma_{0123}a^{0123})+(a^{01}+\gamma_{0123}a^{23})\gamma_{01}+
(a^{02}+\gamma_{0123}a^{31})\gamma_{02}+(a^{03}+\gamma_{0123}a^{12})\gamma_{03}
\]
or taking into account (\ref{e11}) we obtain
\[
\phi=(E^1+iH^1)\gamma_{01}+(E^2+iH^2)\gamma_{02}+(E^3+iH^3)\gamma_{03}=
F_1\gamma_{01}+F_2\gamma_{02}+F_3\gamma_{03}.
\]
In the matrix form we have
\[
\phi=\ar\begin{pmatrix}
0 & 0 & F_3 & F_1-iF_2\\
0 & 0 & F_1+iF_2 & -F_3\\
F_3 & F_1-iF_2 & 0 & 0\\
F_1+iF_2 & -F_3 & 0 & 0
\end{pmatrix}
\]
Then from (\ref{left}) the relation immediately follows between spinors
$\psi\in\C_4$ and $\phi\in\cl_{3,0}$
\begin{equation}\label{S1}
\psi=\phi\frac{1}{2}(1+\gamma_0)\frac{1}{2}(1+i\gamma_1\gamma_2)=
\ar\begin{pmatrix}
0 & 0 & 0 & 0\\
0 & 0 & 0 & 0\\
F_3 & 0 & 0 & 0\\
F_1+iF_2 & 0 & 0 & 0
\end{pmatrix}\sim
\ar\begin{pmatrix}
0\\
F_1\\
F_2\\
F_3
\end{pmatrix}
\end{equation}
Let us consider now an action of the antiautomorphism 
$\cA\rightarrow\widetilde{\cA}$ on the arbitrary element of $\cl_{3,0}$
represented by a formula
\begin{equation}\label{Arb}
\cA=(a^0+\omega a^{123})\e_0+(a^1+\omega a^{23})\e_1+(a^2+\omega a^{31})\e_2
+(a^3+\omega a^{12})\e_3.
\end{equation}
The action of the antiautomorphism $\cA\rightarrow\widetilde{\cA}$ on the
homogeneous element $\cA$ of a degree $k$ is defined by a formula
$\widetilde{\cA}=(-1)^{\frac{k(k-1)}{2}}\cA$. Thus, for the element
(\ref{Arb}) we obtain
\[
\cA\longrightarrow\widetilde{\cA}=(a^0-\omega a^{123})\e_0+
(a^1-\omega a^{23})\e_1+(a^2-\omega a^{31})\e_2+(a^3-\omega a^{12})\e_3.
\]
Therefore, under action of $\cA\rightarrow\widetilde{\cA}$
the element (\ref{e11}) takes a form
\[
\widetilde{(\nabla A)}=(E^1-iH^1)\e_1+(E^2-iH^2)\e_2+(E^3-iH^3)\e_3
\]
and
\[
\widetilde{\phi}=\ar\begin{pmatrix}
0 & 0 & \overset{\ast}{F}_3 & \overset{\ast}{F}_1-i\overset{\ast}{F}_2\\
0 & 0 & \overset{\ast}{F}_1+i\overset{\ast}{F}_2& -\overset{\ast}{F}_3\\
\overset{\ast}{F}_3 & \overset{\ast}{F}_1-i\overset{\ast}{F}_2 & 0 & 0\\
\overset{\ast}{F}_1+i\overset{\ast}{F}_2 & -\overset{\ast}{F}_3 & 0 & 0
\end{pmatrix}.
\]
Hence it immediately follows
\begin{equation}\label{S2}
\widetilde{\psi}=\widetilde{\phi}\frac{1}{2}(1+\gamma_0)\frac{1}{2}
(1+i\gamma_1\gamma_2)=\ar\begin{pmatrix}
0 & 0 & 0 & 0\\
0 & 0 & 0 & 0\\
\overset{\ast}{F}_3 & 0 & 0 & 0\\
\overset{\ast}{F}_1+i\overset{\ast}{F}_2 & 0 & 0 & 0
\end{pmatrix}\sim
\begin{pmatrix}
0\\
\overset{\ast}{F}_1\\
\overset{\ast}{F}_2\\
\overset{\ast}{F}_3
\end{pmatrix}
\end{equation}
From (\ref{S1}) and (\ref{S2}) it follows that the full representation
space $\dS_4\cup\dot{\dS}_4$ is reduced to a 3--dimensional symmetric
space $\Sym_{(2,0)}\cup\Sym_{(0,2)}$. The transition from operator spinors
to $SO(3)$ vectors has been done by several authors (see, for example
\cite{Par00}).
The `vectors' (spintensors) of the
spaces $\Sym_{(2,0)}$ and $\Sym_{(0,2)}$ are
\begin{gather}
f^{00}=\xi^0\otimes\xi^0,\quad f^{01}=f^{10}=\xi^0\otimes\xi^1=
\xi^1\otimes\xi^0,\quad f^{11}=\xi^1\otimes\xi^1,\nonumber\\
f^{\dot{0}\dot{0}}=\xi^{\dot{0}}\otimes\xi^{\dot{0}},\quad
f^{\dot{0}\dot{1}}=f^{\dot{1}\dot{0}}=\xi^{\dot{0}}\otimes\xi^{\dot{1}}=
\xi^{\dot{1}}\otimes\xi^{\dot{0}},\quad
f^{\dot{1}\dot{1}}=\xi^{\dot{1}}\otimes\xi^{\dot{1}}.\label{Sp}
\end{gather}
It is well--known \cite{Rum36,RF68,RF} that spintensors (\ref{Sp})
correspond to a Helmholtz--Silberstein representation $\bF=\bE+i\bH$ and
form a basis of the Majorana--Oppenheimer quantum electrodynamics
\cite{Maj,Opp31,MRB74,Gia85,Rec90,Esp98,Dvo97} in which the electromagnetism
has to be considered as the wave mechanics of the photon.
Moreover, the field $(1,0)\cup(0,1)$ satisfies the Weinberg Theorem.

In such a way, it is easy to see that
\begin{eqnarray}
\text{Weyl neutrino field}&&(1/2,0)\cup(0,1/2),\nonumber\\
\text{Dirac electron--positron field}&&(1/2,0)\oplus(0,1/2),\nonumber\\
\text{Maxwell electromagnetic field}&&(1,0)\cup(0,1)\sim\nonumber\\
&&(1/2,0)\otimes(1/2,0)\cup(0,1/2)\otimes(0,1/2)\label{Max}
\end{eqnarray}
have the same mathematical structure. Namely, all these fields are obtained
from the fundamental field $(1/2,0)$ ($(0,1/2)$) by means of unification,
direct sum and tensor product. From this point of view a division of all
the physical fields into `gauge fields' and `matter fields' has an
artificial character. It is obvious that an origin of such a division
takes its beginning from the invalid description 
(from group theoretical viewpoint)
of electromagnetic field
in quantum field theory (a la Gupta--Bleuler approach) and a groundless
extension of the Yang--Mills idea (as known, a long derivative is equivalent
to a minimal interaction in the lagrangian). Further, as follows from
(\ref{Max}) the Maxwell field has a composite structure that gives rise
to a de Broglie--Jordan neutrino theory of light \cite{Bro32,Jor35},
in which the electromagnetic field is constructed via the two neutrino fields.

As known, the neutrino theory of light has a long and dramatic history
(see excellent review \cite{Dvo99,Per00}). The main obstacle that decelerates
the development of the neutrino theory of light is a Pryce's Theorem
\cite{Pry38} (see also a Berezinskii modification of this Theorem \cite{Ber66}).
In 1938, Pryce claimed the incompatibility of obtaining
transversely--polarized photons and Bose statistics from neutrinos.
However, as it has been recently shown by Perkins \cite{Per00} the
Pryce's Theorem contains an assumption which is unsupported and probably
incorrect. The Perkins arguments based on the experimental similarities
between bosons and quasi--bosons allow to overcome this obstacle
(Pryce's Theorem). In the present paper we show that transversely--polarized
photons follow naturally from composite neutrinos (it is a direct consequence
of the group theoretical consideration). It is obvious that such photons
are quasi--bosons \cite{Per01} that do not satisfy the Bose statistics,
but they satisfy the Lipkin statistics of a composite particle such as
the deuteron and Cooper pairs \cite{Lip73}.
In conclusion, one can say that a correct description of electromagnetic
field is a synthesis of the Majorana--Oppenheimer quantum electrodynamics
and de Broglie--Jordan neutrino theory of light.

\end{document}